\documentclass{emulateapj}
\usepackage{bm}
\usepackage{chngpage}
\usepackage{graphicx}
\usepackage[caption=false]{subfig}
\usepackage{mathrsfs}

\def\apjl{ApJL}

\shorttitle{Electron Cooling}
\shortauthors{Geng et al.}

\begin{document}

\title{LOW-ENERGY SPECTRA OF GAMMA-RAY BURSTS FROM COOLING ELECTRONS}

\author{Jin-Jun Geng\altaffilmark{1,2,3}, Yong-Feng Huang\altaffilmark{1,2}, Xue-Feng Wu\altaffilmark{4},
Bing Zhang\altaffilmark{5,6,7}, Hong-Shi Zong\altaffilmark{3,8}}

\altaffiltext{1}{School of Astronomy and Space Science, Nanjing University, Nanjing 210023, China; gengjinjun@nju.edu.cn, hyf@nju.edu.cn}
\altaffiltext{2}{Key Laboratory of Modern Astronomy and Astrophysics (Nanjing University), Ministry of Education, Nanjing 210023, China}
\altaffiltext{3}{Department of Physics, Nanjing University, Nanjing 210093, China}
\altaffiltext{4}{Purple Mountain Observatory, Chinese Academy of Sciences, Nanjing 210008, China}
\altaffiltext{5}{Department of Physics and Astronomy, University of Nevada Las Vegas, NV 89154, USA}
\altaffiltext{6}{Department of Astronomy, School of Physics, Peking University, Beijing 100871, China}
\altaffiltext{7}{Kavli Institute of Astronomy and Astrophysics, Peking University, Beijing 100871, China}
\altaffiltext{8}{Joint Center for Particle, Nuclear Physics and Cosmology, Nanjing 210093, China}

\begin{abstract}
The low-energy spectra of gamma-ray bursts' (GRBs) prompt emission are closely related to
the energy distribution of electrons, which is further regulated by their cooling processes.
We develop a numerical code to calculate the evolution of the electron distribution with
given initial parameters, in which three cooling processes (i.e., adiabatic, synchrotron
and inverse Compton cooling) and the effect of decaying magnetic field are coherently considered.
A sequence of results are presented by exploring the plausible parameter space
for both the fireball and the Poynting-flux-dominated regime.
Different cooling patterns for the electrons can be identified and they are featured by a specific dominant cooling mechanism.
Our results show that the hardening of the low-energy spectra
can be attributed to the dominance of synchrotron self-Compton cooling within the
internal shock model, or to decaying synchrotron cooling within the Poynting-flux-dominated jet scenario.
These two mechanisms can be distinguished by observing the hard low-energy spectra
of isolated short pulses in some GRBs.
The dominance of adiabatic cooling can also lead to hard low-energy spectra
when the ejecta is moving at an extreme relativistic speed.
The information from the time-resolved low-energy spectra can help to probe the
physical characteristics of the GRB ejecta via our numerical results.
\end{abstract}


\section{INTRODUCTION}

The radiation mechanism responsible for the prompt emission of gamma-ray bursts (GRBs) remains
unidentified since the discovery of GRBs. A typical spectrum of the GRB prompt emission can usually be well fit by the
so-called Band function~\citep{Band93}, which smoothly joins low- and high- energy power laws.
Except for the Band component, the superposition of multiple spectral components was also observed in some GRBs,
such as the thermal component~\citep[e.g.,][]{Ghirlanda03,Ryde05,Ghirlanda07,Ryde10,ZhangBB11}, 
or an additional power-law component~\citep[e.g.,][]{Abdo09,Ackermann11}.
Since the Band function is still an empirical description for the GRB spectra, further studies are
needed to manifest its physical origin.

Synchrotron radiation of electrons has been suggested to be a possible mechanism.
However, one problem (called fast cooling problem, see \citealt{Ghisellini00,ZhangB11})
remains in a simple synchrotron model, i.e., the low-energy spectral index
$\alpha$ ($F_{\nu} \propto \nu^{\alpha}$) is predicted to be $-1/2$ for fast cooling electrons~\citep{Sari98},
which is incompatible with the fact
that the observed value is $\sim 0$ in the majority of GRBs~\citep{Band93,Preece00,ZhangBB11,Nava11,Geng13}.
Modified synchrotron models have been proposed to ease this conflict.
When electrons cool mainly via inverse Compton (IC) scattering in the Klein-Nishina (KN) regime,
it was suggested that the energy loss rate of electrons is roughly $\propto \gamma_{\rm e}^{-1}$ ($\gamma_{\rm e}$ is the electron Lorentz factor),
and the corresponding flux density is $F_{\nu} \sim \nu^0$ (see~\citealt{Derishev01,Bonjak09,Wang09,Nakar09,Fan10,Daigne11}).
Subsequent detailed analytical study on this solution shows that it is impossible to obtain a spectrum with $\alpha > -0.1$
using IC cooling in the KN regime~\citep{Barniol12}.
Recently, by solving the continuity equation of electrons in energy space numerically,
\cite{Uhm14} found that the fast cooling electrons could have a harder energy spectrum
when the surrounding magnetic field is decreasing. Their numerical results show $\alpha$ can
be even harder than $0$ in some certain parameter regime. However, IC cooling of electrons was not
included in their calculations.
It is then crucial to take IC cooling into account when solving the continuity equation of electrons.

Alternative models based on photospheric emission have also been proposed to explain the
GRB prompt emission~\citep[e.g.,][]{Rees05,Peer06,Giannios07,Beloborodov10,Ryde11}.
Indeed, the spectra of some GRBs are found to be consistent with a photospheric component~\citep[e.g.,][]{Ryde10,Larsson11,Peer12}.
However, the photosphere model typically predicts $\alpha \sim 1.4$ \citep{Deng14}.
Some additional effects should be considered to explain the observed index of $\alpha \sim 0$ with
the photosphere model~\citep[e.g.,][]{Lundman13}.
In general, given that the main spectral component of a typical burst is the Band component,
it is still rational to suppose that the emission comes from a non-thermal
mechanism in an optically thin region~\citep{ZhangB12,Veres12,Kumar15}.
So in this study, we work within the framework of synchrotron radiation and focus on the low-energy
spectra of GRBs.

In principle, the profile of synchrotron spectra of GRBs observed is directly determined by the distribution of electrons
(or called the electron spectrum/distribution for short) in the ejecta.
The initial spectrum of the electrons that are accelerated somehow
will soon be modified by cooling processes,
thus electron cooling is a key factor in the prompt emission,
especially when we focus on the low-energy spectra of GRBs.
Although only synchrotron emission is observed by us in the keV --- MeV band,
the electrons actually can be cooled in three ways, which are adiabatic, synchrotron, and IC cooling respectively.
Numerical solution of the electron distribution considering the three processes 
has been presented in previous researches within the internal shock scenario~\citep{Bonjak09,Daigne11}.
They found the majority of observed GRB prompt spectra can be reconciled with a synchrotron origin.
Some useful constraints on the microphysics of internal shocks were presented. 
This numerical approach can also be extended to the Poynting-flux-dominated jet scenario. 
In this study, we investigate electron cooling in different physical situations
and give a clue to distinguish them.

In addition to modeling the time-integrated spectra analytically, the analyses on the time-resolved spectra
of the prompt emission~\citep[e.g.,][]{Lu12,ZhangBB16,Jiang16}
can also provide important information on the radiation process.
On the other hand, the numerical method has the advantage over the analytical method
in that it can incorporate different radiation mechanisms and can follow the evolution of physical
properties in the emitting region.
For example, \cite{Daigne11} predicted that the high-energy ($> 100$ MeV) lightcurve may display a 
prolonged pulse duration due to the IC emission.
\cite{Uhm14} revealed that as a jet expands rapidly from the central engine, 
the magnetic field in the emission region decreases, 
resulting in harder (than the case of constant magnetic field) spectra. 
Therefore, in our study, we develop a numerical code to calculate the evolution of the electron spectra and the
corresponding flux spectra with different parameter sets.
Adiabatic, synchrotron, IC cooling of electrons (also see \citealt{Bonjak09}) and the geometric effect of the emitting shells
are considered properly in our code.
Using this code, we can explore the resulting spectra in the plausible parameter space,
which may provide clues to help relate the observed GRB spectra with the physical processes in the GRB ejecta.
Different kinds of cooling patterns for electrons obtained in different scenarios can also serve as 
the baseline for further explorations.

The structure of this article is as follows. The three main cooling processes considered are briefly described in Section 2.
The constraints from observations on the parameters involved in our calculations are presented in Sections 3.
In Section 4, we derive the conditions under which one particular cooling process will be dominant analytically.
The analytical results are then compared with the numerical results in Section 5, where
we generally study the roles played by different cooling processes in determining the evolution
of the low-energy electron distribution.
Finally, in Section 6, we summarize and discuss our results.
The details of the numerical method and some relevant formulations used are given in
Appendices A and B respectively.

\section{COOLING OF ELECTRONS}

In the co-moving frame of a relativistic jet,
when an electron with a Lorentz factor of $\gamma_{\rm e}^{\prime}$ is moving in the magnetic field of strength $B^{\prime}$,
it will lose energy by the synchrotron radiation at a rate of~\citep{Rybicki79}
\begin{equation}
\dot{\gamma}_{\rm e,syn}^{\prime} = -\frac{\sigma_T B^{\prime 2} \gamma_{\rm e}^{\prime 2}}{6 \pi m_{\rm e} c},
\label{eq:syn}
\end{equation}
where $\sigma_T$, $m_{\rm e}$, and $c$ are the Thomson cross-section, electron mass, and speed of light respectively.
Hereafter, the superscript prime ($\prime$) is used to denote the quantities in the co-moving frame.
The electron also undergoes adiabatic cooling~\citep{Uhm12,Geng14}, i.e.,
\begin{equation}
\dot{\gamma}_{\rm e,adi}^{\prime} = \frac{1}{3} \gamma_{\rm e}^{\prime} \frac{d \ln n_{\rm e}^{\prime}}{d t^{\prime}} = - \frac{2}{3}
\frac{\gamma_{\rm e}^{\prime}}{R} \frac{d R}{d t^{\prime}},
\label{eq:adi}
\end{equation}
where we have taken the co-moving electron number density $n_{\rm e}^{\prime} \propto R^{-2}$ for an expanding shell.

Additionally, the electrons will be cooled by the IC scattering of self-emitted synchrotron photons,
which is referred to as synchrotron self-Compton (SSC) process. The SSC cooling rate is given by~\citep{Blumenthal70,Fan08}
\begin{equation}
\dot{\gamma}_{\rm e,SSC}^{\prime}
= -\frac{1}{m_{\rm e} c^2} \frac{3 \sigma_T c}{4 \gamma_{\rm e}^{\prime 2}} \int_{\nu_{\rm min}^{\prime}}^{\nu_{\rm max}^{\prime}}
\frac{n_{\nu^{\prime}} d \nu^{\prime}}{\nu^{\prime}}
\int_{\nu_{\rm ic,min}^{\prime}}^{\nu_{\rm ic,max}^{\prime}}
h \nu_{\rm ic}^{\prime} d \nu_{\rm ic}^{\prime} F(q,g),
\label{eq:SSC}
\end{equation}
where $F(q,g) = 2 q \ln q + (1 + 2q) (1 - q) + \frac{1}{2} \frac{(4 q g)^2}{1 + 4 q g} (1 - q)$,
$g = \frac{\gamma_{\rm e}^{\prime} h \nu^{\prime}}{m_{\rm e} c^2}$, $w = \frac{h \nu_{\rm ic}^{\prime}}{\gamma_{\rm e}^{\prime} m_{\rm e} c^2}$,
$q = \frac{w}{4 g (1 - w)}$.
The upper limit of the internal integral can be derived as $h \nu_{\rm ic,max}^{\prime} = \gamma_{\rm e}^{\prime} m_{\rm e} c^2 \frac{4 g}{4g + 1}$,
and the lower limit is $\nu_{\rm ic,min}^{\prime} = \nu^{\prime}$.
Overall, the total cooling rate of an electron can be obtained by summing up the processes mentioned above, i.e.,
\begin{equation}
\dot{\gamma}_{\rm e,tot}^{\prime} = \dot{\gamma}_{\rm e,syn}^{\prime} + \dot{\gamma}_{\rm e,adi}^{\prime}
+ \dot{\gamma}_{\rm e,SSC}^{\prime}.
\end{equation}
Heating of low-energy electrons due to synchrotron absorption \citep{Ghisellini91,Gao13} is not considered here,
which may pile up electrons in the low-energy range. In this work, we focus on the cooling processes
in order to investigate their roles clearly. Heating or acceleration of electrons will be incorporated
in our future studies.

The GRB prompt emission comes from a group of electrons, of which
the instantaneous spectrum can be denoted as $d N_{\rm e} / d \gamma_{\rm e}^{\prime}$.
This electron distribution can be obtained by solving the continuity equation of electrons in energy space~\citep{Longair11}
\begin{equation}
\frac{\partial}{\partial t^{\prime}} \left( \frac{d N_{\rm e}}{d \gamma_{\rm e}^{\prime}} \right) + \frac{\partial}{\partial \gamma_{\rm e}^{\prime}}
\left[ \dot{\gamma}_{\rm e,tot}^{\prime} \left( \frac{d N_{\rm e}}{d \gamma_{\rm e}^{\prime}} \right) \right]
= Q(\gamma_{\rm e}^{\prime},t^{\prime}),
\label{eq:continuity}
\end{equation}
where $Q(\gamma_{\rm e}^{\prime},t^{\prime})$ is the source function that describes the electrons injected into the emitting region.
If the bulk Lorentz factor of the jet is $\Gamma$, the co-moving time $t^{\prime}$ can be related to the observer's time by
\begin{equation}
d t_{\rm obs} = (1 + z) \Gamma (1 - \beta) d t^{\prime} \simeq \frac{1 + z}{2 \Gamma} d t^{\prime},
\end{equation}
and the position of the jet head is described by
\begin{equation}
d R = \beta c \Gamma d t^{\prime} = \frac{\beta}{1 - \beta} \frac{c~d t_{\rm obs}}{1 + z},
\end{equation}
where $\beta = \sqrt{1 - \frac{1}{\Gamma^2}}$ is the dimensionless velocity of the jet.

\section{CONSTRAINTS FROM OBSERVATIONS}
On one hand, the characteristics of the emission site for the prompt emission, the relativistic jet (or the relativistic ejecta),
are still under research. On the other hand, the knowledge on the composition of the jet, and some quantities such as
the bulk Lorentz factor of the jet, the emission radius, the strength of $B^{\prime}$, and the Lorentz factor
of the electrons are crucial to model the GRB spectra.
Here, rather than assuming a detailed model, we try to derive some constraints on relevant quantities
from observations. 
These constraints have been derived analytically in prior articles~\citep[e.g.,][]{Kumar08,Beniamini13,Beniamini14,Kumar15}.  
The main logic of our derivation is similar to these works.
With proper ranges of these quantities, we can then analyze the
cooling behavior of electrons and correspondingly perform calculations in the following sections.

Assuming that the bulk Lorentz factor of the GRB ejecta is $\Gamma$, the Lorentz factor of electrons that radiate at the
GRB spectral peak energy $E_{\rm peak}$ is $\gamma_{\rm m}^{\prime}$, then we have
\begin{equation}
E_{\rm peak} = \frac{1}{1+z} \frac{3 h q_{\rm e} B^{\prime}}{4 \pi m_{\rm e} c} \Gamma \gamma_{\rm m}^{\prime 2},
\label{eq:Ep}
\end{equation}
where $h$, $q_{\rm e}$ are the Planck constant and electron charge respectively,
$z$ is the redshift of the burst.
The radiative cooling time for an electron of $\gamma_{\rm e}^{\prime}$ in the observer frame is
\begin{equation}
t_{\rm c} = \frac{3 \pi m_{\rm e} c (1+z)}{\sigma_T B^{\prime 2} \gamma_{\rm e}^{\prime} \Gamma (1+Y)},
\end{equation}
where $Y$ is the Compton-Y parameter.
The dynamical time of the jet can be expressed as $t_{\rm d} \sim R (1+z) / (2 \Gamma^2 c)$.

For typical parameters, the magnetic field strength in the emission region is strong enough
that the electrons are in the fast cooling regime (also see Equation (\ref{eq:Fast Cooling})).
The fast cooling condition requires that $t_{\rm c} (\gamma_{\rm m}^{\prime}) \leq t_{\rm d}$, which implies
\begin{equation}
\frac{B^{\prime 2} \gamma_{\rm m}^{\prime}}{\Gamma} \geq \frac{6 \pi m_{\rm e} c^2}{\sigma_T R (1+Y)}.
\label{eq:tc-td}
\end{equation}
Taking typical values of $E_{\rm peak} \simeq 500$~keV, and $R \simeq 10^{15}$~cm, we can get,
\begin{eqnarray}
B^{\prime} \Gamma \gamma_{\rm m}^{\prime 2} &=& 2.9 \times 10^{13}~(1+z) \left( \frac{E_{\rm peak}}{500~\mathrm{keV}} \right)~\mathrm{G},
\label{eq:Epeak} \\
\Gamma \gamma_{\rm m}^{\prime} &\leq& 3.3 \times 10^7~(1+z)^{2/3} (1+Y)^{1/3}  \\ \nonumber
 & & \left( \frac{E_{\rm peak}}{500~\mathrm{keV}} \right)^{2/3}
\left( \frac{R}{10^{15}~\mathrm{cm}} \right)^{1/3},
\label{eq:Fast Cooling}
\end{eqnarray}
by using Equations (\ref{eq:Ep}) and (\ref{eq:tc-td}).

On the other hand, the specific flux at $E_{\rm peak}$ in the observer frame can be expressed as
~\citep[e.g.,][]{Beniamini13,Beniamini14,Kumar15a}
\begin{equation}
F_{\nu_{\rm obs}} = N_{\rm e} \frac{\sqrt{3} q_{\rm e}^3 B^{\prime} \Gamma}{m_{\rm e} c^2}  \frac{1+z}{4 \pi D_L^2},
\label{eq:Fnu}
\end{equation}
where $N_{\rm e}$ is the total (already corrected for $4\pi$ solid angle) number of electrons with
$\gamma_{\rm e}^{\prime} > \gamma_{\rm m}^{\prime}$,
and $D_{L}$ is the luminosity distance of the burst.
Then we can estimate the number of electrons needed to produce a given observed flux
by combining Equations (\ref{eq:Epeak}) and (\ref{eq:Fnu}),
\begin{equation}
N_{\rm e} = 1.9 \times 10^{39} \gamma_{\rm m}^{\prime 2} (1+z)^{-2} \left( \frac{F_{\nu_{\rm obs}}}{1~\mathrm{mJy}} \right) \left( \frac{E_{\rm peak}}{500~\mathrm{keV}} \right)^{-1}
\left( \frac{D_L}{10^{28}~\mathrm{cm}} \right)^{2}.
\label{eq:Ne}
\end{equation}
The corresponding average injection rate of electrons is
\begin{eqnarray}
N_{\rm inj}^{\prime} &\simeq& \frac{N_{\rm e}}{\delta t_{\rm c}^{\prime} (\gamma_{\rm m})} =  \frac{N_{\rm e}}{2 \Gamma \delta t_{\rm c} /(1+z)} 
= 2.1 \times 10^{57} ~\Gamma^{-2} \gamma_{\rm m}^{\prime -1} \\ \nonumber
&&
\times
(1+Y) \left( \frac{F_{\nu_{\rm obs}}}{1~\mathrm{mJy}} \right)
\left( \frac{E_{\rm peak}}{500~\mathrm{keV}} \right)
\left( \frac{D_L}{10^{28}~\mathrm{cm}} \right)^{2}~\mathrm{s}^{-1},
\end{eqnarray}
where we have used $t_{\rm c}^{\prime} (\gamma_{\rm m}^{\prime})$ as the injection timescale in order to maintain the electron distribution
and the intensity of the radiation flux within this period.
One can see that in a synchrotron model, the typical value of $N_{\rm inj}^{\prime}$ is almost model independent
and may be compared/verified with further detailed simulation results.

Now, we consider two leading models respectively to obtain the plausible range of the key parameter $\gamma_{\rm m}^{\prime}$.
First, if the jet is magnetically dominated, i.e., a relativistic Poynting-flux-dominated jet
\footnote{In this article, by saying the scenario of Poynting-flux-dominated jet,
we mean the regime invoking a large emission radius (the magnetization parameter
is not necessarily very large, see \citealt{ZhangB11}), which is consistent with a Poynting-flux-dominated regime.
Note that besides the synchrotron mechanism, emission from Poynting-flux-dominated jet
has also been discussed in the photosphere context~\citep[e.g.,][]{Drenkhahn02,Metzger11}.}
~\citep[e.g.,][]{ZhangB11,Beniamini14,Kumar15a},
its isotropic equivalent magnetic luminosity is
$L_B \simeq \frac{B^{\prime 2}}{8 \pi} \Gamma^2 4 \pi R^2 c$.
The kinetic energy power of the accelerated electrons is
$L_{\rm e} \simeq N_{\rm inj}^{\prime} m_{\rm e} c^2 \gamma_{\rm m}^{\prime} \Gamma^2$.
In principle, the ratio of $L_{\rm e}$ to $L_B$,
\begin{eqnarray}
\label{eq:eta_e}
\eta_{\rm e} &=& \frac{L_{\rm e}}{L_B} = 1.3 \times 10^{-16} \gamma_{\rm m}^{\prime 4} (1+z)^{-2} (1+Y) \\ \nonumber
&&
\times \left( \frac{F_{\nu_{\rm obs}}}{1~\mathrm{mJy}} \right)
\left( \frac{E_{\rm peak}}{500~\mathrm{keV}} \right)^{-1}
\left( \frac{R}{10^{15}~\mathrm{cm}} \right)^{-2}
\left( \frac{D_L}{10^{28}~\mathrm{cm}} \right)^{2}
\end{eqnarray}
should be less than $1$ since the jet is magnetically dominated.
The upper limit of $\gamma_{\rm m}^{\prime}$ can thus be derived from Equation (\ref{eq:eta_e}).
However, one should note that $\eta_{\rm e}$ may still be slightly larger than 1 in the realistic case, since the
magnetic field $B^{\prime}$ that cools electrons here may be smaller than the average magnetic field of the ejecta.

In the framework of the internal shock model~\citep[e.g.,][]{Rees94,Daigne11},
the dominant energy of the jet should be the kinetic energy carried by protons.
If we consider that there are $\eta_{\rm p}$ protons for every accelerated electron
and assume that the accelerated protons remain non-relativistic, then
we get the kinetic energy power of protons as $L_{\rm p} \simeq \eta_{\rm p} N_{\rm inj}^{\prime} m_{\rm p} c^2 \Gamma^2$
~\citep{Bonjak09,Daigne11,Beniamini13},
and the ratio between $L_B$ and $L_{\rm p}$ is
\begin{eqnarray}
\label{eq:xiB}
\xi_B &=& \frac{L_B}{L_{\rm p}} = 4.1 \times 10^{12} \gamma_{\rm m}^{\prime -3} \eta_{\rm p}^{-1} (1+z)^{2} (1+Y)^{-1} \\ \nonumber
&&\times \left( \frac{F_{\nu_{\rm obs}}}{1~\mathrm{mJy}} \right)^{-1}
\left( \frac{E_{\rm peak}}{500~\mathrm{keV}} \right)
\left( \frac{R}{10^{15}~\mathrm{cm}} \right)^{2}
\left( \frac{D_L}{10^{28}~\mathrm{cm}} \right)^{-2}.
\end{eqnarray}
Here, $\xi_B$ can be equivalently treated as the familiar magnetization parameter $\sigma$ at the emission radius $R$.
On the other hand, the ratio between $L_{\rm e}$ and $L_{\rm p}$ is
\begin{equation}
\xi_{\rm e} = \frac{L_{\rm e}}{L_{\rm p}} = \frac{\gamma_{\rm m}^{\prime} m_{\rm e}}{\eta_{\rm p} m_{\rm p}}.
\end{equation}
This gives an upper limit of $\gamma_{\rm m}^{\prime}$, i.e.,
$\gamma_{\rm m}^{\prime} = \xi_{\rm e} \frac{\eta_{\rm p} m_{\rm p}}{m_{\rm e}} <  \frac{\eta_{\rm p} m_{\rm p}}{m_{\rm e}}$
\citep{Barniol12,Beniamini13,Kumar15a} since $\xi_{\rm e}$ should be less than 1.
Combining this limit with Equation (\ref{eq:xiB}), we obtain the plausible range of $\xi_B$,
\begin{eqnarray}
\label{eq:xiBrange}
1 &>& \xi_B > 6.6 \times 10^2 \eta_{\rm p}^{-4} (1+z)^{2} (1+Y)^{-1} \\ \nonumber
& & \times \left( \frac{F_{\nu_{\rm obs}}}{1~\mathrm{mJy}} \right)^{-1}
\left( \frac{E_{\rm peak}}{500~\mathrm{keV}} \right)
\left( \frac{R}{10^{15}~\mathrm{cm}} \right)^{2}
\left( \frac{D_L}{10^{28}~\mathrm{cm}} \right)^{-2}.
\end{eqnarray}

Consequently, the range of $\gamma_{\rm m}^{\prime}$ could be derived from Equations (\ref{eq:xiB}) and (\ref{eq:xiBrange}).
One may notice that the free parameter $\eta_{\rm p}$ is crucial to determine the ranges of other quantities in the internal shock model.
Previous studies indicate that $\eta_{\rm p} \ge 10$ should be satisfied for the internal shock model
to explain the GRB spectra (see \citealt{Kumar15} for a review).
Whether this requirement can be fulfilled within the simulation of collisionless ion-electron shocks
is still under debate~\citep{Sironi15}. This issue goes beyond the scope of our current study.
In this work, we admit $\eta_{\rm p} \ge 10$ first and see whether the low-energy spectra
can be explained naturally.

\section{DIFFERENT REGIMES}

With the estimates on the ranges of the key parameters shown above, we now discuss the possible cooling
behaviors of electrons in the emission region of a GRB analytically.
Conventionally, we first take the synchrotron radiation as the main cooling process for electrons, since
the spectra of the observed prompt emission resemble the synchrotron spectra.
However, it is possible that the electrons could also lose energy largely through the other two processes.
For instance, the adiabatic cooling rate for an electron of $\gamma_{\rm e}^{\prime}$
will dominate the synchrotron cooling rate if
\begin{equation}
\frac{\dot{\gamma}_{\rm e,adi}^{\prime}}{\dot{\gamma}_{\rm e,syn}^{\prime}} \ge 1,
\end{equation}
which further gives
\begin{eqnarray}
\label{eq:adi/syn}
\Gamma^3  \gamma_{\rm m}^{\prime 4} \gamma_{\rm e}^{\prime -1} &\ge& \frac{4 \pi m_{\rm e} \sigma_T R (1+z)^2 E_{\rm peak}^2}{9 h^2 q_{\rm e}^2} \\ \nonumber
&\simeq& 5.3 \times 10^{22} (1+z)^2 \left( \frac{E_{\rm peak}}{500~\mathrm{keV}} \right)^2 \left( \frac{R}{10^{15}~\mathrm{cm}} \right)
\end{eqnarray}
by using Equations (\ref{eq:syn}), (\ref{eq:adi}) and (\ref{eq:Epeak}).
For electrons of $\gamma_{\rm e}^{\prime} \le 10^3$, one would find this situation occurs
when $\Gamma \ge 10^3$ and $\gamma_{\rm m}^{\prime} \ge 10^4$, or when $R$ is significantly smaller than $10^{15}$~cm.
We will see that adiabatic cooling does dominate in some cases in the following calculations.

Moreover, SSC cooling is dominant if
\begin{equation}
\frac{\dot{\gamma}_{\rm e,SSC}^{\prime}}{\dot{\gamma}_{\rm e,syn}^{\prime}} \ge 1.
\end{equation}
The corresponding physical requirements are not straightforward since $\dot{\gamma}_{\rm e,SSC}^{\prime}$ involves
double integral in Equation (\ref{eq:SSC}).
Before giving accurate numerical results, some simple estimates can be done primarily.
The magnetic energy density in the co-moving frame of the ejecta is
\begin{equation}
U_{B}^{\prime} = \frac{B^{\prime 2}}{8 \pi},
\end{equation}
while the radiation energy density in the co-moving frame can be calculated as
\begin{equation}
U_{\gamma}^{\prime} \simeq \frac{N_{\rm e} m_{\rm e} c^2 \dot{\gamma}_{\rm m,syn}^{\prime}}{4 \pi R^2 c}.
\end{equation}

If scattering between an electron of $\gamma_{\rm e}^{\prime}$ with photons is always in the Thomson regime,
it is well known that $\dot{\gamma}_{\rm e,SSC}^{\prime} / \dot{\gamma}_{\rm e,syn}^{\prime}$
can be approximated as $U_{\gamma}^{\prime} / U_{B}^{\prime}$.
We define $\gamma_T^{\prime}$ as the Lorentz factor of the electrons above which the scattering with the $E_{\rm peak}$ photons
is in the KN regime~\citep{Wang09,Nakar09}, i.e.,
\begin{equation}
\gamma_T^{\prime} = \frac{\Gamma m_{\rm e} c^2}{E_{\rm peak} (1+z)} \simeq \Gamma (1+z)^{-1} \left( \frac{E_{\rm peak}}{500~\mathrm{keV}} \right)^{-1}.
\end{equation}
$\gamma_T^{\prime}$ will be smaller than $\gamma_{\rm m}^{\prime}$ only if $\Gamma$ is not very large.
So, for electrons within the range of $\gamma_T^{\prime} < \gamma_{\rm e}^{\prime} < \gamma_{\rm m}^{\prime}$,
IC cooling is in the KN regime.

We also define $h \nu_{\rm KN}^{\prime} = m_{\rm e} c^2 / \gamma_{\rm e}^{\prime}$ as the critical photon energy of which the IC scattering
between the electron of $\gamma_{\rm e}^{\prime}$ is in the KN regime.
The scattering between the electron with photons of frequency $\nu^{\prime} < \nu_{\rm KN}^{\prime}$ can effectively cool the electron.
Assuming a low-energy photon spectrum of $\nu F_{\nu} \propto \nu^{\delta}$~$(h \nu < E_{\rm peak})$,
then we have
\begin{equation}
\frac{\dot{\gamma}_{\rm e,SSC}^{\prime}}{\dot{\gamma}_{\rm e,syn}^{\prime}}
\simeq \frac{U_{\gamma}^{\prime} (\nu^{\prime} < \nu_{\rm KN}^{\prime})}
{U_B^{\prime}} \simeq \left( \frac{\gamma_{\rm e}^{\prime}}{\gamma_T^{\prime}} \right)^{-\delta} \frac{U_{\gamma}^{\prime}}{U_{B}^{\prime}}
= \left( \frac{\gamma_{\rm e}^{\prime}}{\gamma_T^{\prime}} \right)^{-\delta} N_{\rm e} \frac{\sigma_T \gamma_{\rm m}^{\prime 2}}{3 \pi R^2}.
\label{eq:SSC/syn}
\end{equation}

Therefore, the condition for the dominance of SSC cooling turns to be
\begin{eqnarray}
\label{eq:SSC>syn}
1.3 \times 10^{-16} \left( \frac{\gamma_{\rm e}^{\prime}}{\gamma_T^{\prime}} \right)^{-\delta} \gamma_{\rm m}^{\prime 4} (1+z)^{-2}
\left( \frac{F_{\nu_{\rm obs}}}{1~\mathrm{mJy}} \right) \\ \nonumber
\left( \frac{E_{\rm peak}}{500~\mathrm{keV}} \right)^{-1}
\left( \frac{R}{10^{15}~\mathrm{cm}} \right)^{-2}
\left( \frac{D_L}{10^{28}~\mathrm{cm}} \right)^{2}
> 1
\end{eqnarray}
by substituting Equation (\ref{eq:Ne}) into Equation (\ref{eq:SSC/syn}).
If the system is steady ($\partial / \partial t = 0$),
the lower limit of $\delta$ is 0.5 when electrons are in fast cooling due to the synchrotron
radiation ($d N_{\rm e} / d \gamma_{\rm e}^{\prime} \propto \gamma_{\rm e}^{\prime -2}$ for
$\gamma_{\rm e}^{\prime} < \gamma_{\rm m}^{\prime}$),
while the upper limit is 1 when electrons are cooled by the SSC radiation
($d N_{\rm e} / d \gamma_{\rm e}^{\prime} \propto \gamma_{\rm e}^{\prime -1}$ for
$\gamma_{\rm e}^{\prime} < \gamma_{\rm m}^{\prime}$, \citealt{Wang09}).
One will then find that for $\gamma_{\rm m}^{\prime} \ge 10^4$, the condition in Equation (27) can be met
at least for electrons of $\gamma_{\rm e}^{\prime} \simeq \gamma_T^{\prime}$.
Moreover, for even smaller emission radius (e.g., $R \approx 10^{14}$~cm), this condition will be relaxed significantly.
So it is essential to consider SSC cooling during the evolution of the electron distribution.
In summary, Equations (\ref{eq:adi/syn}) and (\ref{eq:SSC>syn}) obtained are useful explicit criteria
on judging how an electron cools with given relevant parameters.

\section{NUMERICAL CALCULATIONS}

The main task is to solve the continuity equation of electrons in the energy space, i.e., Equation (\ref{eq:continuity}),
which is also called the advection equation with a source term.
This kind of partial differential equations can be efficiently solved by
the constrained interpolation profile (CIP) method~\citep{Yabe91,Yabe01}.
Detailed discretization procedure can be found in Appendix \ref{app:Method}.
In principle, the final results are determined by the initial and boundary conditions for Equation (\ref{eq:continuity}).
On the other hand, we have already obtained the plausible range for relevant parameters according to the estimates in Section 3.
As we mentioned before, we intend to give an overview of the evolution of the electron distribution
under different physical conditions.
So we explore the parameter space by performing several groups of calculations
to investigate the roles played by different radiation mechanisms in different cases.
In this paper, we adopt the assumption that the co-moving magnetic field in the jet is decaying with radius
as proposed in \citep{Uhm14}, i.e.,
\begin{equation}
B^{\prime} = B_0^{\prime} \left( \frac{R}{R_0} \right)^{-q},
\end{equation}
where $B_0^{\prime}$ is the magnetic strength at $R_0$, and $R_0$ is the radius
where the jet begins to produce the first photon that observed by us.

The injected electrons is assumed to be a power-law
$Q (\gamma_{\rm e}^{\prime},t^{\prime}) = Q_0 (t^{\prime}) (\gamma_{\rm e}^{\prime} / \gamma_{\rm m}^{\prime})^{-p}$
for $\gamma_{\rm e}^{\prime} > \gamma_{\rm m}^{\prime}$, where $Q_0$ is related to the
injection rate by $N_{\rm inj}^{\prime} = \int_{\gamma_{\rm m}^{\prime}}^{\gamma_{\rm max}^{\prime}}
Q (\gamma_{\rm e}^{\prime},t^{\prime}) d \gamma_{\rm e}^{\prime}$
\footnote{$\gamma_{\rm max}^{\prime}$ is the maximum Lorentz factor of electrons and is
given by the approximation $\gamma_{\rm max}^{\prime} \simeq 10^8 \left( \frac{B^{\prime}}{1~\mathrm{G}} \right)^{-0.5}$
\citep{Dai99,Huang00}. So authors may notice that $\gamma_{\rm max}^{\prime}$ is evolving
with time in the results of some calculations.}.
Then, there are eight free parameters in total in our calculations, i.e.,
$\Gamma$, $\gamma_{\rm m}^{\prime}$, $B_0^{\prime}$, $p$,
$N_{\rm inj}^{\prime}$, $q$, $R_0$ and $\eta_{\rm e}$ (or $\eta_{\rm p}$).
Particularly, $q = 1$ is commonly adopted for all calculations unless explicitly stated since the toroidal magnetic field
in the ejecta decreases as $R^{-1}$.
As we will see, this treatment does not markedly impact our main conclusions.
Also, $p = 2.8$ is commonly adopted since the evolution of the low-energy electron distribution is
nearly unaffected by $p$ in fast cooling cases.
Therefore, there are still six free parameters left.
Below, we choose reasonable values for these parameters and perform
a sequence of calculations to represent various physical conditions.
With the electron spectra being derived numerically, we can then calculate the corresponding synchrotron radiation spectra
according to Appendix \ref{app:Radiation}.

\subsection{Testing Calculations}

\begin{figure}
\centering
    \subfloat{\includegraphics[width=0.42\linewidth]{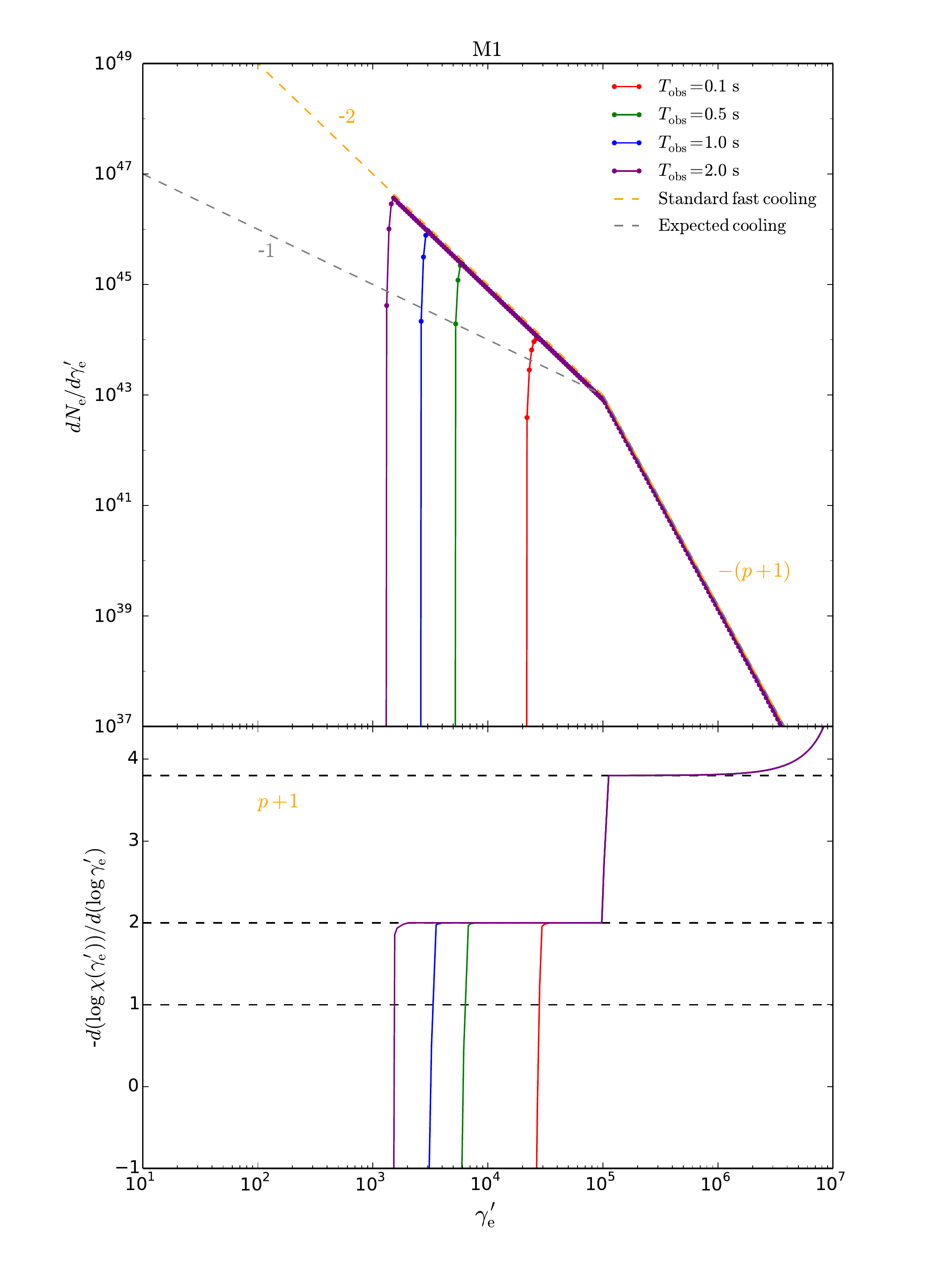}}
    \subfloat{\includegraphics[width=0.42\linewidth]{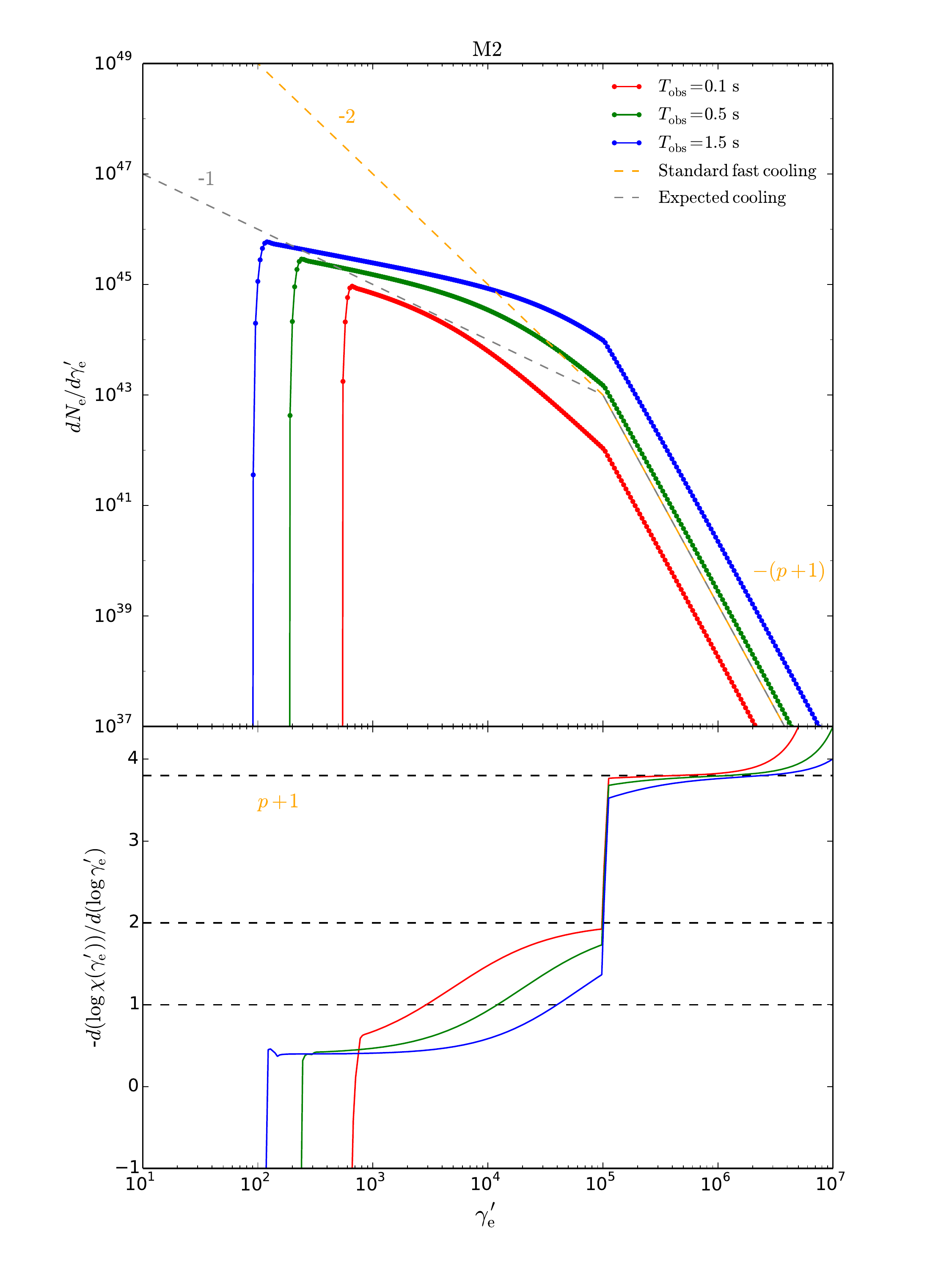}} \\
    \subfloat{\includegraphics[width=0.42\linewidth]{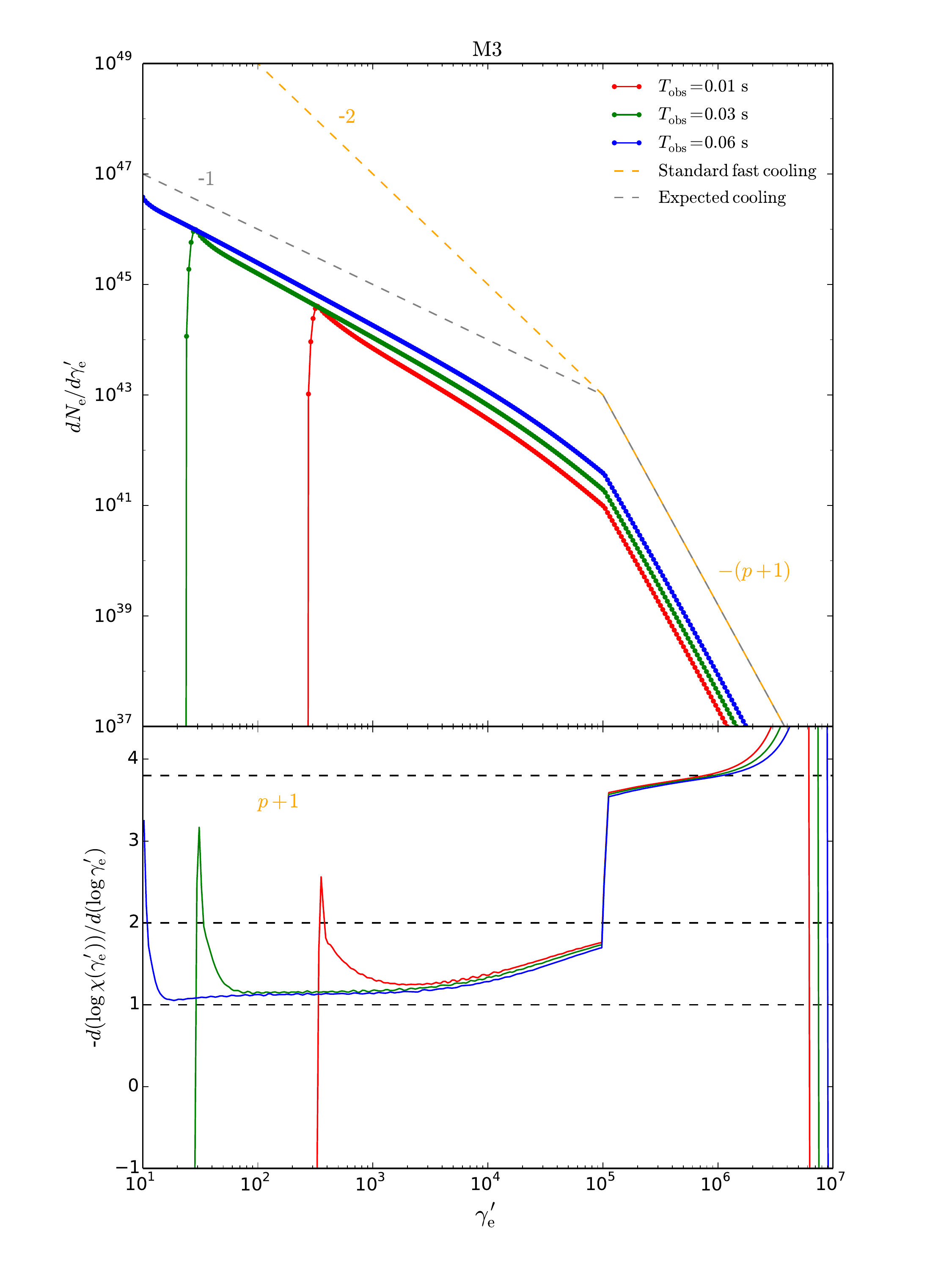}}
    \subfloat{\includegraphics[width=0.42\linewidth]{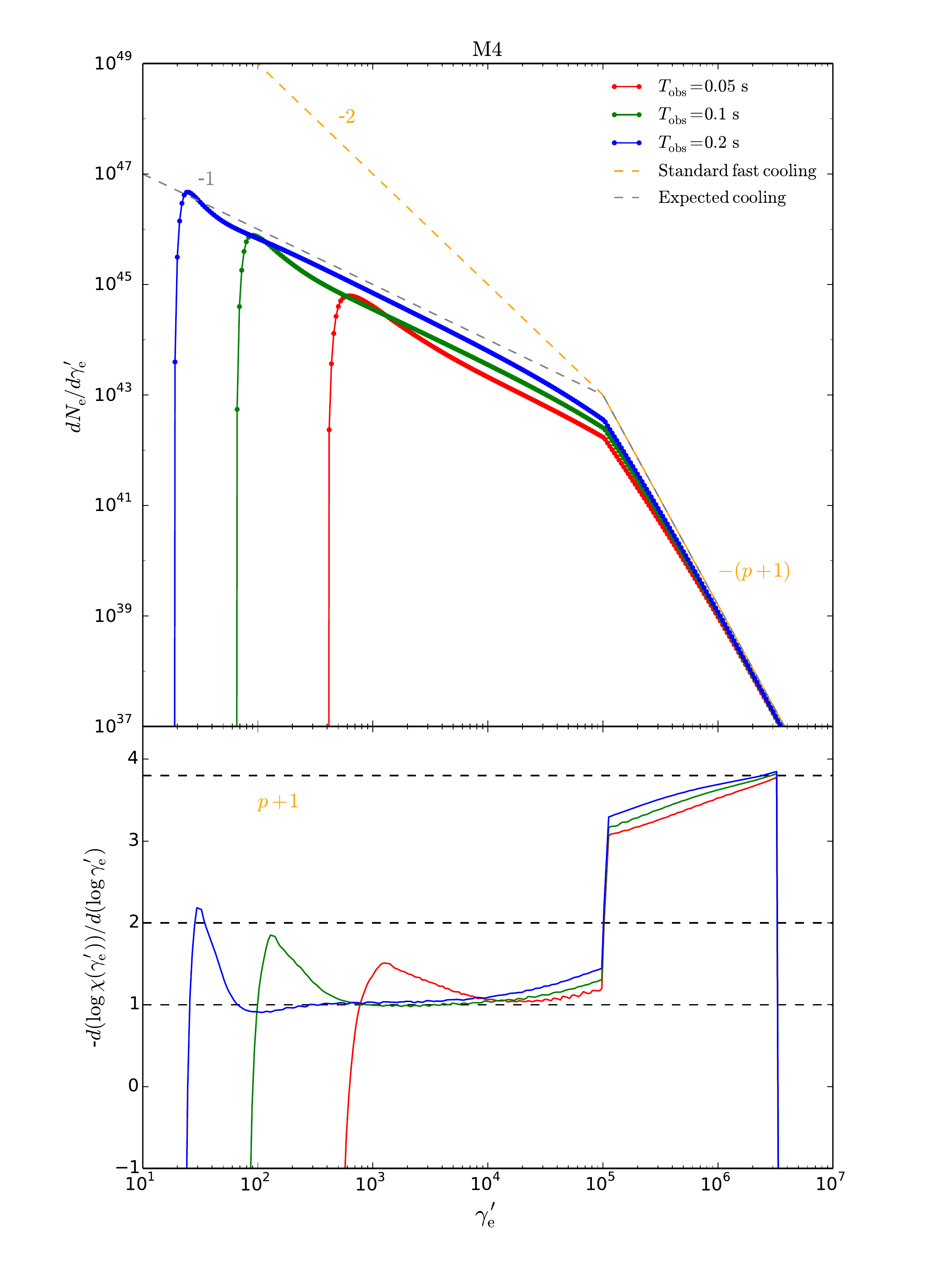}}
    \caption{The evolution of the electron energy spectrum for four cases in the testing calculations (see Table \ref{TABLE:testing}).
    The orange dashed lines are the standard fast cooling pattern, i.e., $d N_{\rm e} / d \gamma_{\rm e}^{\prime} \propto \gamma_{\rm e}^{\prime -2}$,
    and the grey dashed lines present the expected cooling pattern according to the observations, i.e.,
    $d N_{\rm e} / d \gamma_{\rm e}^{\prime} \propto \gamma_{\rm e}^{\prime -1}$. The epochs shown for each case are different
    since the electron cooling timescales are different. The lower panel of each case is the negative spectral index
    of the electron spectrum, i.e., $\chi (\gamma_{\rm e}^{\prime}) = d N_{\rm e} / d \gamma_{\rm e}^{\prime}$. In the standard calculation, M1,
    the electron spectrum shows a typical broken power-law profile, with the spectral indices being $-(p+1)$ above $\gamma_{\rm m}^{\prime}$
    and being -2 below $\gamma_{\rm m}^{\prime}$. In other cases, the spectral indices below $\gamma_{\rm m}^{\prime}$ are
    notably harder than -2.}
    \label{fig:testing1}
\end{figure}

\begin{figure}
\centering
    \subfloat{\includegraphics[width=0.45\linewidth]{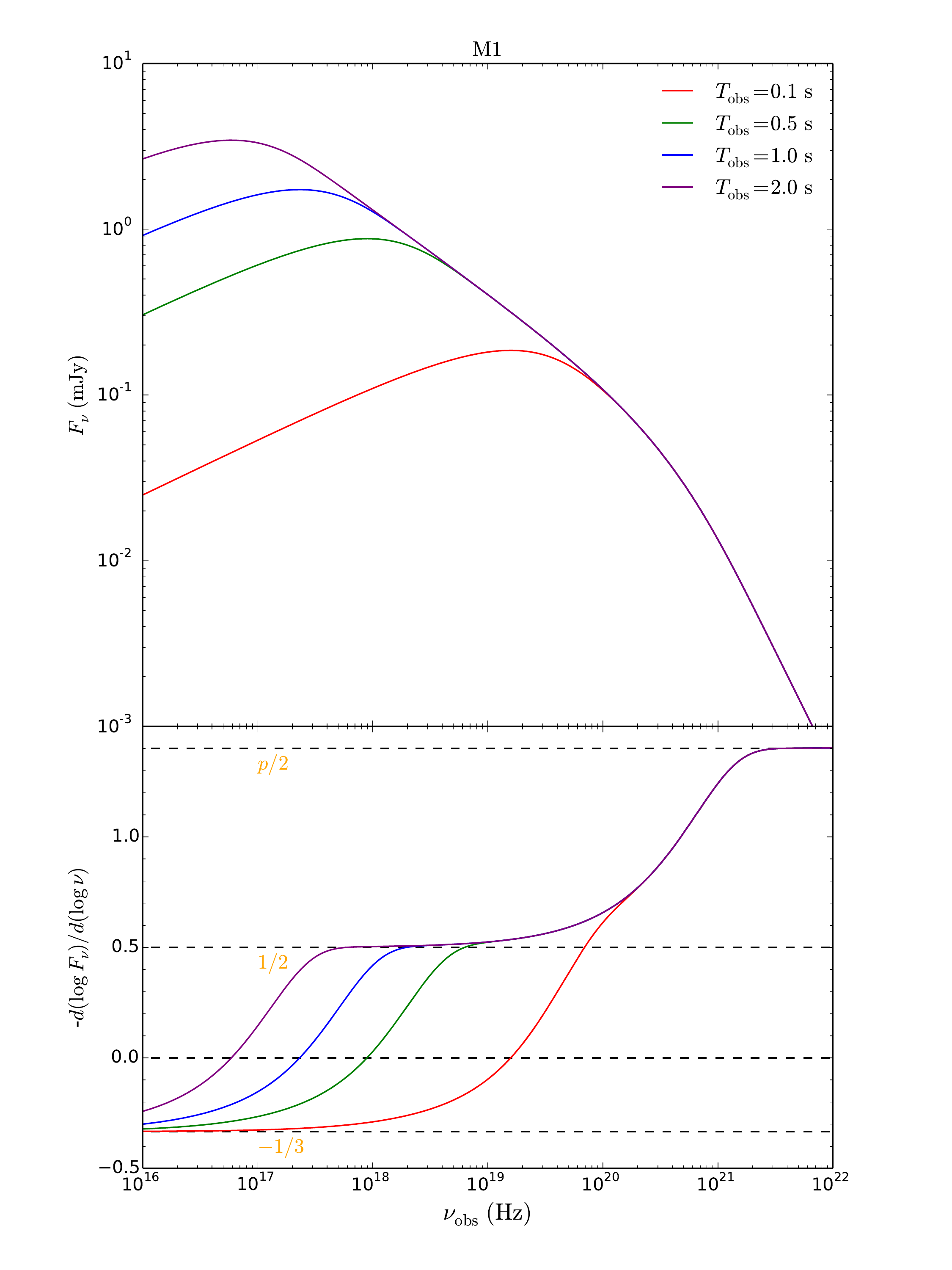}}
    \subfloat{\includegraphics[width=0.45\linewidth]{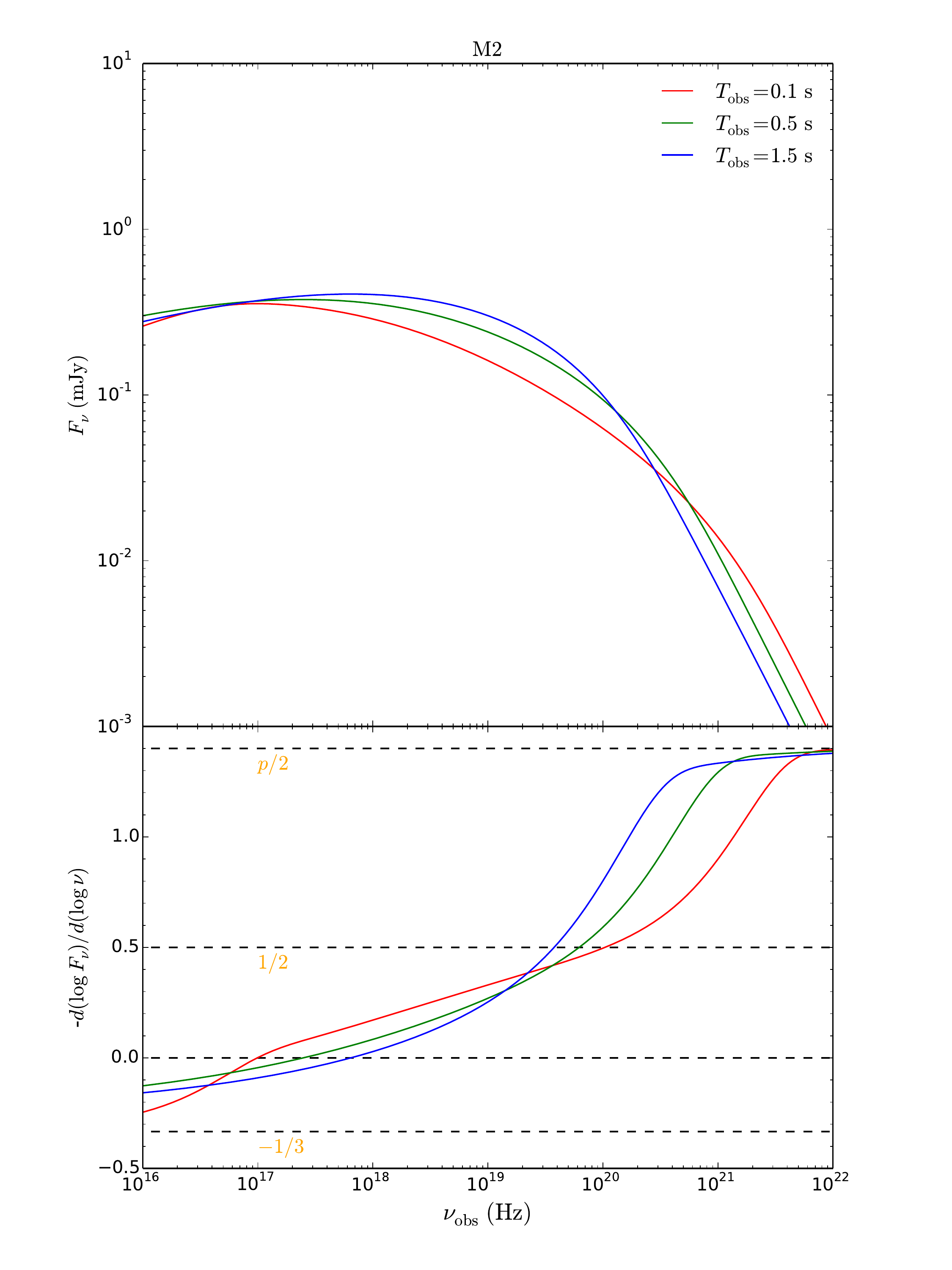}} \\
    \subfloat{\includegraphics[width=0.45\linewidth]{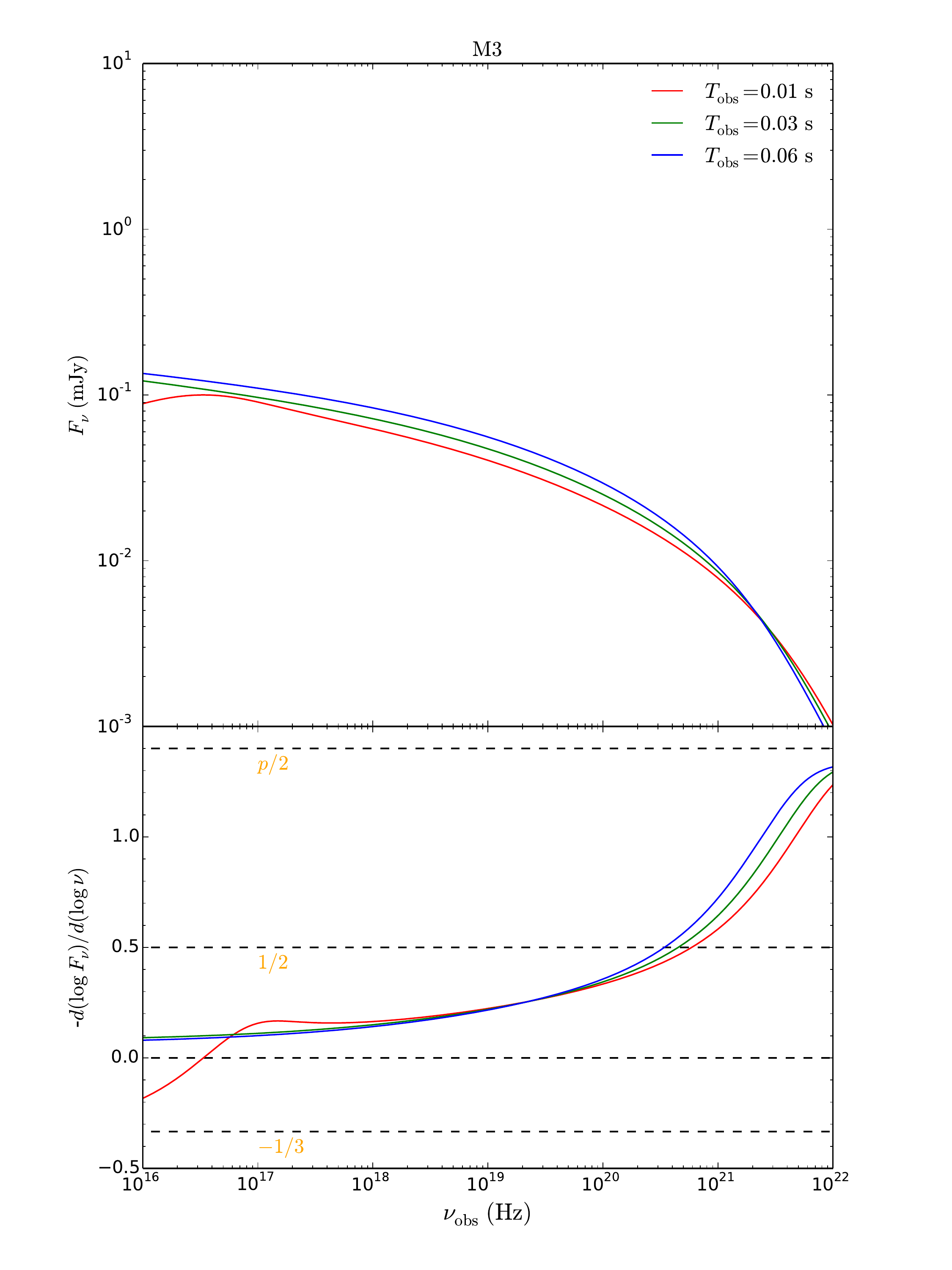}}
    \subfloat{\includegraphics[width=0.45\linewidth]{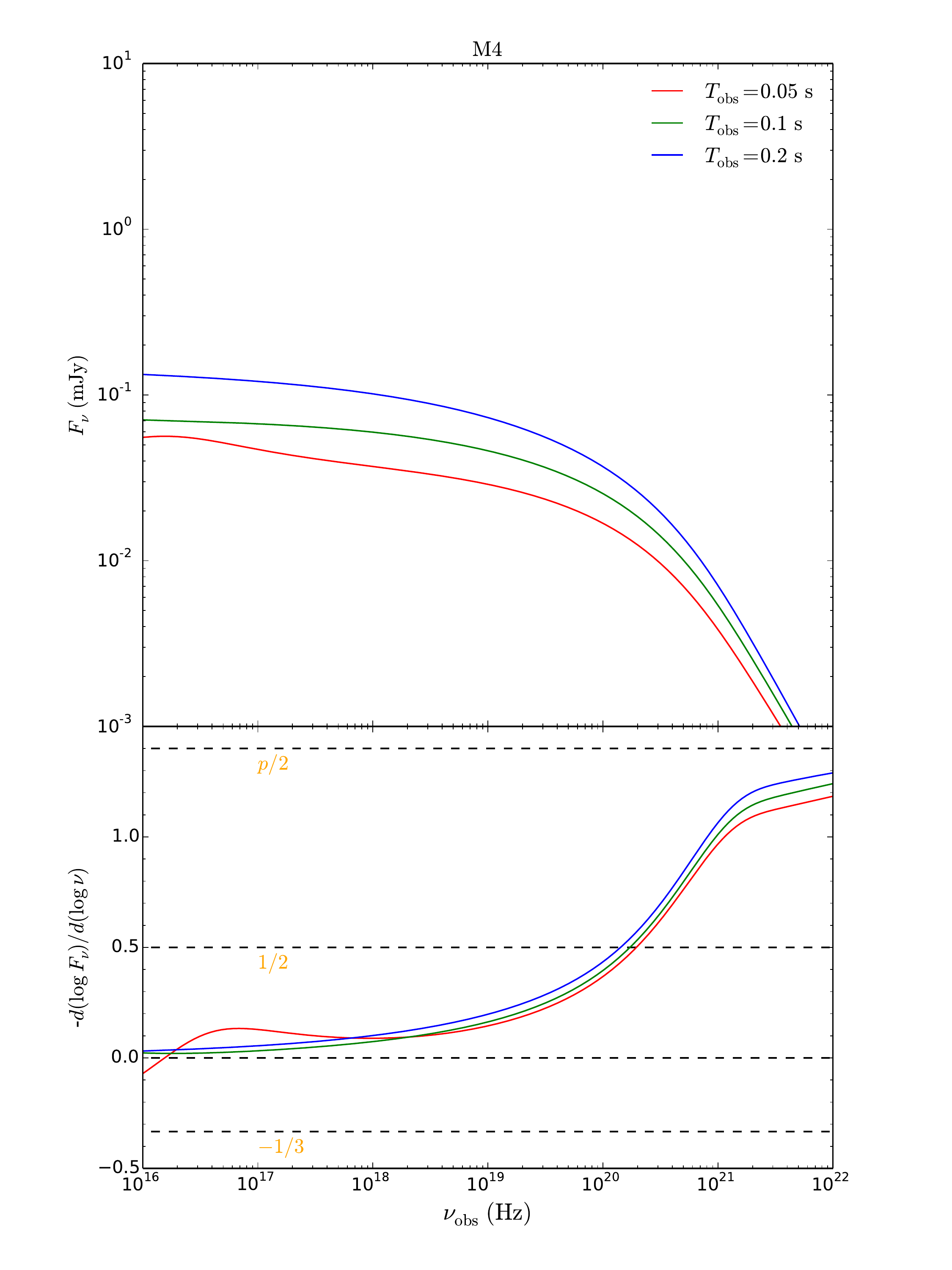}}
    \caption{The corresponding synchrotron flux-density spectra $F_{\nu}$ from the electrons with the energy distribution presented in Figure \ref{fig:testing1}.
    The lower panel of each case shows the negative local spectral indices ($-\alpha$, $F_{\nu} \propto \nu^{\alpha}$) of $F_{\nu}$ in the upper panel.
    In the standard calculation, M1, it gives the typical fast cooling spectrum $F_{\nu} \propto \nu^{-1/2}$
    below $\nu_{\rm m}$ ($\nu_{\rm m} \propto \gamma_{\rm m}^{\prime 2}$).
    For even lower frequency, the spectrum is $F_{\nu} \propto \nu^{1/3}$, which is the profile
    of the low frequency part of a single electron's synchrotron spectrum.
    In other cases, the spectral indices below $\nu_{\rm m}$ can be harder than $-1/2$
    and can approach 0.}
    \label{fig:testing2}
\end{figure}

\begin{figure}
\centering
    \subfloat{\includegraphics[width=0.45\linewidth]{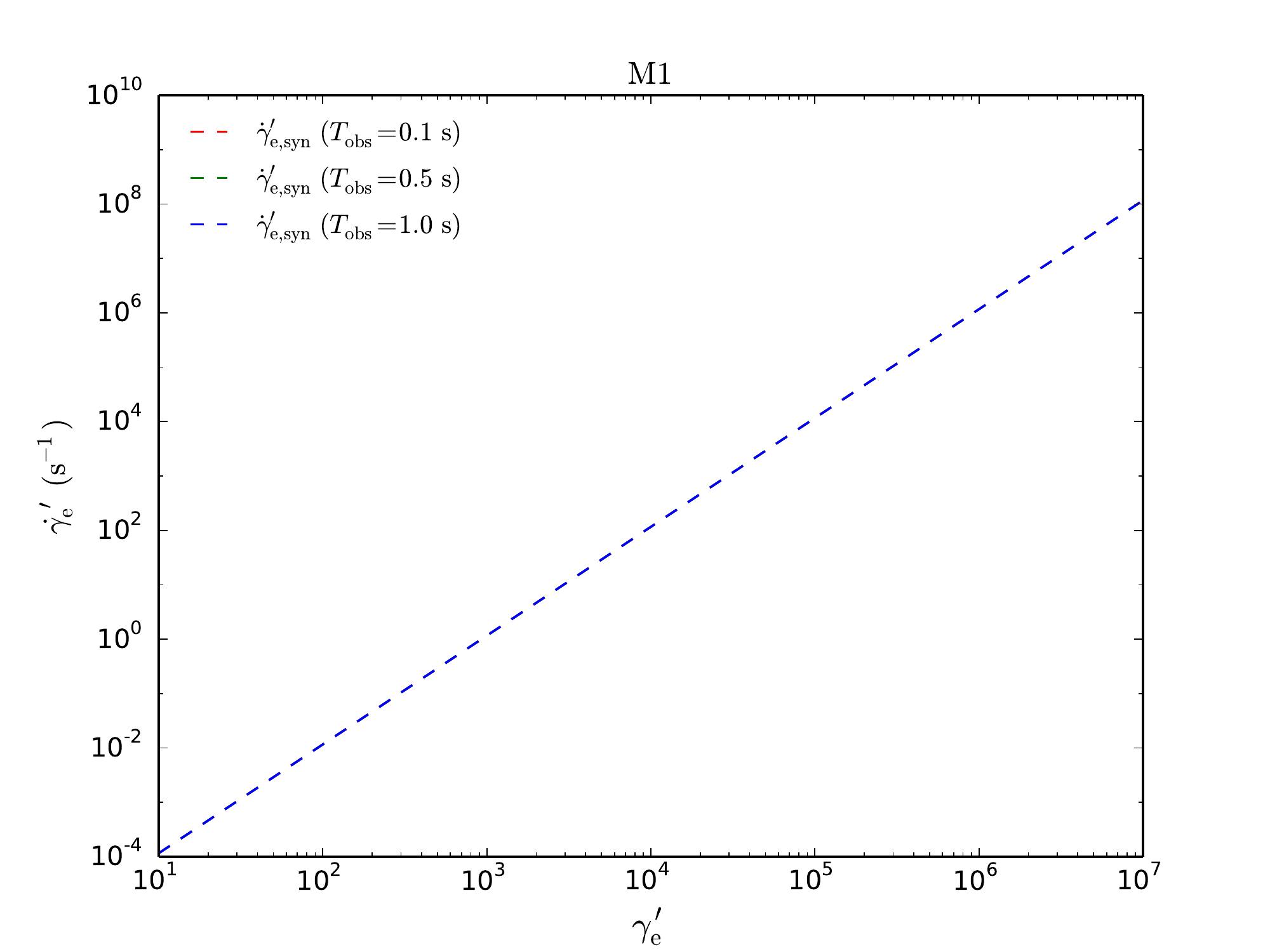}}
    \subfloat{\includegraphics[width=0.45\linewidth]{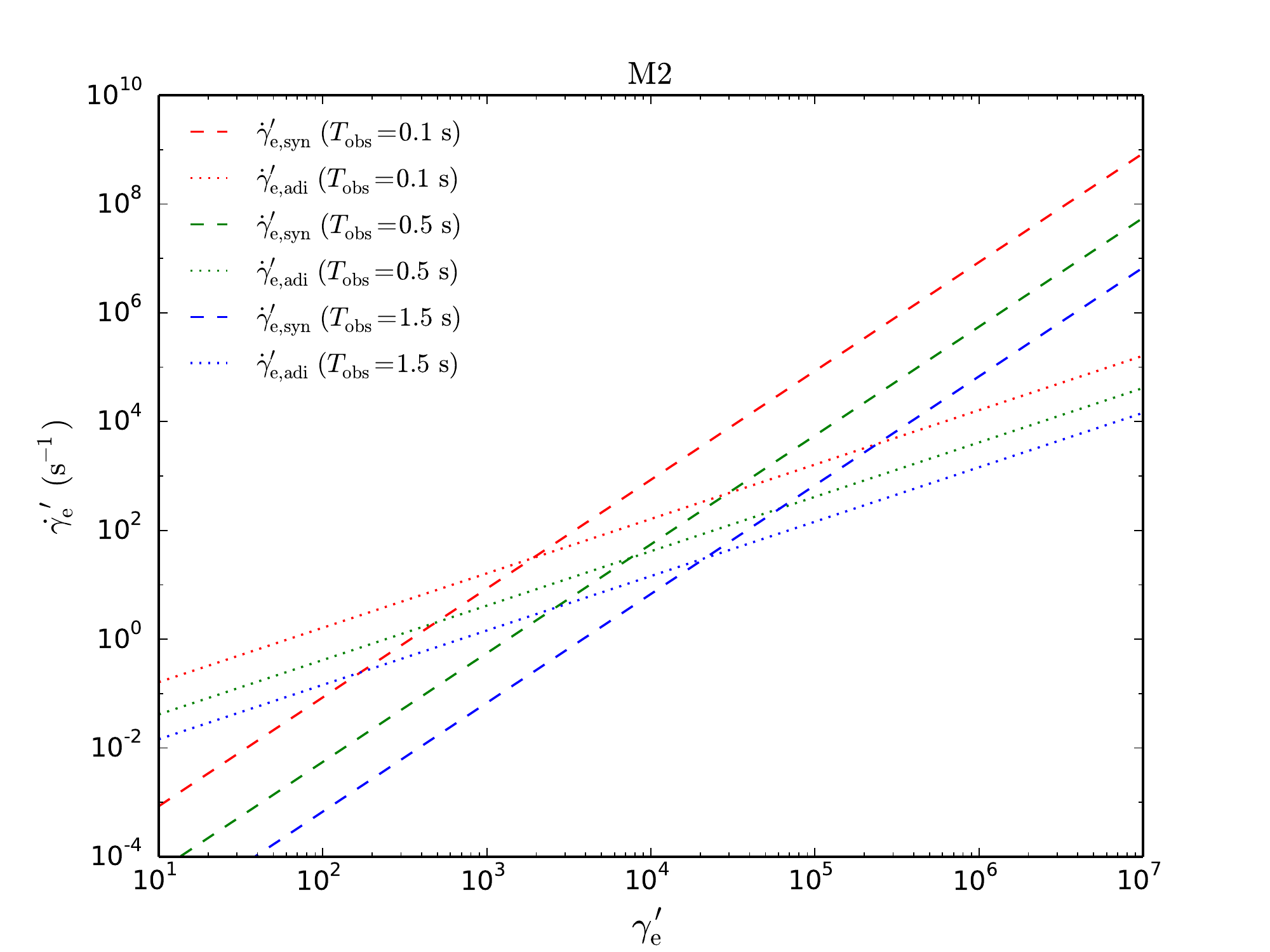}} \\
    \subfloat{\includegraphics[width=0.45\linewidth]{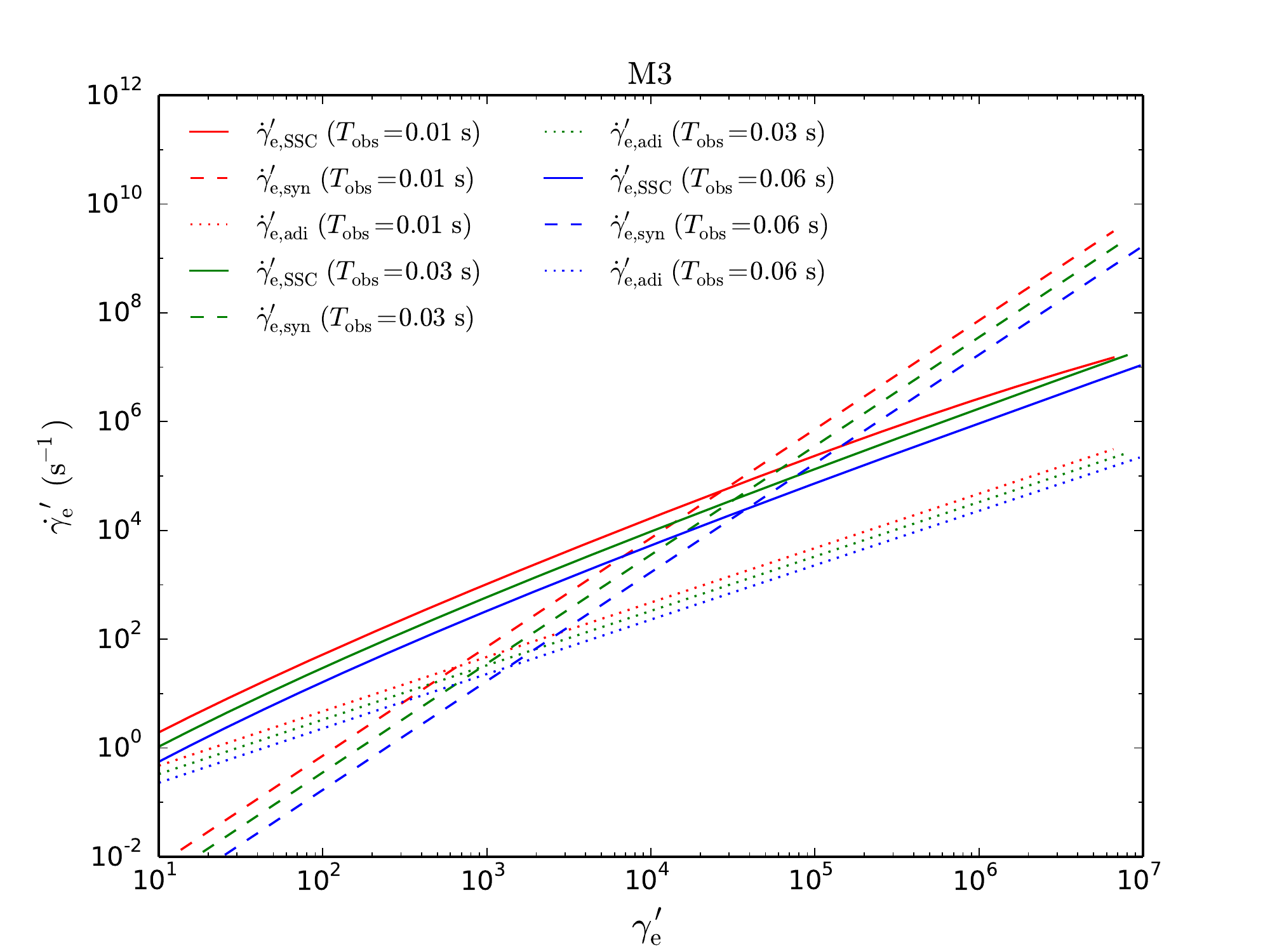}}
    \subfloat{\includegraphics[width=0.45\linewidth]{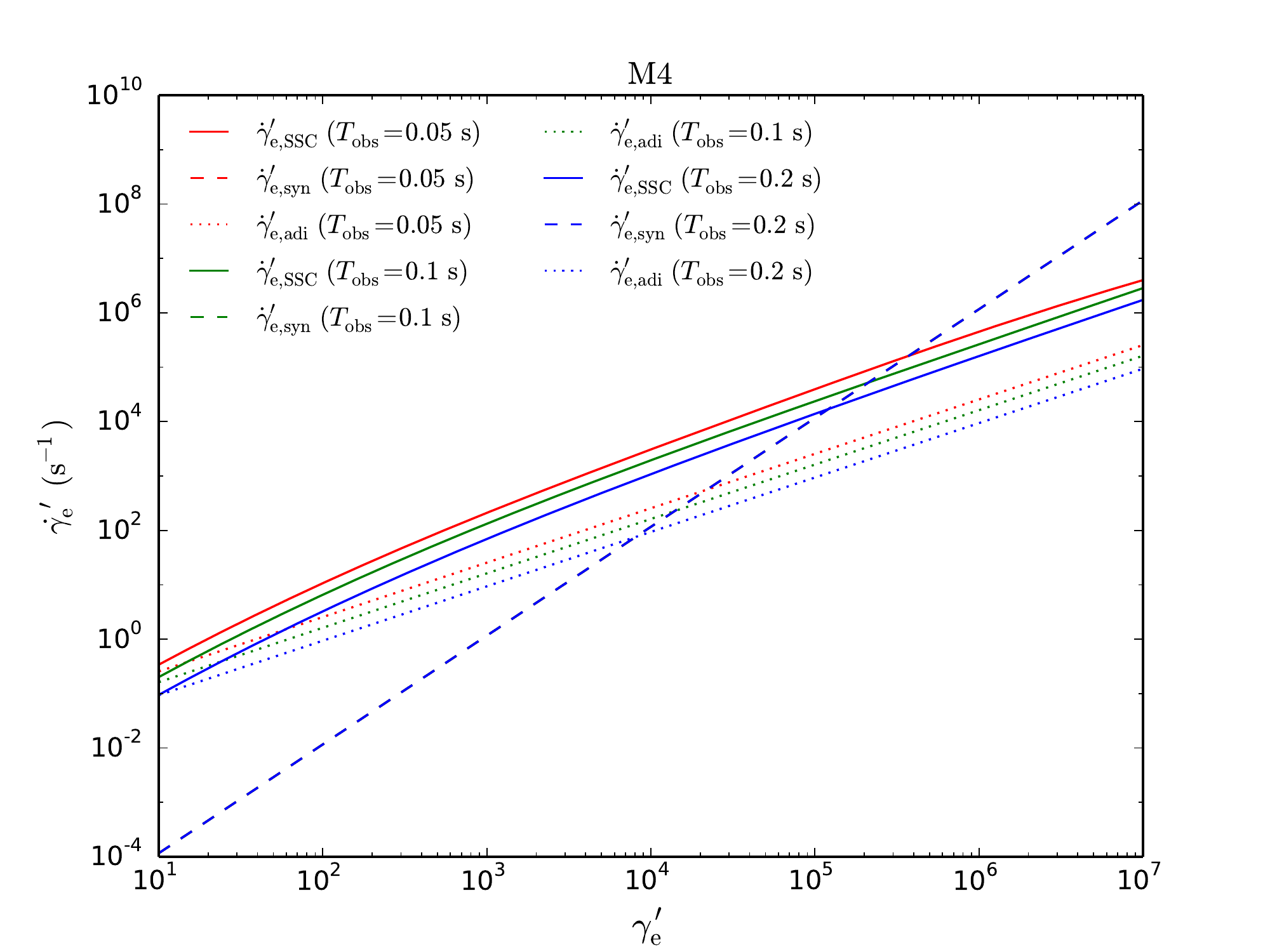}}
    \caption{The co-moving cooling rates of different cooling mechanisms for the electrons with the energy distribution presented in Figure \ref{fig:testing1}.
    Different colours (red, green, blue) denote the cooling rates at different epochs, and different line styles (solid, dashed, dotted)
    present different cooling mechanisms (SSC, synchrotron, adiabatic) respectively.
    Note that in M1, synchrotron cooling rates are time-independent since $B^{\prime}$ is a constant in this case.}
    \label{fig:testing3}
\end{figure}

In this section, we first check the significance of SSC cooling in the evolution of the electron
distribution in the testing calculations.
Four calculations are performed, which are named in form of ``M$i$'' ($i = 1,...,4$).
In M1, we set $q = 0$, ignore SSC and adiabatic cooling so that this should give
a ``standard'' evolution pattern ($d N_{\rm e} / d \gamma_{\rm e}^{\prime} \propto \gamma_{\rm e}^{\prime -2}$)
for electrons under synchrotron cooling only.
The values for other parameters (shown in Table \ref{TABLE:testing}) are taken as the same as those in \cite{Uhm14}
in order to compare the results directly.
In M2, we set $q = 1$ to achieve the similar results with the decaying $B^{\prime}$ case
shown in \cite{Uhm14}. In M3, we introduce the SSC cooling process and compare the results with M2.
This should be more close to the realistic situation. At last, we set $q = 0$ to ignore the effect of decaying $B^{\prime}$
in M4, where the result can clearly show the role played by SSC cooling.
The resultant electron distributions and radiation spectra are shown in Figure \ref{fig:testing1} and Figure \ref{fig:testing2}
respectively for these four calculations.
Moreover, we present the cooling rates of different radiation mechanisms in Figure \ref{fig:testing3}.
It is not surprising that the indices of the low-energy electron spectra are always strictly $-2$ for
$\gamma_{\rm e}^{\prime} < \gamma_{\rm m}^{\prime}$ in M1,
and the indices in M2 turn harder along with $B^{\prime}$'s decreasing as proposed by \cite{Uhm14}.
In M3 and M4, it is interesting to find that the indices of the low-energy electron spectra are approaching $-1$ with the elapsing time,
and the electrons with $\gamma_{\rm e}^{\prime} < \gamma_{\rm m}^{\prime}$ are being cooled mainly via SSC process
as shown in Figure \ref{fig:testing3}.
The asymptotic value of $-1$ is consistent with what is predicted theoretically as mentioned.
Another natural result is that the electrons are cooled much faster after considering the SSC process.
For example, in M3, the minimum Lorentz factor of electrons has already reached $\simeq 20$
at 0.03 s in the observer frame, while it takes 1.5 s for electrons to cool to $\gamma_{\rm e}^{\prime} = 100$ in M2.
With these testing calculations, we see that SSC cooling can play an important role
in determining the electron distribution, at least for in cases considered in previous researches.

\subsection{Cases in Different Scenarios}

Next, we perform numerical calculations by taking the parameters in plausible ranges
for GRBs, corresponding to different physical scenarios,

\subsubsection{The Poynting-flux-dominated Jet}
\label{subsec:PJet}
Let us first consider the case that the jet is a Poynting-flux dominated.
According to Equation (\ref{eq:eta_e}), we can have the upper limit of $\gamma_{\rm m}^{\prime}$ ($\eta_{\rm e} = 1$)
by taking the following set of typical parameters: $F_{\nu_{\rm obs}} = 1~\mathrm{mJy}$,
$E_{\rm peak} = 500~\mathrm{keV}$, $z = 1$ and $Y = 0$.
In this scenario, we only use the upper limit of $\gamma_{\rm m}^{\prime}$ in the calculations.
One may note that $\eta_{\rm e} = 1$ means that $L_{\rm e} = L_B$, which seems to be contrary
to the fact of a Poynting-flux-dominated jet.
However, we would see that the corresponding results are representative.
In addition, we perform two groups of calculations, in which $R_0$ is taken
to be $10^{15}$~cm (called Group PJR15) and $10^{14}$~cm (called Group PJR14), respectively.
Then, in each group, we assume a series of values for $\Gamma$,
and the corresponding $B_0^{\prime}$ can be obtained via Equation (\ref{eq:Epeak}).
For $R_0 = 10^{15}$~cm, we perform five calculations named in form of ``PJR15$\Gamma$N'' with N denoting the value of $\Gamma$,
which ranges from 50 to $10^3$ (see details in Table \ref{TABLE:I}).
For $R_0 = 10^{14}$~cm, we perform four calculations named in form of ``PJR14$\Gamma$N'',
in which $\Gamma$ ranges from $10^2$ to $10^3$ (see details in Table \ref{TABLE:II}).

\begin{figure}
   \centering
   \includegraphics[scale=0.7]{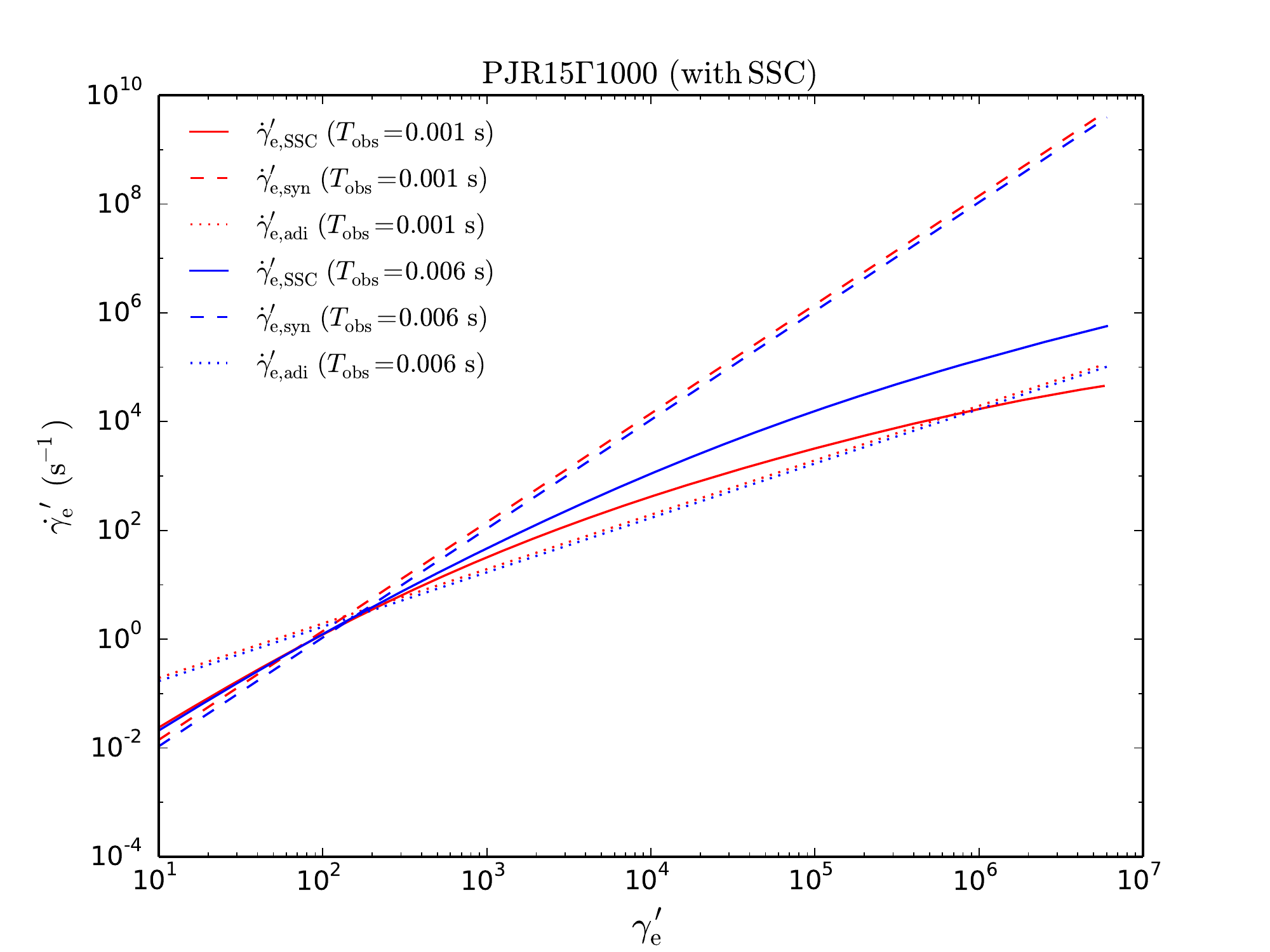}
   \caption{The co-moving cooling rates of different cooling mechanisms for the electrons with the energy distribution
   in calculation PJR15$\Gamma$1000. The cooling rates at two epochs (0.001 s, red; 0.006 s, blue) are shown, and the
   SSC, synchrotron, adiabatic cooling mechanisms are illustrated as the solid, dashed and dotted lines respectively.
   In this case, the cooling of electrons with $\gamma_{\rm e}^{\prime} > 10^2$ is dominated by the synchrotron radiation,
   whereas electrons with $\gamma_{\rm e}^{\prime} < 10^2$ are mainly cooled by adiabatic cooling.}
   \label{fig:MI1SSC}
\end{figure}

\begin{figure}
\begin{adjustwidth}{-2cm}{-2cm}
\centering
    \subfloat{\includegraphics[width=0.3\paperwidth]{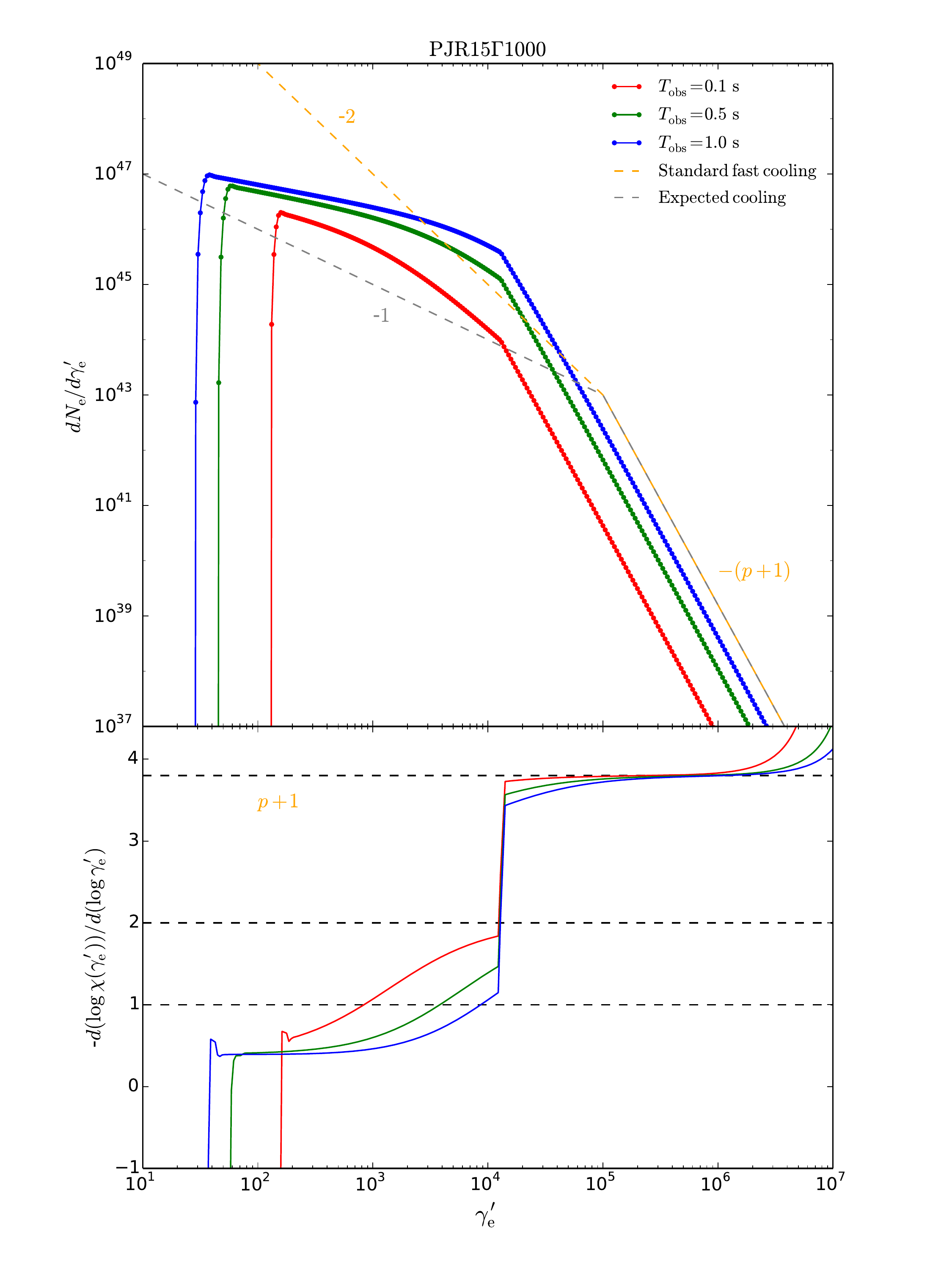}}
    \subfloat{\includegraphics[width=0.3\paperwidth]{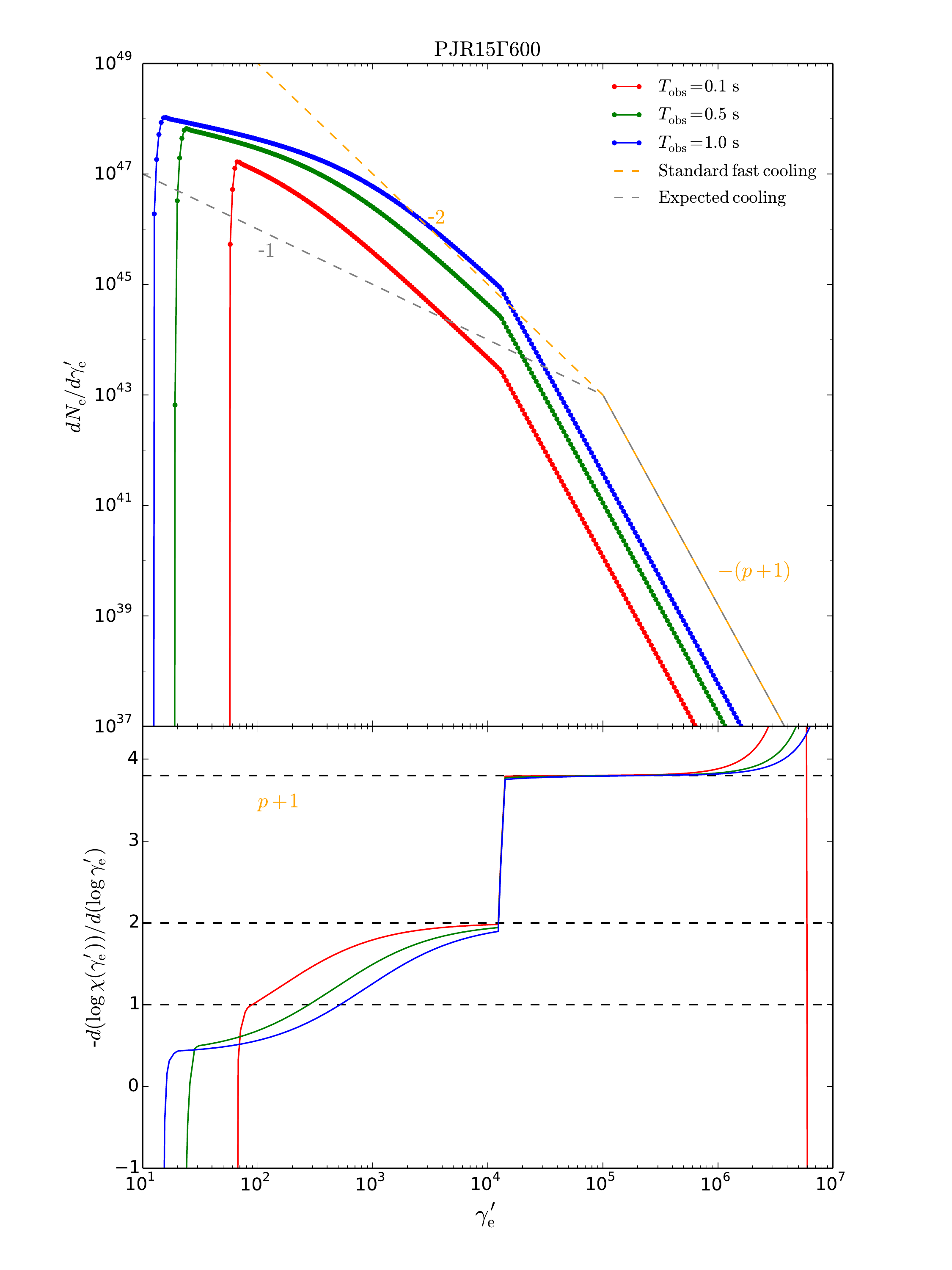}}
    \subfloat{\includegraphics[width=0.3\paperwidth]{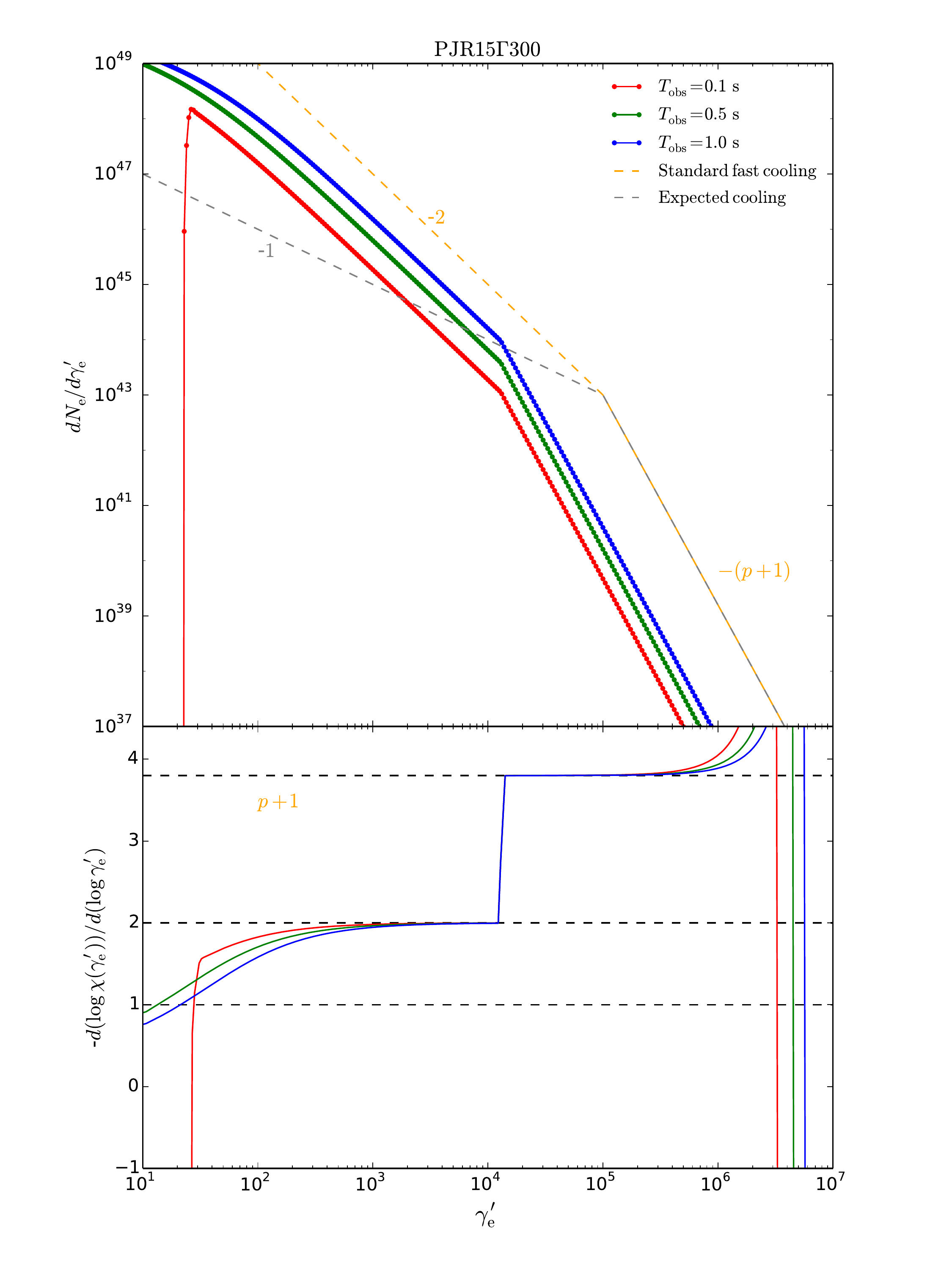}} \\
    \subfloat{\includegraphics[width=0.3\paperwidth]{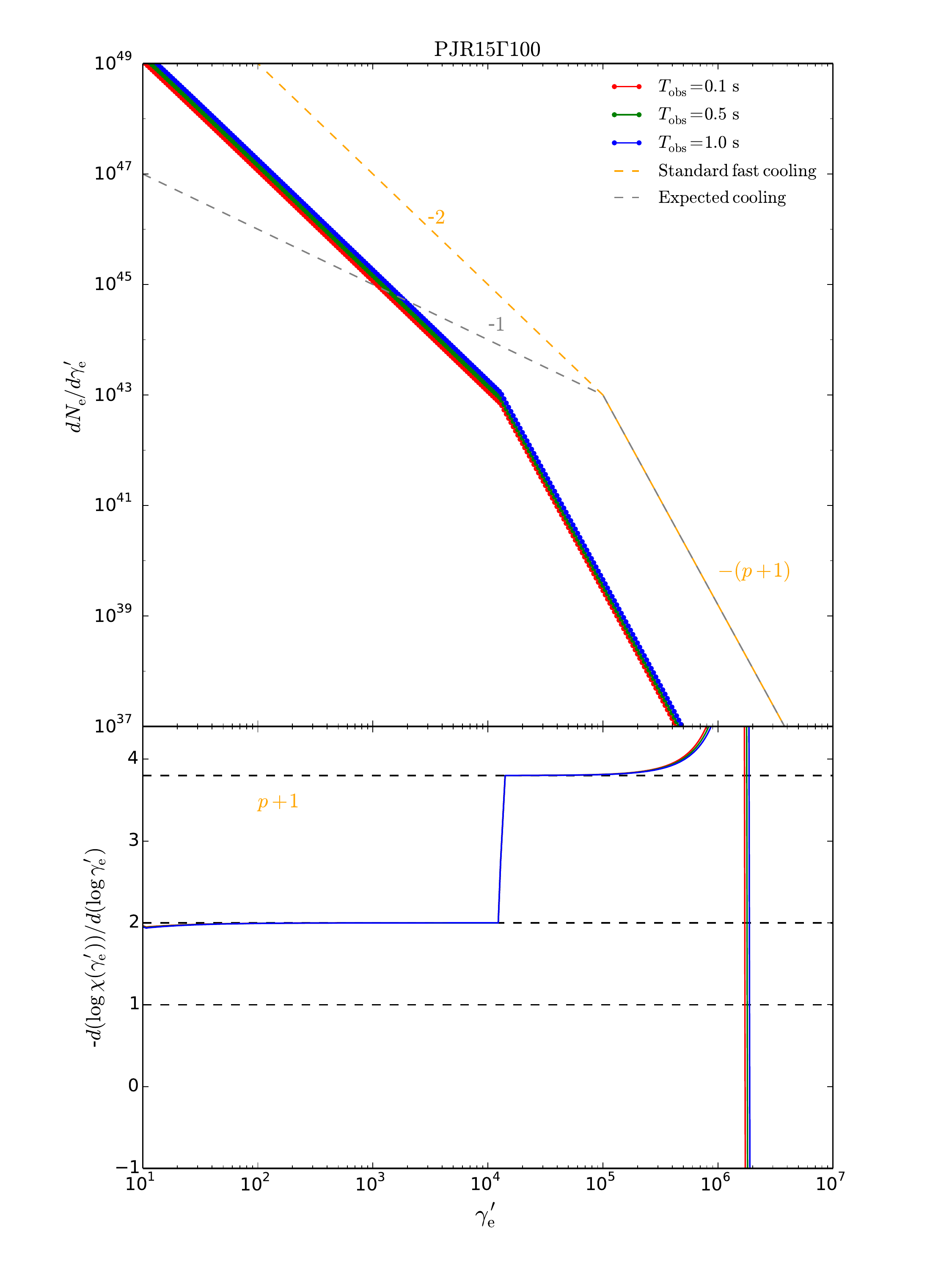}}
    \subfloat{\includegraphics[width=0.3\paperwidth]{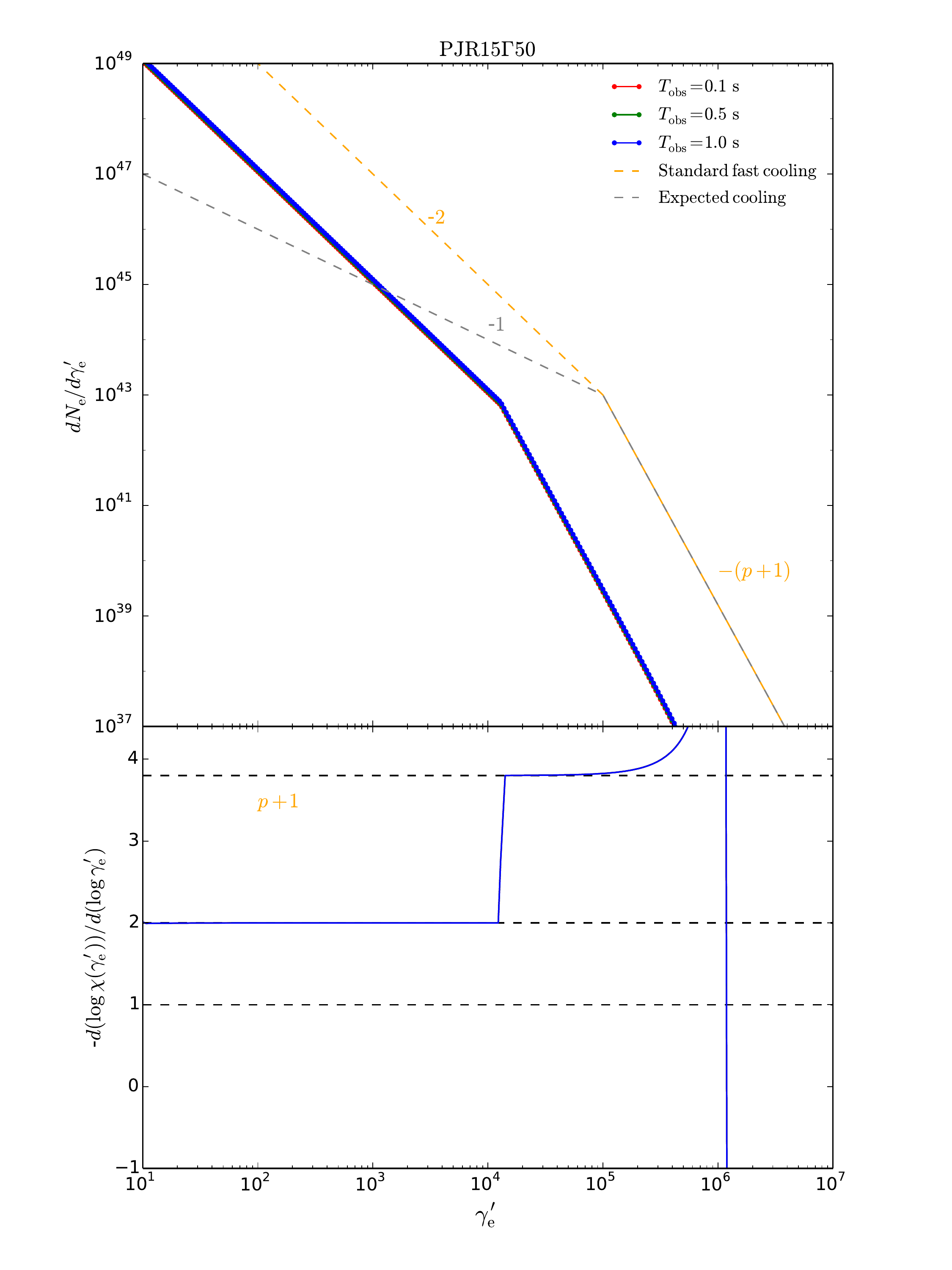}}
    \caption{The evolution of the electron energy spectrum for the five cases in calculations of Group PJR15 (see Table \ref{TABLE:I}). \label{fig:MI-electron}}
\end{adjustwidth}
\end{figure}

\begin{figure}
\begin{adjustwidth}{-2cm}{-2cm}
\centering
    \subfloat{\includegraphics[width=0.3\paperwidth]{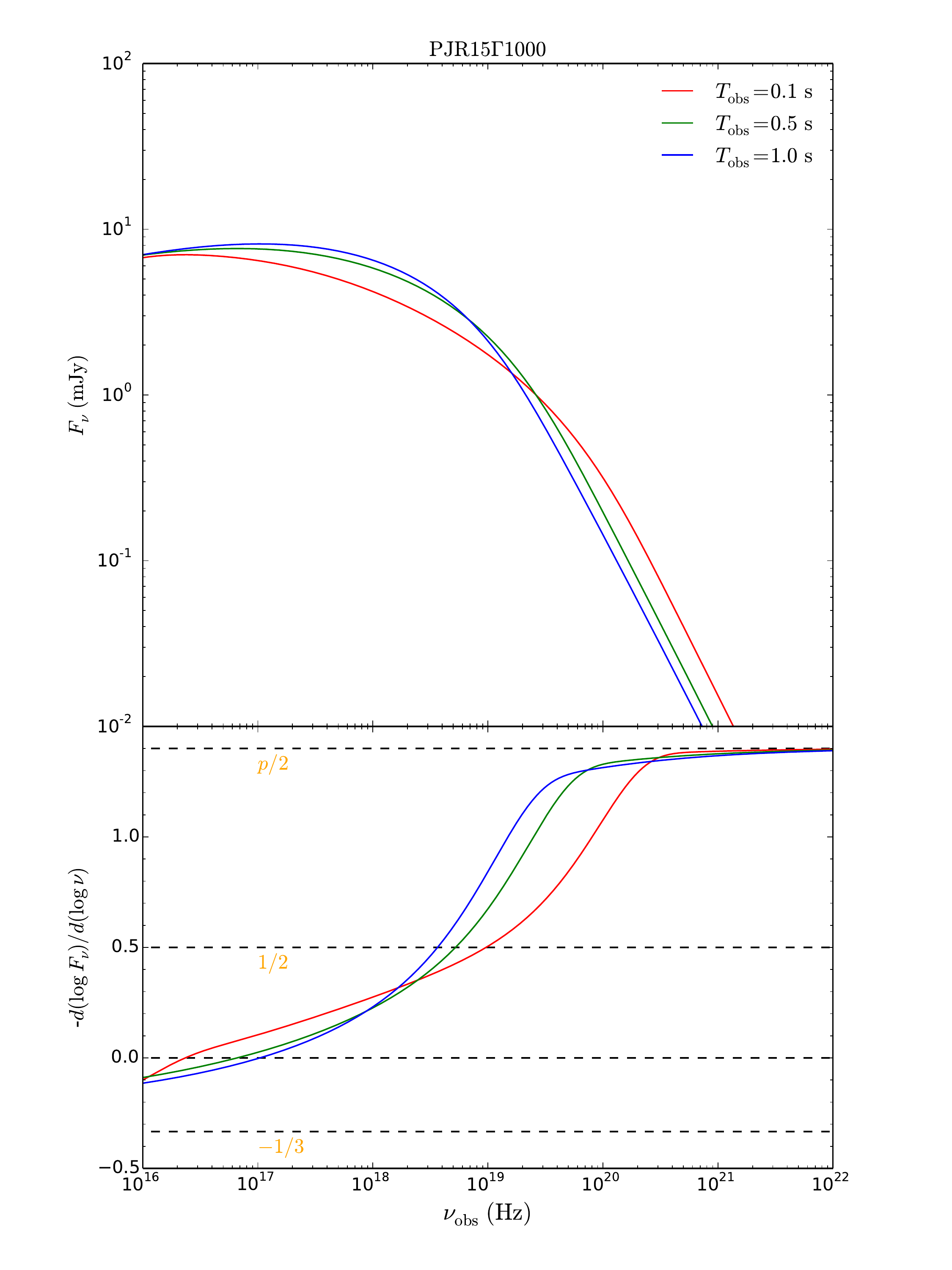}}
    \subfloat{\includegraphics[width=0.3\paperwidth]{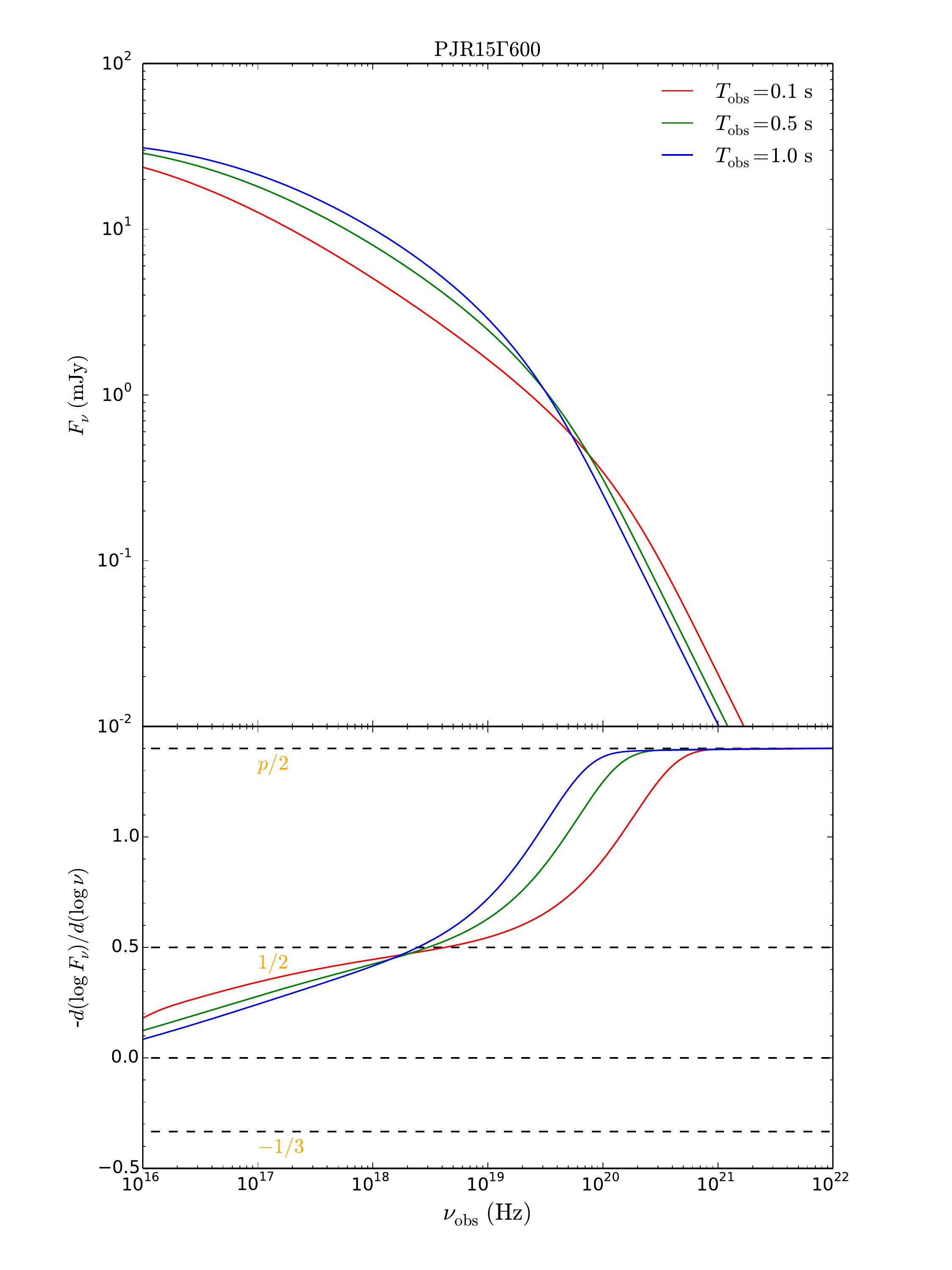}}
    \subfloat{\includegraphics[width=0.3\paperwidth]{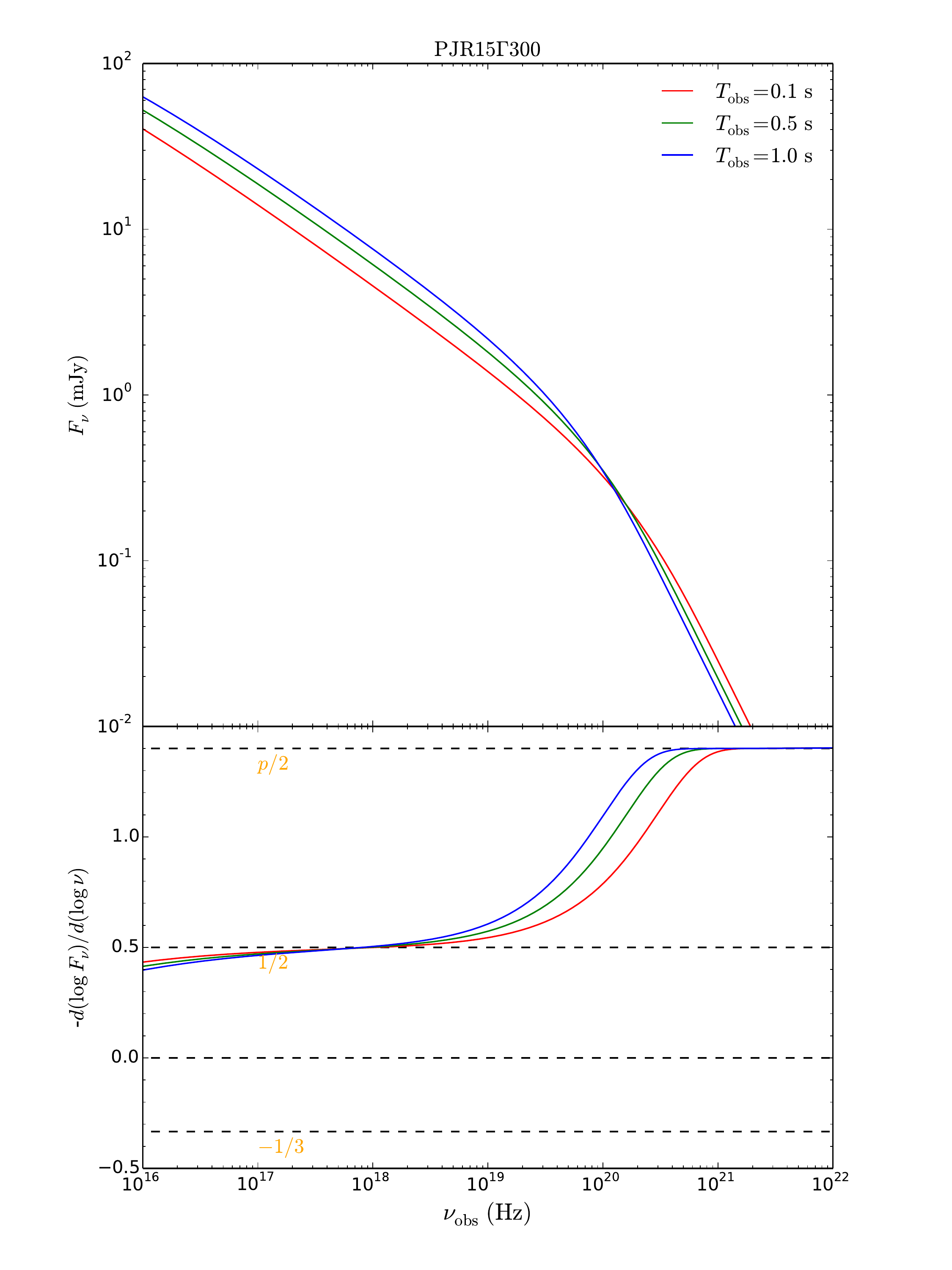}} \\
    \subfloat{\includegraphics[width=0.3\paperwidth]{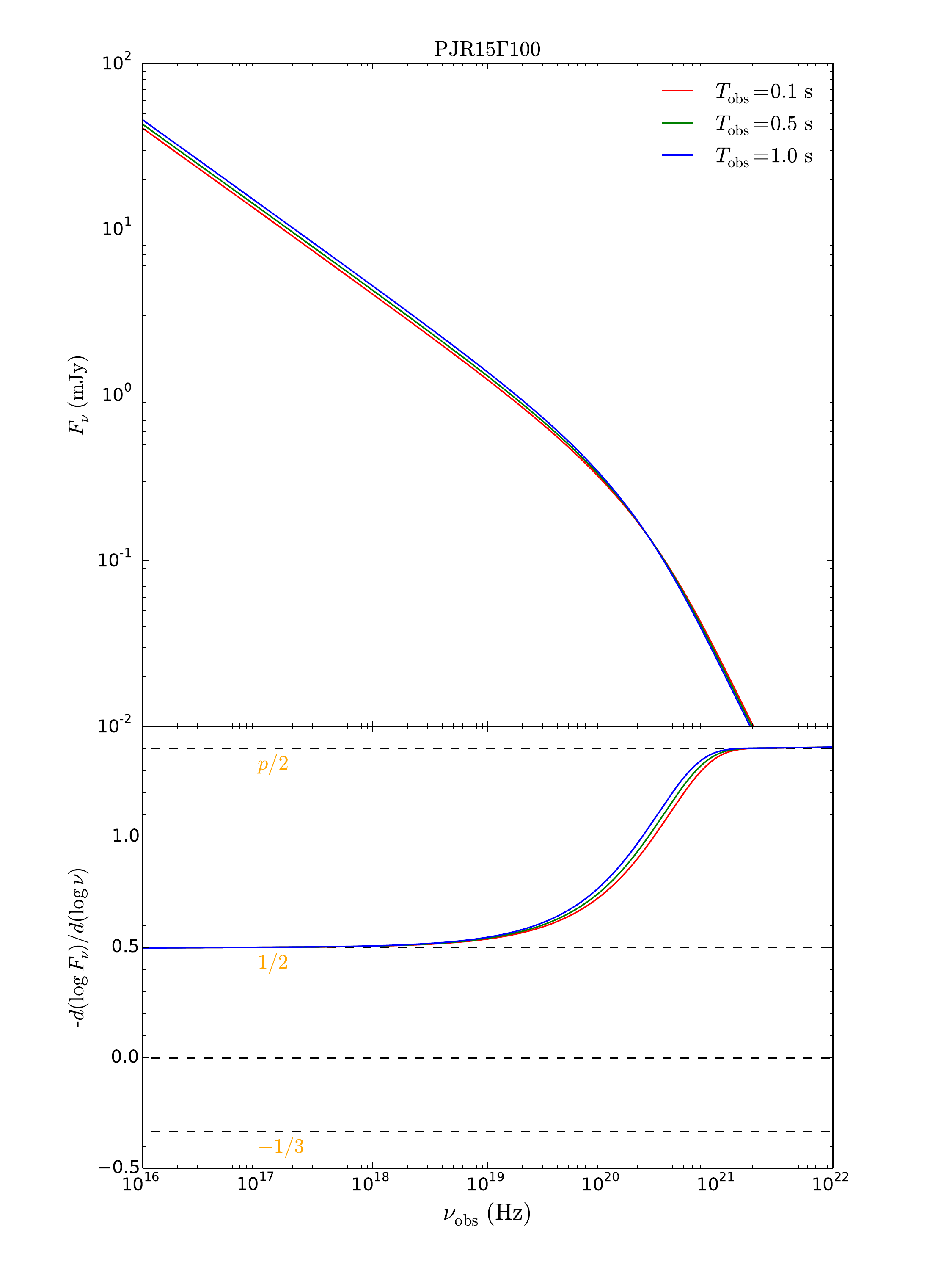}}
    \subfloat{\includegraphics[width=0.3\paperwidth]{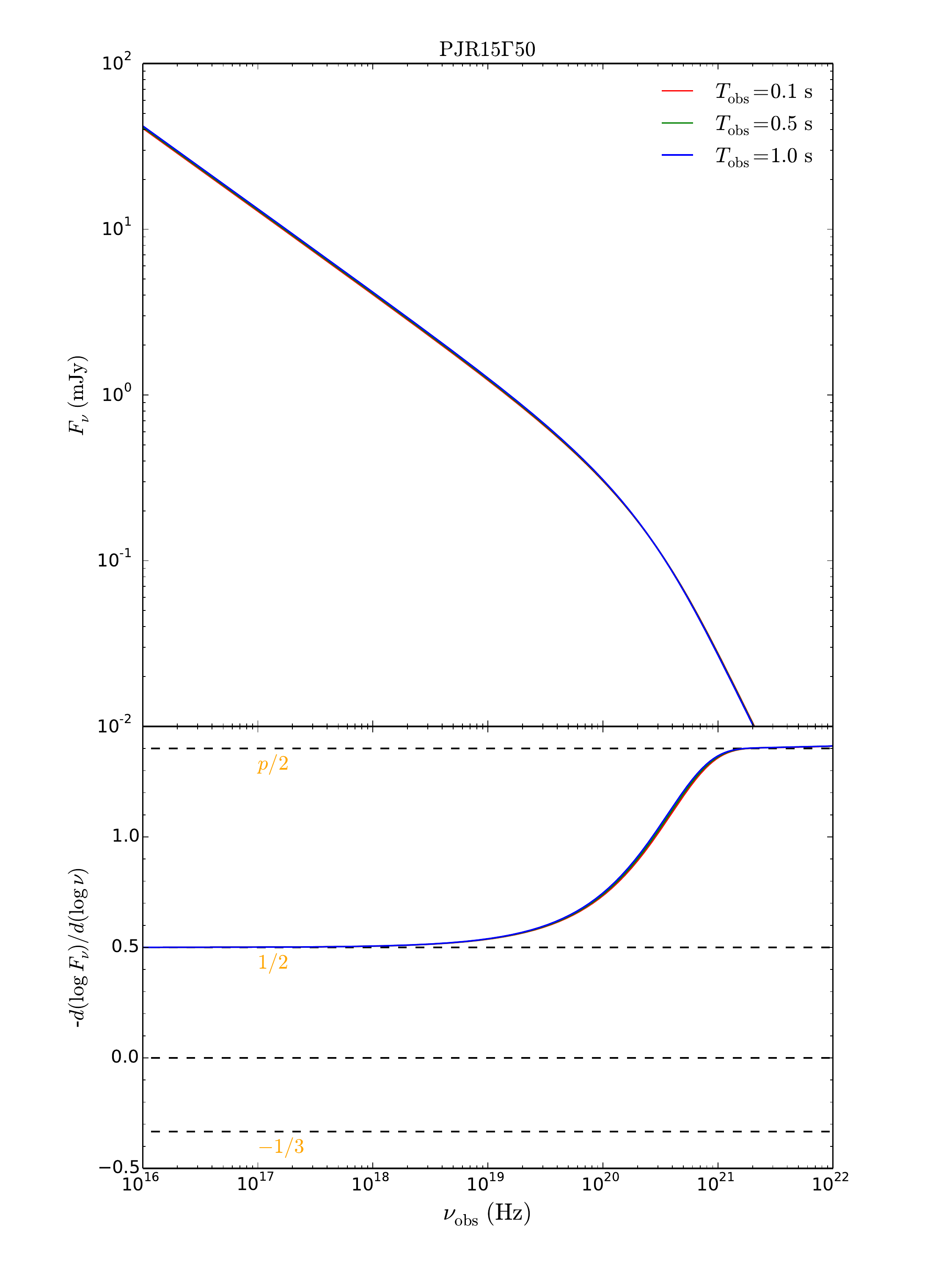}}
    \caption{The corresponding synchrotron flux-density spectra $F_{\nu}$ from the electrons with the energy distribution presented in Figure \ref{fig:MI-electron}.
    \label{fig:MI-spectra}}
\end{adjustwidth}
\end{figure}

\begin{figure}
\begin{adjustwidth}{-2cm}{-2cm}
\centering
    \subfloat{\includegraphics[width=0.3\paperwidth]{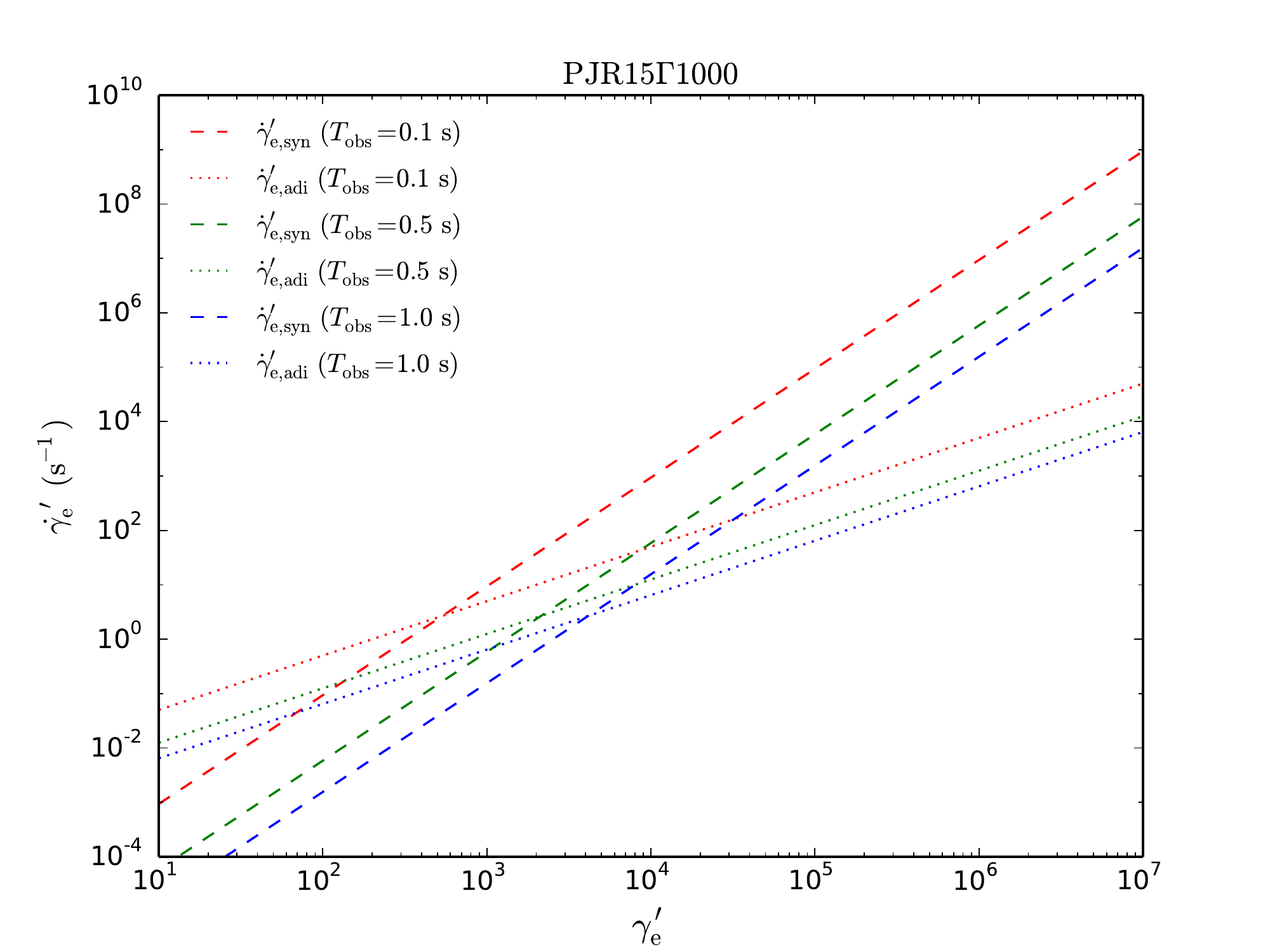}}
    \subfloat{\includegraphics[width=0.3\paperwidth]{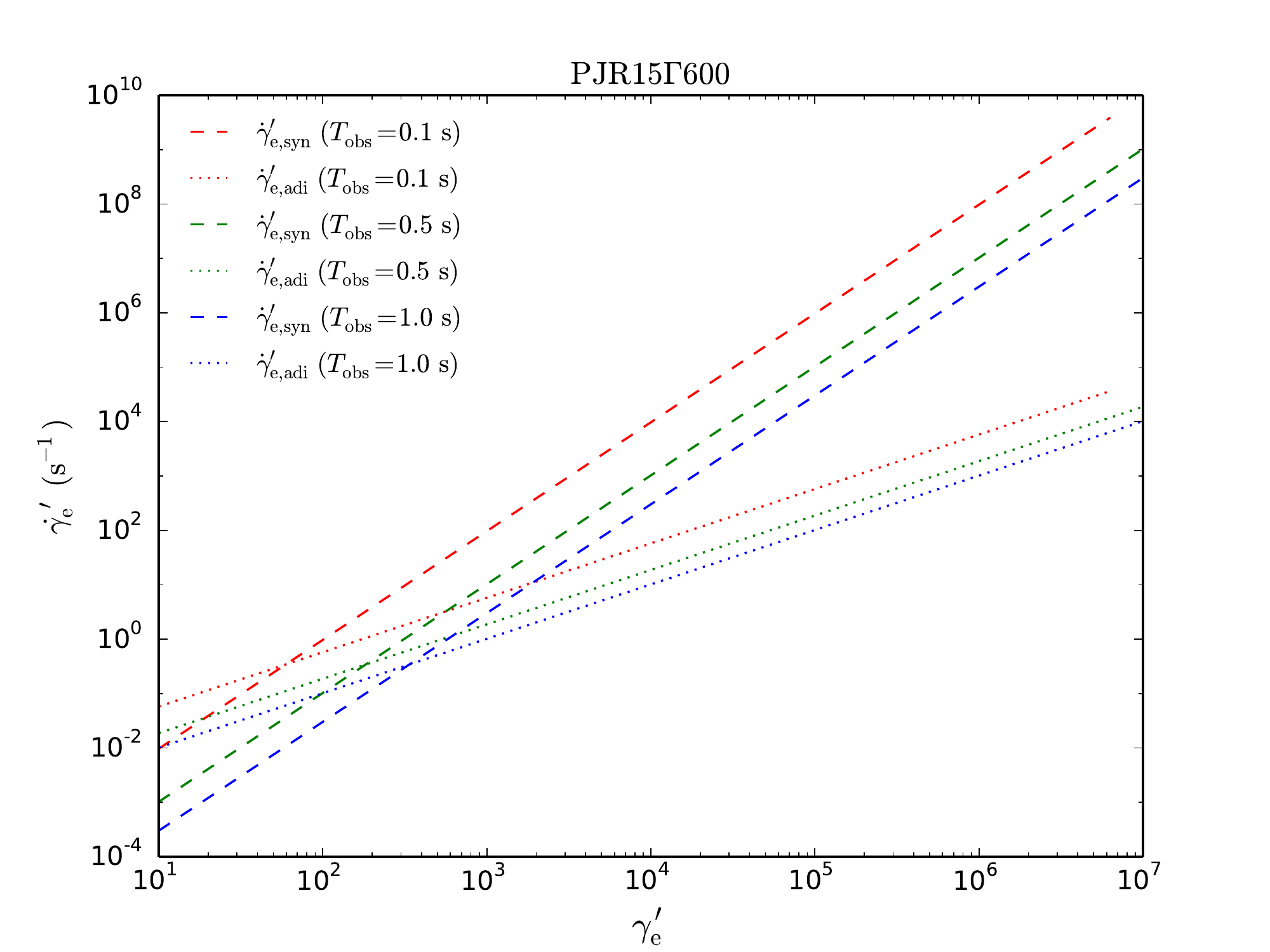}}
    \subfloat{\includegraphics[width=0.3\paperwidth]{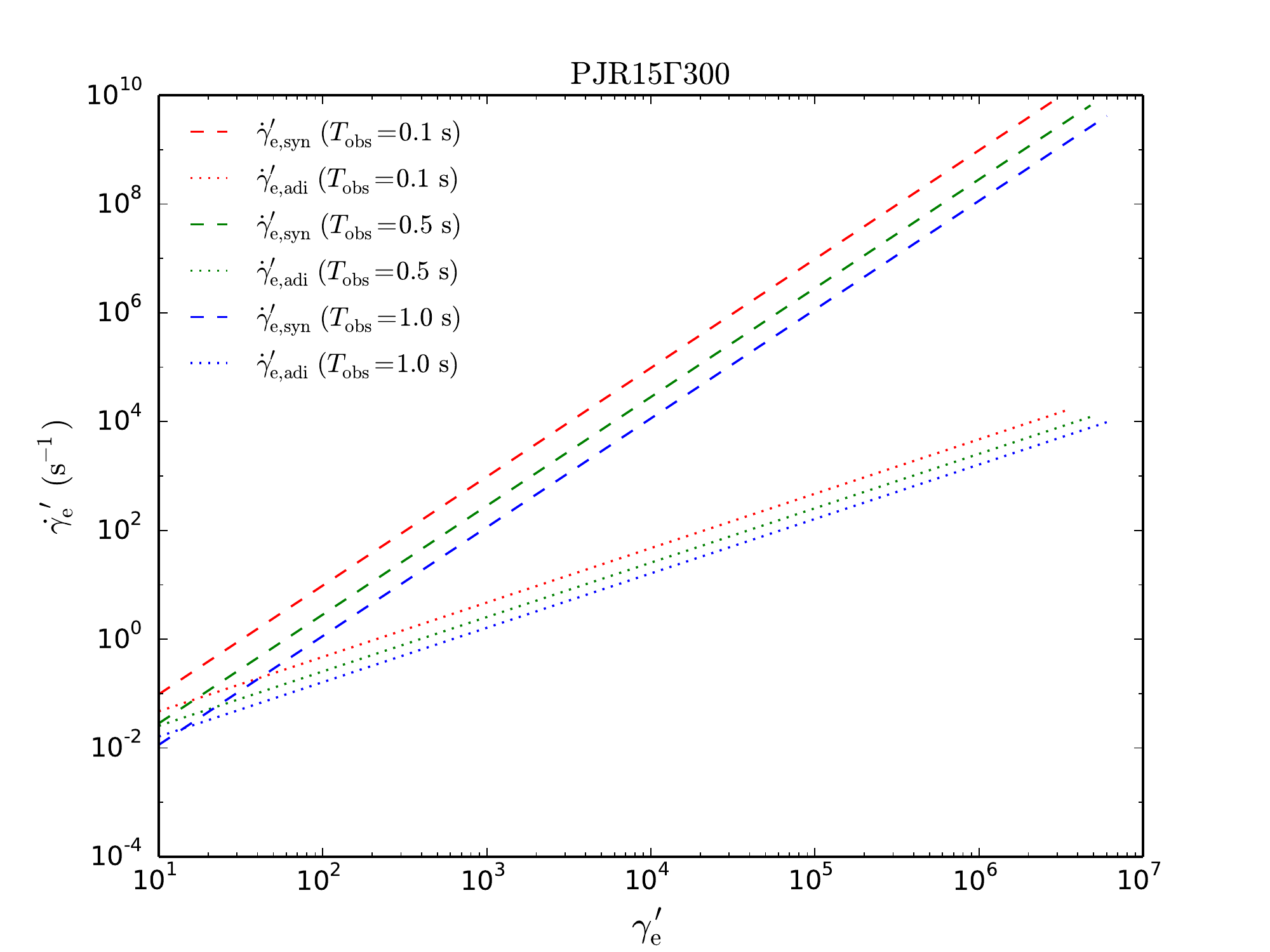}} \\
    \subfloat{\includegraphics[width=0.3\paperwidth]{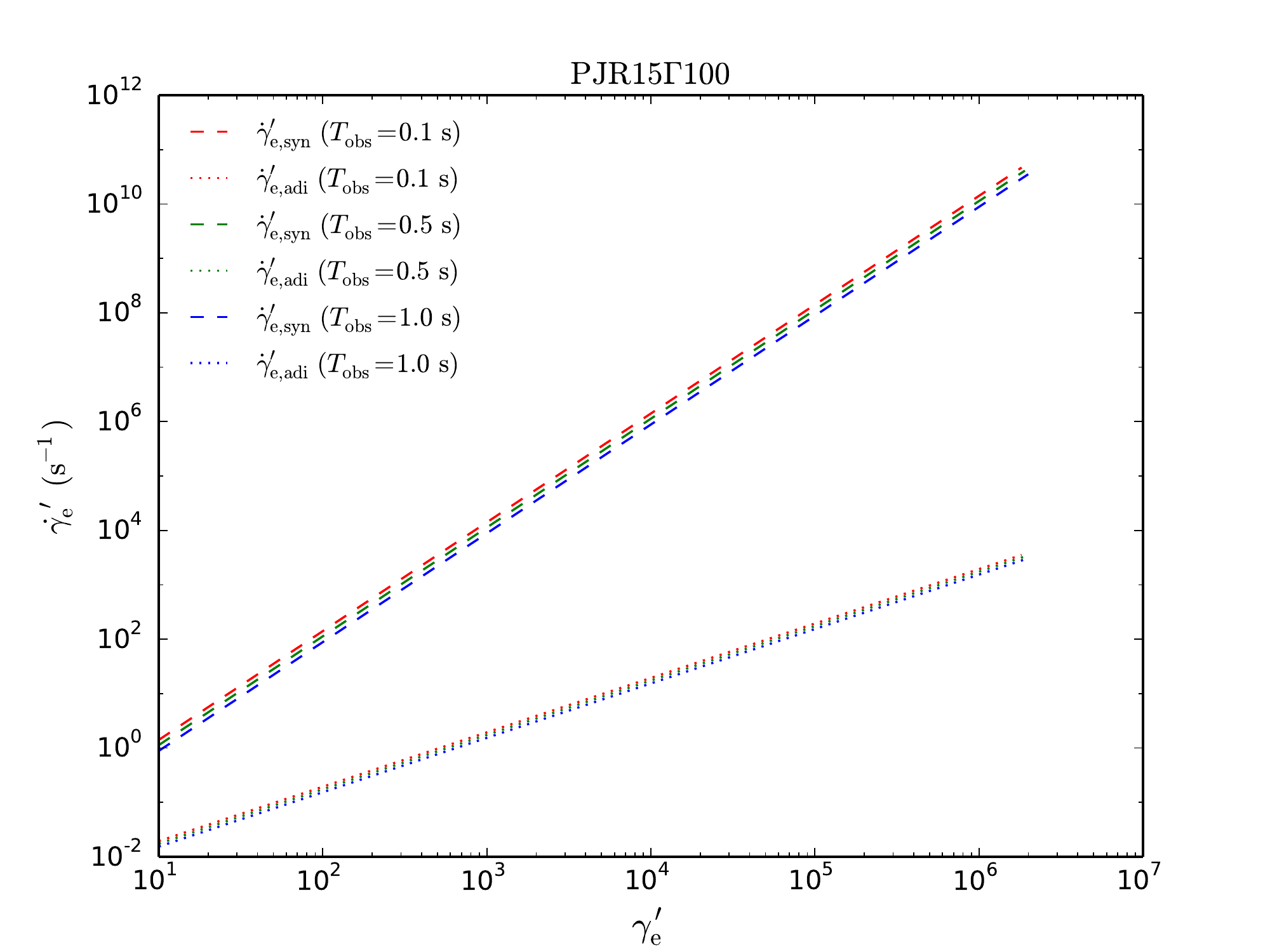}}
    \subfloat{\includegraphics[width=0.3\paperwidth]{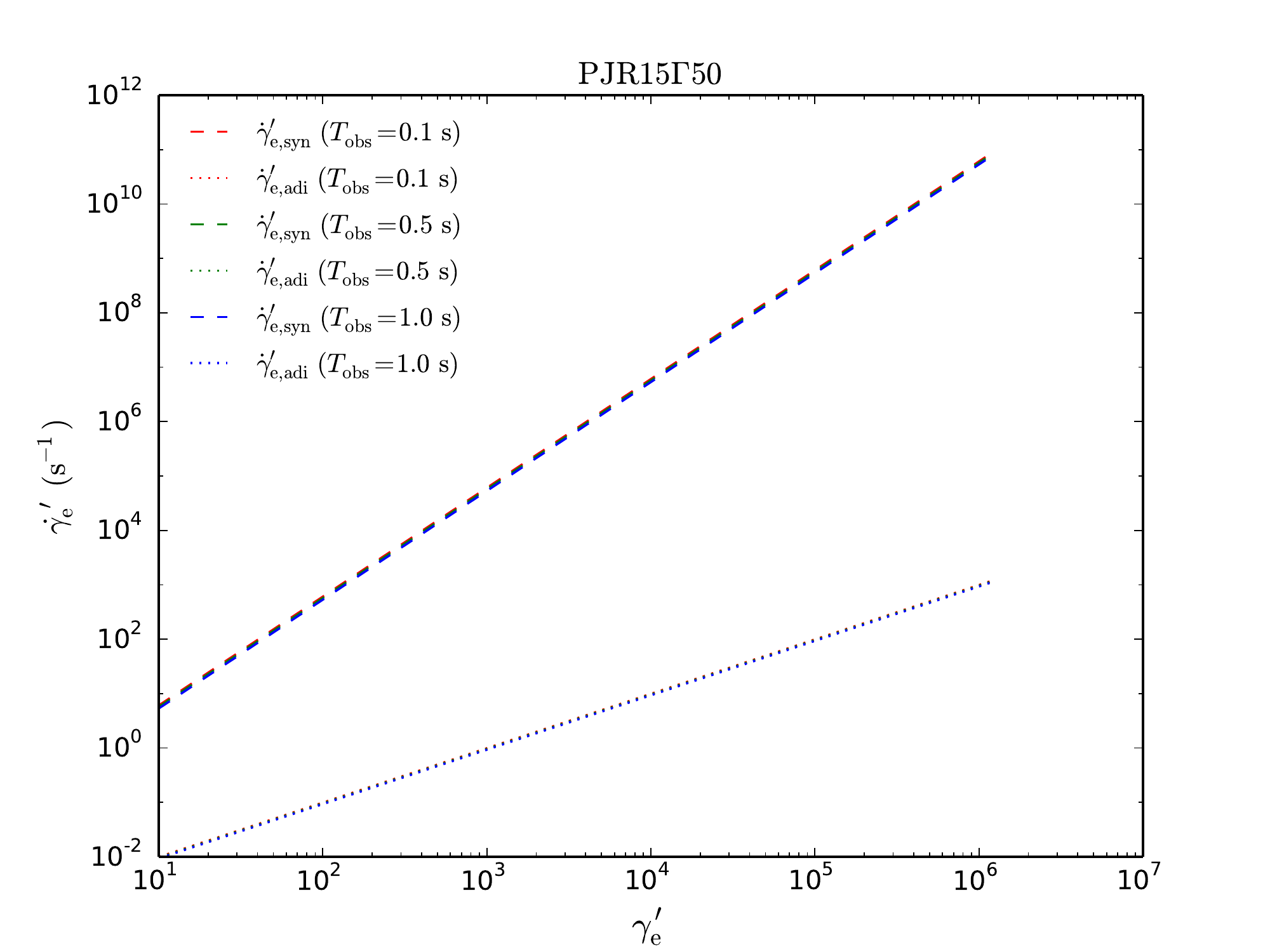}}
    \caption{The co-moving cooling rates of different cooling mechanisms for the electrons with the energy distribution presented in Figure \ref{fig:MI-electron}.
    \label{fig:MI-rate}}
\end{adjustwidth}
\end{figure}

From the results of PJR15$\Gamma$1000,
we notice that the electron cooling is dominated by the synchrotron radiation for $\gamma_{\rm e}^{\prime} > 10^2$,
whereas the adiabatic expansion dominates the cooling of electrons with $\gamma_{\rm e}^{\prime} < 10^2$ (see Figure \ref{fig:MI1SSC}).
It indicates that SSC cooling is not significant in this case,
which also holds for other cases with even greater $B_0^{\prime}$ in Group PJR15.
On the other hand, calculating the SSC cooling rates is extremely time-consuming since it is difficult to achieve
the integral convergence in our code. So, we do not include the SSC cooling effect in this group of calculations.
The corresponding results for electron distributions, flux spectra, and cooling rates are shown in
Figures \ref{fig:MI-electron}, \ref{fig:MI-spectra} and \ref{fig:MI-rate}, respectively.
In the results of PJR15$\Gamma$1000 and PJR15$\Gamma$600, it is seen that along with the decaying of $B^{\prime}$, adiabatic cooling
becomes more and more dominant for low-energy electrons since $\dot{\gamma}_{\rm e,syn}^{\prime} \propto
B^{\prime 2} \propto R^{-2}$ and $\dot{\gamma}_{\rm e,adi}^{\prime} \propto R^{-1}$.
The decreasing of $\dot{\gamma}_{\rm e,adi}^{\prime}$ along $R$ leads to an electron spectrum
that is harder than $-1$, just like what is done by the decreasing $\dot{\gamma}_{\rm e,syn}^{\prime}$ (see results of M2).
The low-energy indices of the flux density spectra in PJR15$\Gamma$1000 ($\alpha$, $F_{\nu} \propto \nu^{\alpha}$) can be larger than 0
and $\alpha$ in PJR15$\Gamma$600 are within [-0.5,0].
In results of PJR15$\Gamma$300, PJR15$\Gamma$100 and PJR15$\Gamma$50, the cooling of all electrons is dominated by the synchrotron radiation only. In Figure \ref{fig:MI-electron},
the decaying of $B^{\prime}$ leads the electron spectra to become harder than -2 as expected in PJR15$\Gamma$300.
However, this does not occur in PJR15$\Gamma$100 or PJR15$\Gamma$50.
Comparing with the results of \cite{Uhm14}, the electron spectrum seems to become harder
when the electrons are cooled at a decreasing cooling rate.
A history/experience of decreasing cooling rate for an electron is the key factor to result in a hard spectrum.
For relatively large $B^{\prime}$ and $R$, the cooling timescale of electrons is significantly smaller than
the dynamical timescale. As a consequence, the history of the decreasing cooling rate of an electron
is too short to take effect.

\begin{figure}
\centering
    \subfloat{\includegraphics[width=0.45\linewidth]{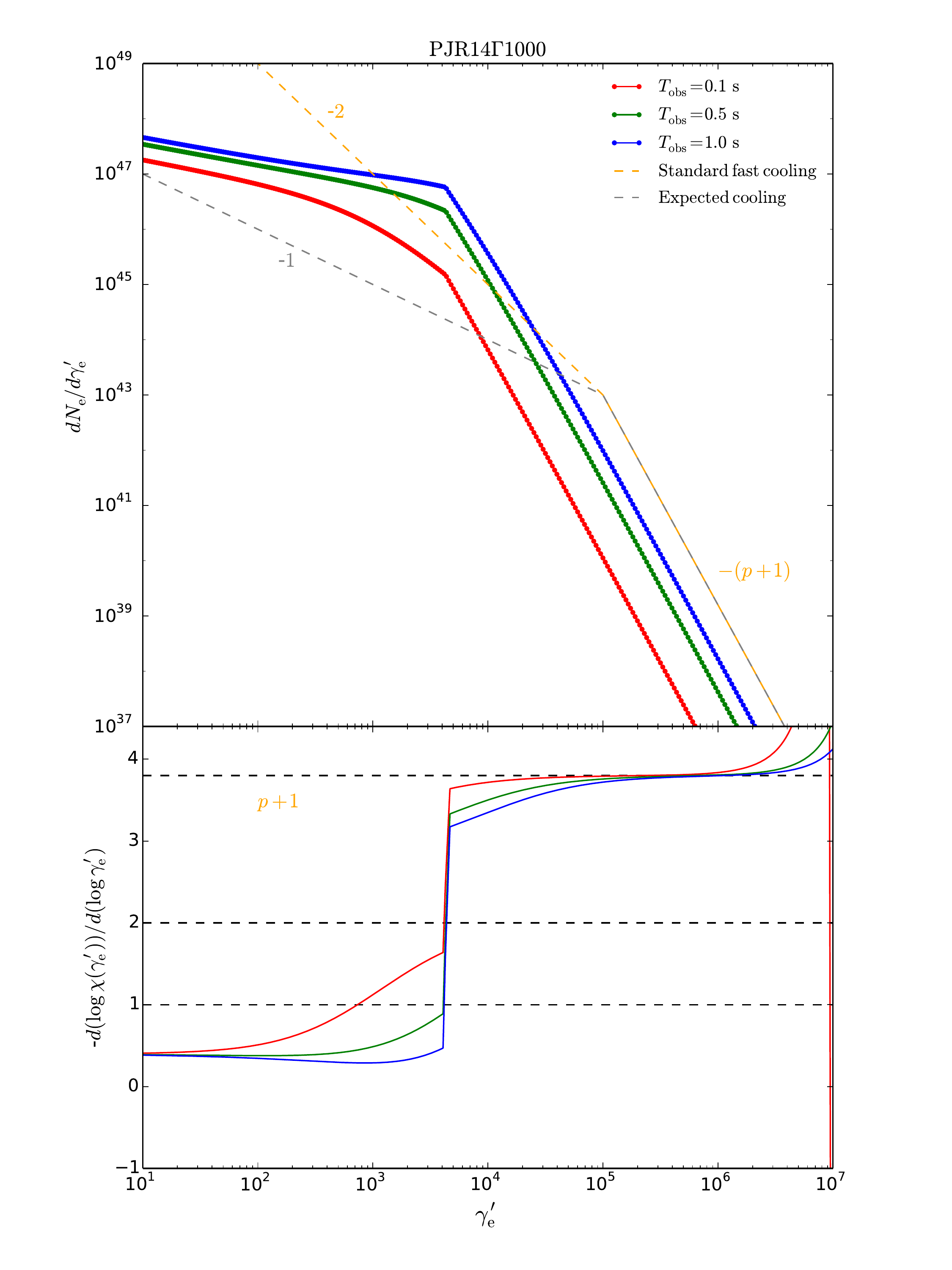}}
    \subfloat{\includegraphics[width=0.45\linewidth]{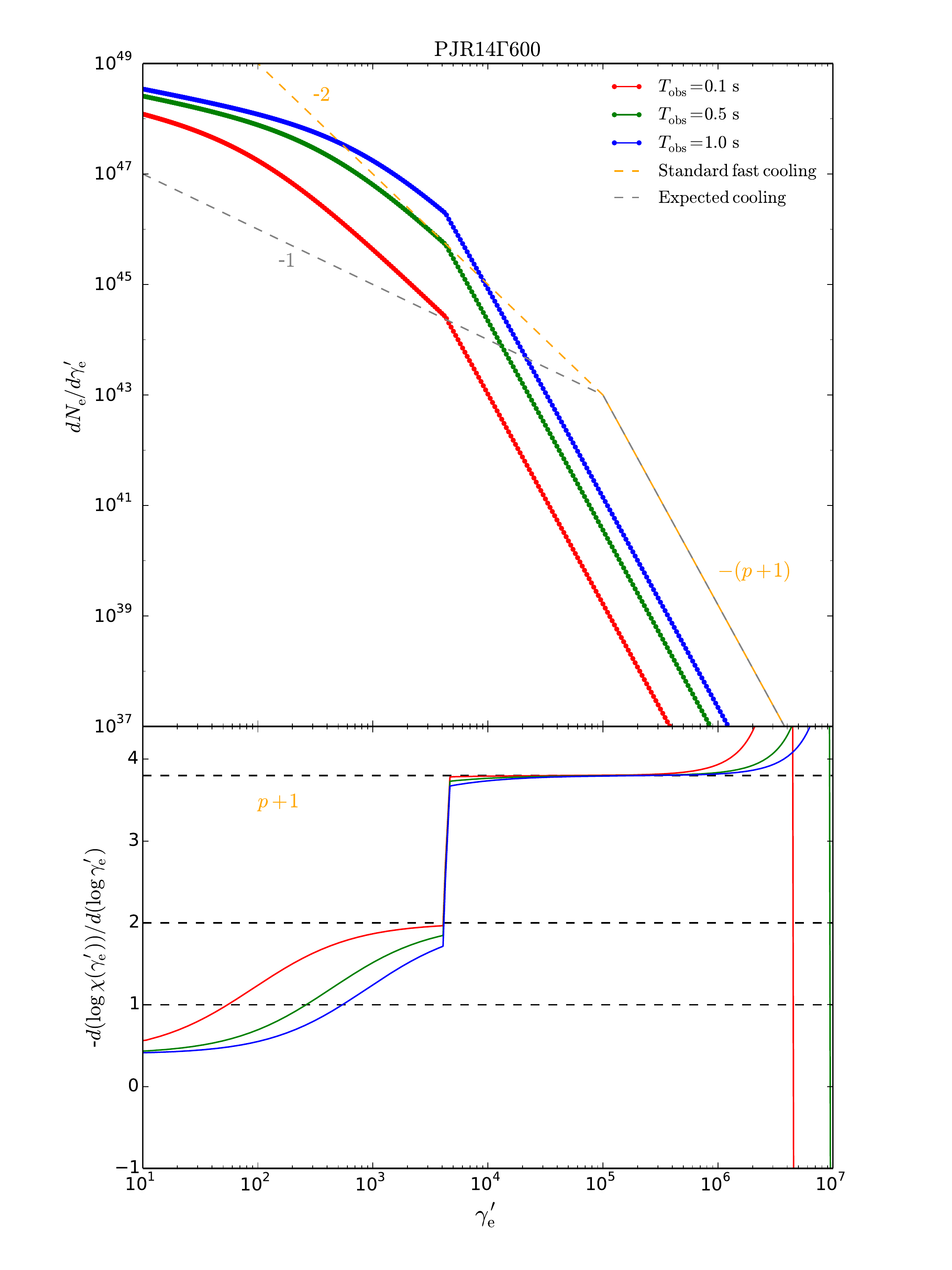}} \\
    \subfloat{\includegraphics[width=0.45\linewidth]{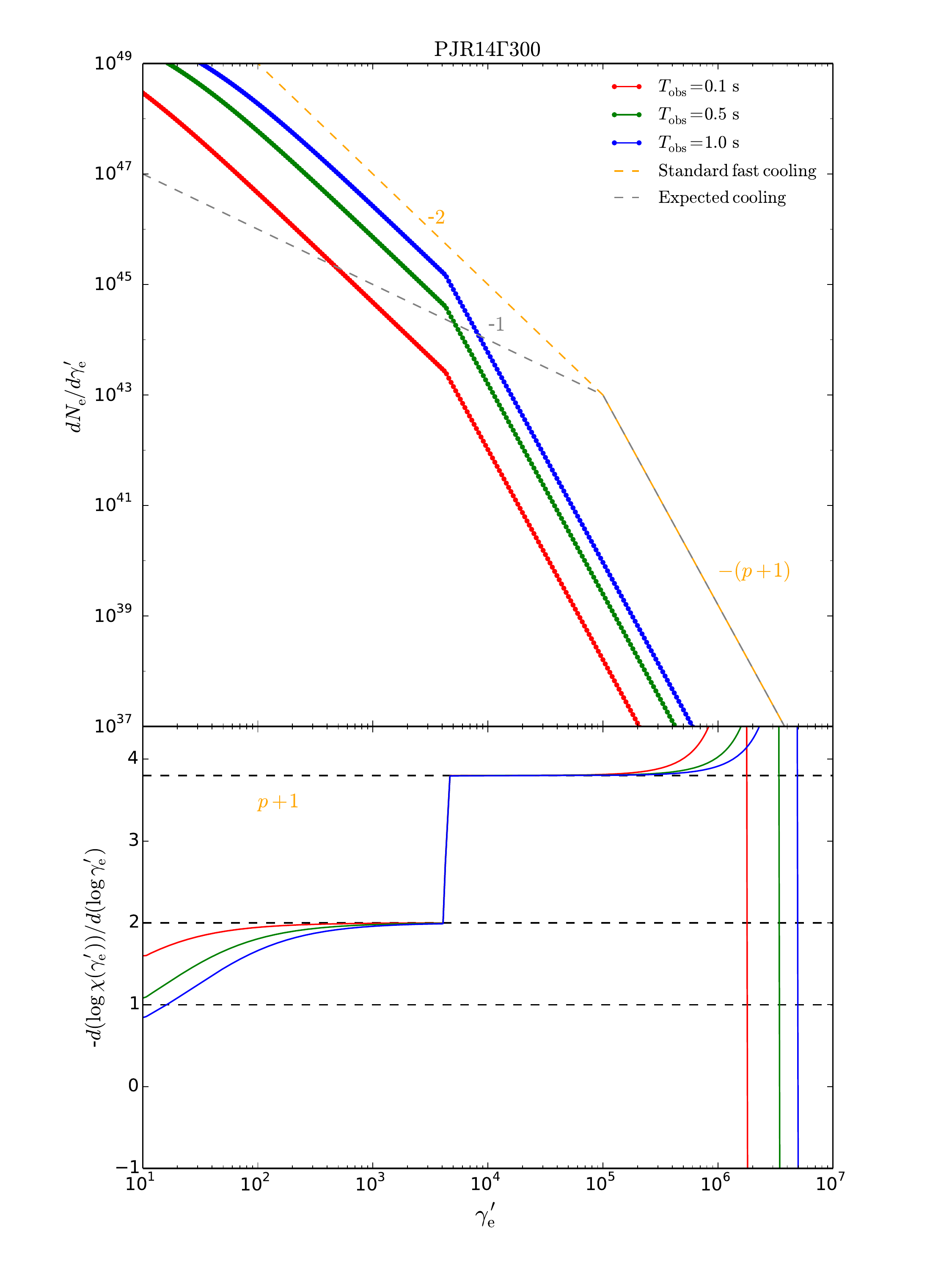}}
    \subfloat{\includegraphics[width=0.45\linewidth]{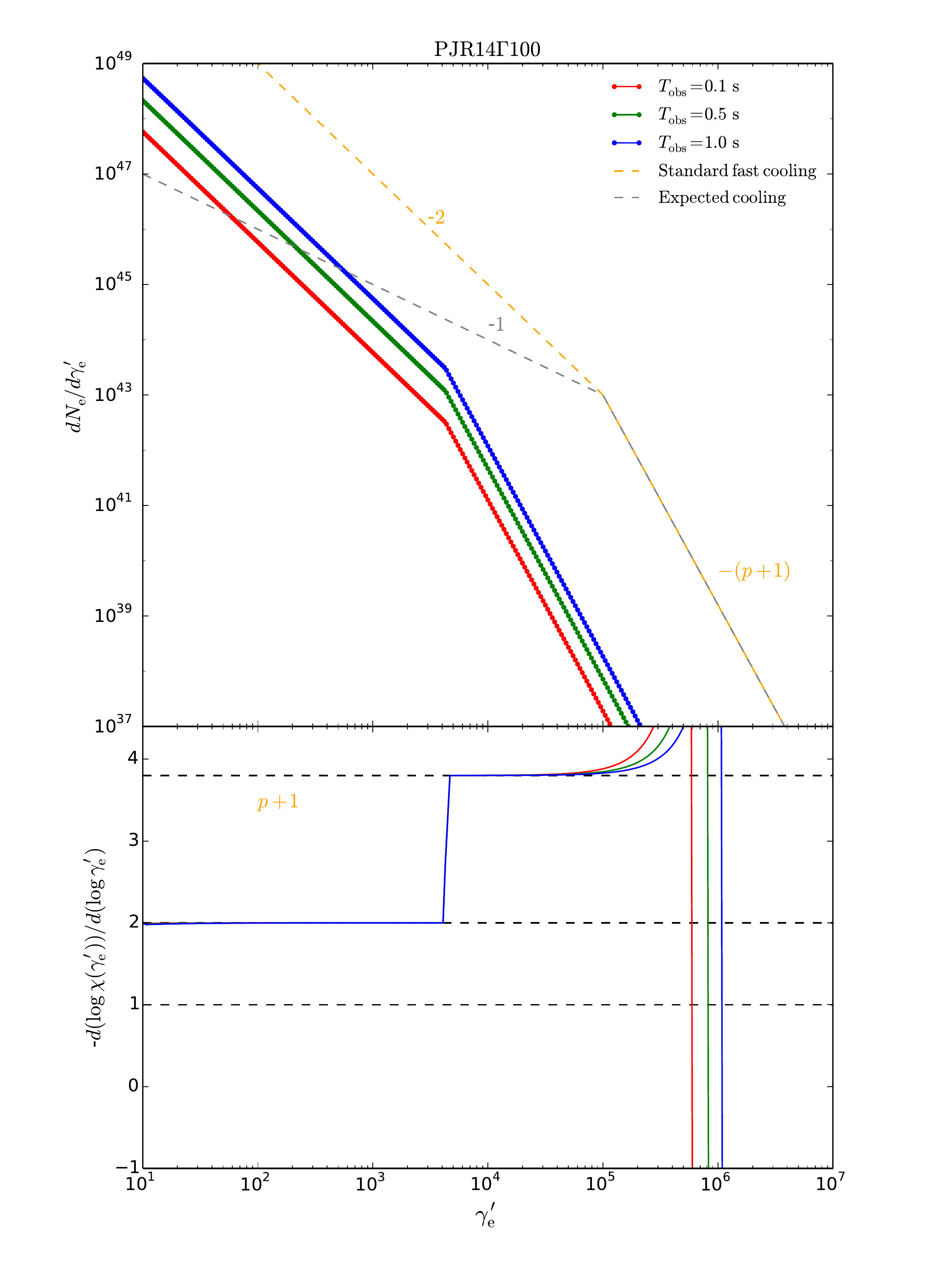}}
    \caption{The evolution of the electron energy spectrum for the four cases in calculations of Group PJR14 (see Table \ref{TABLE:II}).}
    \label{fig:MII-electron}
\end{figure}

\begin{figure}
\centering
    \subfloat{\includegraphics[width=0.45\linewidth]{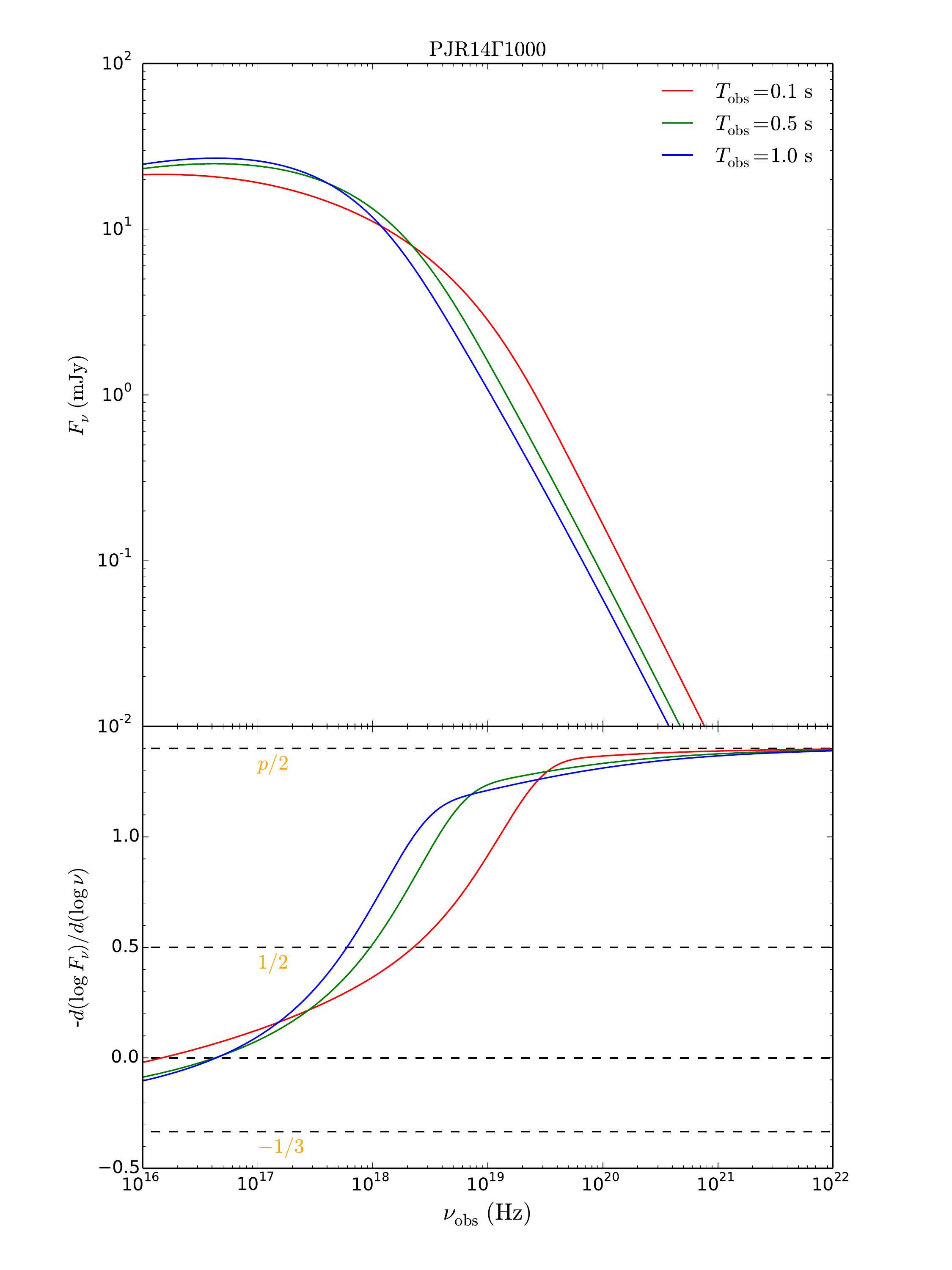}}
    \subfloat{\includegraphics[width=0.45\linewidth]{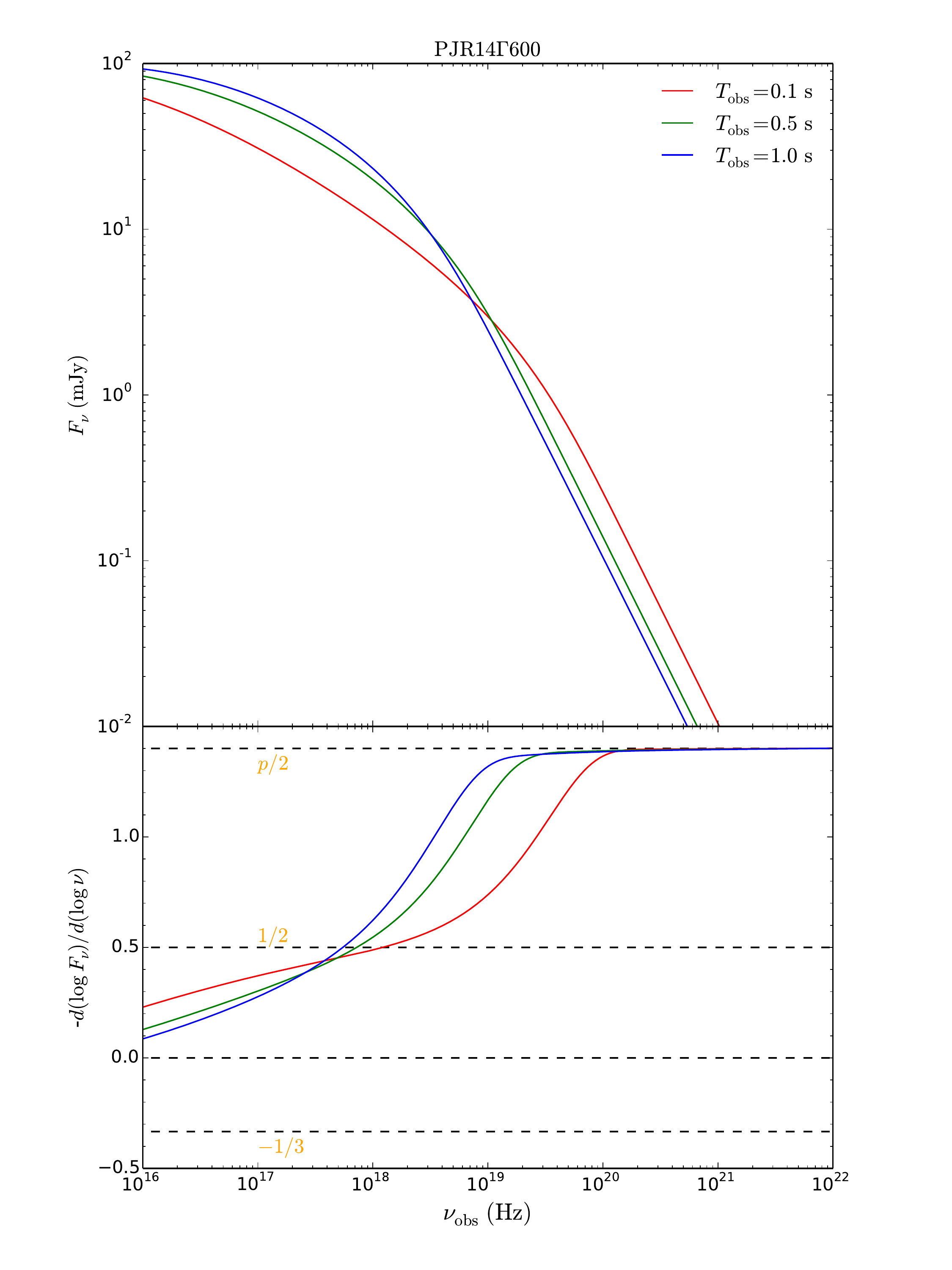}} \\
    \subfloat{\includegraphics[width=0.45\linewidth]{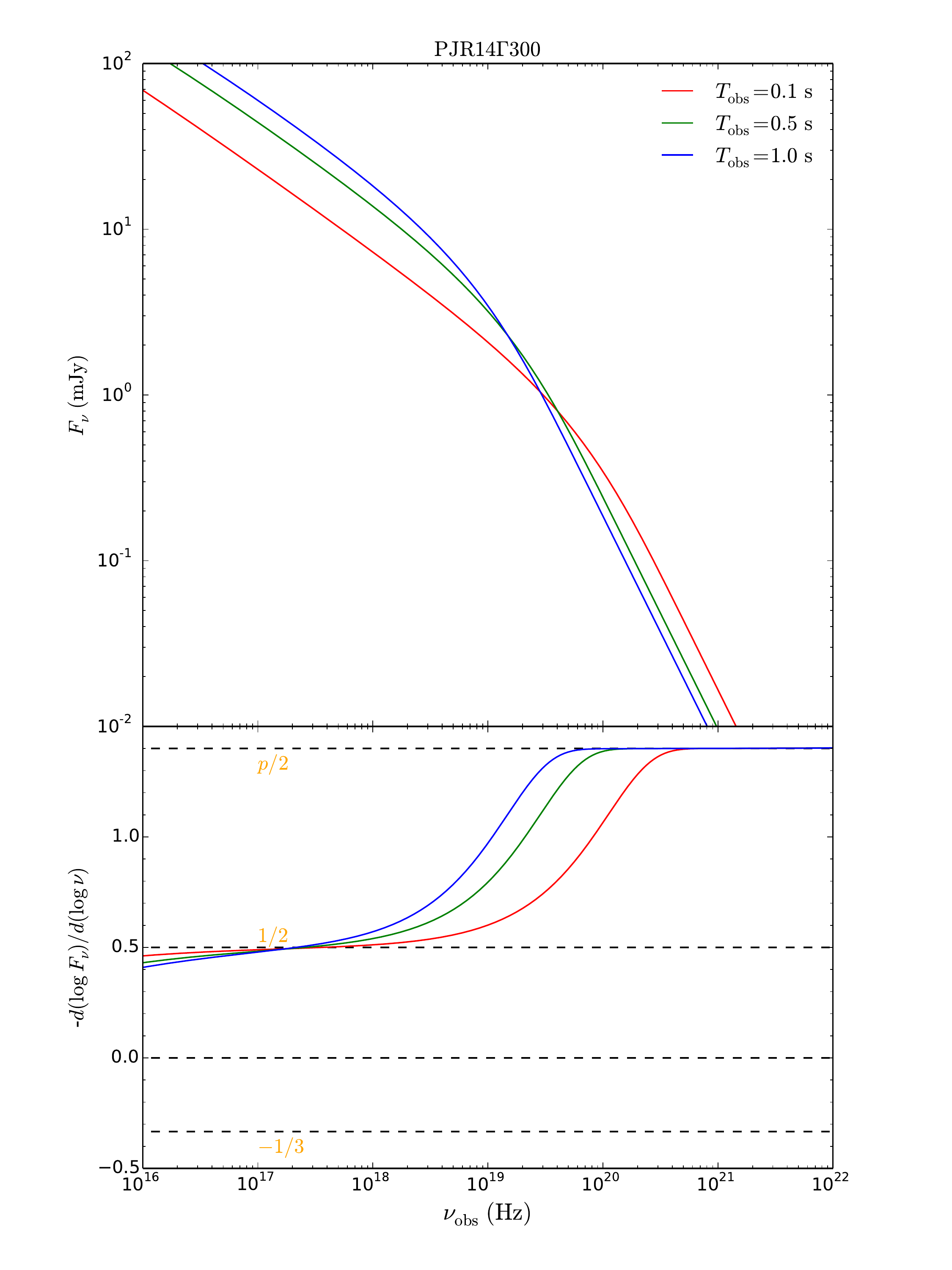}}
    \subfloat{\includegraphics[width=0.45\linewidth]{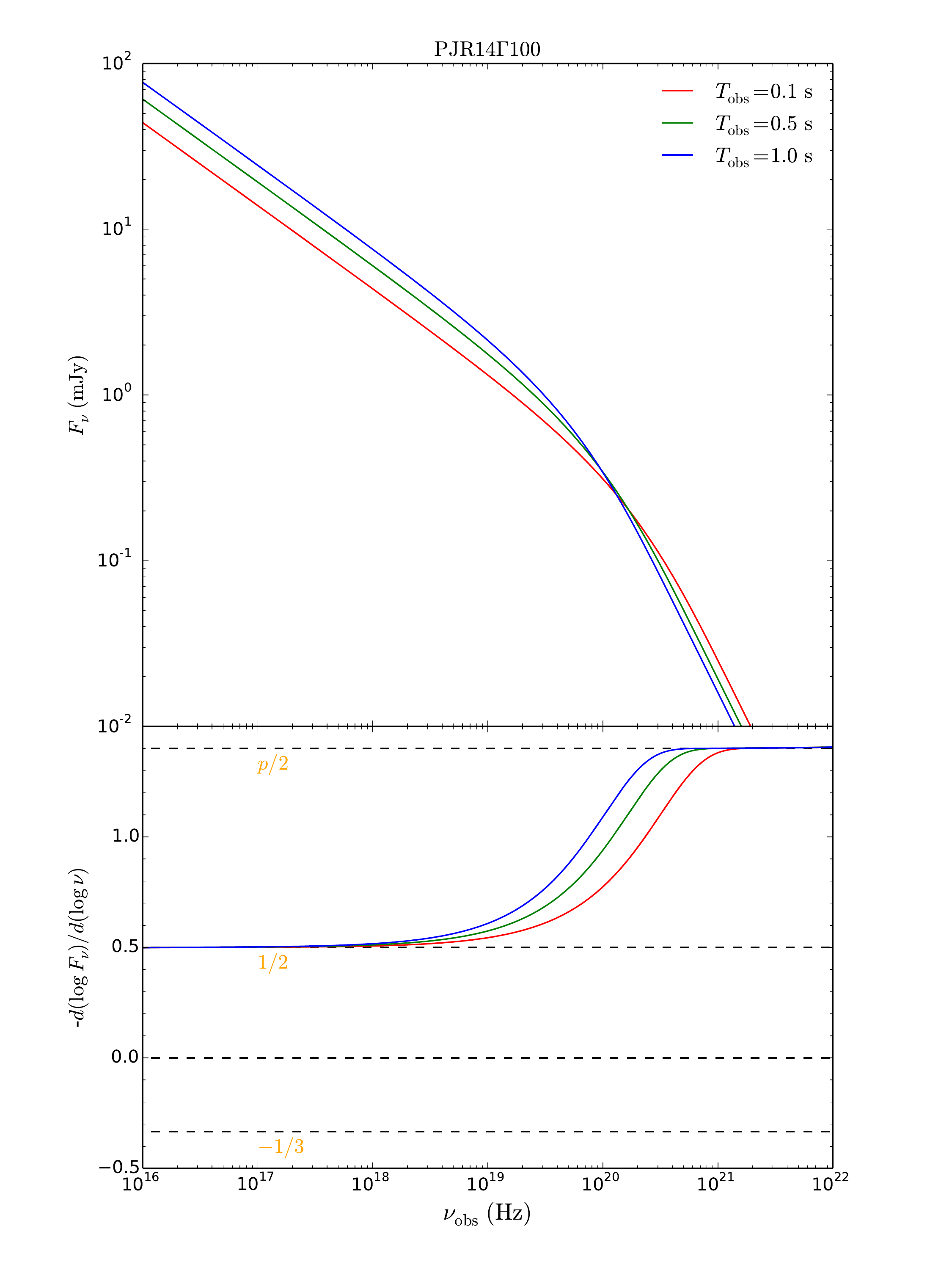}}
    \caption{The corresponding synchrotron flux-density spectra $F_{\nu}$ from the electrons with the energy distribution presented in Figure \ref{fig:MII-electron}.}
    \label{fig:MII-spectra}
\end{figure}

\begin{figure}
\centering
    \subfloat{\includegraphics[width=0.45\linewidth]{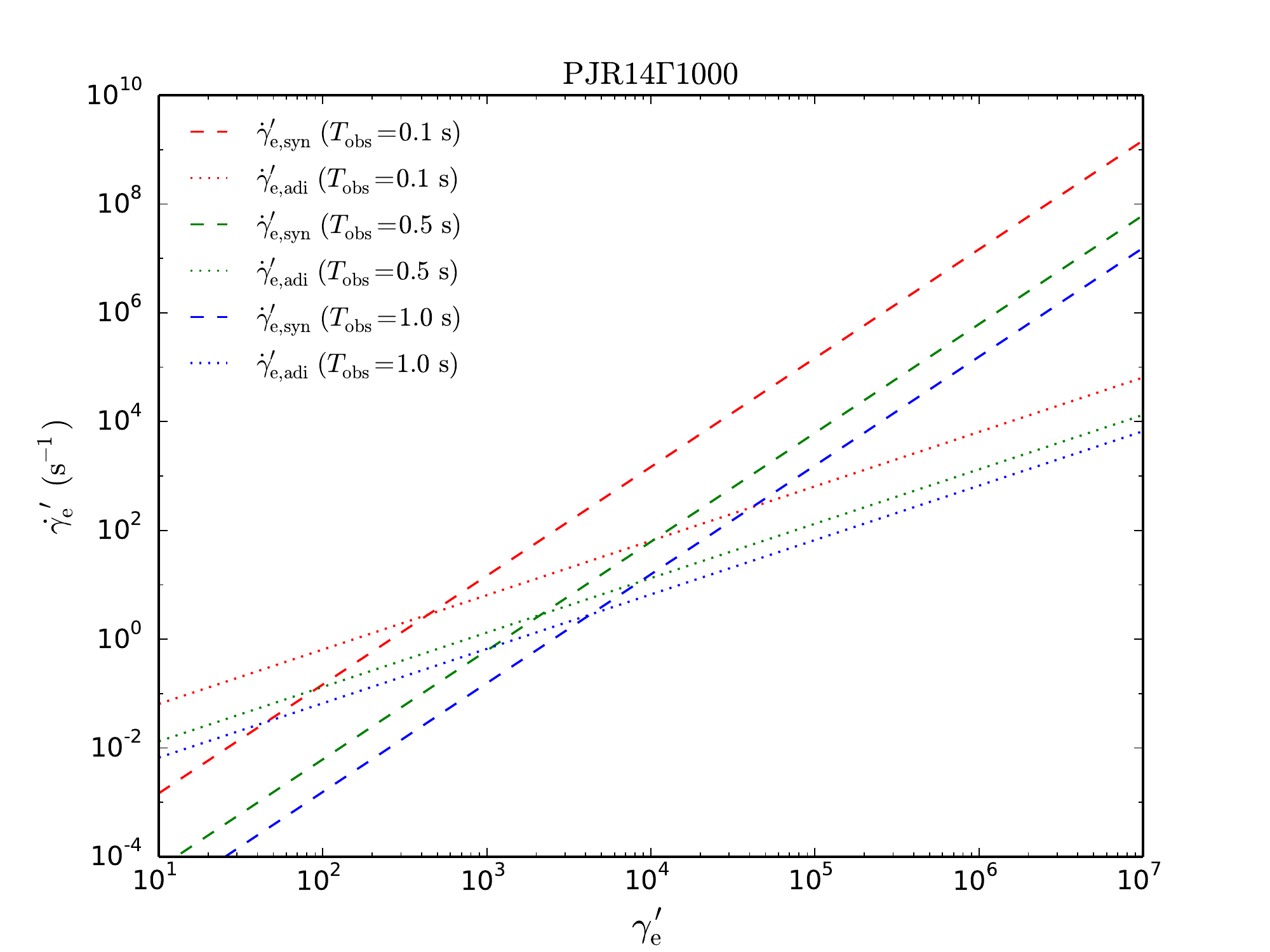}}
    \subfloat{\includegraphics[width=0.45\linewidth]{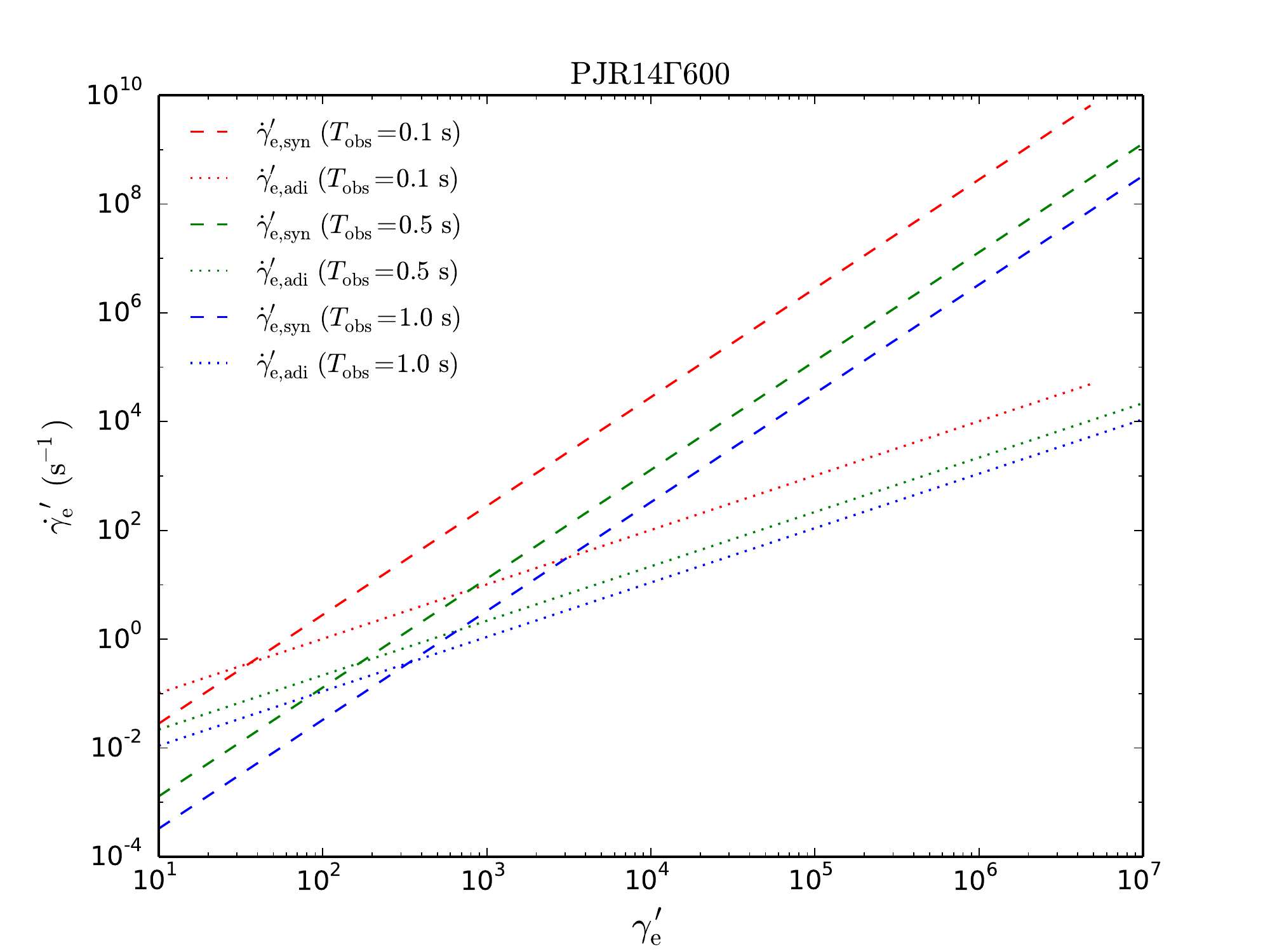}} \\
    \subfloat{\includegraphics[width=0.45\linewidth]{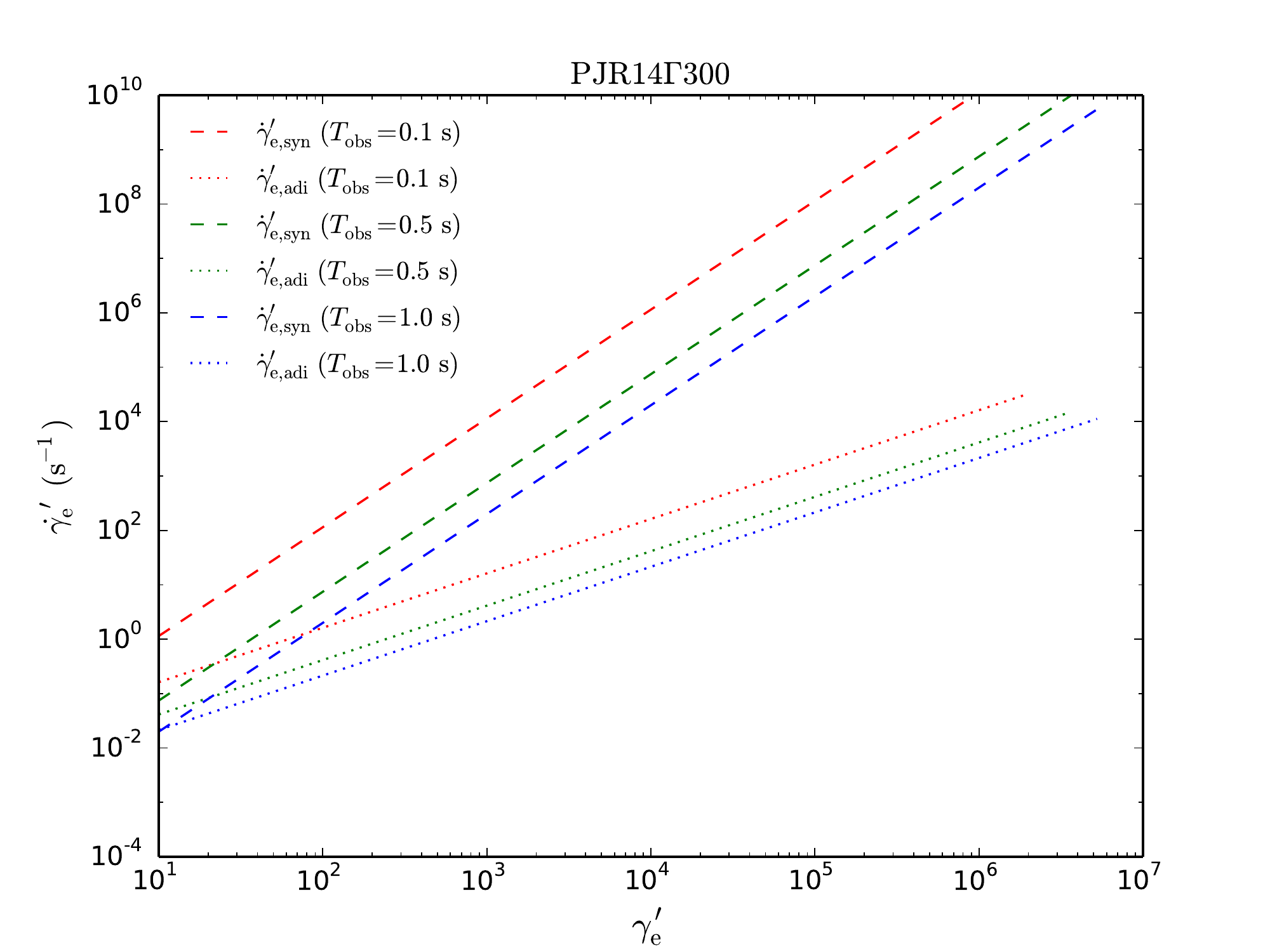}}
    \subfloat{\includegraphics[width=0.45\linewidth]{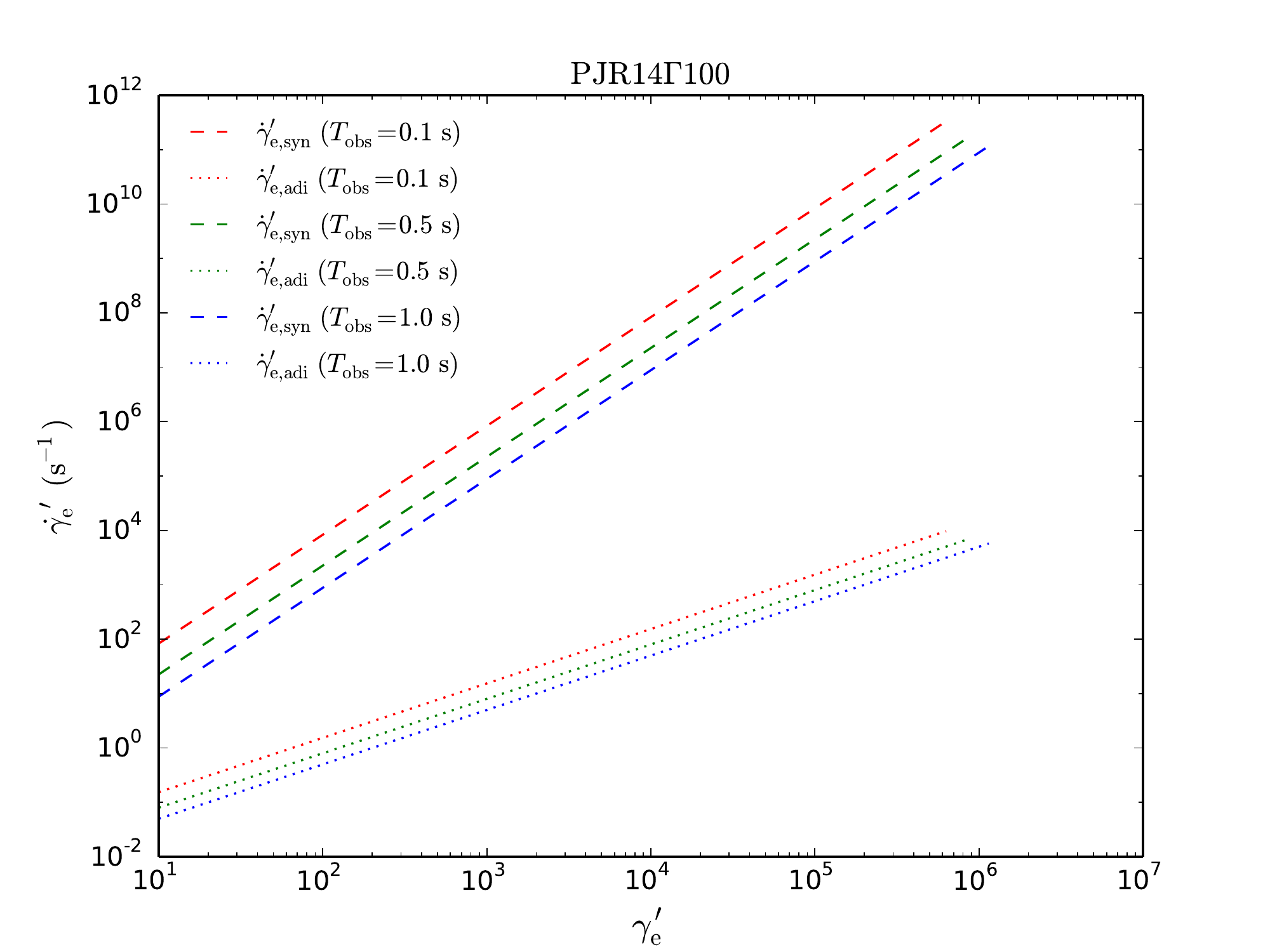}}
    \caption{The co-moving cooling rates of different cooling mechanisms for the electrons with the energy distribution
    presented in Figure \ref{fig:MII-electron}.}
    \label{fig:MII-rate}
\end{figure}

In calculations of Group PJR14, the results for electron distributions, flux spectra, and cooling rates are shown in
Figures \ref{fig:MII-electron}, \ref{fig:MII-spectra} and \ref{fig:MII-rate}, respectively.
In the results of PJR14$\Gamma$1000, it can be seen that adiabatic cooling is dominant for low-energy electrons
due to the relatively large $\Gamma$ and small $B^{\prime}$, $R_0$.
The indices of the low-energy electron spectra in PJR14$\Gamma$1000 are harder than $-1$ and
$\alpha$ also becomes harder than 0 at frequency $\sim 1~\mathrm{keV}$ when $t_{\rm obs} > 0.5$~s.
In PJR14$\Gamma$600, synchrotron cooling is dominant at early times,
while adiabatic cooling becomes dominant at late stages.
The low-energy electron spectra and the flux spectra are harder than the standard ones,
but not so hard as those in PJR14$\Gamma$1000.
In the results of PJR14$\Gamma$300 and PJR14$\Gamma$100, we can find that synchrotron cooling is always dominant
and the low-energy electron spectra and the flux spectra are just similar to the standard ones.
Here again, in PJR14$\Gamma$300 and PJR14$\Gamma$100, the cooling timescale of an electron under relatively large $B^{\prime}$
is significantly smaller than the dynamical timescale,
and the mechanism of ``decreasing synchrotron cooling rate'' cannot work to make the electron spectra significantly harder than $-2$.

For Poynting-flux-dominated jets considered here, we confirm that the decreasing synchrotron cooling rate (decreasing $B^{\prime}$)
will lead to hard electron spectra and flux spectra, according to the results of our calculation Groups PJR15 and PJR14.
The $\gamma_{\rm m}^{\prime}$ used in PJR15 and PJR14 are one or two orders smaller than those used in \citep{Uhm14},
to meet the physical condition of $L_{B} \ge L_{\rm e}$.
Although $\gamma_{\rm m}^{\prime}$ used here is only the upper limit,
the results from Calculations PJR15 and PJR14 should be representative for three kinds of cooling patterns in this scenario,
i.e., hard electron spectra caused by decreasing $\dot{\gamma}_{\rm e,adi}^{\prime}$,
or decreasing $\dot{\gamma}_{\rm e,syn}^{\prime}$, and normal spectra under large $B^{\prime}$.
Moreover, we notice that $B^{\prime}$ should not be too large in this scenario,
otherwise the synchrotron cooling timescale would be much shorter than the dynamical timescale
and the effect of decaying $B^{\prime}$ is weakened.
This is also the reason that we did not explore the regime of $L_{B} \gg L_{\rm e}$ (when SSC cooling is not important).
In other words, if we want to use the decreasing $\dot{\gamma}_{\rm e,syn}^{\prime}$ to work for hard electron spectra,
there should be a lower limit $\zeta$ for the extent of fast cooling, i.e., $\zeta \le t_{\rm c} / t_{\rm d} < 1$.
The ``unsuccessful'' results from PJR15$\Gamma$100--PJR15$\Gamma$50 and PJR14$\Gamma$300--PJR14$\Gamma$100 together indicate that
$\zeta$ should be larger than $10^{-5}$.
Since $B^{\prime} \propto t_{\rm c}^{-1/2}$, the proper range of $B^{\prime}$ for decaying $\dot{\gamma}_{\rm e,syn}^{\prime}$
to work is likely to be within roughly 2 orders of magnitude.
If this mechanism is true for GRB spectral hardening,
the narrow range of $B^{\prime}$ indicates a potential way to probe the magnetic strength
of the GRB jet using its spectral characteristics.

\subsubsection{The Internal Shock Scenario}
\begin{figure}
   \centering
   \includegraphics[scale=0.5]{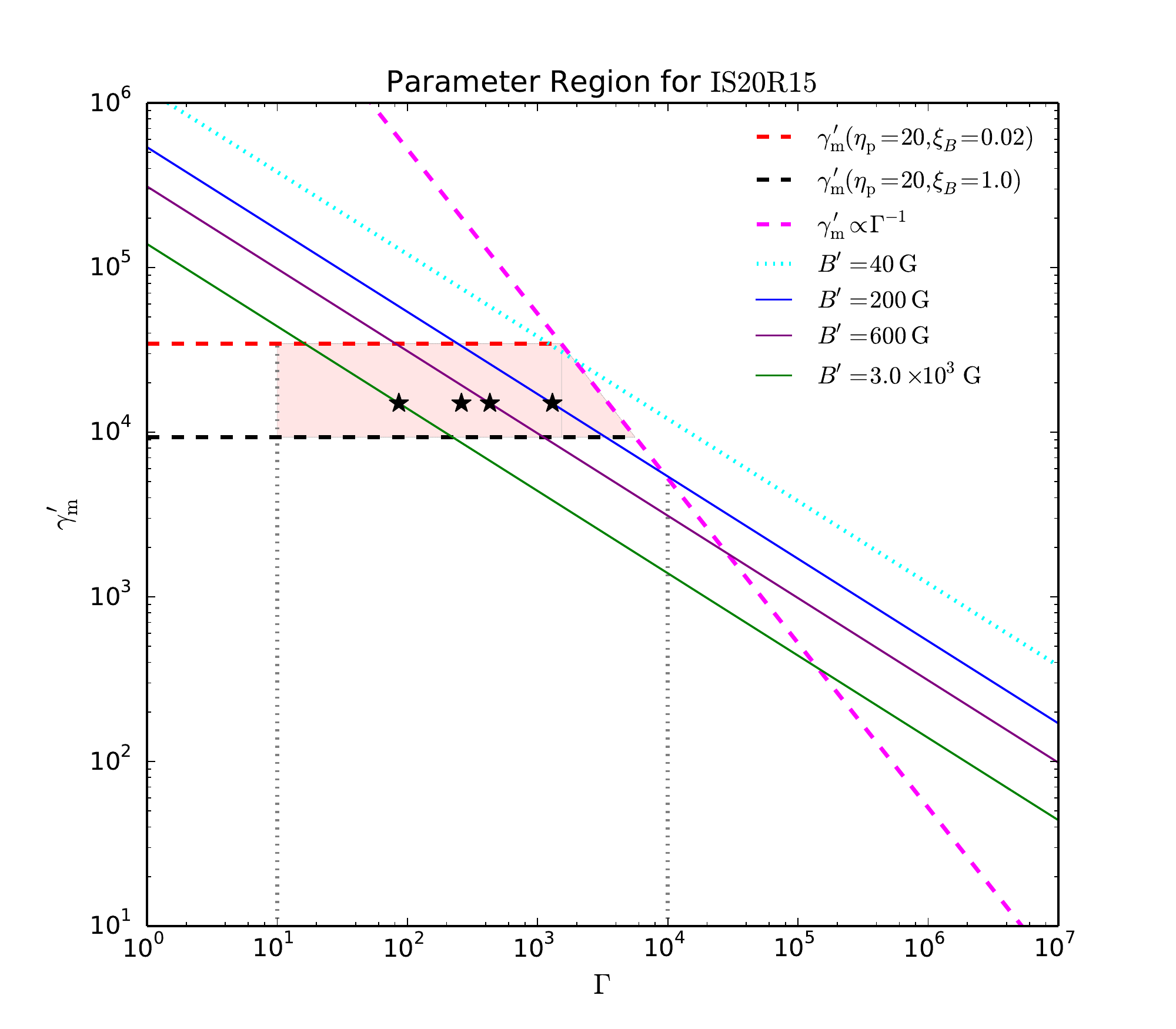}
   \caption{The parameter space for the internal shock model when $R_0 = 10^{15}$~cm and $\eta_{\rm p} = 20$ are adopted.
   The shadow region is the plausible region for the $\gamma_{\rm m}^{\prime}$--$\Gamma$ couple
   constrained from Equations (\ref{eq:Fast Cooling}) and (\ref{eq:xiBrange}).
   The magenta dashed line presents the condition for fast cooling, while the grey dashed and
   the red dashed lines show the lower limit
   ($\gamma_{\rm m}^{\prime} < \eta_{\rm p} m_{\rm p} / m_{\rm e}$)
   and the upper limit (deduced by $\xi_B \le 1$ with Equation (\ref{eq:xiB})) of $\gamma_{\rm m}^{\prime}$, respectively.
   $\Gamma$ is assumed to range from 10 to $10^4$ according to previous researches on GRB jets~\citep{Liang10,Liang13}.
   The solid lines are the relationship between $\gamma_{\rm m}^{\prime}$ and $\Gamma$ using Equation (\ref{eq:Epeak}),
   when different values for $B^{\prime}$ (denoted by different colours) are adopted
   and $E_{\rm peak}$ is set to be $500~\mathrm{keV}$ typically.
   Particularly, the cyan dotted line represents the minimum value of $B^{\prime}$ for $\gamma_{\rm m}^{\prime}$--$\Gamma$
   couples to overlap with the shadow region.
   The positions of the parameters in the four calculations of Group IS20R15 (see Table \ref{TABLE:A}) are marked by star symbols.}
   \label{fig:groupA}
\end{figure}

\begin{figure}
   \centering
   \includegraphics[scale=0.5]{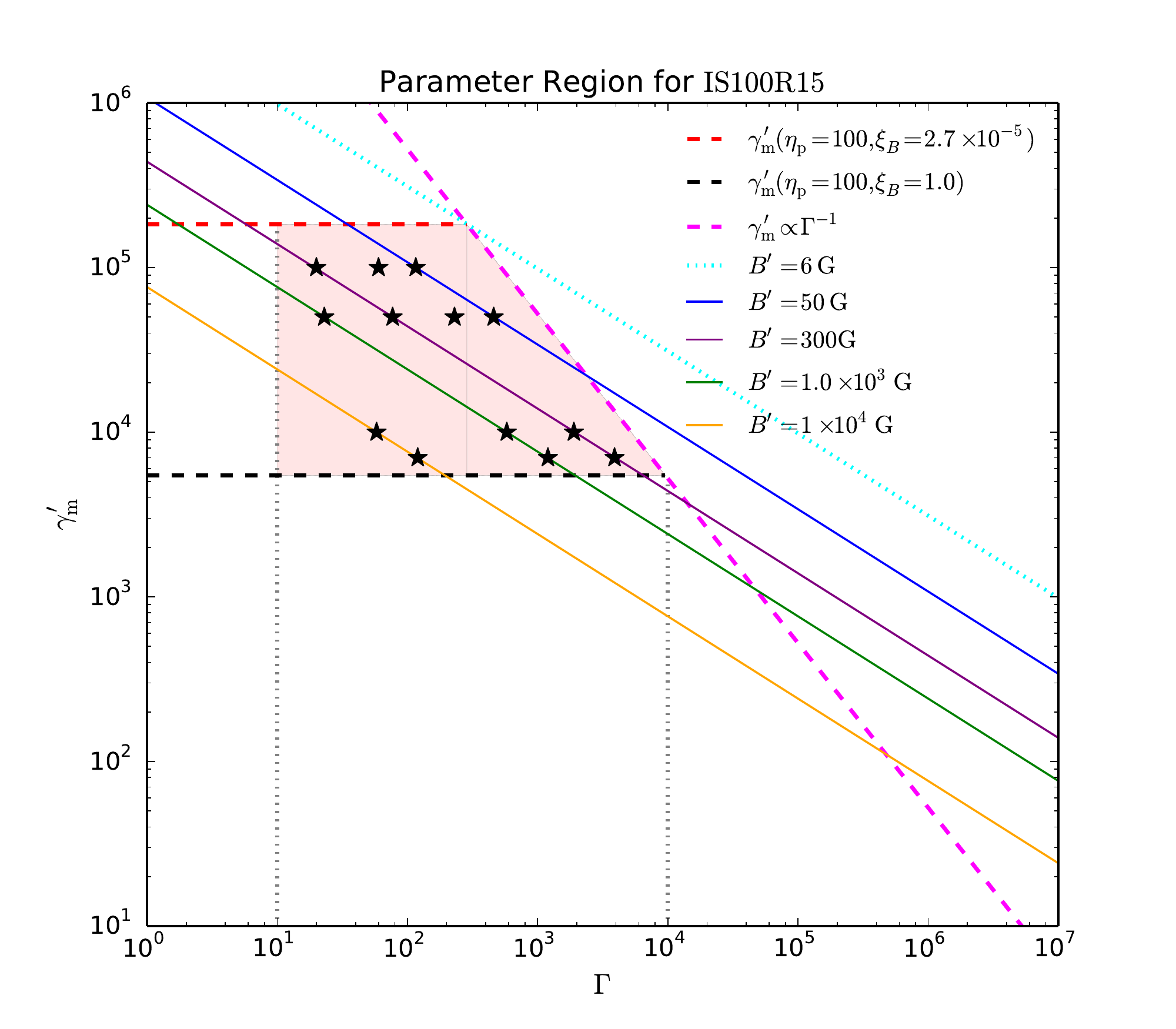}
   \caption{The parameter space for the internal shock model when $R_0 = 10^{15}$~cm and $\eta_{\rm p} = 100$ are adopted.
   The meanings of lines are similar to those explained in Figure \ref{fig:groupA}.
   The positions of the parameters in the thirteen calculations of Group IS100R15 (see Table \ref{TABLE:B}) are marked by star symbols.}
   \label{fig:groupB}
\end{figure}

\begin{figure}
   \centering
   \includegraphics[scale=0.5]{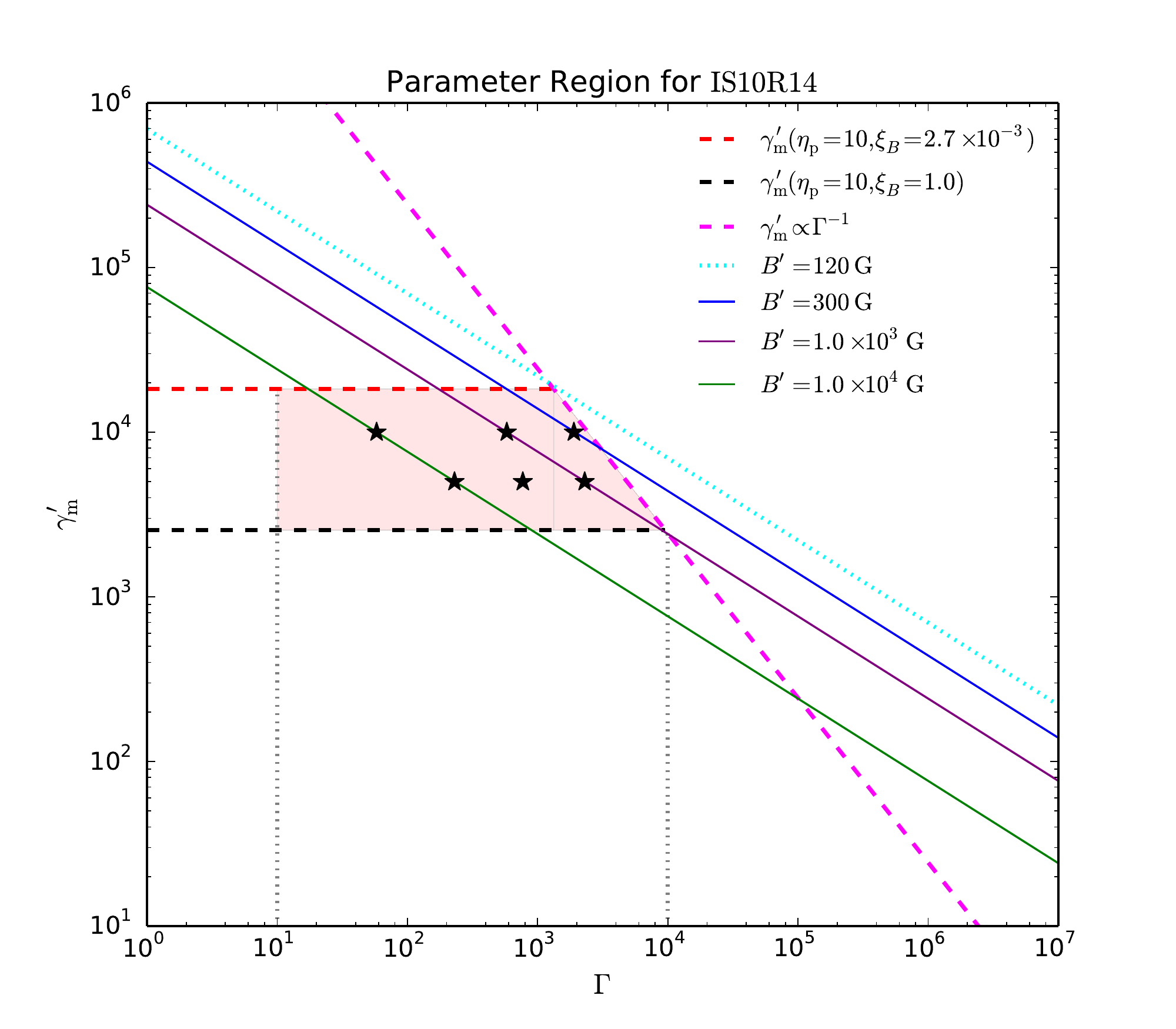}
   \caption{The parameter space for the internal shock model when $R_0 = 10^{14}$~cm and $\eta_{\rm p} = 10$ are adopted.
   The meanings of lines are similar to those explained in Figure \ref{fig:groupA}.
   The positions of the parameters in the six calculations of Group IS10R14 (see Table \ref{TABLE:C}) are marked by star symbols.}
   \label{fig:groupC}
\end{figure}

\begin{figure}
   \centering
   \includegraphics[scale=0.5]{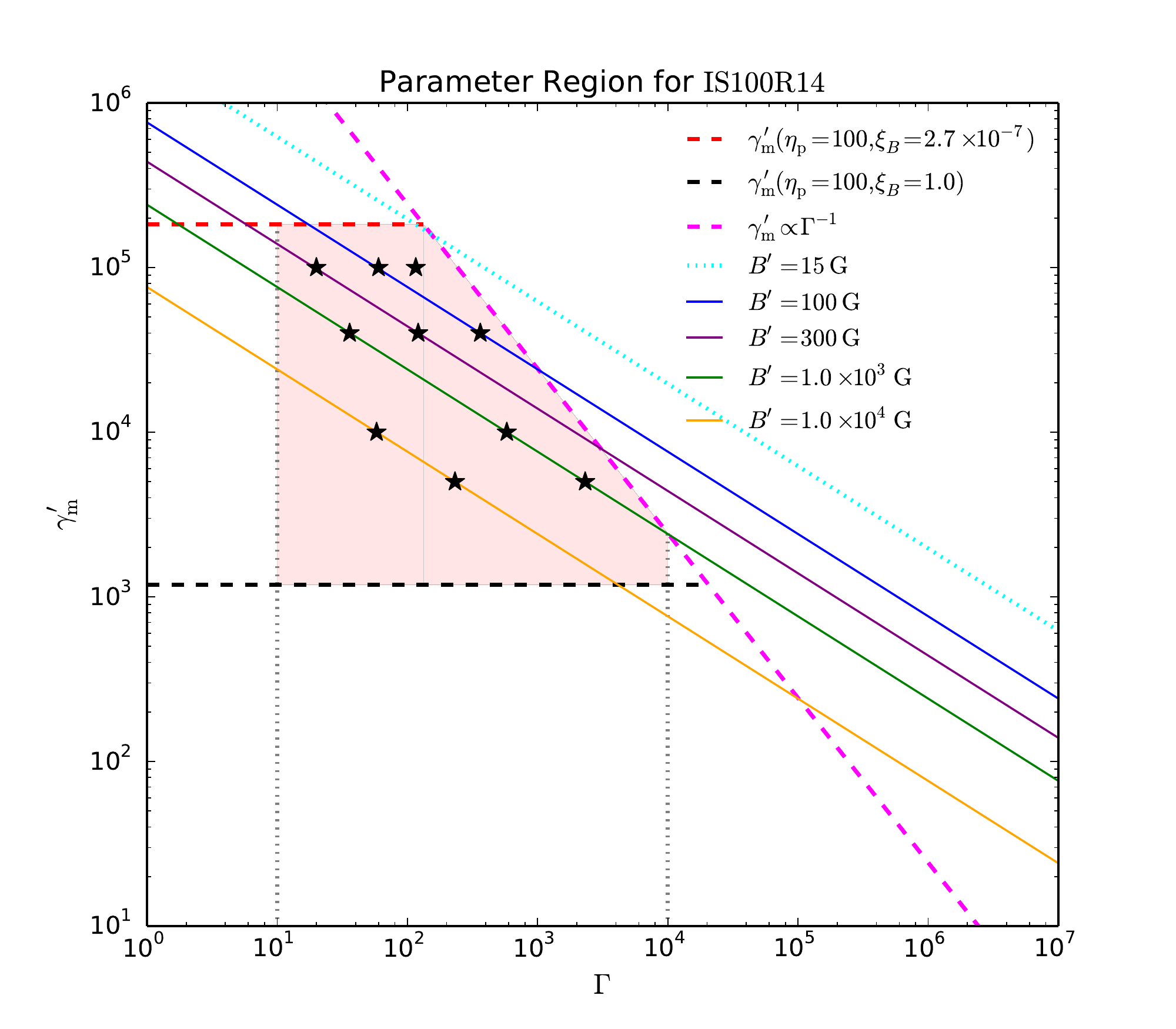}
   \caption{The parameter space for the internal shock model when $R_0 = 10^{14}$~cm and $\eta_{\rm p} = 100$ are adopted.
   The meanings of lines are similar to those explained in Figure \ref{fig:groupA}.
   The positions of the parameters in the ten calculations of Group IS100R14 (see Table \ref{TABLE:D}) are marked by star symbols.}
   \label{fig:groupD}
\end{figure}

Now, we turn to the internal shock model. In this scenario, since we still have little knowledge
about the emission radii of GRBs and the crucial parameter $\eta_{\rm p}$, we perform four groups of calculations
to try to cover various possibilities.
For $R_0 = 10^{15}$~cm,
we consider two situations, i.e., $\eta_{\rm p} = 20$\footnote{Here, we use $\eta_{\rm p} =20$
rather than $\eta_{\rm p} = 10$, due to the fact that $\eta_{\rm p} = 10$ would slightly violate
the underlying condition of $\xi_B + \xi_{\rm e} \le 1$.} or $\eta_{\rm p} = 100$.
For $\eta_{\rm p} = 20$, we show the allowed parameter region in the
$\gamma_{\rm m}^{\prime} - \Gamma$ diagram (see Figure \ref{fig:groupA}) by combing the restrictions given in Equations (\ref{eq:Fast Cooling})
and (\ref{eq:xiB}). In this group of calculations, called Group IS20R15,
four calculations are performed and the positions of the corresponding parameters in the parameter space
are marked as black stars in Figure \ref{fig:groupA}.
Detailed parameter values in each calculation are listed in Table \ref{TABLE:A}
and each calculation is named in form of ``IS20R15N'', with N denoting the value of $\Gamma$.
Similarly, in the calculations of Group IS100R15 ($\eta_{\rm p} = 100$),
we perform thirteen calculations of which the corresponding information can be seen in Table \ref{TABLE:B} and Figure \ref{fig:groupB}.
These thirteen calculations are classified into four subgroups according to different values used for $\gamma_{\rm m}^{\prime}$.
Each calculation is named in form of ``IS20R15W$\Gamma$N'' with letter ``W'' (e.g., A, B etc.) distinguishing different subgroups
and N denoting the value of $\Gamma$.
When $R_0 = 10^{14}$~cm is adopted, we also consider two situations, i.e., $\eta_{\rm p} = 10$ or $\eta_{\rm p} = 100$.
Six calculations are performed in Group IS10R14 (see Table \ref{TABLE:C} and Figure \ref{fig:groupC}), while ten calculations are performed
in Group IS100R14 (see Table \ref{TABLE:D} and Figure \ref{fig:groupD}).

\begin{figure}
\centering
    \subfloat{\includegraphics[width=0.45\linewidth]{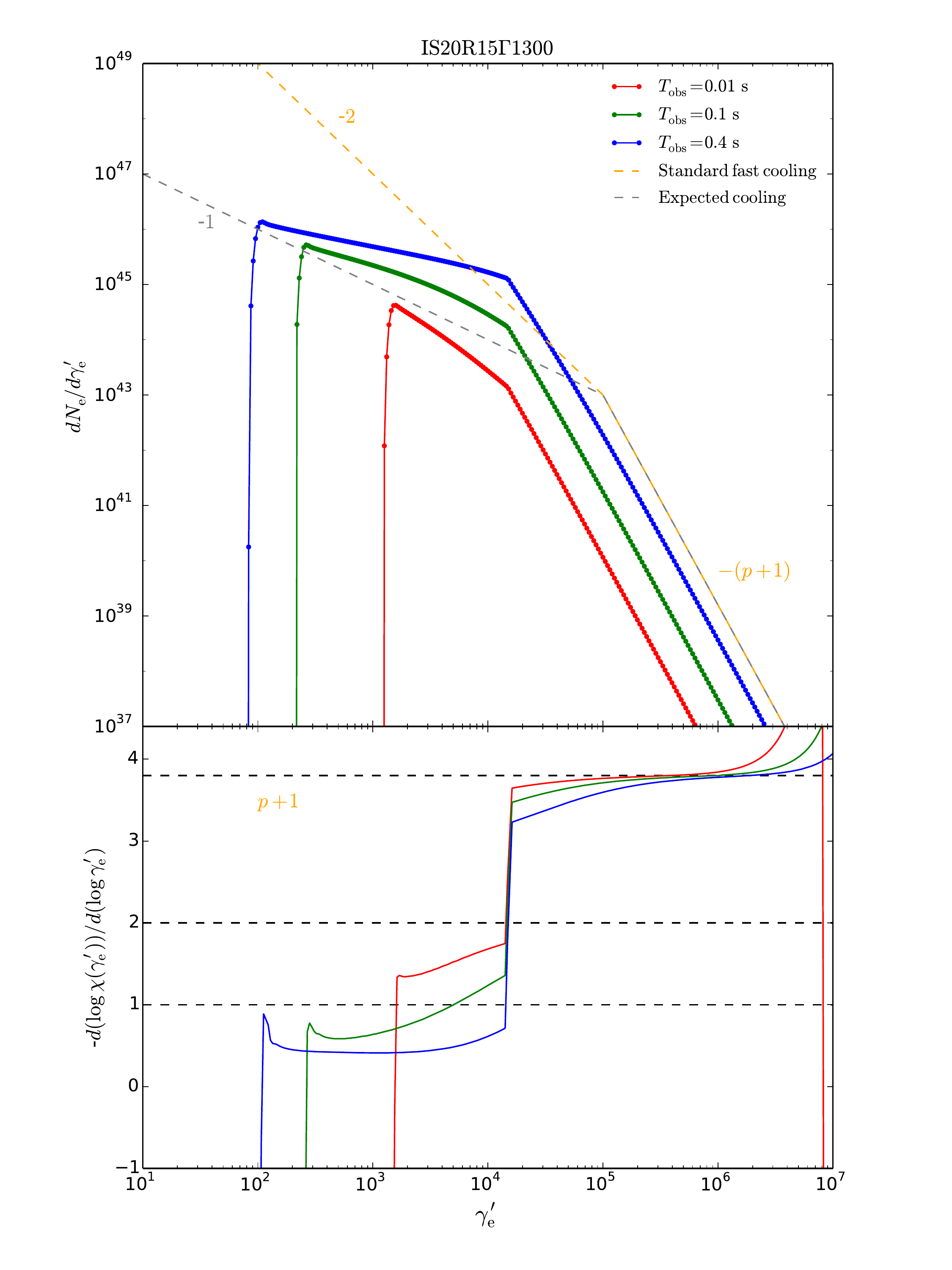}}
    \subfloat{\includegraphics[width=0.45\linewidth]{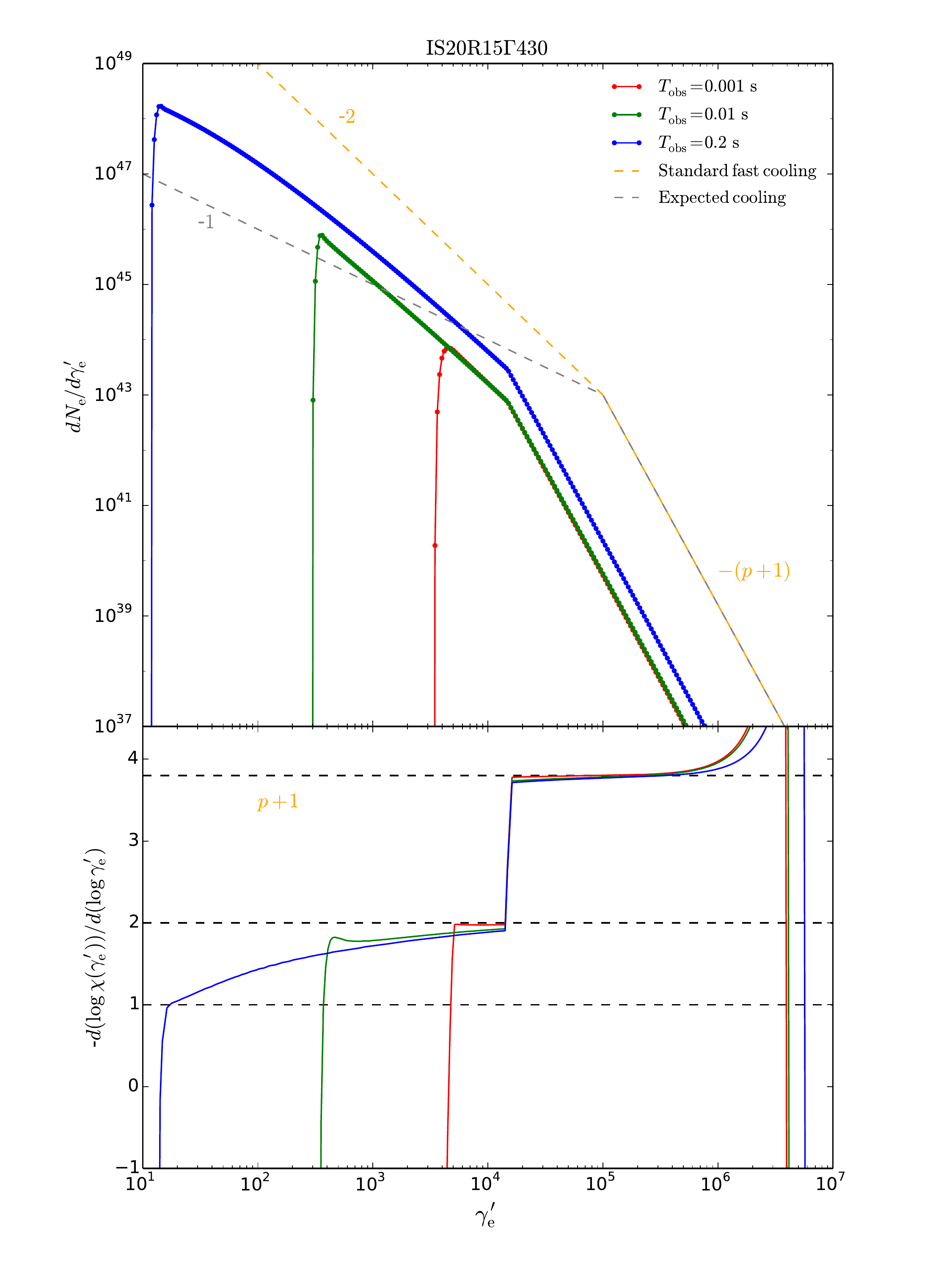}} \\
    \subfloat{\includegraphics[width=0.45\linewidth]{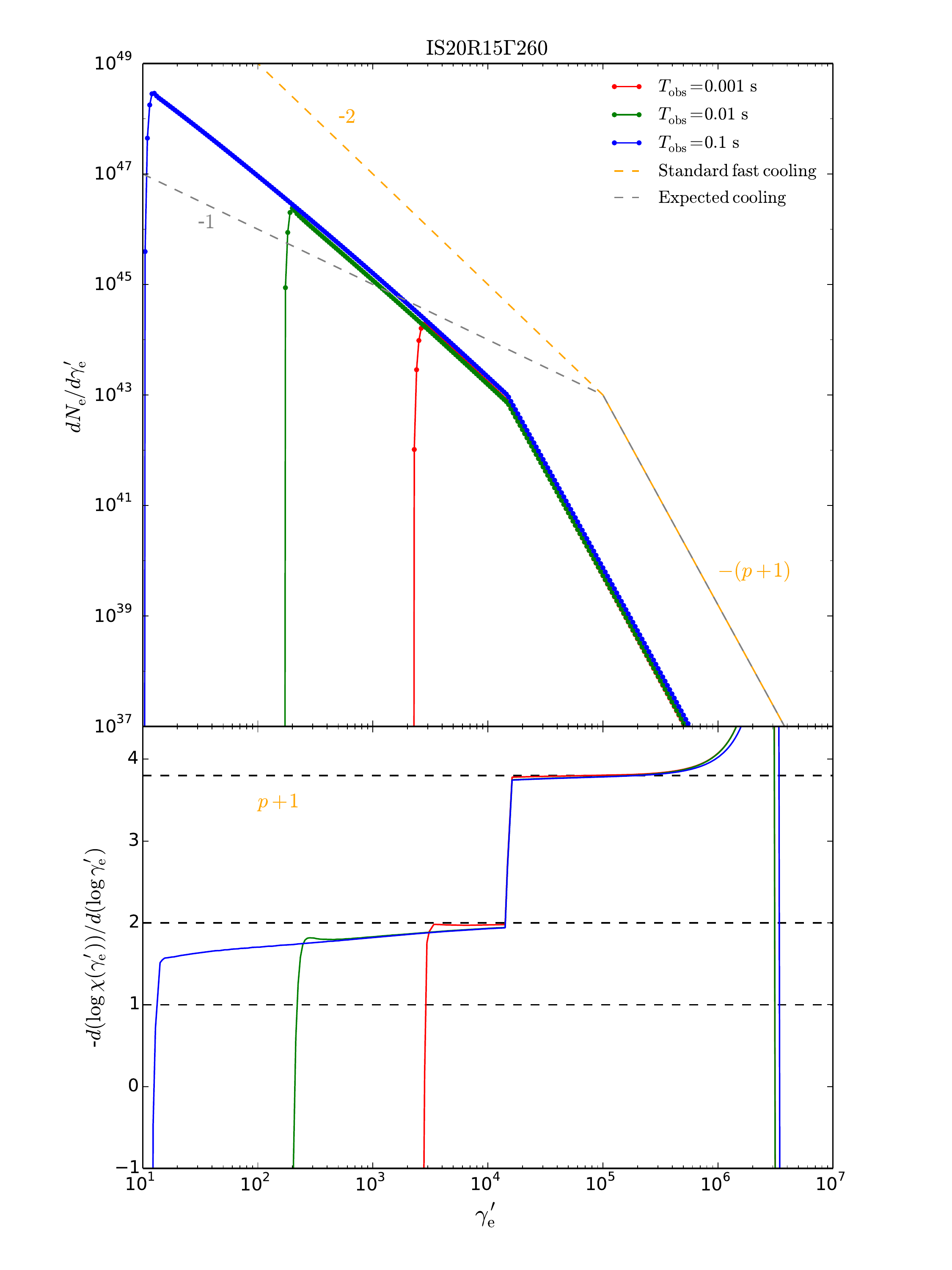}}
    \subfloat{\includegraphics[width=0.45\linewidth]{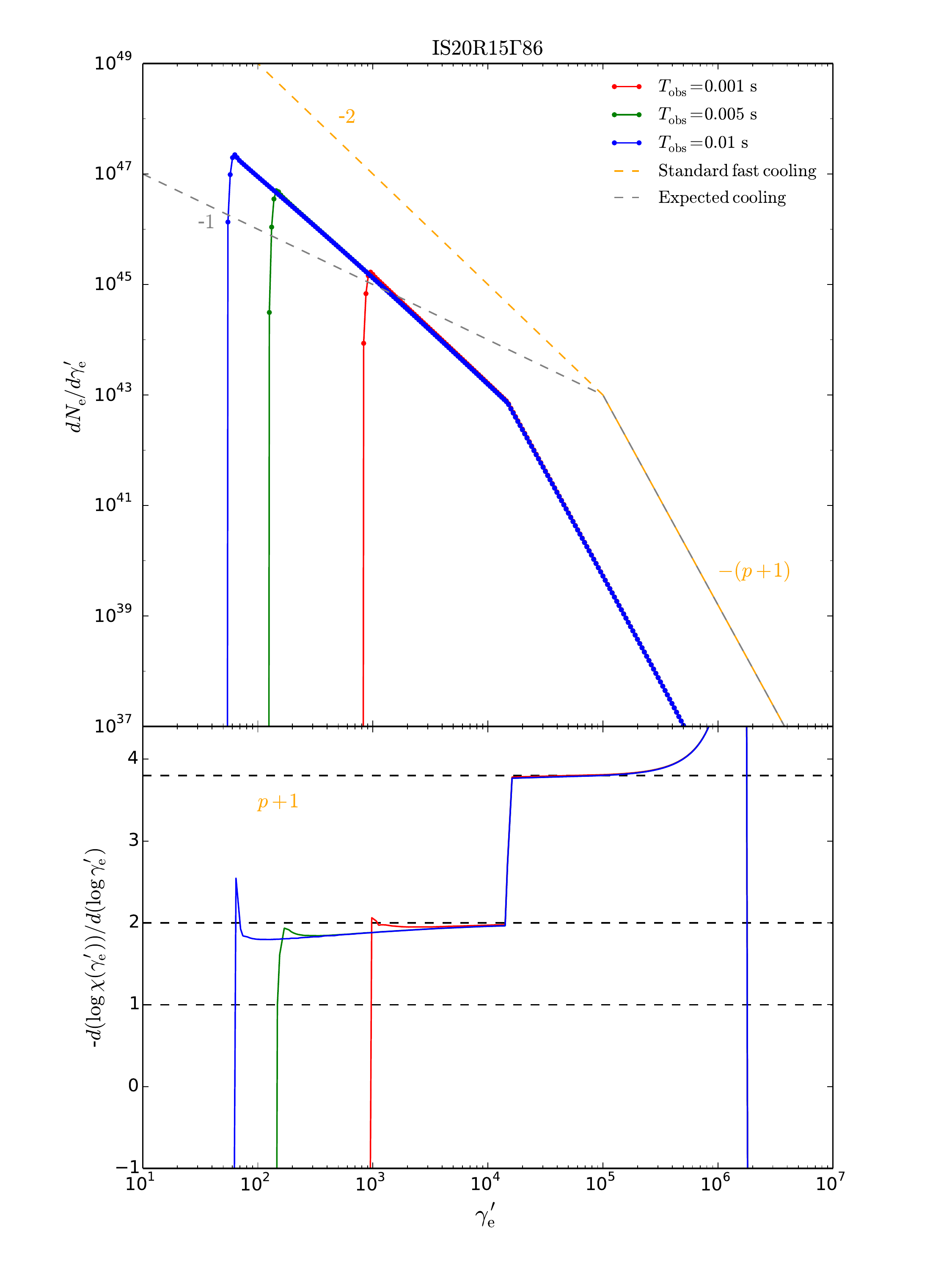}}
    \caption{The evolution of the electron energy spectrum for the four cases of Group IS20R15 (see Table \ref{TABLE:A}).}
    \label{fig:MA-electron}
\end{figure}

\begin{figure}
\centering
    \subfloat{\includegraphics[width=0.45\linewidth]{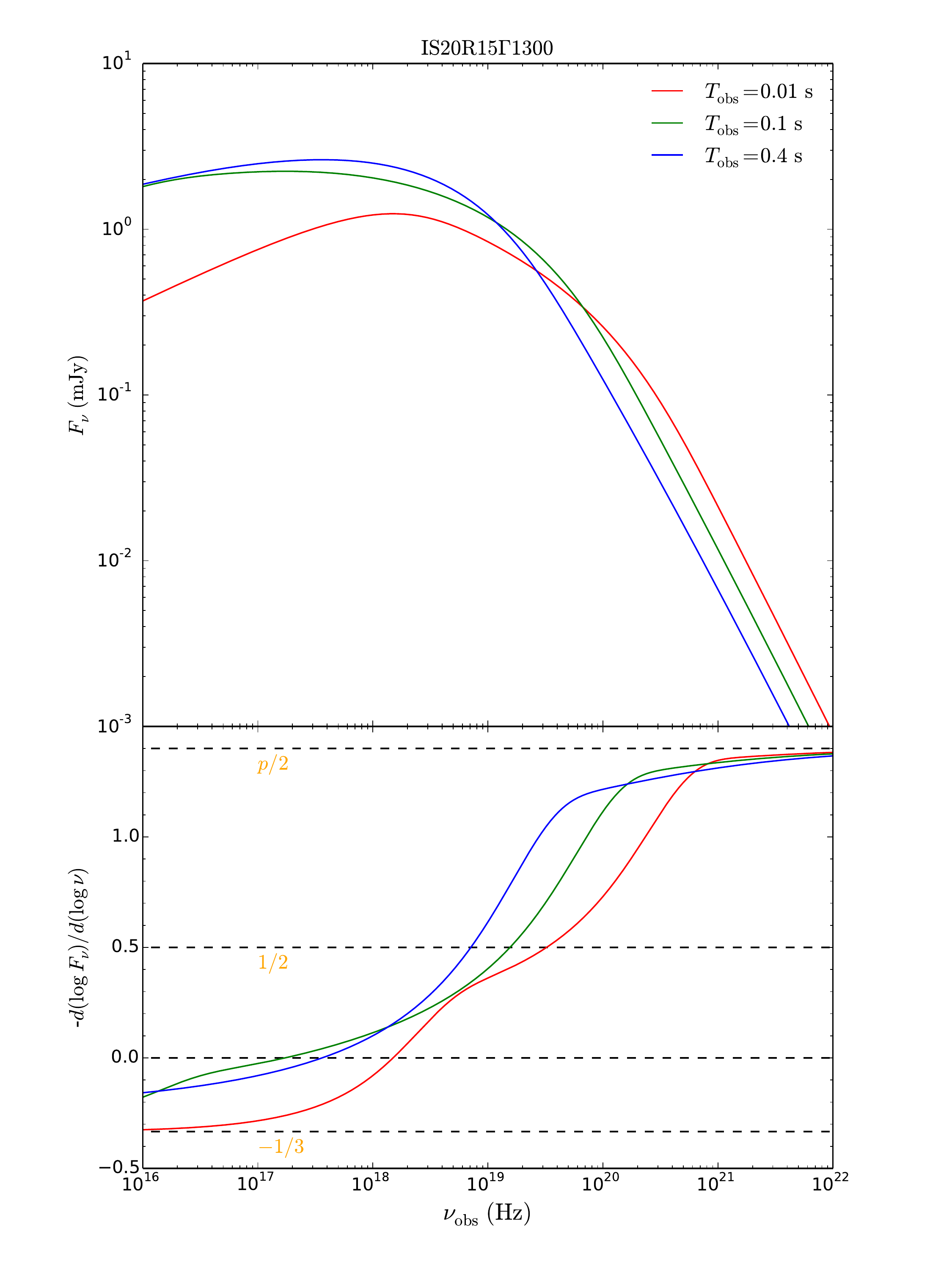}}
    \subfloat{\includegraphics[width=0.45\linewidth]{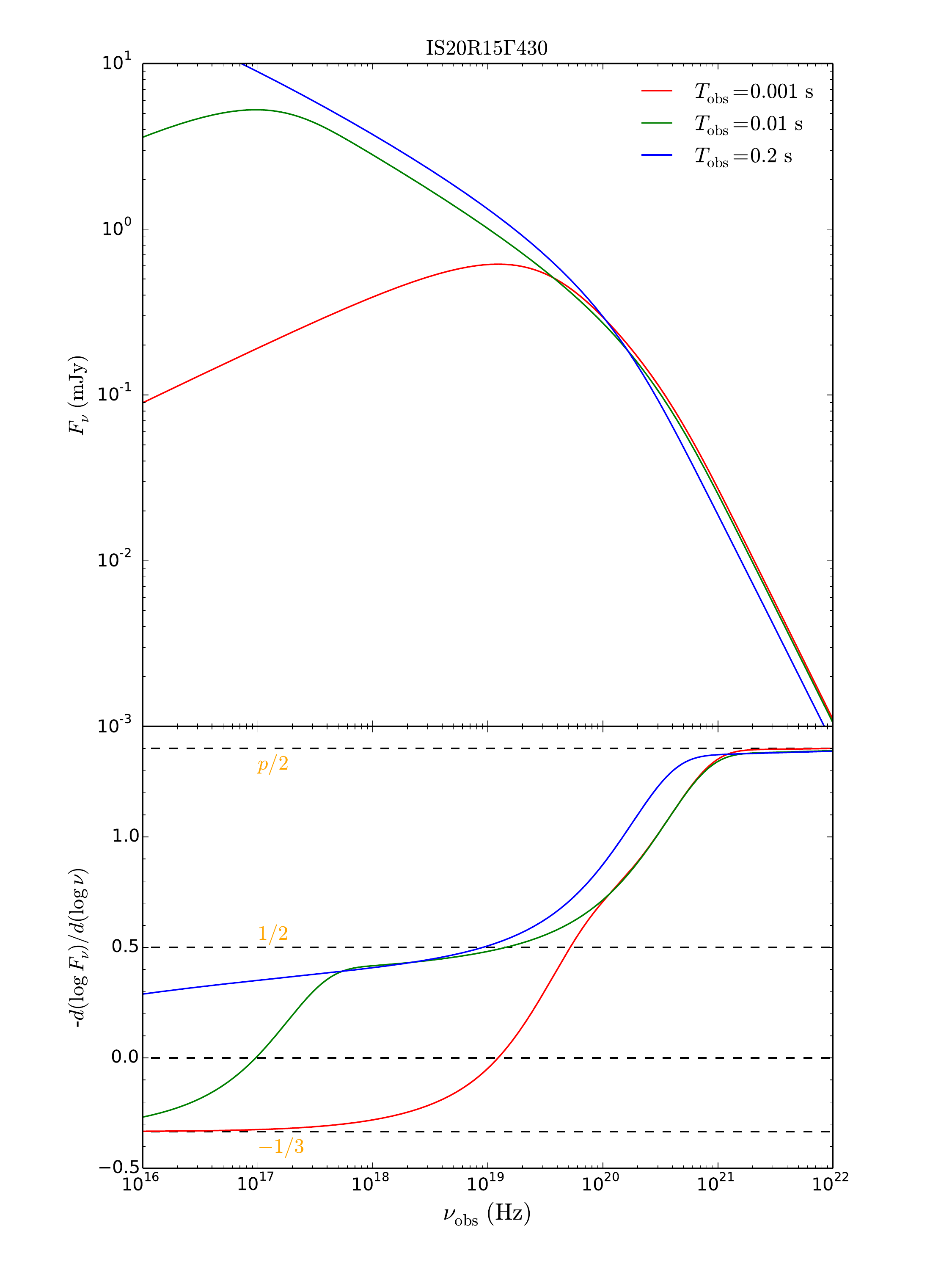}} \\
    \subfloat{\includegraphics[width=0.45\linewidth]{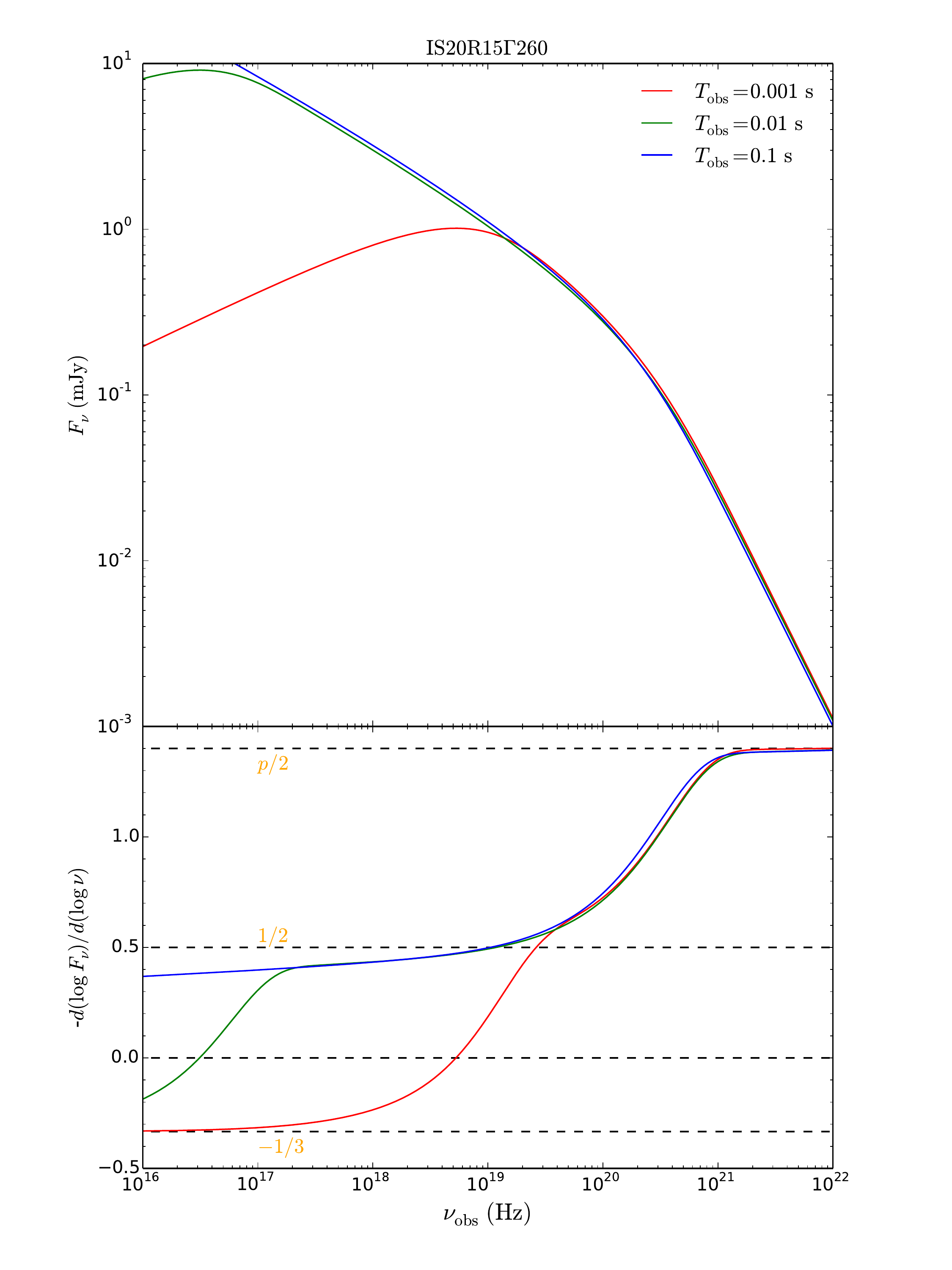}}
    \subfloat{\includegraphics[width=0.45\linewidth]{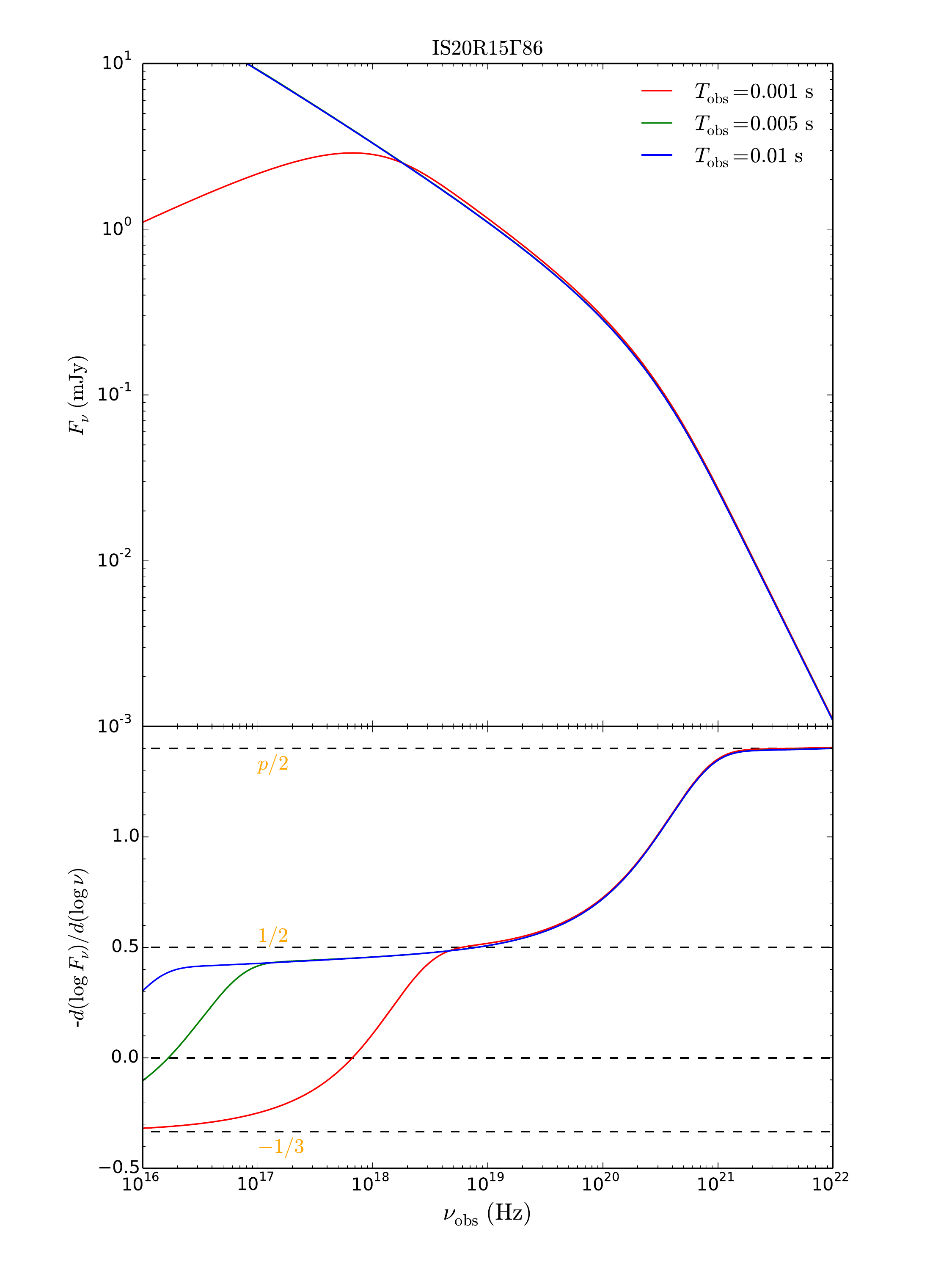}}
    \caption{The corresponding synchrotron flux-density spectra $F_{\nu}$ from the electrons with the energy distribution presented in Figure \ref{fig:MA-electron}.}
    \label{fig:MA-spectra}
\end{figure}

\begin{figure}
\centering
    \subfloat{\includegraphics[width=0.45\linewidth]{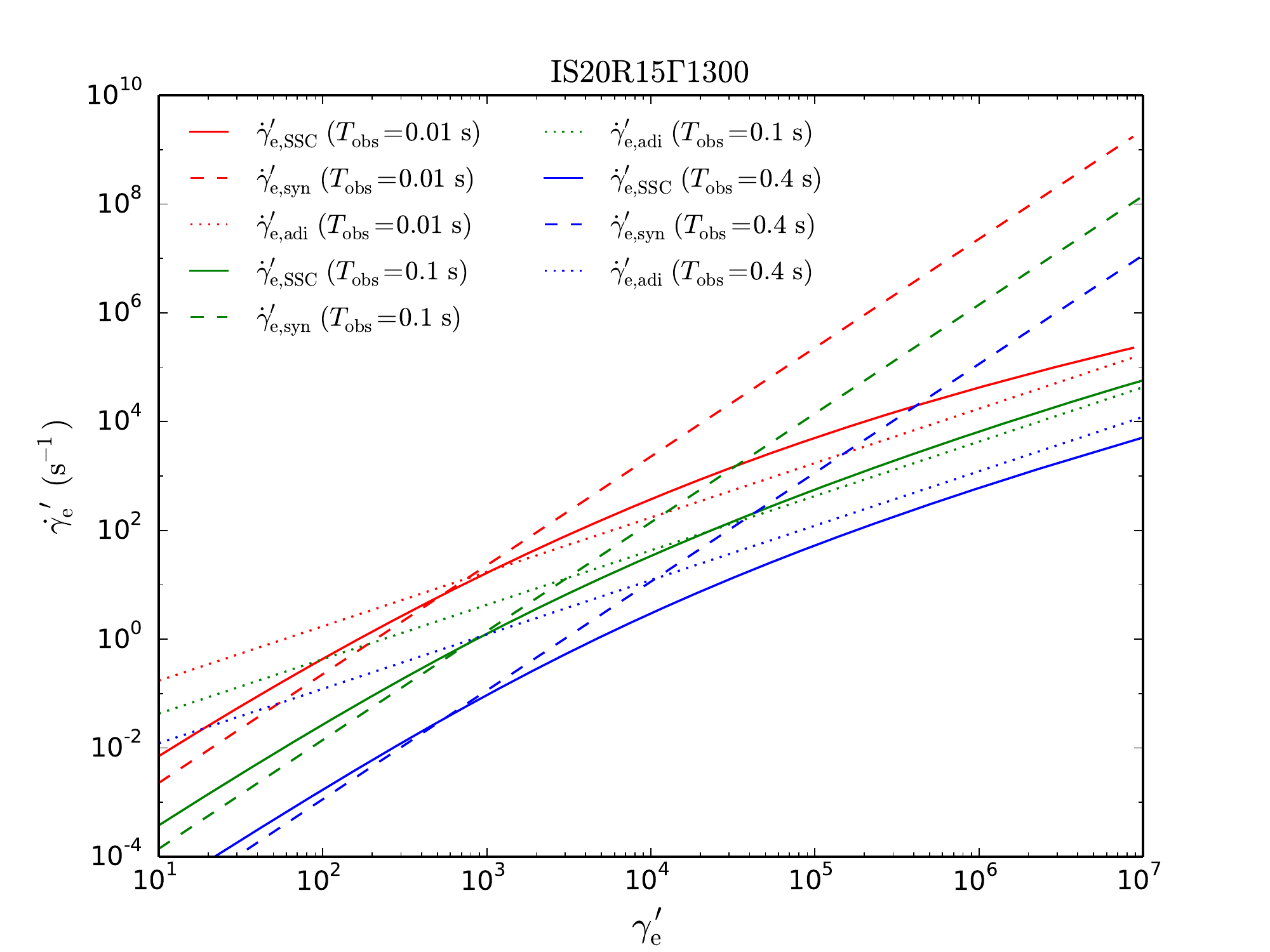}}
    \subfloat{\includegraphics[width=0.45\linewidth]{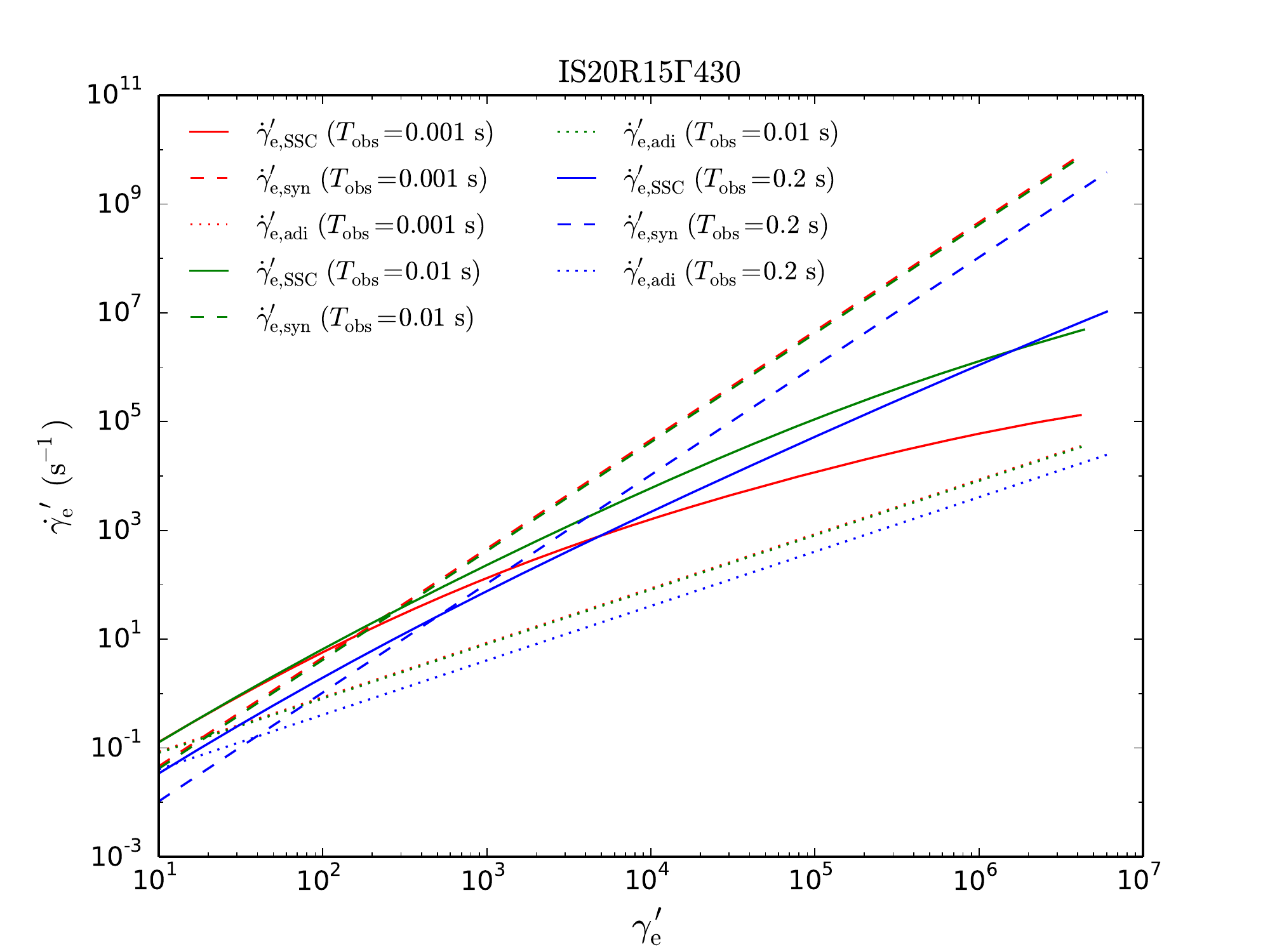}} \\
    \subfloat{\includegraphics[width=0.45\linewidth]{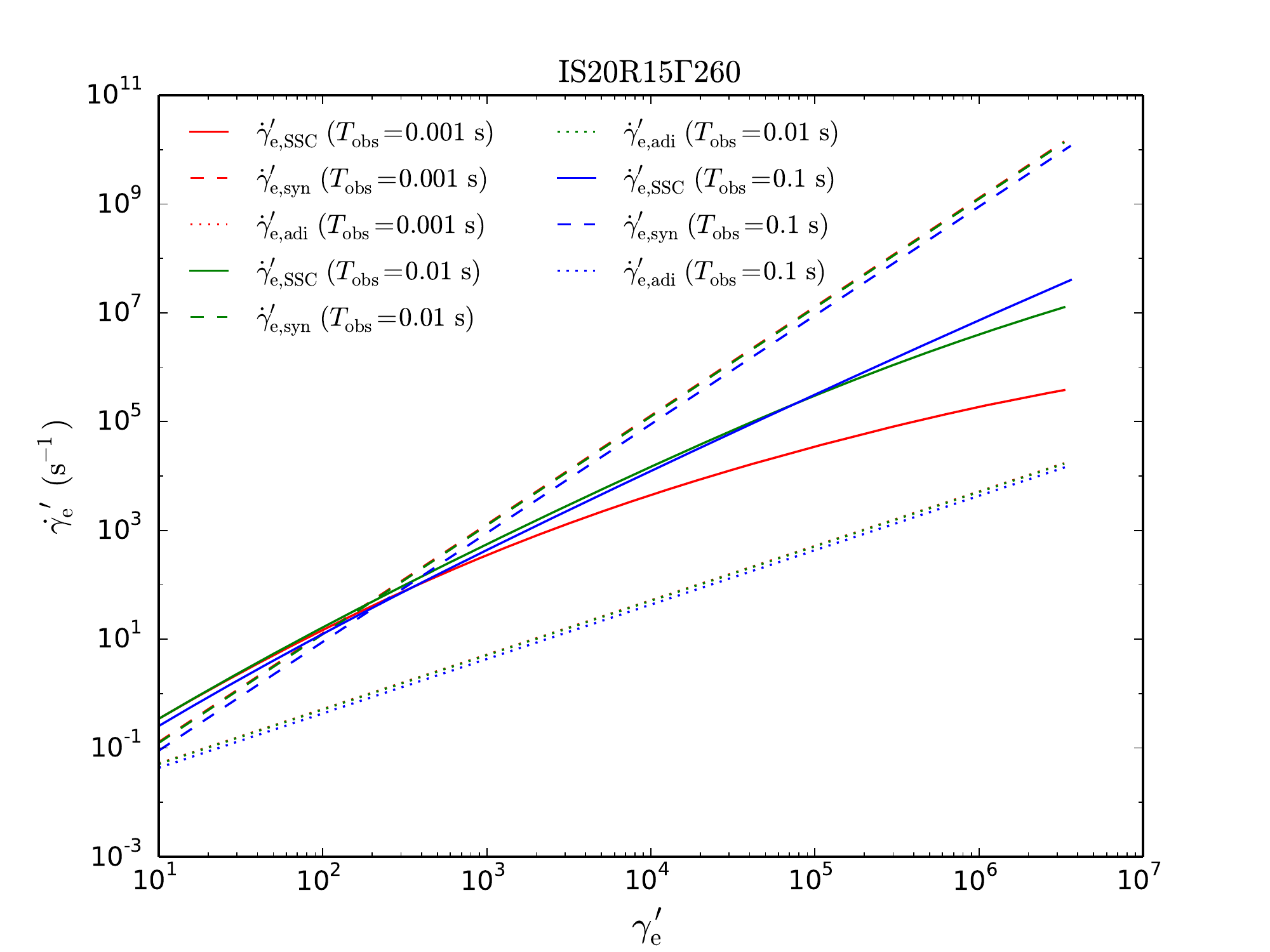}}
    \subfloat{\includegraphics[width=0.45\linewidth]{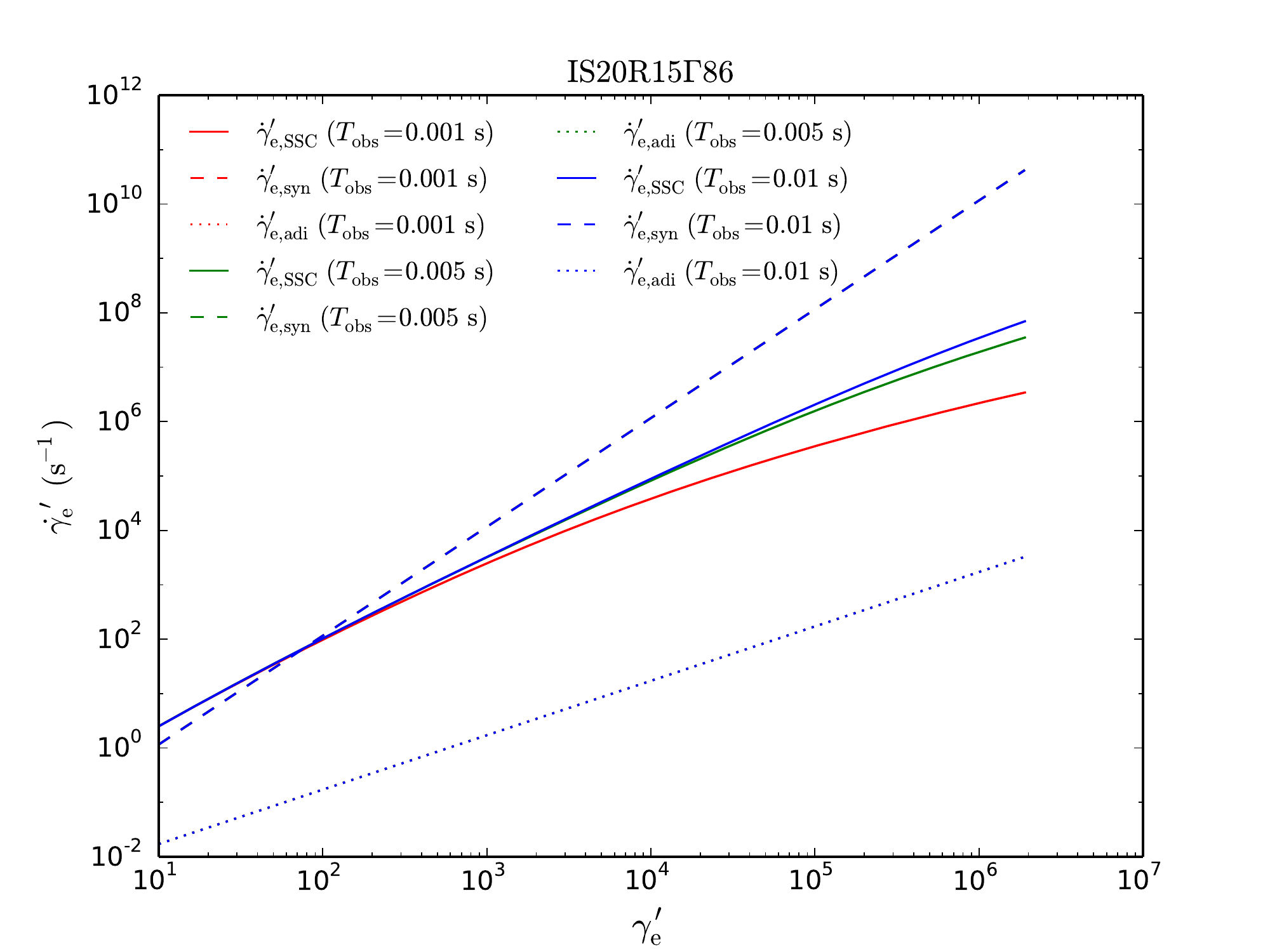}}
    \caption{The co-moving cooling rates of different cooling mechanisms for the electrons with the energy distribution
    presented in Figure \ref{fig:MA-electron}.}
    \label{fig:MA-rate}
\end{figure}

In the calculations of Group IS20R15, we notice that $\gamma_{\rm m}^{\prime} = 1.5 \times 10^4$
and $L_{\rm e} \approx L_B$. The results for electron distributions, flux spectra, and cooling rates are shown
in Figures \ref{fig:MA-electron}, \ref{fig:MA-spectra}, and \ref{fig:MA-rate} respectively.
From these results, we find that the electron spectra and the flux spectra can be hard enough
to match the observations only when adiabatic cooling is dominant (in IS20R15$\Gamma$1300).
For other three cases (IS20R15$\Gamma$430--IS20R15$\Gamma$86), $\dot{\gamma}_{\rm e,SSC}^{\prime}$ is only comparable to $\dot{\gamma}_{\rm e,syn}^{\prime}$
for electrons of $\gamma_{\rm e}^{\prime} \le 10^2$, and the indices of the low-energy electron spectra are slightly harder than $-2$.

In the calculations of Group IS100R15, we can see $\gamma_{\rm m}^{\prime}$ ranges from $7 \times 10^3$ to $10^5$ 
and $L_{\rm e}/L_B$ ranges from 0.08 to 3000.
According to the results (see Figures \ref{fig:MB-electron}, \ref{fig:MB-spectra} and \ref{fig:MB-rate}),
we find:
1, for Subgroup IS100R15A ($L_{\rm e}/L_B$ = 3000), $\dot{\gamma}_{\rm e,SSC}^{\prime}$
is always dominant for electrons of $\gamma_{\rm e}^{\prime} < \gamma_{\rm m}^{\prime}$ and the
resulting indices of electron spectra are $\sim -1.3$;
2, for Subgroup IS100R15B ($L_{\rm e}/L_B$ = 200), $\dot{\gamma}_{\rm e,SSC}^{\prime}$
is dominant for electrons of $\gamma_{\rm e}^{\prime} < 10^4$ and the resulting indices of electron spectra are $\sim -1.4$,
except for IS100R15B$\Gamma$460, in which $\dot{\gamma}_{\rm e,adi}^{\prime}$ becomes dominant at the late time
and the low-energy electron indices can be even harder than $-1$;
3, for Subgroup IS100R15C ($L_{\rm e}/L_B$ = 0.3), $\dot{\gamma}_{\rm e,syn}^{\prime}$
is always larger than $\dot{\gamma}_{\rm e,SSC}^{\prime}$.
The electron spectra resemble the standard ones in IS100R15C$\Gamma$580 and IS100R15C$\Gamma$58, while the
electron indices in IS100R15C$\Gamma$1900 are becoming harder than $-1$ due to the dominance of adiabatic cooling;
4, for Subgroup IS100R15D ($L_{\rm e}/L_B$ = 0.08), $\dot{\gamma}_{\rm e,syn}^{\prime}$
is always larger than $\dot{\gamma}_{\rm e,SSC}^{\prime}$.
The electron spectra resemble the standard ones in IS100R15D$\Gamma$1200 and IS100R15D$\Gamma$120, while the
electron indices in IS100R15D$\Gamma$3900 are becoming even harder than $0$ due to the dominance of adiabatic cooling.

In the calculations of Group IS10R14, we find (see Figures \ref{fig:MC-electron}, \ref{fig:MC-spectra} and \ref{fig:MC-rate}):
1, for Subgroup IS10R14A ($L_{\rm e}/L_B$ = 30), $\dot{\gamma}_{\rm e,SSC}^{\prime}$
is larger than $\dot{\gamma}_{\rm e,syn}^{\prime}$ for electrons of $\gamma_{\rm e}^{\prime} < 3 \times 10^3$,
and the resulting indices of electron spectra are $\sim -1.5$ for IS10R14A$\Gamma$580 and IS10R14A$\Gamma$58.
For IS10R14A$\Gamma$1900, $\dot{\gamma}_{\rm e,adi}^{\prime}$ is always dominant for electrons of $\gamma_{\rm e}^{\prime} < \gamma_{\rm m}^{\prime}$
and the electron indices are becoming harder than 0;
2, for Subgroup IS10R14B ($L_{\rm e}/L_B$ = 2), $\dot{\gamma}_{\rm e,SSC}^{\prime}$
is slightly larger than $\dot{\gamma}_{\rm e,syn}^{\prime}$ for electrons of $\gamma_{\rm e}^{\prime} < 3 \times 10^2$,
and the resulting indices of electron spectra are $\sim -1.8$ for IS10R14B$\Gamma$770 and IS10R14B$\Gamma$230.
For IS10R14B$\Gamma$2300, $\dot{\gamma}_{\rm e,adi}^{\prime}$ is becoming increasingly dominant for $\gamma_{\rm e}^{\prime} < \gamma_{\rm m}^{\prime}$
and the electron indices are becoming harder than 0.

In the calculations of Group IS100R14, $\gamma_{\rm m}^{\prime}$ ranges from $5 \times 10^3$ to $10^5$ 
and $L_{\rm e}/L_B$ ranges from 2 to $3 \times 10^5$.
According to the results (see Figures \ref{fig:MD-electron}, \ref{fig:MD-spectra} and \ref{fig:MD-rate}),
we find:
1, for Subgroups IS100R14A and IS100R14B ($L_{\rm e}/L_B$ is $3 \times 10^5$ or $8.5 \times 10^3$),
$\dot{\gamma}_{\rm e,SSC}^{\prime}$ is always much larger than both $\dot{\gamma}_{\rm e,syn}^{\prime}$
and $\dot{\gamma}_{\rm e,adi}^{\prime}$ for electrons of $\gamma_{\rm e}^{\prime} < \gamma_{\rm m}^{\prime}$,
the resulting indices of electron spectra are $\sim -1$;
2, for Subgroup IS100R14C ($L_{\rm e}/L_B$ = 30),
$\dot{\gamma}_{\rm e,SSC}^{\prime}$ is larger than both $\dot{\gamma}_{\rm e,syn}^{\prime}$
and $\dot{\gamma}_{\rm e,adi}^{\prime}$ for $\gamma_{\rm e}^{\prime} <  3 \times 10^3$,
the resulting indices of electron spectra are $\sim -1.5$.
3, for Subgroup IS100R14D ($L_{\rm e}/L_B$ = 2), since $\dot{\gamma}_{\rm e,adi}^{\prime}$ is becoming increasingly dominant for
low-energy electrons, the indices of electron spectra are getting harder than 0.
The results of IS100R14D$\Gamma$230 are similar to those of IS10R14B$\Gamma$230 since their parameters
are actually the same. So we have not shown them as subfigures in the relevant
figures of this calculation group.

To sum up, in cases of a large $\Gamma$ ($\Gamma > 10^3$), adiabatic cooling is
the most dominant process for low-energy electrons' cooling, making the electron indices
harder than $-1$, or even 0.
For cases of $L_{\rm e} / L_B > 1$, and when $\Gamma$ is not too large, SSC cooling
dominates over synchrotron cooling, resulting in electron spectra with indices ranging
from $\sim -1$ to $-2$ with the decreasing of $L_{\rm e} / L_B$.
When $L_{\rm e} / L_B \le 1$ is met while $\Gamma$ is still not too large, synchrotron cooling will
then take over and the resulting spectra slightly deviate from the standard ones.
These three kinds of cooling patterns are generally consistent with previous 
numerical results in \cite{Bonjak09} and \cite{Daigne11} 
\footnote{In our results, spectral indices $\alpha$ are 
strictly smaller than 0 for cases when SSC cooling dominates.
This is consistent with the limit of $\alpha > -0.1$ given in \cite{Barniol12}
with detailed analytical study. While the results in \cite{Daigne11}
violate this limit slightly (see their Figure 2),
this difference does not affect much on the consensus on the effect of SSC cooling.}.
The conditions for the dominance of adiabatic cooling or SSC cooling in these results
are consistent with the analyses in Section 4.
In general, the combination of a large $\gamma_{\rm m}^{\prime}$ and small $B^{\prime}$, $R$ will favor the
dominance of adiabatic cooling or SSC cooling, rather than
synchrotron cooling according to Equations (\ref{eq:adi/syn}) and (\ref{eq:SSC>syn}).

In reality, the observed minimum variability timescales of GRBs can be used to derive the internal shock radii,
which typically gives $R_0 = 10^{14}$~cm or smaller. We have only explored the region of $R_0 \ge 10^{14}$~cm
in our calculations. However, according to Equations (\ref{eq:SSC/syn}) and (\ref{eq:SSC>syn}), one will realize that
a smaller $R_0$ will enhance the SSC cooling rate due to the increase of the radiation energy density,
i.e., $U_{\gamma}^{\prime} \propto R^{-2}$.
Therefore, for a smaller $R_0$, SSC cooling will be more significant.
Also, we can expect that the results of a smaller $R_0$ can be well differentiated by values of $L_{\rm e} / L_B$ and $\Gamma$.

\section{DISCUSSION AND CONCLUSIONS}
We have developed a code to solve the continuity equation of electrons
in GRB ejecta, and analyzed the roles played by three cooling mechanisms
(synchrotron, SSC, adiabatic cooling) and the effect of decaying magnetic field
in determining the electron/flux spectra in both the fireball and the Poynting-flux-dominated regimes.
By exploring the parameter space and calculating the corresponding
electron spectra, flux spectra, and electron cooling rates, we find
that the hardening of the electron spectra can be attributed to the
synchrotron radiation with a decaying $B^{\prime}$, or the dominance of adiabatic cooling
or the dominance of SSC cooling.
Therefore, it is essential to coherently consider them together in future studies of the GRB prompt emission.
The numerical method, as proposed in this paper, has the advantage over the analytical method
in solving equations involving several factors simultaneously.
According to our results, the low-energy spectra of GRBs could be explained by the synchrotron
radiation from either a Poynting-flux-dominated jet or an internal shock,
although some shortcomings may exist for the two scenarios.

SSC cooling of electrons in the KN regime has been proposed to solve the fast cooling problem previously.
In this paper, our analyses confirm that SSC cooling is crucial in the internal shock scenario
when we take parameters deduced from typical observational characteristics.
A sequence of numerical calculations further reveal that SSC cooling
will result in electron spectra with low-energy indices ranging from $-2$ to $\sim -1$.
The physical condition for SSC cooling to be dominant is $L_{\rm e}/L_B > 1$.
In order to match the observations, i.e., $d N_{\rm e} / \gamma_{\rm e}^{\prime} \sim \gamma_{\rm e}^{\prime -1}$,
$F_{\nu} \sim \nu^0$, the condition should be more strict, i.e., $L_{\rm e}/L_B \ge 10^4$
according to the results from the calculation Groups IS100R15 and IS100R14.
This condition means $\xi_{\rm e}/\xi_B \ge 10^4$ within the internal shock model.
It is still hard to understand why the energy fraction in the magnetic field could be so small.
However, a possible explanation is that the $B^{\prime}$ that cools the electrons here
is much smaller than the magnetic field in the acceleration region.
As proposed by \cite{Rossi03}, \cite{PeerZhang06}, \cite{Lemoine13}, and \cite{Zhao14},
the magnetic field may accumulate in a small region (but carries the majority of the
magnetic energy created by the shock) just near the shock front,
and the magnetic field behind the shock decays rapidly with the distance from the front.
This possibility is hinted in particle-in-cell simulations of shocks (e.g., \citealt{Medvedev05,Chang08}).
Since the electrons may be cooled in the downstream region behind the shock,
then the $B^{\prime}$ they stream through should be smaller than those near the shock.
Moreover, to achieve the electron spectra with low-energy indices of $\sim -1$,
$\eta_{\rm p} \ge 10^2$ is preferred to take $\gamma_{\rm m}^{\prime} \ge 10^4$.
It indicates that only $\sim$ 1\% electrons are accelerated
and they carry $\sim$ 10\% of the energy when they cross the shock.
This should correspond to the cases in which the relativistic electron-ion shock is
of low magnetizations ($\sigma \le 10^{-3}$) according to the simulations of \cite{Sironi11}
and \cite{Sironi15}.

Our numerical results also support the idea that the hardening of the electron spectra
purely by SSC cooling in KN regime can be only up to, but not equal to $-1$ \citep{Barniol12}.
So the fast cooling problem could not be fully solved by SSC cooling,
since we have some GRB spectra with $\alpha > 0$.
However, for this small fraction of GRBs, they may be well explained when
adiabatic cooling is the dominant cooling process for low-energy electrons.
This requires that the ejecta is moving at an extremely relativistic speed, i.e., $\Gamma > 10^3$.
In the framework of the internal shock model, this condition may be naturally fulfilled since the initial energy
of the ejecta is mainly turned into kinetic energy.
On the other hand, the lower limit for $\Gamma$ in three GRBs set by the {\it Fermi} team
is $\sim 1000$ (\citealt{Abdo09,Abdo09Nat,Abdo09Sci}).
Thus it is possible that adiabatic cooling may be dominant in a few GRBs.

For a Poynting-flux-dominated jet ($L_{B} \ge L_{\rm e}$),
we confirm that the low-energy indices of the electron spectra can be harder than $\sim -1$
when $B^{\prime}$ is under decaying with $R$ \citep{Uhm14}, as shown in the results of PJR15$\Gamma$600 and PJR14$\Gamma$600.
By using the constraints from observations and performing a sequence of calculations (see Section~\ref{subsec:PJet}), 
we further reveal that $B^{\prime}$ should not be too large in this scenario,
otherwise the synchrotron cooling timescale will be much shorter than the dynamical timescale
and the effect of the decaying $B^{\prime}$ is weakened.
This feature provides a way to identify the effect of SSC cooling from the effect of decaying $B^{\prime}$.

Although spectra hardening could be achieved by two mechanisms, i.e., SSC cooling in KN regime
or the effect of decaying $B^{\prime}$, they may be distinguished in observations.
For SSC cooling in KN regime, the hardening of electron spectra is a result of the ``intrinsic'' 
scale relation of $\dot{\gamma}_{\rm e,tot}^{\prime} \simeq \dot{\gamma}_{\rm e,SSC}^{\prime}
\propto \gamma_{\rm e}^{\prime -1}$~\citep{Derishev01,Bonjak09,Wang09,Nakar09,Fan10,Daigne11}.
It can be seen that the hardening of electron spectra could be established within a rather
short timescale, e.g., $t_{\rm obs} < 0.1$~s via SSC cooling as in the results of Subgroups IS100R14A and IS100R14B. 
In contrast, decaying $B^{\prime}$ is an ``external'' way,
and hardening of the electron spectra due to this effect
needs a longer time, e.g., $t_{\rm obs} \ge 0.5$~s according to the results of PJR15$\Gamma$600 and PJR14$\Gamma$600.
Therefore, if a flux spectrum of form $F_{\nu} \sim \nu^0$ is observed in
a single, distinguishable, short pulse ($t_{\rm obs} < 0.1$~s) for a particular GRB,
then it is more likely that the SSC mechanism should be the dominant cooling process.
On the other hand, spectral lags and $E_{\rm peak}$ evolution patterns are found to be 
related with broad pulses (with durations of seconds, or called slow component, \citealt{Gao12}) rather than quick variabilities
\citep{ZhangB11,Uhm16a}. 
According to $R \sim \Gamma^2 c \delta t_{\rm obs}$, the emission radius is large for large $\delta t_{\rm obs}$,
which would weaken the effect of SSC cooling. The large emission radius is also contrived within the internal shock model. 
Thus, for these broad pulses, the effect of decaying $B^{\prime}$ should be more important if the evolution of 
spectral hardening (and softening) could be well matched with numerical results.

No general consensus on GRB jet properties (e.g., jet composition, emission radius) has been reached in the community. 
The information from the time-resolved low-energy spectra can help to probe
the physical characteristics of the GRB ejecta via our numerical results.
As mentioned above, SSC cooling in KN regime works in the scenario of internal shocks 
(baryon-dominated jet), while the effect of decaying $B^{\prime}$ mainly happens in 
the scenario of Poynting-flux-dominated jet. 
Once the time-resolved low-energy spectra hardening is affirmed
to be due to a specific mechanism, the jet composition could also be inferred simultaneously.
Furthermore, an overall comparison of the results among the four calculation groups within the 
internal shock model indicates that $\alpha \sim 0$ 
is more likely to be achieved at a small emission radius ($\le 10^{14}$~cm).
This value is well consistent with the results from other methods
\citep{Rees94,Gupta08,Kumar15}. 

There is another way to solve the fast synchrotron cooling problem.
Except for the cooling processes of electrons, their acceleration processes should also
be important in determining the final electron distribution.
A hard electron energy distribution with index $\sim -1$ may be produced by a slow heating
process~\citep[e.g.,][]{Ghisellini99,Stern04,Asano09}.
This possibility was confirmed by \cite{Xu17} recently, who considered the second-order Fermi acceleration
in the turbulent reconnection~\citep{ZhangB11}.
In future studies, we will incorporate the acceleration term into Equation (\ref{eq:continuity})
and investigate the effect in more details.

Some parameters are fixed in our calculations, such as $N_{\rm inj}^{\prime}$, $\Gamma$ etc.
Although these parameters may be variable in reality,
our results can still be a useful baseline for further detailed explorations.
Actually, the variation of these parameters can be easily taken into account
in our numerical code, by setting the boundary conditions as time-dependent functions.
Furthermore, more physical processes could be considered instantly.
For example, the scattering of electrons by various external radiation fields~\citep{Yan16}
may also be an important cooling mechanism.
Additionally, the electrons that are not efficiently accelerated (the low-energy electrons of Maxwellian distribution)
may provide considerable low-energy seed photons.
For the complete test of a GRB model, it is necessary to consider the dynamics of the
GRB jet at much earlier stages.
For example, a Poynting-flux-dominated jet may undergo accelerating when it is emitting~\citep{Uhm16b},
while the final Lorentz factor of merged ejecta is determined by the momentum of early shells.
In the future, we will improve our code to simulate the dynamical processes of GRBs more physically,
deriving the light curves and spectra at the same time.
Such complete theoretical fitting to the observed spectra and light curves will
give more clues on the characteristics of the emission region.
We note that the Hard X-ray Modulation Telescope (HXMT)~\citep{Li07,Xie15} launched by China recently can cover
an energy range of 1~---~250 keV. It will be efficient in collecting the time-resolved low-energy spectra
of GRBs in the near future. Our methodology proposed here will be helpful to make use of the HXMT data.

\acknowledgments
We thank the anonymous referee for valuable suggestions.
We also thank Z. Lucas Uhm, Mi-Xiang Lan, Liang Li and Bing Li for helpful discussion. 
J.J.G thanks Pengfei Chen for useful help on the numerical methods.  
This work is partially supported by
the National Postdoctoral Program for Innovative Talents (Grant No. BX201700115),
China Postdoctoral Science Foundation funded project (Grant No. 2017M620199),
the National Natural Science Foundation of China (Grants No. 11473012, 11673068 and 11725314),
the National Basic Research Program of China (``973'' Program, Grant No. 2014CB845800),
and by the Strategic Priority Research Program of the Chinese Academy of Sciences
``Multi-waveband Gravitational Wave Universe'' (Grant No. XDB23040000).

\appendix

\section{Numerical Method}
\label{app:Method}
In this appendix, we present the discretization procedure to solve Equation (\ref{eq:continuity}),
i.e., the CIP method (see \citealt{Yabe01} for a general review).
For an one-dimensional non-linear equation
\begin{equation}
\frac{\partial f}{\partial t} + \frac{\partial u f}{\partial x} = g,
\end{equation}
it is expedient to separate the solution procedure into two fractional steps.
One is the advection phase,
\begin{eqnarray}
\frac{\partial f}{\partial t} + u \frac{\partial f}{\partial x} &=& 0, \\
\frac{\partial \hat{f}}{\partial t} + u \frac{\hat{f}}{\partial x} &=& 0,
\end{eqnarray}
and the other is the non-convection phase,
\begin{eqnarray}
\frac{\partial f}{\partial t} &=& g - f \frac{\partial u}{\partial x} = G, \label{eq:nonconv1} \\
\frac{\partial \hat{f}}{\partial t} &=& \hat{G} - \hat{f} \frac{\partial u}{\partial x}, \label{eq:nonconv2}
\end{eqnarray}
where $\hat{f} = \partial f / \partial x$, $\hat{G} = \partial G / \partial x$ stands for the spatial derivative of $f$ and $G$.
$\hat{f}$ should be solved together with $f$ in the CIP method, which is crucial to obtain the
propagation of the spatial derivative during the evolution.
If we assume that the profile between two adjacent points can be interpolated by the cubic polynomial
$F(x) = a x^3 + b x^2 + c x + d$, then the solution at grid $i$ can be evolved from step $n$ to step $n+1$ by
\begin{eqnarray}
f_i^{n+1} &=& a_i \xi^3 + b_i \xi^2 + \xi \hat{f}_i^* + f_i^*, \label{eq:fn+1} \\
\hat{f}_i^{n+1} &=& 3 a_i \xi^2 + 2 b_i \xi + \hat{f}_i^*, \label{eq:fhn+1}
\end{eqnarray}
where $a_i = \frac{\hat{f}_i^* + \hat{f}_{i-1}^*}{\Delta x^2} - 2 \frac{f_i^*-f_{i-1}^*}{\Delta x^3}$,
$b_i = 3 \frac{f_{i-1}^*-f_i^*}{\Delta x^2} + \frac{2 \hat{f}_i^*+\hat{f}_{i-1}^*}{\Delta x}$,
$\xi = - u \Delta t$, and $\Delta t$ is the time step. Note that if $u < 0$ (just the case for electron cooling), one should replace $i -1$ with $i + 1$,
and $\Delta x$ with $- \Delta x$.
$f_i^*$ and $\hat{f}_i^*$ in Equations (\ref{eq:fn+1}) and (\ref{eq:fhn+1}) are the the intermediate solution from the non-convection phase,
and can be obtained through centered finite-difference of (\ref{eq:nonconv1}) and (\ref{eq:nonconv2}), i.e.,
\begin{eqnarray}
f_i^* &=& f_i^n + G_i \Delta t, \\
\hat{f}_i^* &=& \hat{f}_i^n + \frac{G_{i+1} - G_{i-1}}{2 \Delta x} \Delta t - \hat{f}_i^n \frac{u_{i+1} - u_{i-1}}{2 \Delta x} \Delta t \\
&=& \hat{f}_i^n + \frac{(f_{i+1}^* - f_{i+1}^n)-(f_{i-1}^* - f_{i-1}^n)}{2 \Delta x} - \hat{f}_i^n \frac{u_{i+1} - u_{i-1}}{2 \Delta x} \Delta t.
\end{eqnarray}

For the specific case in this paper, since the thermal Lorentz factor $\gamma_{\rm e}^{\prime}$ ranges from 10 to $10^7$,
it is necessary to solve Equation (\ref{eq:continuity}) in the logarithm space of $\gamma_{\rm e}^{\prime}$.
We determine $\log_{10} \gamma_{\rm e}^{\prime} = x$, so that
$\frac{d N_{\rm e}}{d x} = \ln 10 \gamma_{\rm e}^{\prime} \frac{d N_{\rm e}}{d \gamma_{\rm e}^{\prime}}$,
and the Equation (\ref{eq:continuity}) is transformed to
\begin{equation}
\frac{\partial}{\partial t^{\prime}} \left( \frac{d N_{\rm e}}{d x} \right) + \frac{\partial}{\partial x}
\left[ \frac{d x}{d t^{\prime}} \left( \frac{d N_{\rm e}}{d x} \right) \right]
= Q(x,t^{\prime}) \gamma_{\rm e}^{\prime} \ln 10,
\end{equation}
where $\frac{d N_{\rm e}}{d x}$ is actually what we want to solve in our code.
In all the calculations, we set the range of $x$ to be [1,8], and the total grids number $N_{\rm num}$ to be 401.
The time step for every evolution is determined by the Courant condition
\begin{equation}
\Delta t^{\prime} \le \frac{\Delta \gamma_{\rm e}^{\prime}}{\dot{\gamma}_{\rm e,tot}^{\prime}} \vert_{\rm min}
= \frac{x_{\rm max}-x_{\rm min}}{N_{\rm num}-1} \ln 10 \frac{\gamma_{\rm max}^{\prime}}{\dot{\gamma}_{\rm max,tot}^{\prime}},
\end{equation}
where $\gamma_{\rm max}^{\prime}$ is the maximum thermal Lorentz factor and can be
given by the approximation
$\gamma_{\rm max}^{\prime} \simeq 10^8 \left( \frac{B^{\prime}}{1~\mathrm{G}} \right)^{-0.5}$
\citep{Dai99,Huang00}.
However, the uncertainty of the real value of $\gamma_{\rm max}^{\prime}$ has little influence on the
evolution of the electron distribution and the spectra we focus on.

\section{Formulations for Radiation}
\label{app:Radiation}
In the co-moving frame, the synchrotron radiation power at frequency $\nu^{\prime}$ is~\citep{Rybicki79}
\begin{equation}
P^{\prime}(\nu^{\prime}) = \frac{\sqrt{3} q_{\rm e}^3 B^{\prime}}{m_{\rm e} c^2} \int_{\gamma_{\rm e,min}^{\prime}}^{\gamma_{\rm e,max}^{\prime}}
\left( \frac{d N_{\rm e}}{d \gamma_{\rm e}^{\prime}} \right) F \left( \frac{\nu^{\prime}}{\nu_{\rm c}^{\prime}} \right) d \gamma_{\rm e}^{\prime},
\end{equation}
where $\nu_{\rm c}^{\prime} = 3 q_{\rm e} B^{\prime} \gamma_{\rm e}^{\prime 2} /(4 \pi m_{\rm e} c)$,
$F(x) = x \int_{x}^{+\infty} K_{5/3}(k) dk$, and $K_{5/3}(k)$ is the Bessel function.
The synchrotron seed photon spectra can then be calculated as~\citep{Fan08}
\begin{equation}
n_{\nu^{\prime}} \simeq \frac{T^{\prime}}{h \nu^{\prime}} \frac{\sqrt{3} q_{\rm e}^3 B^{\prime}}{m_{\rm e} c^2}
\int_{\gamma_{\rm e,min}^{\prime}}^{\gamma_{\rm e,max}^{\prime}} n^{\prime}(\gamma_{\rm e}^{\prime}) F \left( \frac{\nu^{\prime}}{\nu_{\rm c}^{\prime}} \right)
d \gamma_{\rm e}^{\prime},
\end{equation}
where $T^{\prime} \approx \Delta / c$ is the time that the synchrotron radiation photons stay within the
ejecta, $n^{\prime}(\gamma_{\rm e}^{\prime}) = \frac{d N_{\rm e} / d \gamma_{\rm e}^{\prime}}{4 \pi \Delta R^2}$
is the co-moving electron number density,
and $\Delta \approx R / \Gamma$ is the co-moving width of the ejecta.
One should notice that $\Delta$ does not appear in our calculations since
the $\Delta$ in $T^{\prime}$ and $n^{\prime}(\gamma_{\rm e}^{\prime})$ are canceled out.
Here, for simplicity, we have considered only the single scattering case for SSC cooling and ignored the multiple scattering process.

If we ignore the effect of the equal-arrival-time surface (EATS, \citealt{Waxman97,Granot99,Huang07,Geng17}),
then the observed spectral flux can be expressed as
\begin{equation}
F_{\nu_{\rm obs}} = \frac{(1+z) \Gamma P^{\prime}(\nu^{\prime} (\nu_{\rm obs}))}{4 \pi D_L^2},
\end{equation}
where $\nu^{\prime} = (1+z) \nu_{\rm obs} / \mathcal{D}$,
and $\mathcal{D} = 1/[\Gamma (1-\beta \cos \theta)]$ is the Doppler factor.
In this work, the luminosity distance $D_L$ is obtained by adopting a flat $\Lambda$CDM universe,
in which $H_0 = 71$~km~s$^{-1}$, $\Omega_{\rm m} = 0.27$, and $\Omega_{\Lambda} = 0.73$.
For all the calculations in this work, the burst is assumed to be at a cosmological redshift $z = 1$.
If we take the EATS effect into account, the observed spectral flux should be
\begin{equation}
F_{\nu_{\rm obs}} = \frac{1+z}{4 \pi D_L^2} \int_0^{\theta_j} P^{\prime}(\nu^{\prime} (\nu_{\rm obs})) \mathcal{D}^3 \frac{\sin \theta}{2} d \theta,
\end{equation}
where $\theta_j$ is the half-opening angle of the jet.
The integration of $\theta$ is performed over an elliptical surface (or a sequence of $R_{\theta}$),
which is determined by~\citep{Geng16}
\begin{equation}
t_{\rm obs} = (1+z) \int_0^{R_{\theta}} \frac{1-\beta \cos \theta}{\beta c} dr \equiv {\rm const},
\end{equation}
from which $R_{\theta}$ can be derived for a given $\theta$.
However, we found that there is little difference for the spectrum calculated
from the case with the EATS effect and the case without in our calculations.
Also the EATS effect has no significant influence on the low-energy indices of the flux spectra.
So we just show the spectra calculated without the EATS effect in this paper.

\begin{deluxetable}{cccccccc}
\tabletypesize{\scriptsize}
\tablewidth{0pt}
\tablecaption{Parameters used in the testing calculations.\label{TABLE:testing}}

\tablehead{%
        \colhead{Model} &
        \colhead{$\Gamma$} &
        \colhead{$\gamma_{\rm m}^{\prime}$~($10^5$)} &
        \colhead{$B_0^{\prime}$~(G)} &
        \colhead{$N_{\rm inj}^{\prime}$~($10^{47}$~s$^{-1}$)}&
        \colhead{$q$} &
        \colhead{Adiabatic} &
        \colhead{SSC}
        }
\startdata
M1      &   300      &   1   &   30   &   1   &  0  & No   &   No  \\
M2      &   300      &   1   &   30   &   1   &  1  & Yes  &   No   \\
M3      &   300      &   1   &   30   &   1   &  1  & Yes  &   Yes    \\
M4      &   300      &   1   &   30   &   1   &  0  & Yes  &   Yes
\enddata
\tablecomments{In this group of calculations, one should note that $R_0 = 10^{15}$~cm is adopted, whereas the
starting radius at which the jet begin to produce emission is $R_{\rm s} = 10^{14}$~cm.
This is only to achieve the same initial conditions as those in \cite{Uhm14}.
However, $R_{\rm s} = R_0$ is commonly used all through this paper.}
\end{deluxetable}

\begin{deluxetable}{ccccccc}
\tabletypesize{\scriptsize}
\tablewidth{0pt}
\tablecaption{Parameters used in the calculations of Group PJR15 ($R_0 = 10^{15}$~cm).\label{TABLE:I}}
\tablehead{%
        \colhead{Model} &
        \colhead{$\Gamma$} &
        \colhead{$\gamma_{\rm m}^{\prime}$} &
        \colhead{$B_0^{\prime}$} &
        \colhead{$N_{\rm inj}^{\prime}$}&
        \colhead{$L_{\rm e}$} &
        \colhead{$L_{B}$} \\
        \colhead{} & \colhead{} &  \colhead{($10^4$)}  &  \colhead{($10^2$~G)} &  \colhead{($10^{47}$~s$^{-1}$)} &
        \colhead{($10^{51}$~erg~s$^{-1}$)} &  \colhead{($10^{51}$~erg~s$^{-1}$)}
        }
\startdata
PJR15$\Gamma$1000     &  1000    &   1.3  &  3.4   &  1.6    &   1.7  &     1.8  \\
PJR15$\Gamma$600      &  600     &   1.3  &  5.7   &  4.5    &   1.7  &     1.8  \\
PJR15$\Gamma$300      &  300     &   1.3  &  11    &  18     &   1.7  &     1.8   \\
PJR15$\Gamma$100      &  100     &   1.3  &  34    &  160   &   1.7  &     1.8   \\
PJR15$\Gamma$50       &  50      &   1.3  &  69    &  650   &   1.7  &     1.8
\enddata
\tablecomments{
From PJR15$\Gamma$1000 to PJR15$\Gamma$50, $\Gamma$ is decreasing while $B_0^{\prime}$ is increasing.
This group of calculations are within the Poynting-flux-dominated jet scenario and correspond to case of $L_{\rm e} \simeq L_B$.}
\end{deluxetable}

\begin{deluxetable}{ccccccc}
\tabletypesize{\scriptsize}
\tablewidth{0pt}
\tablecaption{Parameters used in the calculations of Group PJR14 ($R_0 = 10^{14}$~cm).\label{TABLE:II}}
\tablehead{%
        \colhead{Model} &
        \colhead{$\Gamma$} &
        \colhead{$\gamma_{\rm m}^{\prime}$} &
        \colhead{$B_0^{\prime}$} &
        \colhead{$N_{\rm inj}^{\prime}$}&
        \colhead{$L_{\rm e}$} &
        \colhead{$L_{B}$} \\
        \colhead{} &  \colhead{} &  \colhead{($10^4$)}  &  \colhead{($10^3$~G)} &  \colhead{($10^{47}$~s$^{-1}$)} &
        \colhead{($10^{51}$~erg~s$^{-1}$)} &  \colhead{($10^{51}$~erg~s$^{-1}$)}
        }
\startdata
PJR14$\Gamma$1000   &  1000    &   0.42  &  3.3  &  5      & 1.7   &  1.6  \\
PJR14$\Gamma$600    &   600    &   0.42  &  5.5  &  14    & 1.7   &  1.6  \\
PJR14$\Gamma$300    &   300    &   0.42  &  11   &   56   &  1.7  &   1.6 \\
PJR14$\Gamma$100    &   100    &   0.42  &  33   &  500  &  1.7  &  1.6
\enddata
\tablecomments{
From PJR14$\Gamma$1000 to PJR14$\Gamma$100, $\Gamma$ is decreasing while $B_0^{\prime}$ is increasing.
This group of calculations are within the Poynting-flux-dominated jet scenario and correspond to case of $L_{\rm e} \simeq L_B$.
It differs from Table \ref{TABLE:I} mainly on the values of $R_0$ and $\gamma_{\rm m}^{\prime}$.}
\end{deluxetable}

\begin{deluxetable}{cccccccc}
\tabletypesize{\scriptsize}
\tablewidth{0pt}
\tablecaption{Parameters used in the calculations of Group IS20R15 ($\eta_{\rm p} = 20$, $R_0 = 10^{15}$~cm).\label{TABLE:A}}
\tablehead{%
        \colhead{Model} &
        \colhead{$\Gamma$} &
        \colhead{$\gamma_{\rm m}^{\prime}$} &
        \colhead{$B_0^{\prime}$} &
        \colhead{$N_{\rm inj}^{\prime}$}&
        \colhead{$L_{\rm e}$} &
        \colhead{$L_{B}$} &
        \colhead{$L_{\rm p}$} \\
        \colhead{} &  \colhead{} &  \colhead{($10^4$)}  &  \colhead{($10^2$~G)} &  \colhead{($10^{47}$~s$^{-1}$)} &
        \colhead{($10^{51}$~erg~s$^{-1}$)} &  \colhead{($10^{51}$~erg~s$^{-1}$)} & \colhead{($10^{51}$~erg~s$^{-1}$)}
        }
\startdata
IS20R15$\Gamma$1300   &  1300  &  1.5    &  2    &   0.83     &  1.7  &  1.0  &  4.2 \\
IS20R15$\Gamma$430    &  430   &  1.5    &  6    &   7.6       &  1.7  &  1.0  &  4.2 \\
IS20R15$\Gamma$260    &  260   &  1.5    &  10  &    21       &  1.7  &  1.0  &  4.2 \\
IS20R15$\Gamma$86     &  86    &  1.5    &  30  &   190      &  1.7  &  1.0  &  4.2
\enddata
\tablecomments{
This group of calculations are within the internal shock model and correspond to the case of $L_{\rm e}/L_B \approx 1$.}
\end{deluxetable}

\begin{deluxetable}{cccccccc}
\tabletypesize{\scriptsize}
\tablewidth{0pt}
\tablecaption{Parameters used in the calculations of Group IS100R15 ($\eta_{\rm p} = 100$, $R_0 = 10^{15}$~cm).\label{TABLE:B}}
\tablehead{%
        \colhead{Model} &
        \colhead{$\Gamma$} &
        \colhead{$\gamma_{\rm m}^{\prime}$} &
        \colhead{$B_0^{\prime}$} &
        \colhead{$N_{\rm inj}^{\prime}$}&
        \colhead{$L_{\rm e}$} &
        \colhead{$L_{B}$} &
        \colhead{$L_{\rm p}$} \\
        \colhead{} &  \colhead{} &  \colhead{($10^4$)}  &  \colhead{($10^2$~G)} &  \colhead{($10^{47}$~s$^{-1}$)} &
        \colhead{($10^{51}$~erg~s$^{-1}$)} &  \colhead{($10^{51}$~erg~s$^{-1}$)} & \colhead{($10^{51}$~erg~s$^{-1}$)}
        }
\startdata
IS100R15A$\Gamma$120  &  120   &  10    &  0.5  &   16      & 1.7  & $5.4 \times 10^{-4}$  &  3.2 \\
IS100R15A$\Gamma$60   &  60    &  10    &  1     &    58     &  1.7 & $5.4 \times 10^{-4}$  &  3.2 \\
IS100R15A$\Gamma$20   &  20    &  10    &  3     &    530   &  1.7 & $5.4 \times 10^{-4}$  &  3.2  \\ \hline

IS100R15B$\Gamma$460   &  460    &  5  &  0.5  &   2     & 1.7   & $8.0 \times 10^{-3}$  &  6.4 \\
IS100R15B$\Gamma$230   &  230    &  5  &  1     &   8     & 1.7   & $8.0 \times 10^{-3}$  &  6.4 \\
IS100R15B$\Gamma$77    &  77     &  5  &  3     &   71   & 1.7   & $8.0 \times 10^{-3}$  &  6.4 \\
IS100R15B$\Gamma$23    &  23     &  5  &  10   &  800  & 1.7   & $8.0 \times 10^{-3}$  &  6.4 \\  \hline

IS100R15C$\Gamma$1900   &  1900   &  1  &  3      &   0.58    & 1.7   & 5  &  32 \\
IS100R15C$\Gamma$580    &  580    &  1  &  10    &   6.2      & 1.7   & 5  &  32 \\
IS100R15C$\Gamma$58     &  58     &  1  &  100  &   620     & 1.7   & 5  &  32  \\ \hline

IS100R15D$\Gamma$3900  &  3900    &  0.7   &   3     &   0.2    & 1.7  & 21  &  46 \\
IS100R15D$\Gamma$1200  &  1200    &  0.7   &  10    &   2.1    & 1.7  & 21  &  46 \\
IS100R15D$\Gamma$120   &  120     &  0.7   &  100  &  210    & 1.7  & 21   & 46  \\
\enddata
\tablecomments{
This group of calculations are within the internal shock model and include four subgroups, of which
the values of $\gamma_{\rm m}^{\prime}$ are different and $L_{\rm e}/L_B$ ranges from 0.08 to 3000.}
\end{deluxetable}

\begin{deluxetable}{cccccccc}
\tabletypesize{\scriptsize}
\tablewidth{0pt}
\tablecaption{Parameters used in the calculations of Group IS10R14 ($\eta_{\rm p} = 10$, $R_0 = 10^{14}$~cm).\label{TABLE:C}}
\tablehead{%
        \colhead{Model} &
        \colhead{$\Gamma$} &
        \colhead{$\gamma_{\rm m}^{\prime}$} &
        \colhead{$B_0^{\prime}$} &
        \colhead{$N_{\rm inj}^{\prime}$}&
        \colhead{$L_{\rm e}$} &
        \colhead{$L_{B}$} &
        \colhead{$L_{\rm p}$} \\
        \colhead{} &  \colhead{} &  \colhead{($10^4$)}  &  \colhead{($10^2$~G)} &  \colhead{($10^{47}$~s$^{-1}$)} &
        \colhead{($10^{51}$~erg~s$^{-1}$)} &  \colhead{($10^{51}$~erg~s$^{-1}$)} & \colhead{($10^{51}$~erg~s$^{-1}$)}
        }
\startdata
IS10R14A$\Gamma$1900  &  1900      &  1   &  3      &   0.58    & 1.7  & 0.05   &  3.1 \\
IS10R14A$\Gamma$580   &  580       &  1   &  10    &   6.2      & 1.7  & 0.05   &  3.1 \\
IS10R14A$\Gamma$58    &  58        &  1   &  100  &   620     & 1.7  & 0.05   &  3.1  \\ \hline

IS10R14B$\Gamma$2300  &  2300  &  0.5  &  10   &   0.79  & 1.7  & 0.8  &  6.3 \\
IS10R14B$\Gamma$770   &  770   &  0.5  &  30   &   7.1    & 1.7  & 0.8  &  6.3 \\
IS10R14B$\Gamma$230   &  230   &  0.5  &  100  &  79     & 1.7  & 0.8  &  6.3 \\
\enddata
\tablecomments{
This group of calculations are within the internal shock model and include two subgroups, of which
the values of $\gamma_{\rm m}^{\prime}$ are different and $L_{\rm e}/L_B$ ranges from 2 to 30.}
\end{deluxetable}

\begin{deluxetable}{cccccccc}
\tabletypesize{\scriptsize}
\tablewidth{0pt}
\tablecaption{Parameters used in the calculations of Group IS100R14 ($\eta_{\rm p} = 100$, $R_0 = 10^{14}$~cm).\label{TABLE:D}}
\tablehead{%
        \colhead{Model} &
        \colhead{$\Gamma$} &
        \colhead{$\gamma_{\rm m}^{\prime}$} &
        \colhead{$B_0^{\prime}$} &
        \colhead{$N_{\rm inj}^{\prime}$}&
        \colhead{$L_{\rm e}$} &
        \colhead{$L_{B}$} &
        \colhead{$L_{\rm p}$} \\
        \colhead{} &  \colhead{} &  \colhead{($10^4$)}  &  \colhead{($10^2$~G)} &  \colhead{($10^{47}$~s$^{-1}$)} &
        \colhead{($10^{51}$~erg~s$^{-1}$)} &  \colhead{($10^{51}$~erg~s$^{-1}$)} & \colhead{($10^{51}$~erg~s$^{-1}$)}
        }
\startdata
IS100R14A$\Gamma$120  &  120   &  10   &  0.5  &   16     & 1.7   & $5.4 \times 10^{-6}$  &  3.2 \\
IS100R14A$\Gamma$60   &  60    &  10   &  1     &   58     &  1.7  & $5.4 \times 10^{-6}$  &  3.2 \\
IS100R14A$\Gamma$20   &  20    &  10   &  3     &   530   &  1.7  & $5.4 \times 10^{-6}$  &  3.2  \\ \hline

IS100R14B$\Gamma$360  &  360   &   4    &  1     &   4     & 1.7  & $2.0 \times 10^{-4}$  &  7.9 \\
IS100R14B$\Gamma$120  &  120   &   4    &  3     &   36   & 1.7  & $2.0 \times 10^{-4}$  &  7.9 \\
IS100R14B$\Gamma$36   &  36    &   4    &  10   &  400  & 1.7 &  $2.0 \times 10^{-4}$  &  7.9 \\  \hline

IS100R14C$\Gamma$580 &  580    &  1    & 10     &   6.2     & 1.7  & 0.05  &  31 \\
IS100R14C$\Gamma$58  &  58  &  1    &  100  &   620    & 1.7  & 0.05  &  31 \\ \hline

IS100R14D$\Gamma$2300  &  2300    &  0.5    &  10    &  0.78   & 1.7  & 0.81  &  63 \\
IS100R14D$\Gamma$230   &  230     &  0.5    &  100  &  78      & 1.7  & 0.81  &  63
\enddata
\tablecomments{
This group of calculations are within the internal shock model and include four subgroups, of which
the values of $\gamma_{\rm m}^{\prime}$ are different and $L_{\rm e}/L_B$ ranges from 2 to $3 \times 10^5$.}
\end{deluxetable}

\clearpage

\clearpage

\clearpage

\clearpage

\clearpage

\clearpage

\clearpage

\clearpage

\clearpage

\clearpage

\clearpage

\clearpage

\clearpage

\clearpage

\begin{figure}
\begin{adjustwidth}{-2cm}{-2cm}
\centering
    \subfloat{\includegraphics[width=0.25\paperwidth]{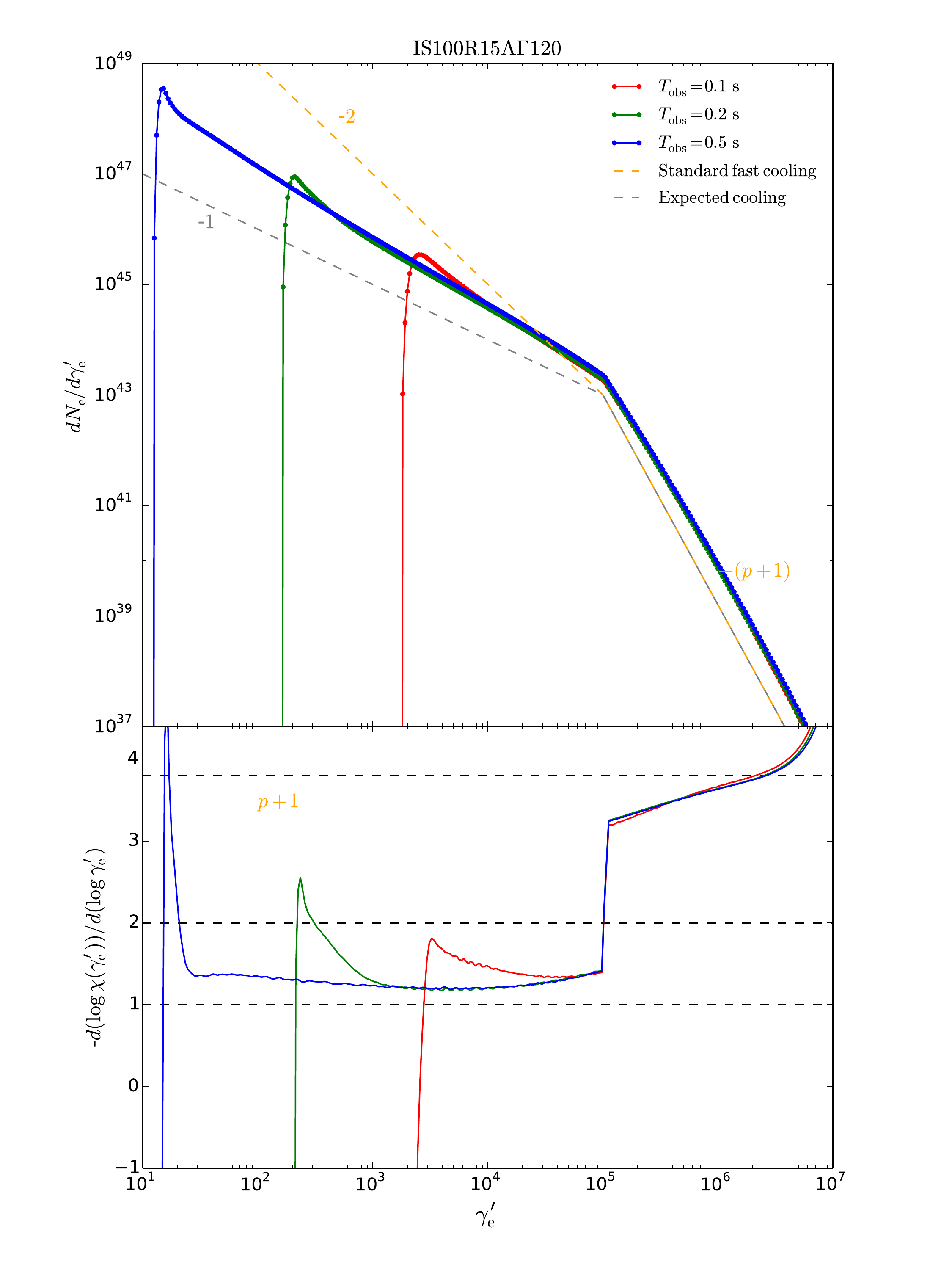}}
    \subfloat{\includegraphics[width=0.25\paperwidth]{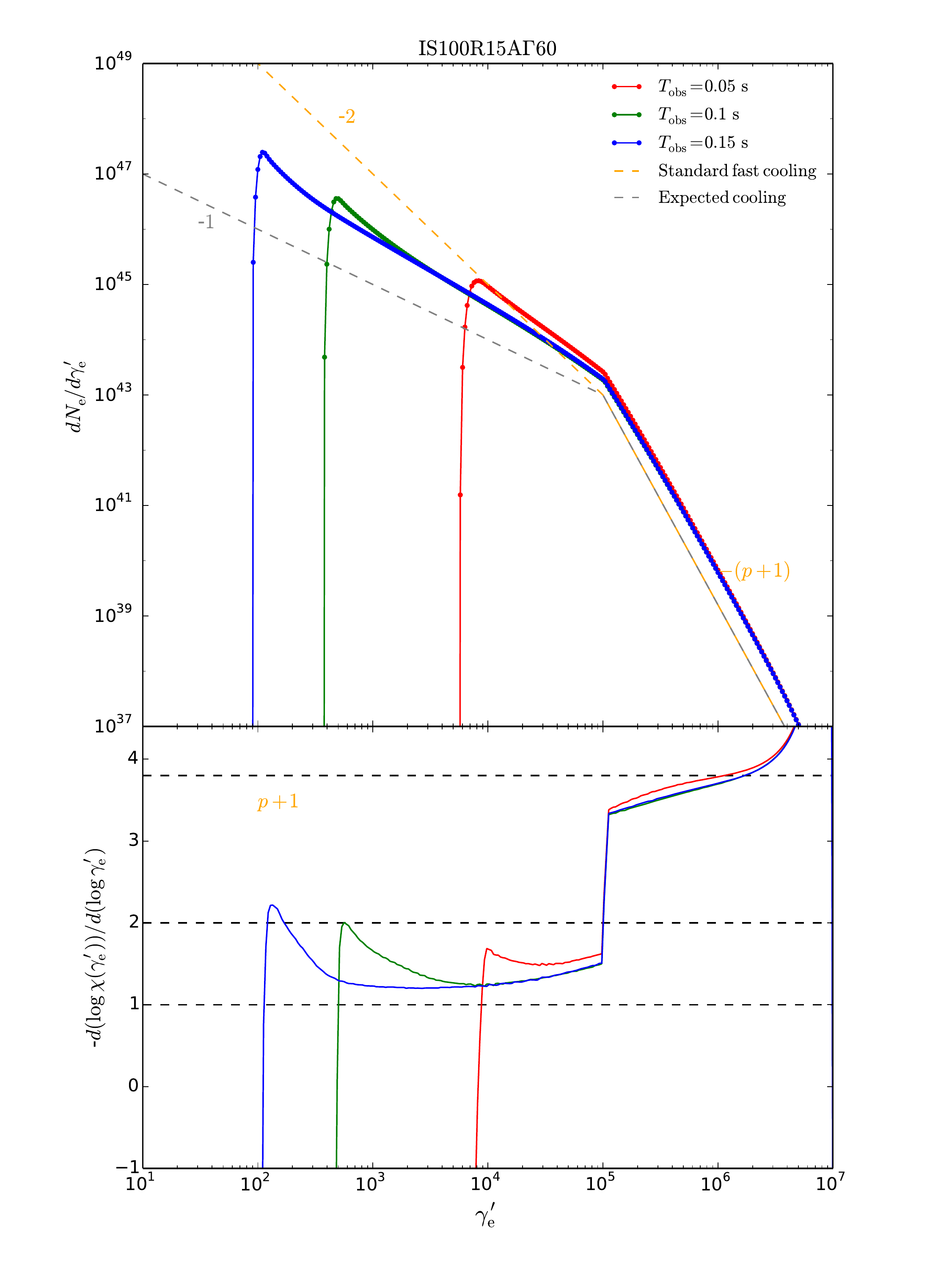}}
    \subfloat{\includegraphics[width=0.25\paperwidth]{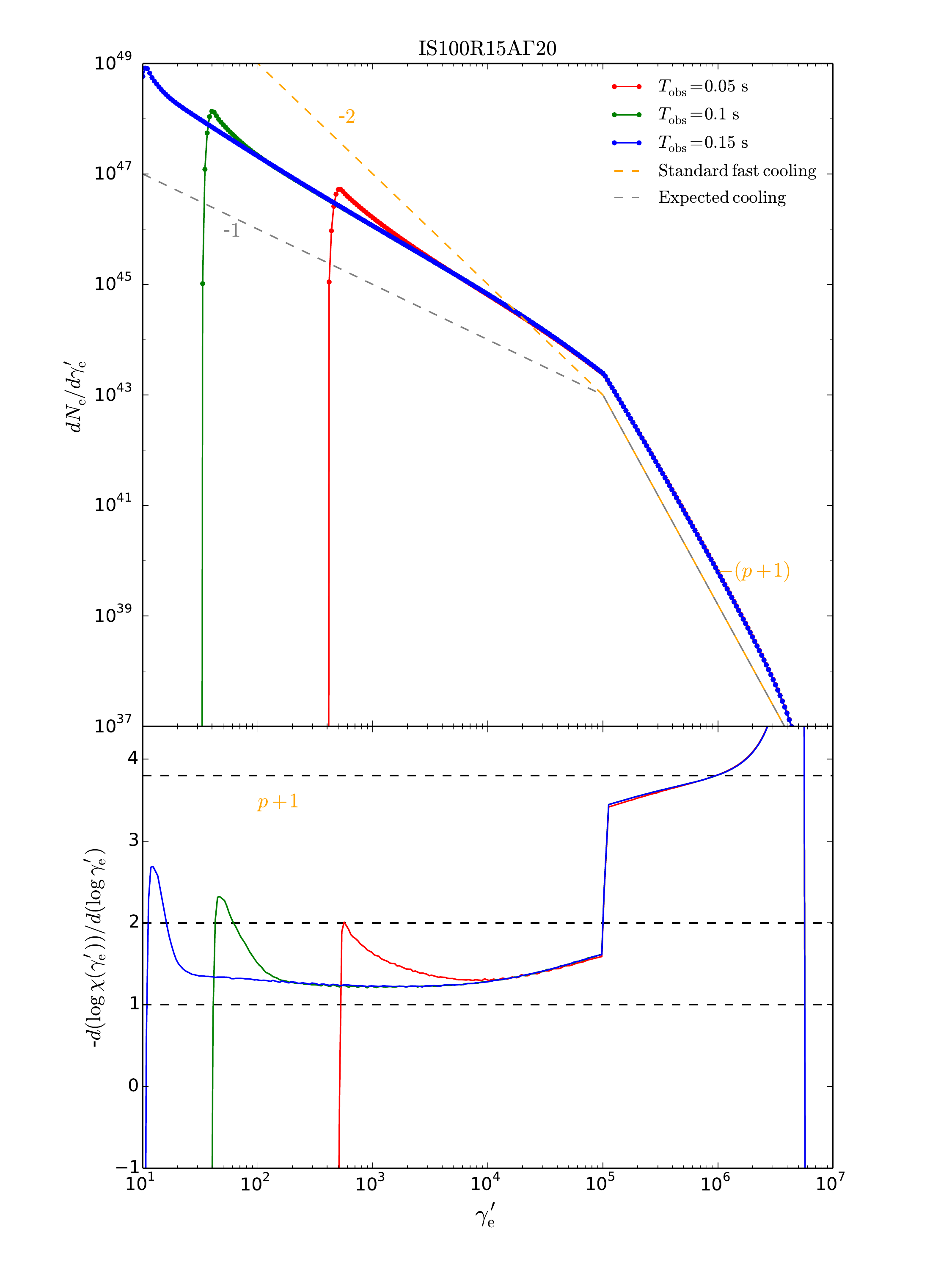}} \\
    \subfloat{\includegraphics[width=0.25\paperwidth]{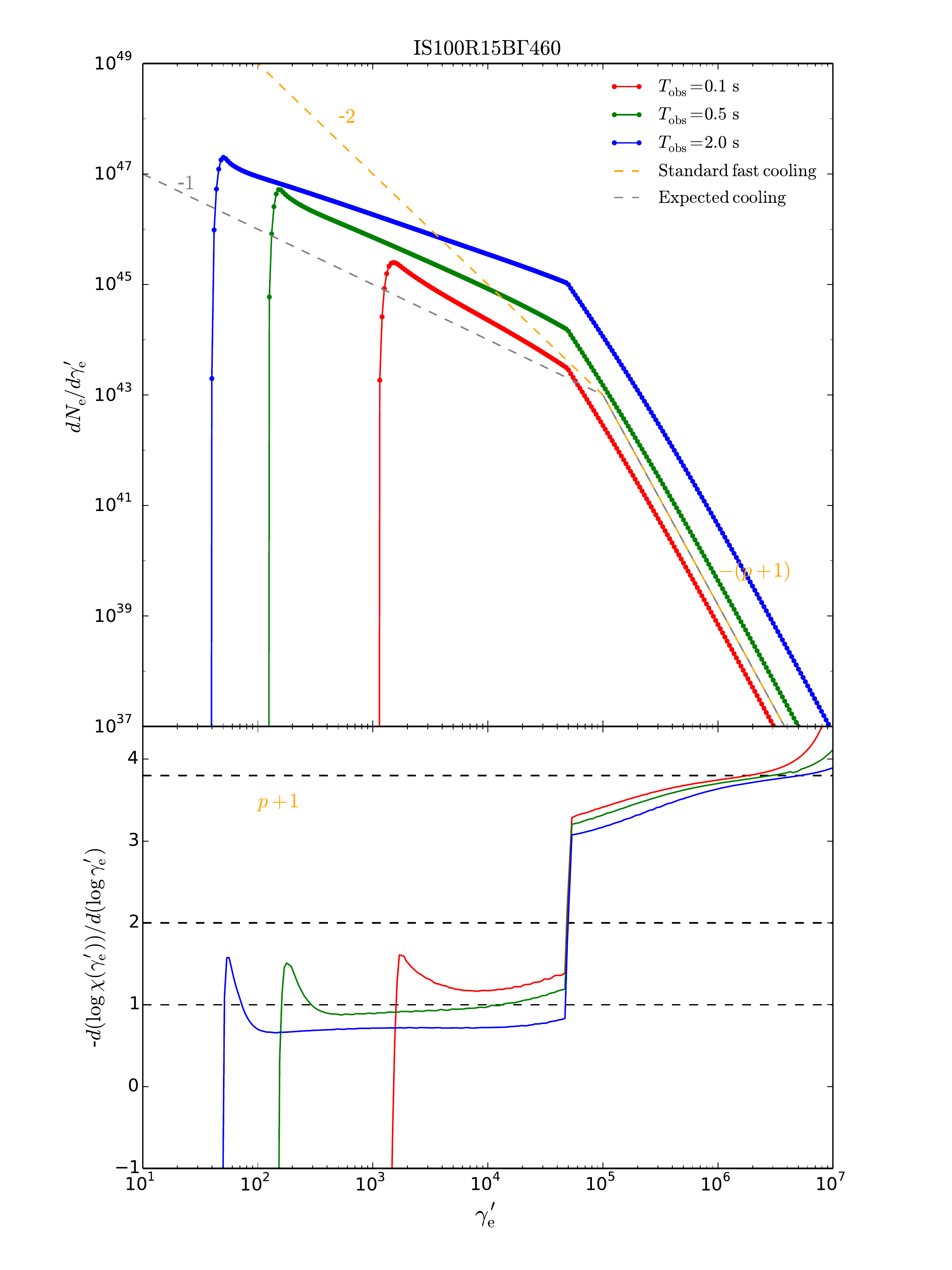}}
    \subfloat{\includegraphics[width=0.25\paperwidth]{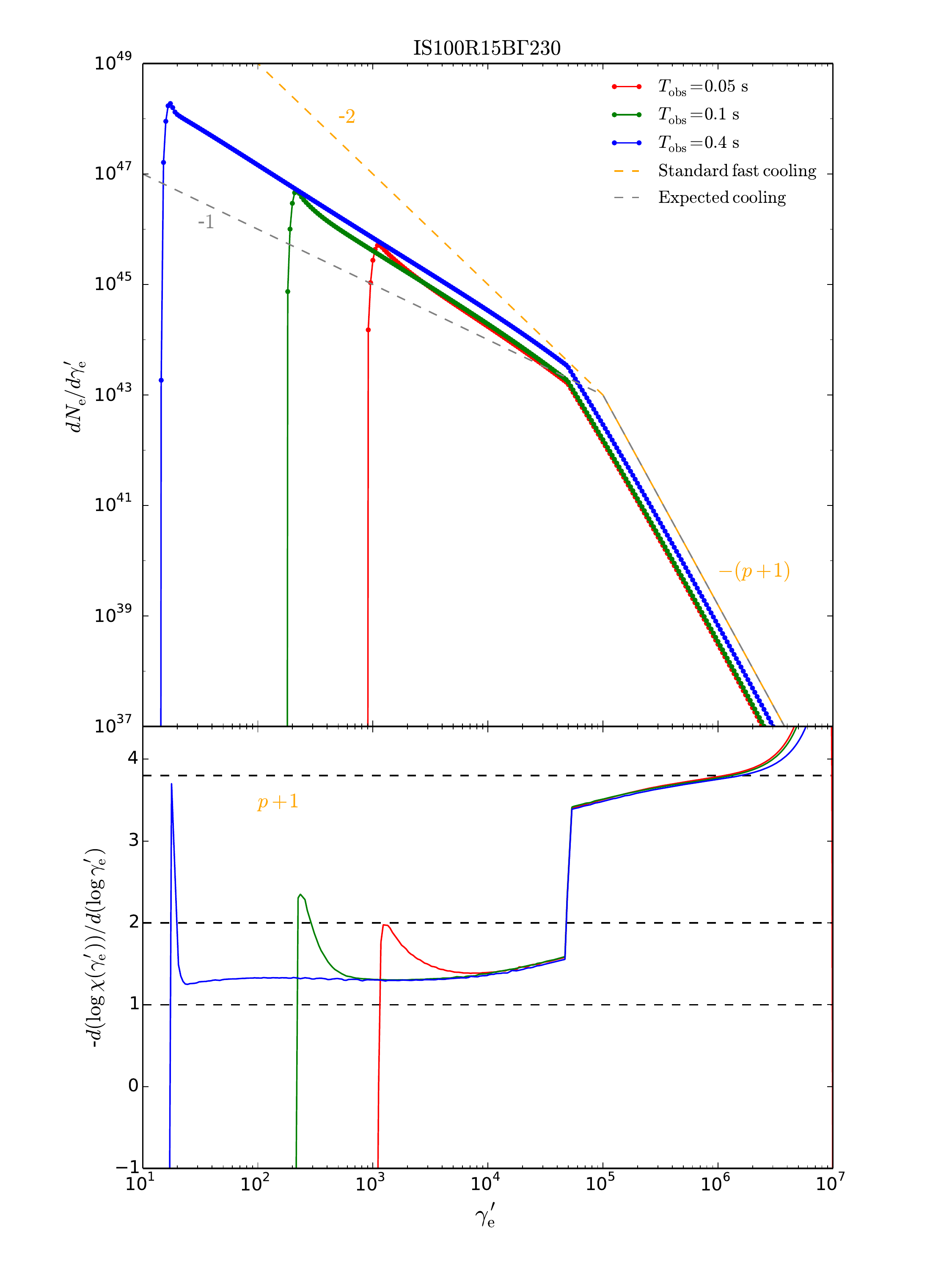}}
    \subfloat{\includegraphics[width=0.25\paperwidth]{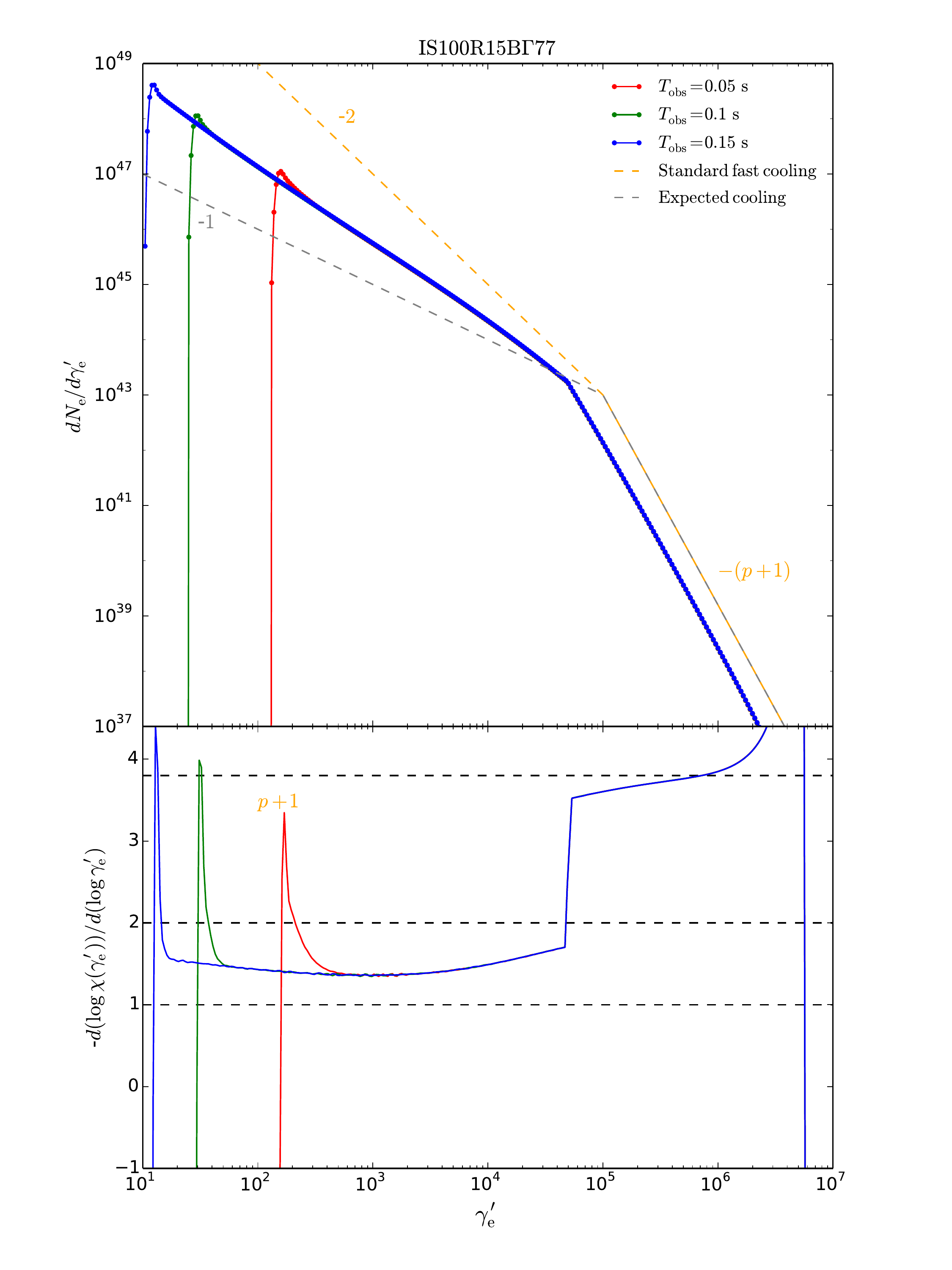}} \\
    \subfloat{\includegraphics[width=0.25\paperwidth]{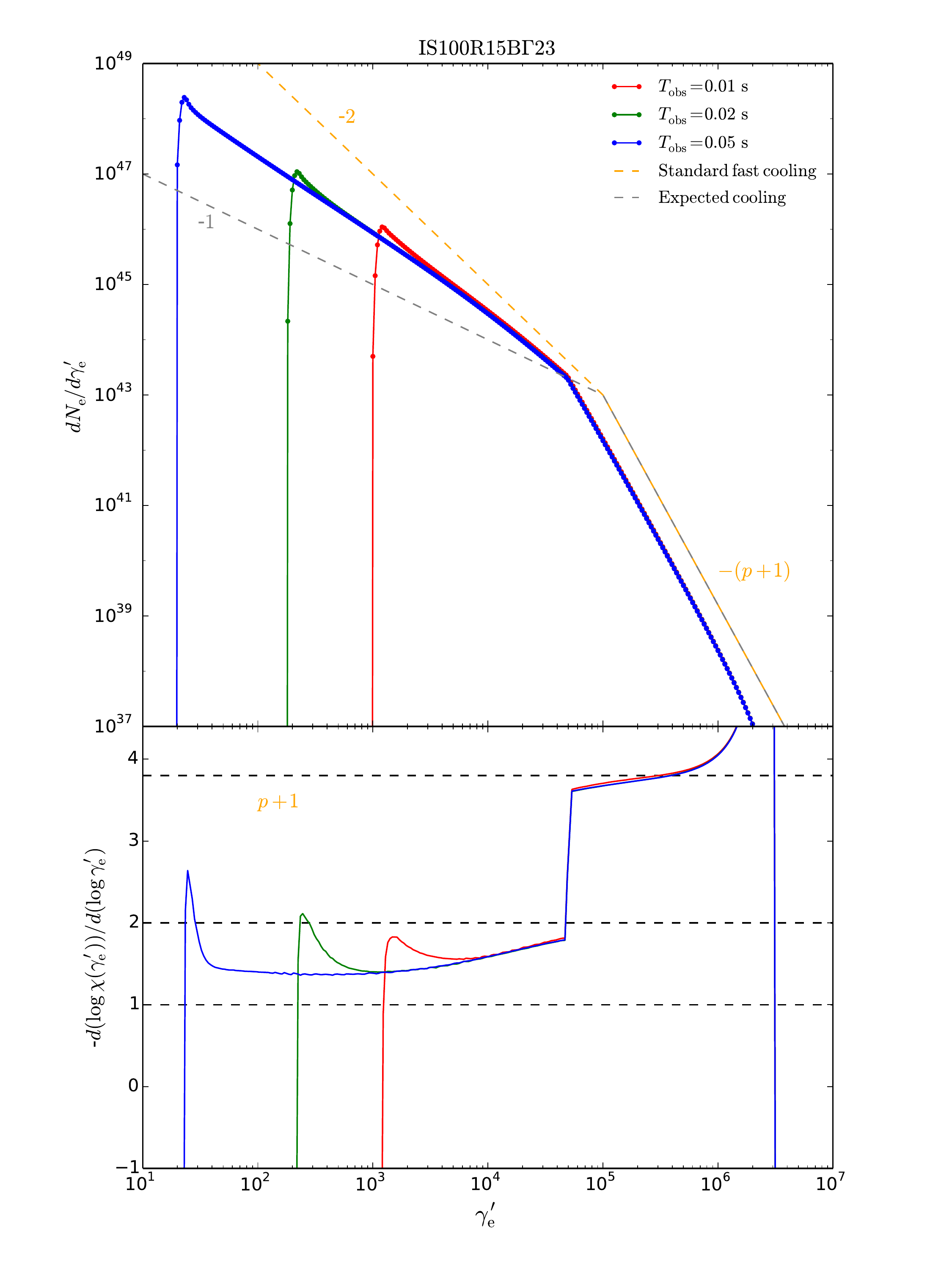}}
    \subfloat{\includegraphics[width=0.25\paperwidth]{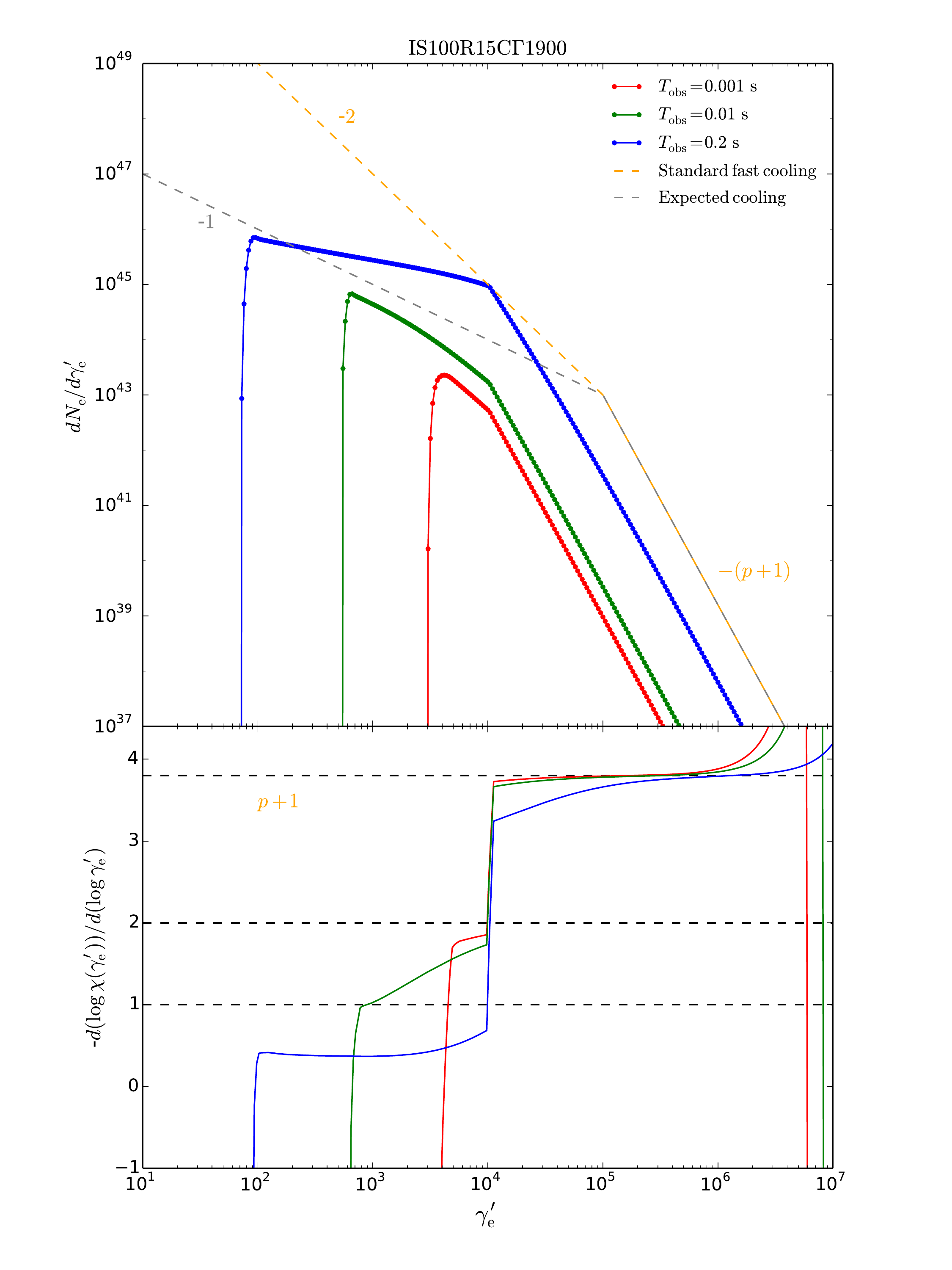}}
    \subfloat{\includegraphics[width=0.25\paperwidth]{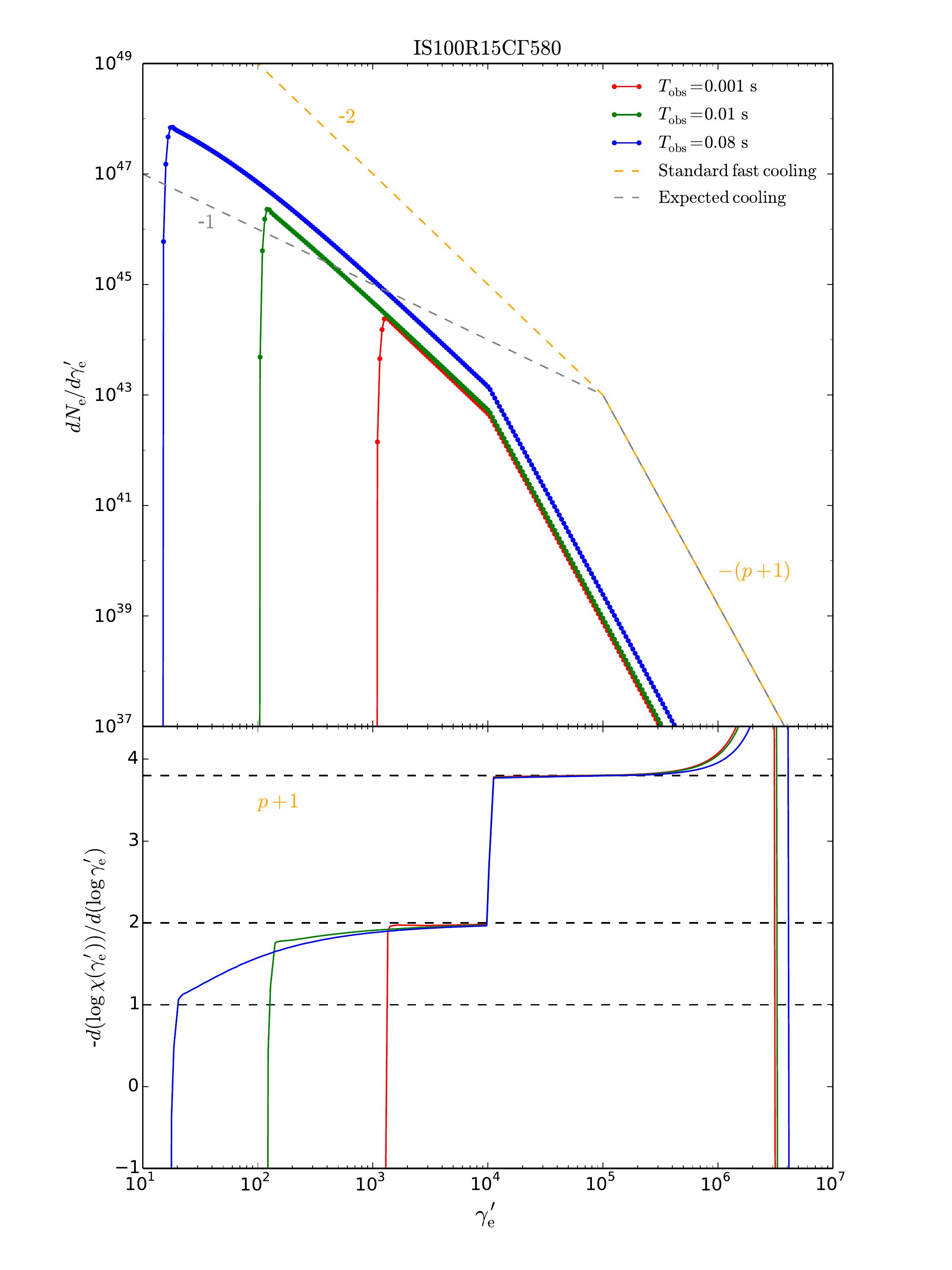}}
    \caption{The evolution of the electron energy spectrum for the thirteen cases in Group IS100R15 (see Table \ref{TABLE:B}). \label{fig:MB-electron}}
\end{adjustwidth}
\end{figure}
\clearpage
\begin{figure}
\ContinuedFloat
\begin{adjustwidth}{-2cm}{-2cm}
\centering
    \subfloat{\includegraphics[width=0.25\paperwidth]{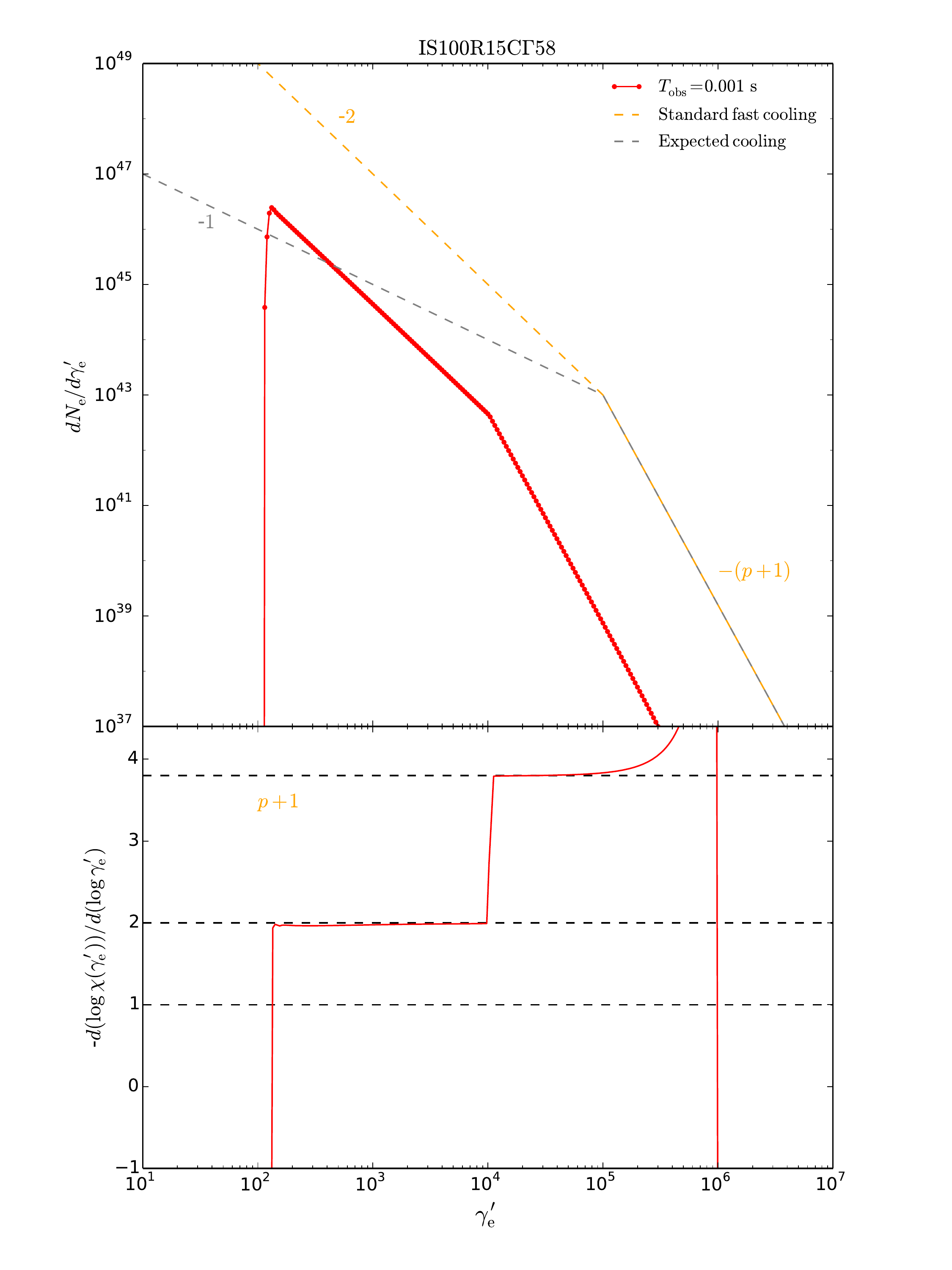}}
    \subfloat{\includegraphics[width=0.25\paperwidth]{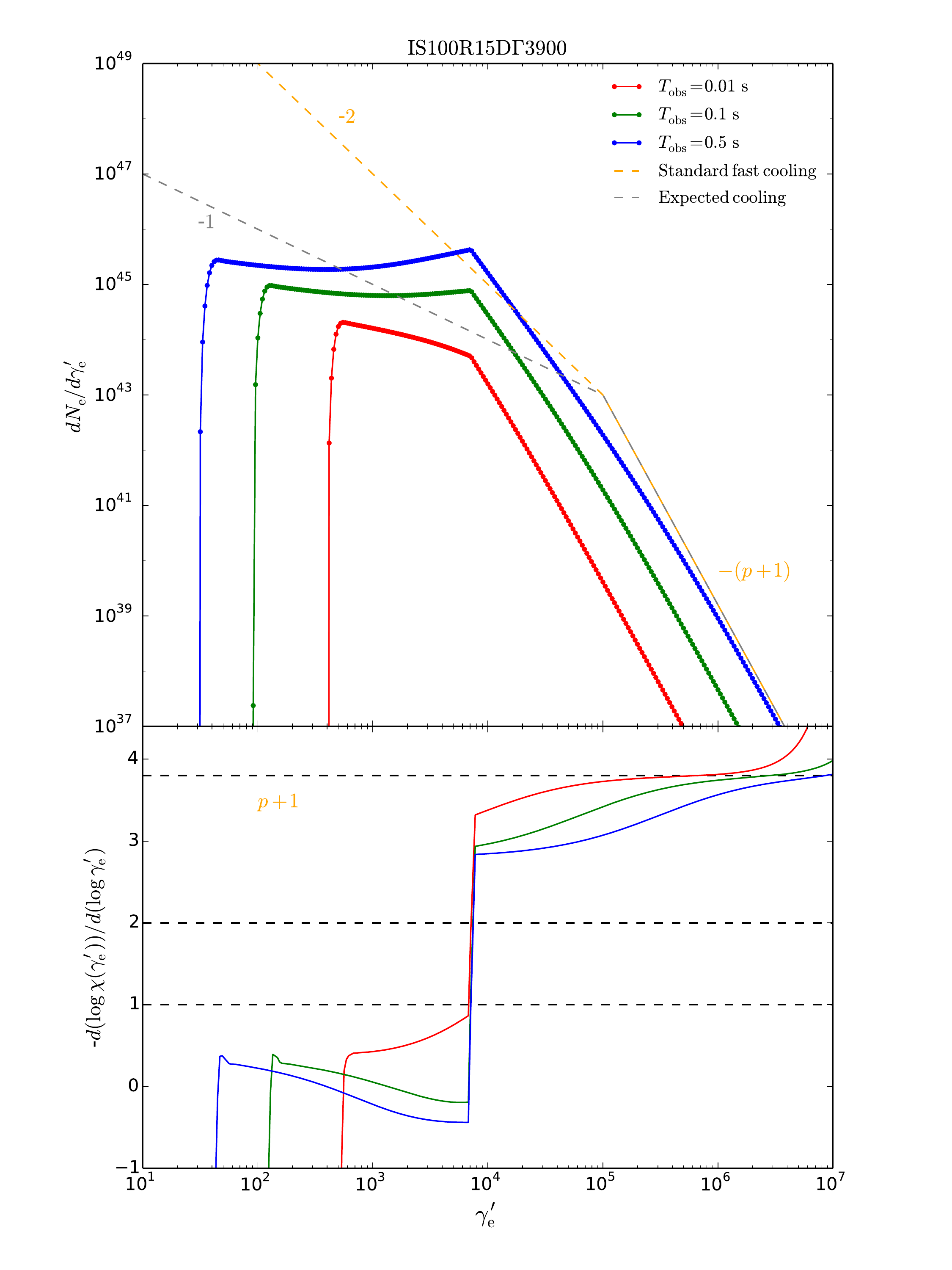}}
    \subfloat{\includegraphics[width=0.25\paperwidth]{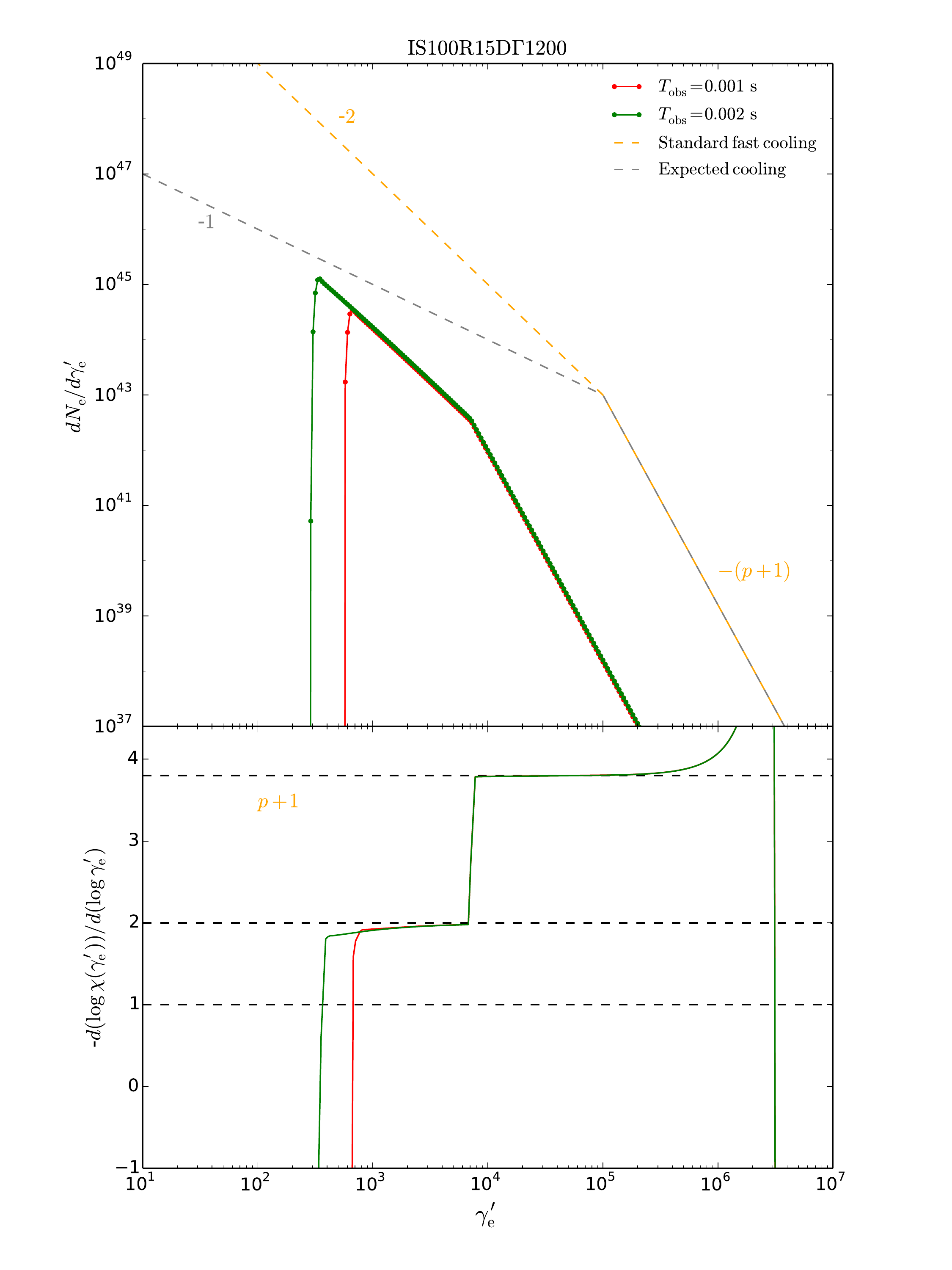}} \\
    \subfloat{\includegraphics[width=0.25\paperwidth]{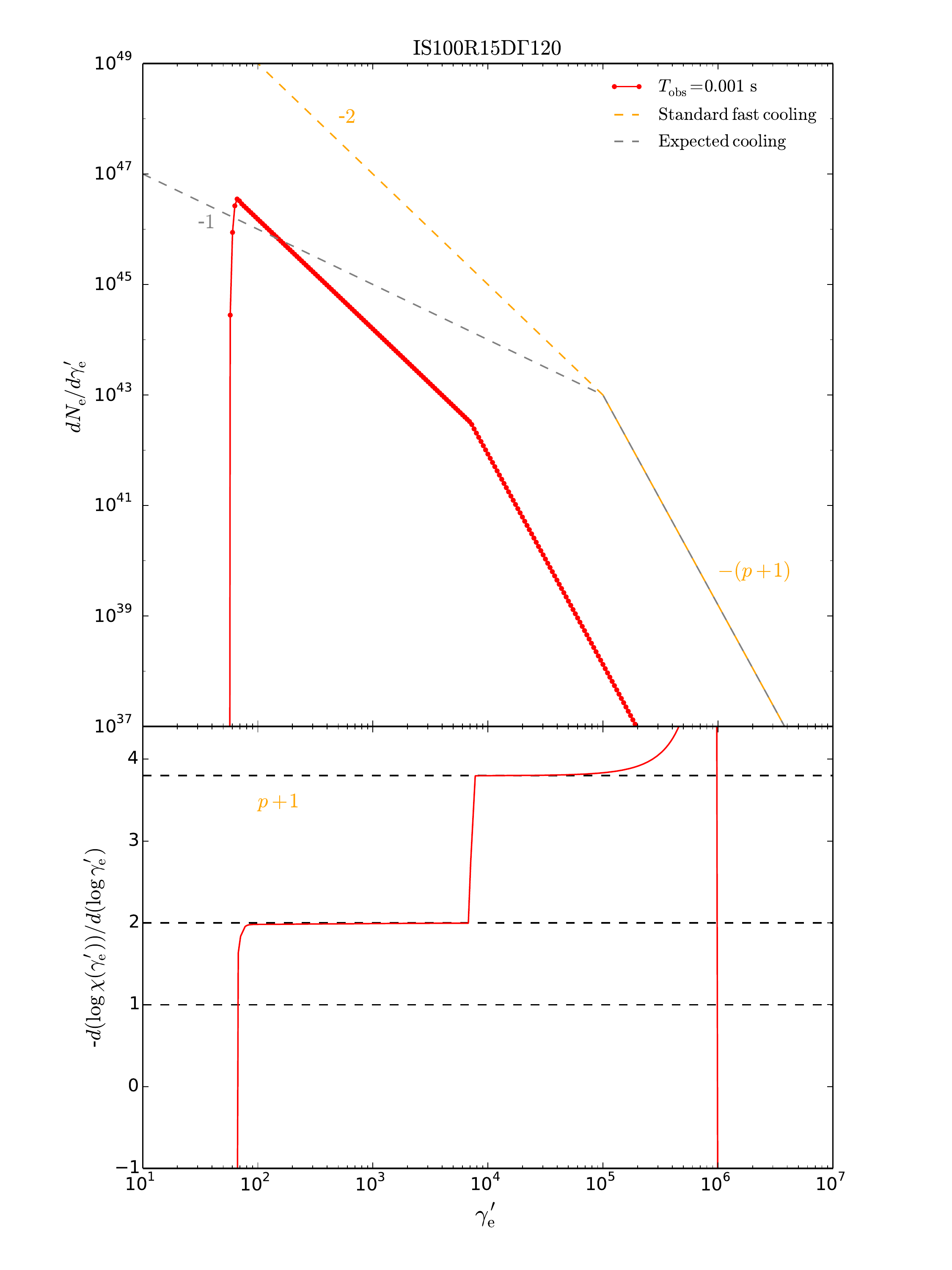}}
   \caption{(continued.)}
\end{adjustwidth}
\end{figure}

\clearpage

\begin{figure}
\begin{adjustwidth}{-2cm}{-2cm}
\centering
    \subfloat{\includegraphics[width=0.25\paperwidth]{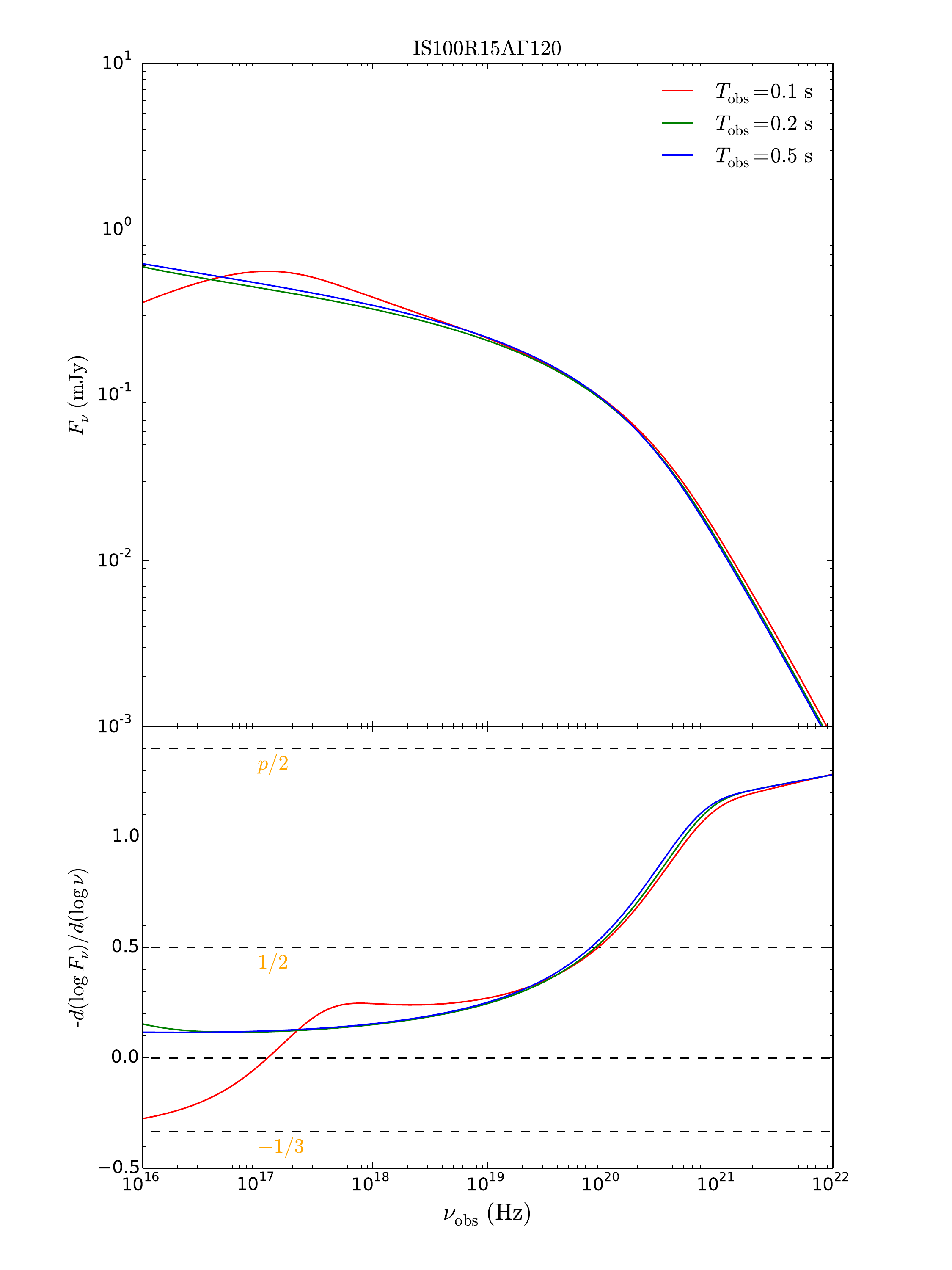}}
    \subfloat{\includegraphics[width=0.25\paperwidth]{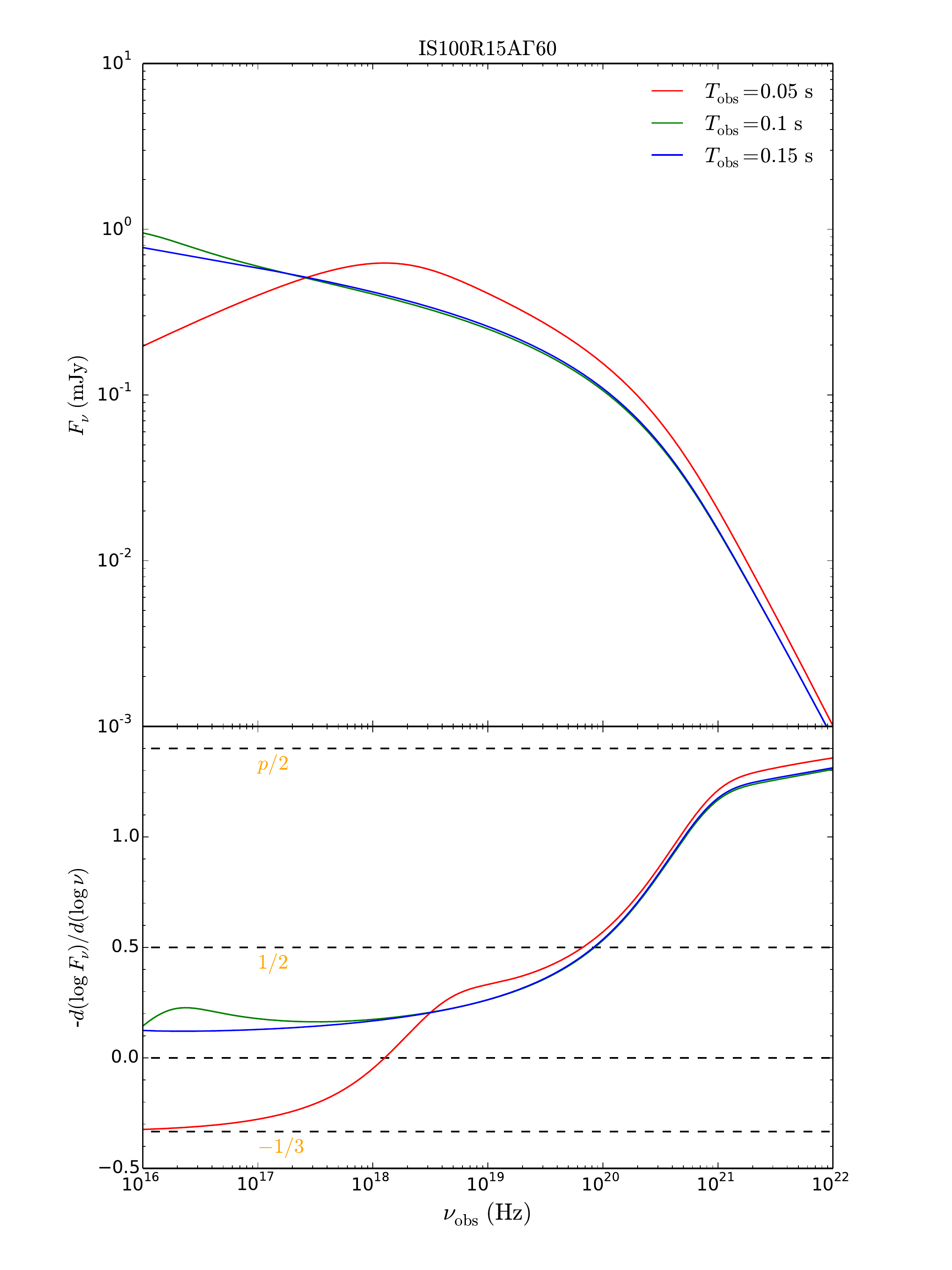}}
    \subfloat{\includegraphics[width=0.25\paperwidth]{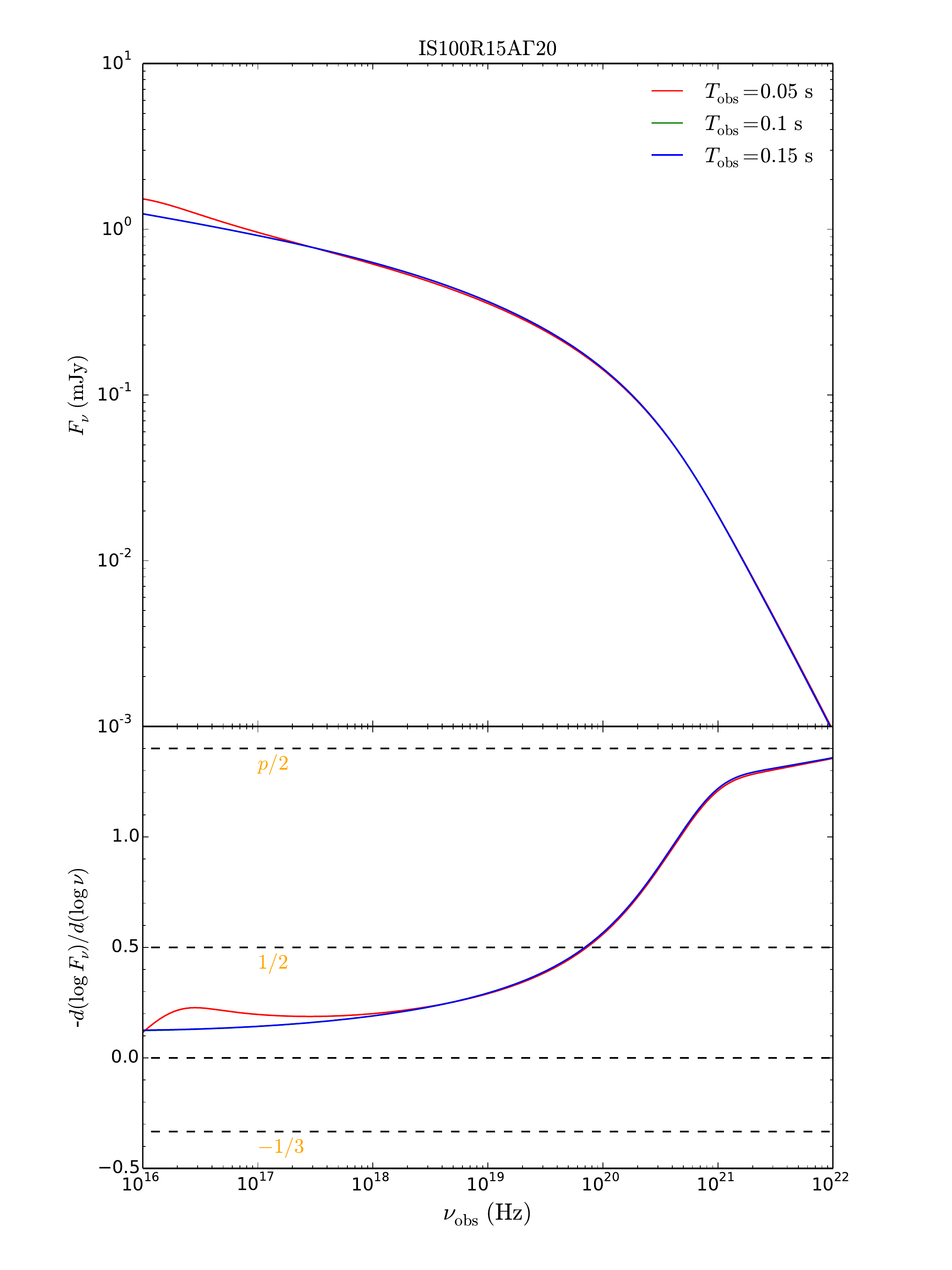}} \\
    \subfloat{\includegraphics[width=0.25\paperwidth]{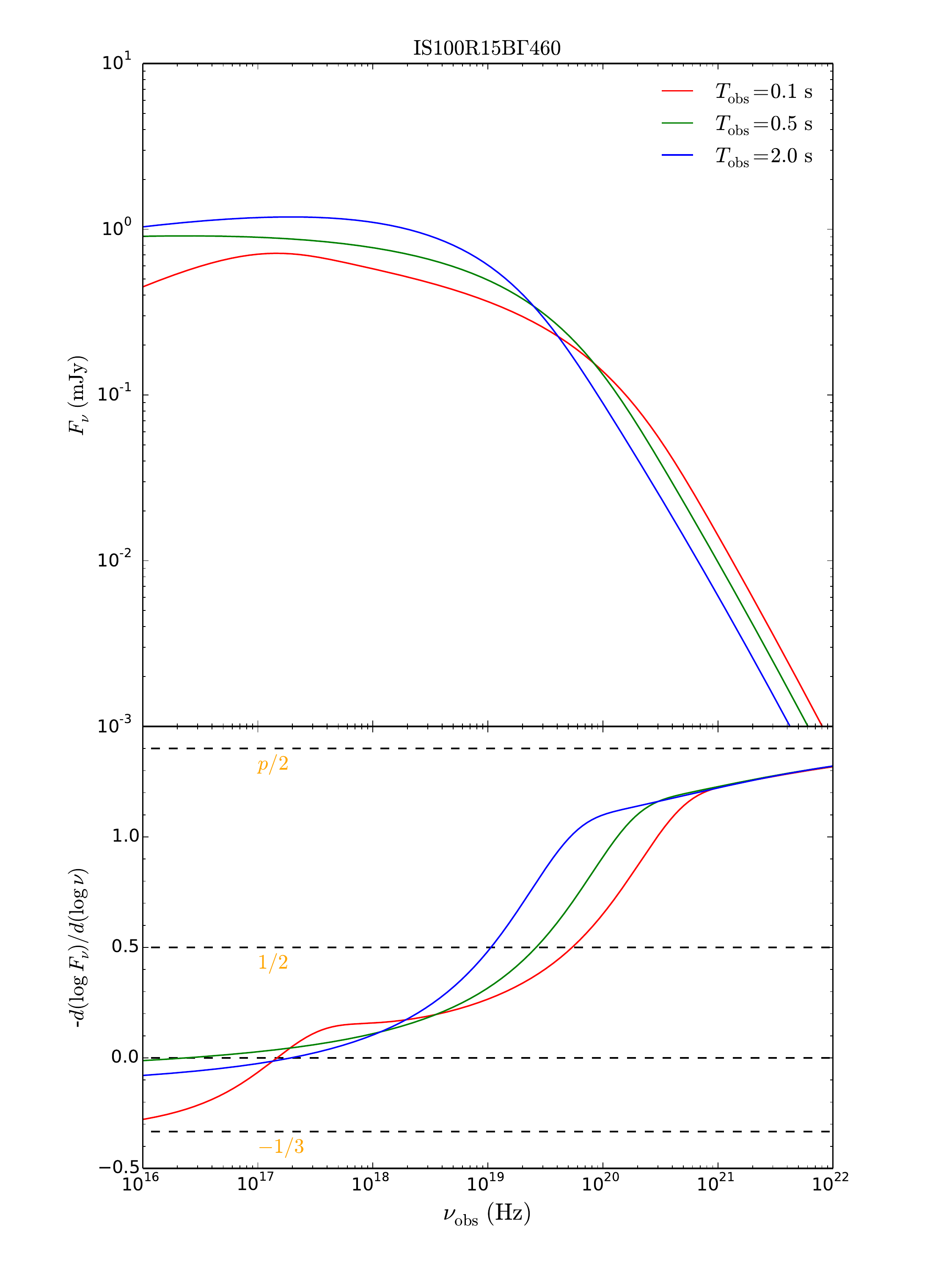}}
    \subfloat{\includegraphics[width=0.25\paperwidth]{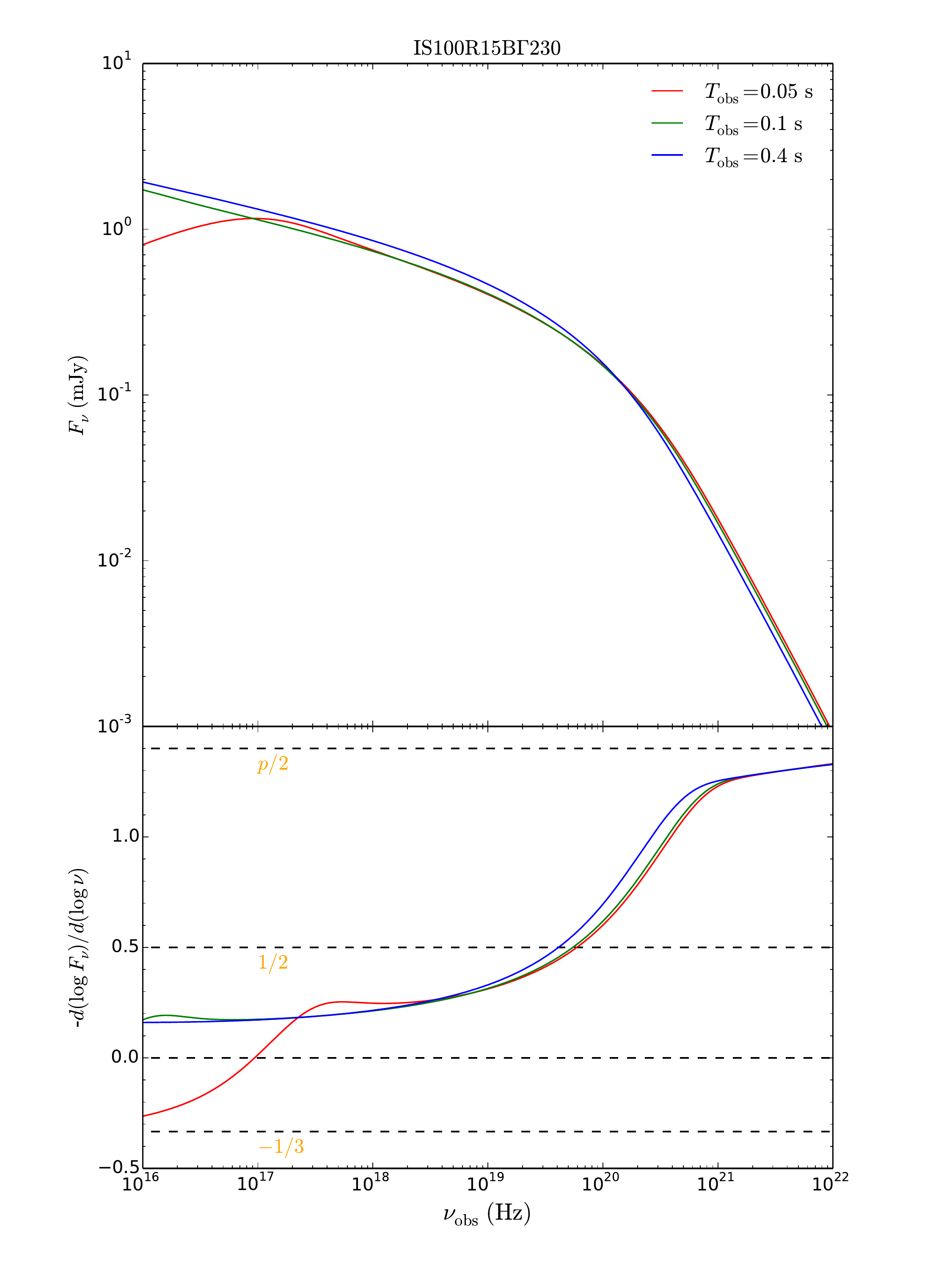}}
    \subfloat{\includegraphics[width=0.25\paperwidth]{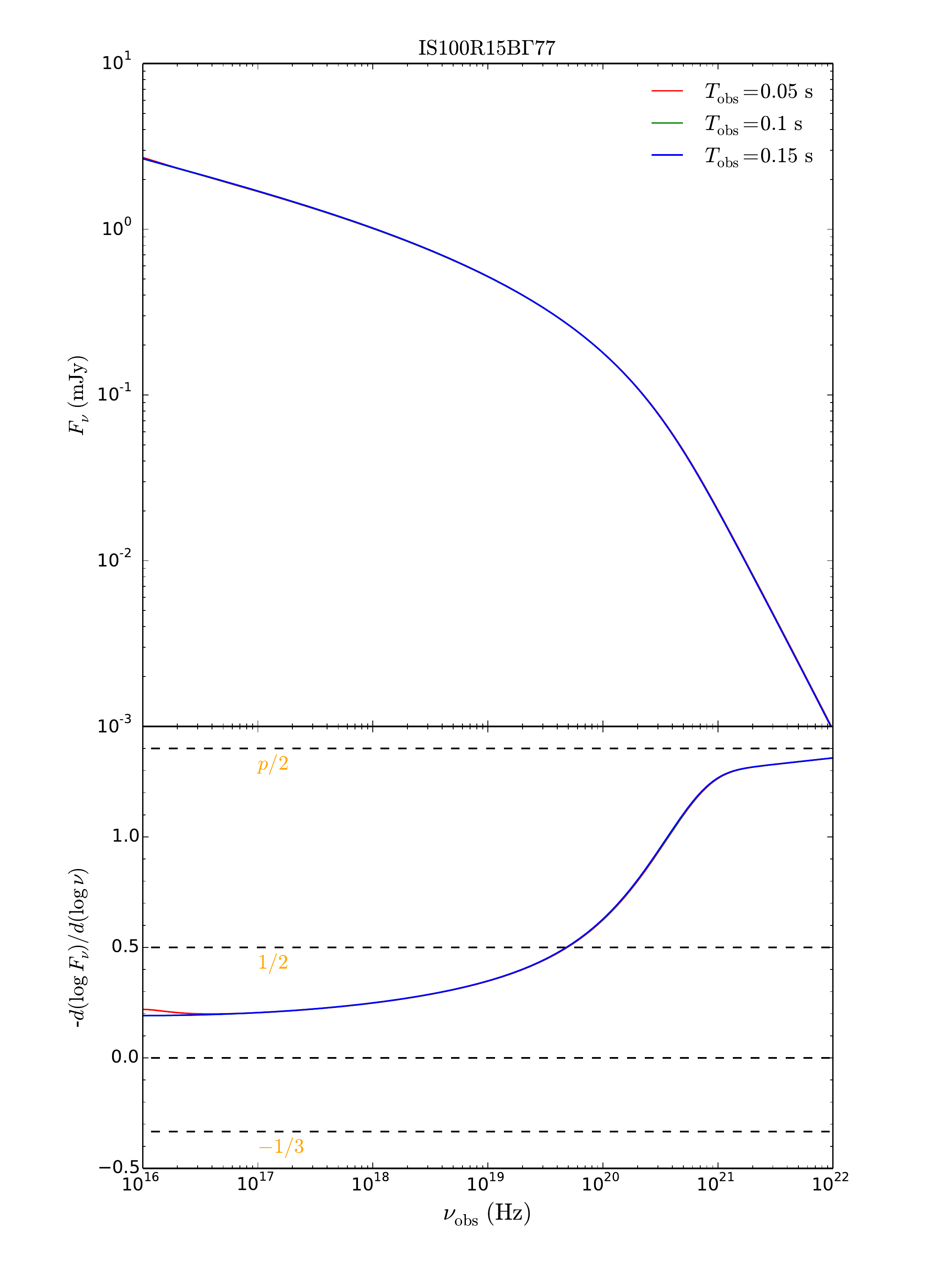}} \\
    \subfloat{\includegraphics[width=0.25\paperwidth]{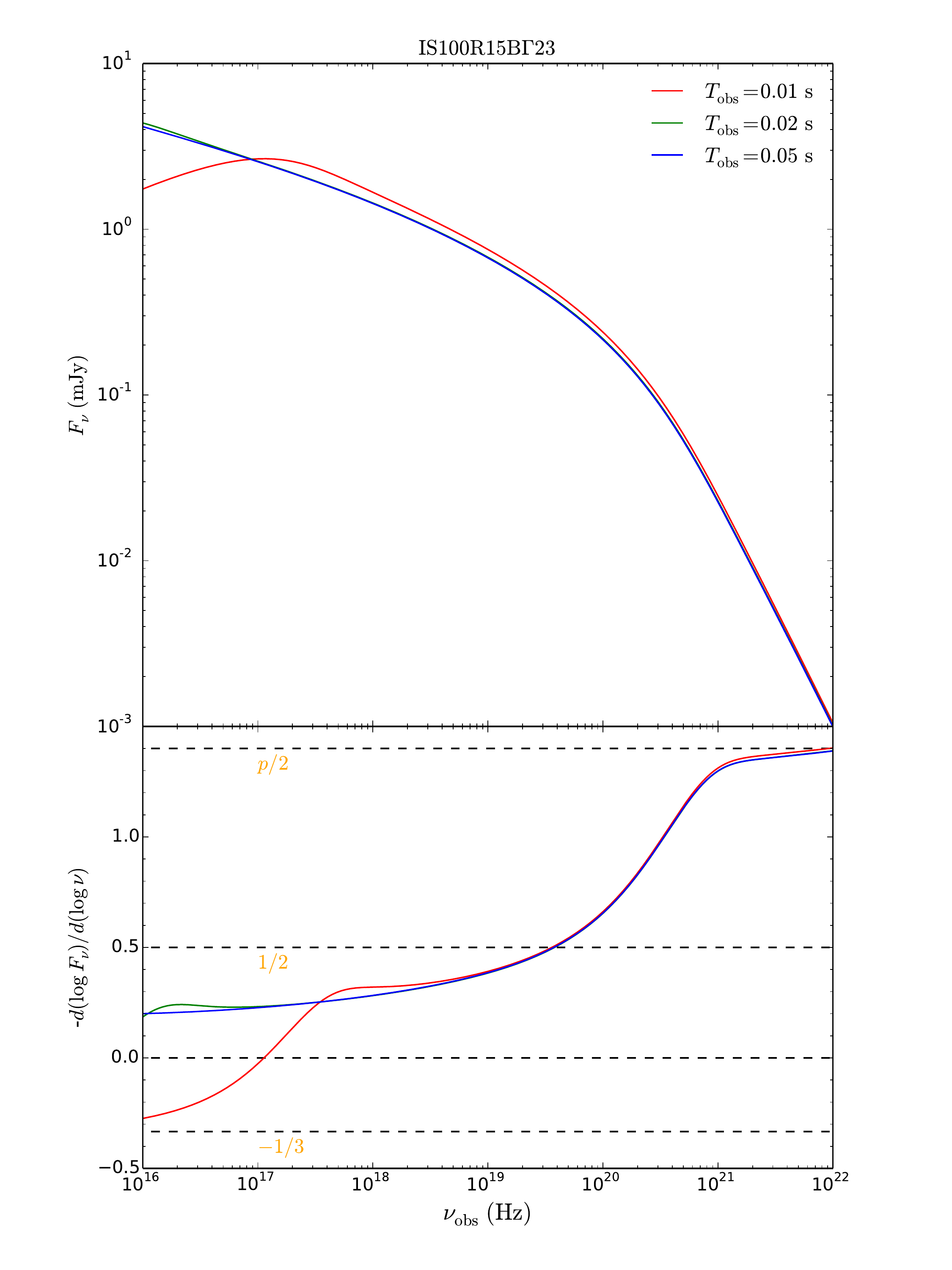}}
    \subfloat{\includegraphics[width=0.25\paperwidth]{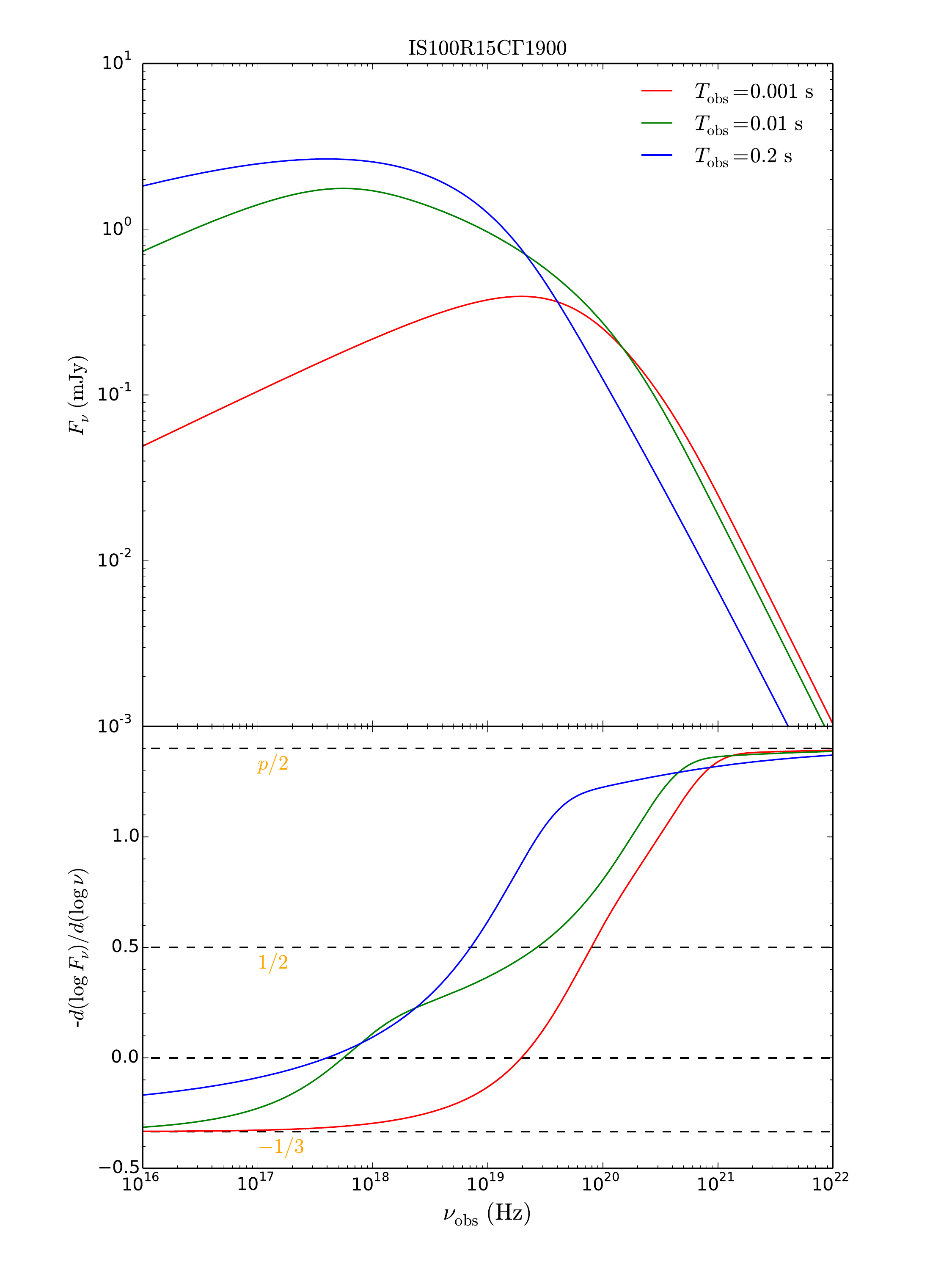}}
    \subfloat{\includegraphics[width=0.25\paperwidth]{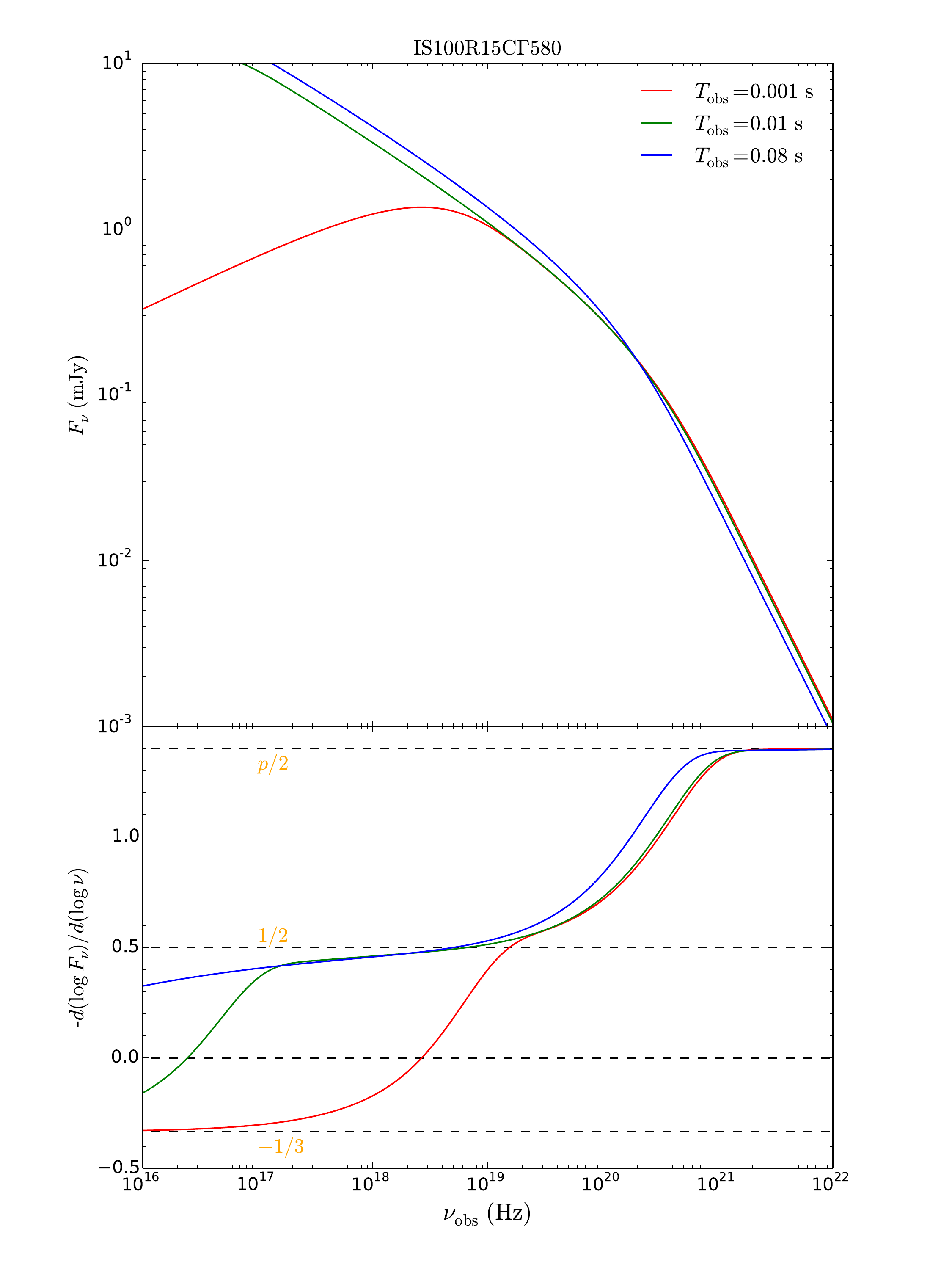}}
    \caption{The corresponding synchrotron flux-density spectra $F_{\nu}$ from the electrons with the energy distribution presented in Figure
    \ref{fig:MB-electron}.\label{fig:MB-spectra}}
\end{adjustwidth}
\end{figure}
\clearpage
\begin{figure}
\ContinuedFloat
\begin{adjustwidth}{-2cm}{-2cm}
\centering
    \subfloat{\includegraphics[width=0.25\paperwidth]{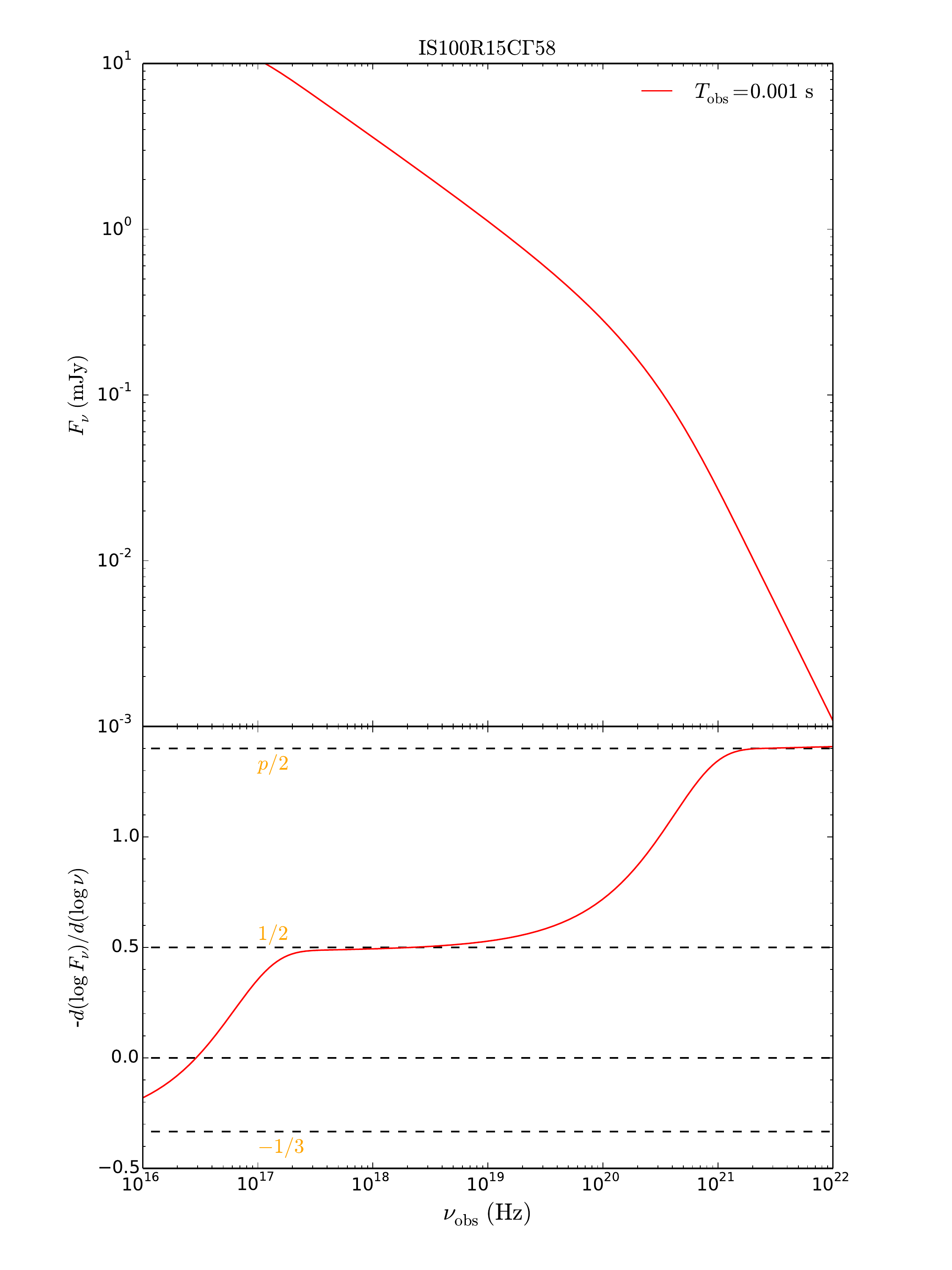}}
    \subfloat{\includegraphics[width=0.25\paperwidth]{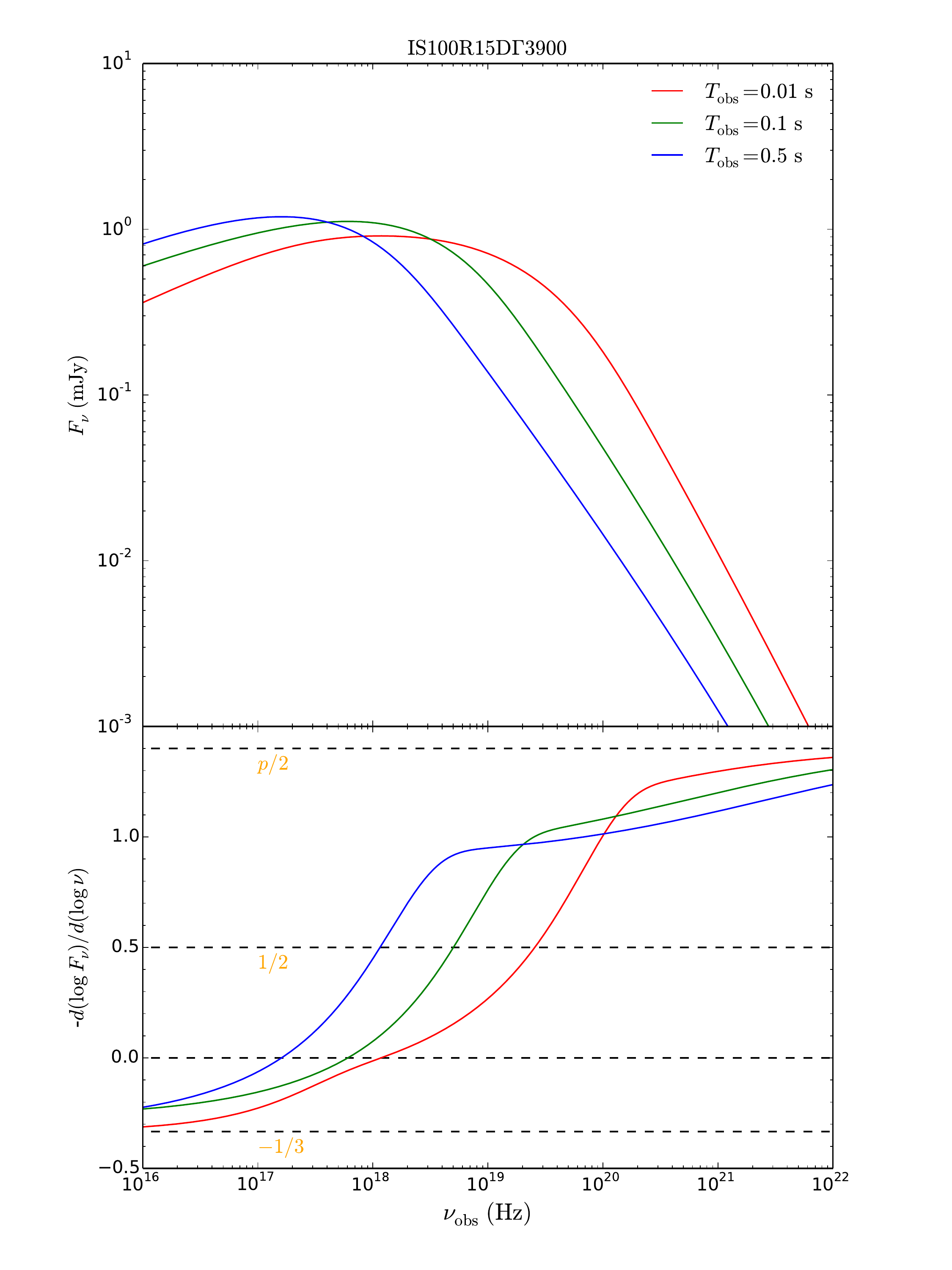}}
    \subfloat{\includegraphics[width=0.25\paperwidth]{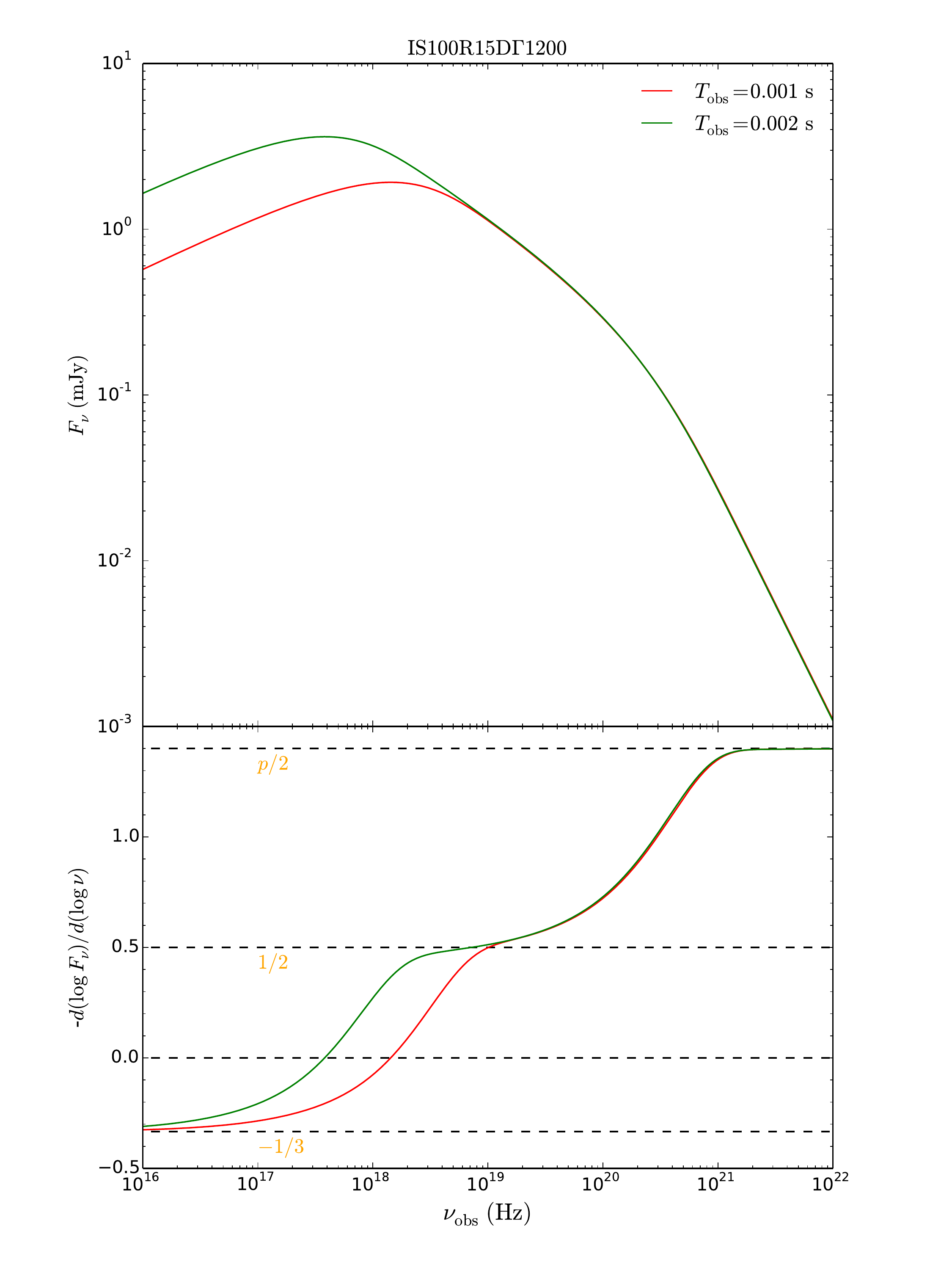}} \\
    \subfloat{\includegraphics[width=0.25\paperwidth]{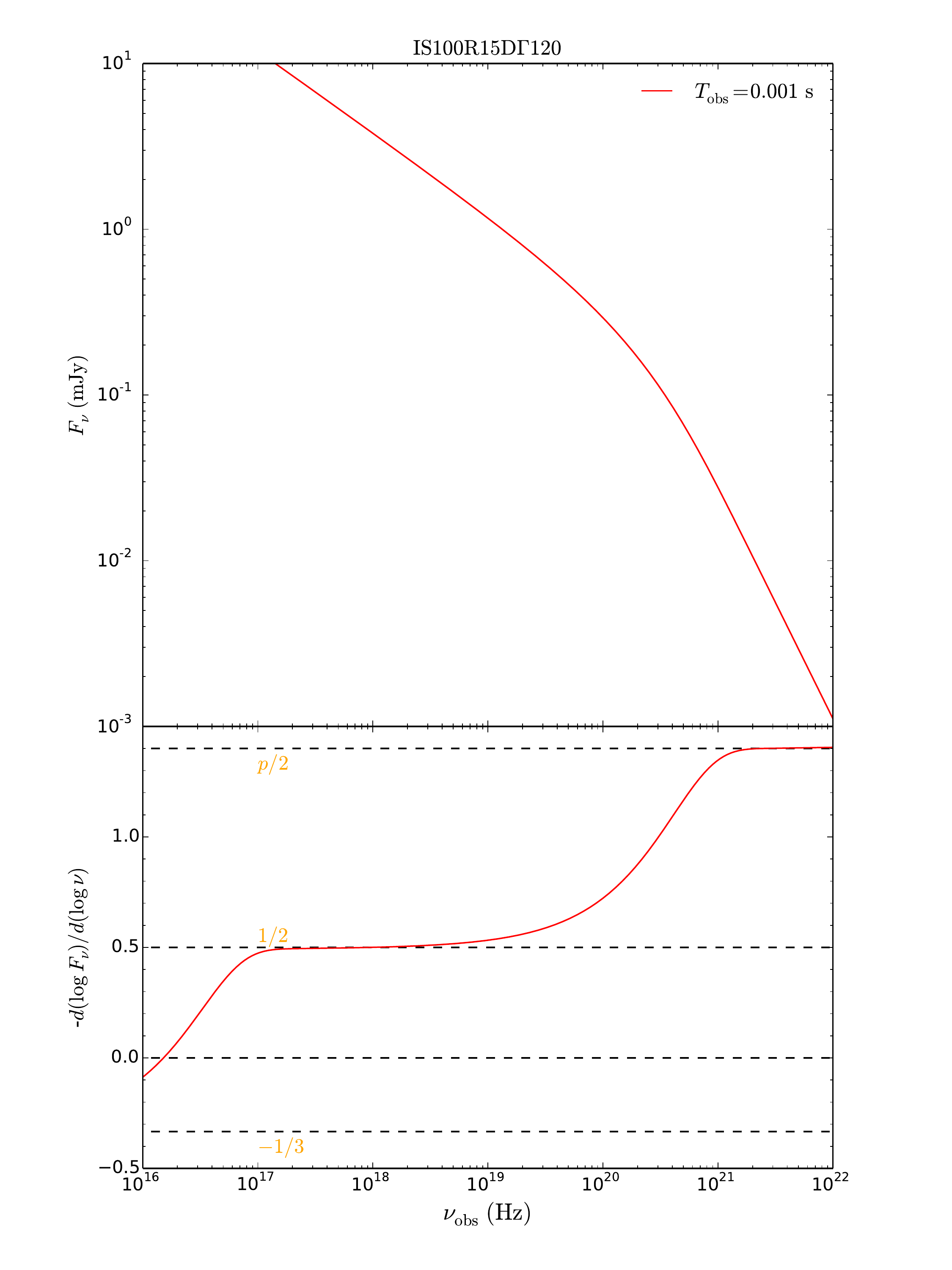}}
   \caption{(continued.)}
\end{adjustwidth}
\end{figure}

\clearpage

\begin{figure}
\begin{adjustwidth}{-2cm}{-2cm}
\centering
    \subfloat{\includegraphics[width=0.25\paperwidth]{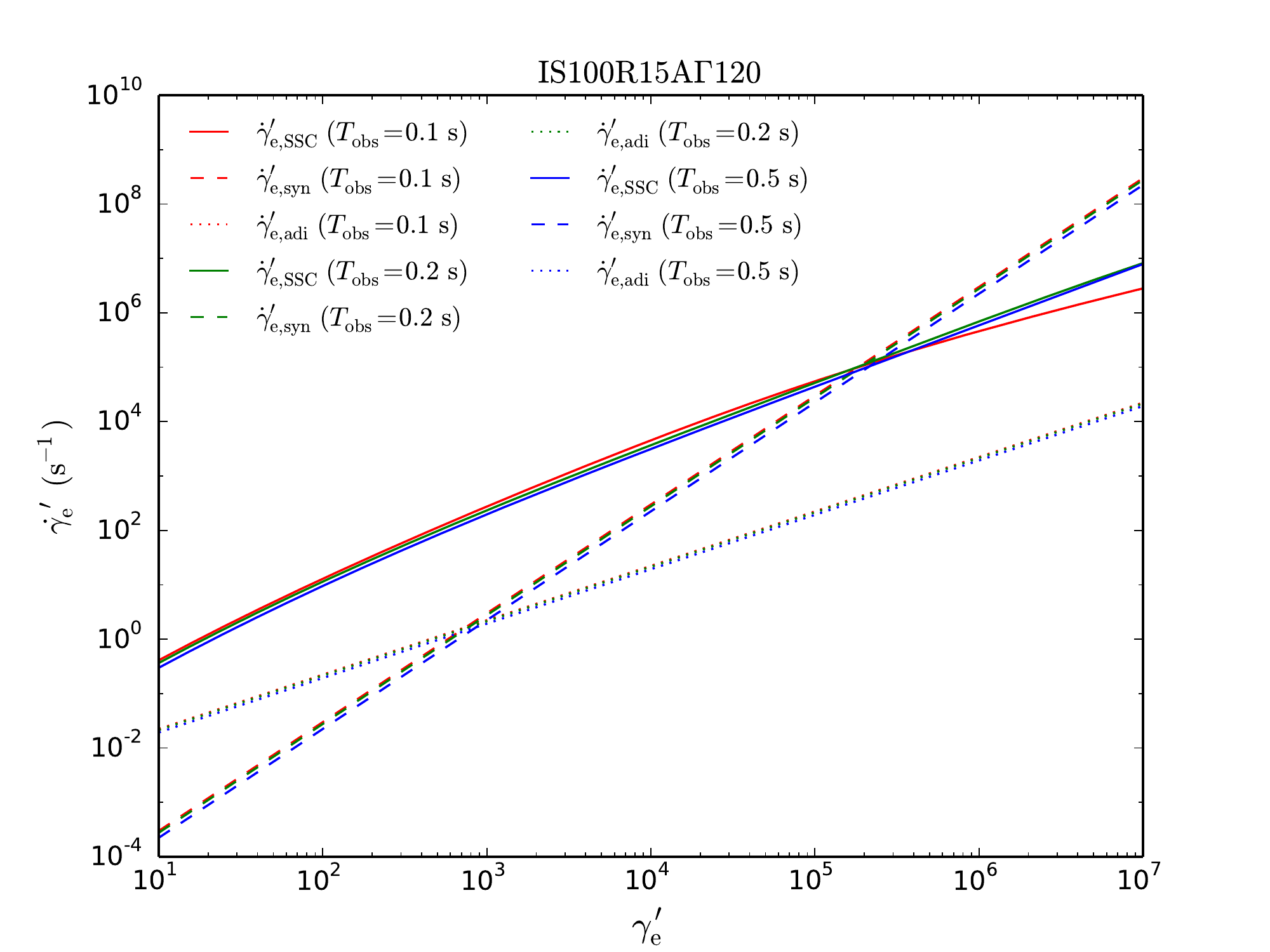}}
    \subfloat{\includegraphics[width=0.25\paperwidth]{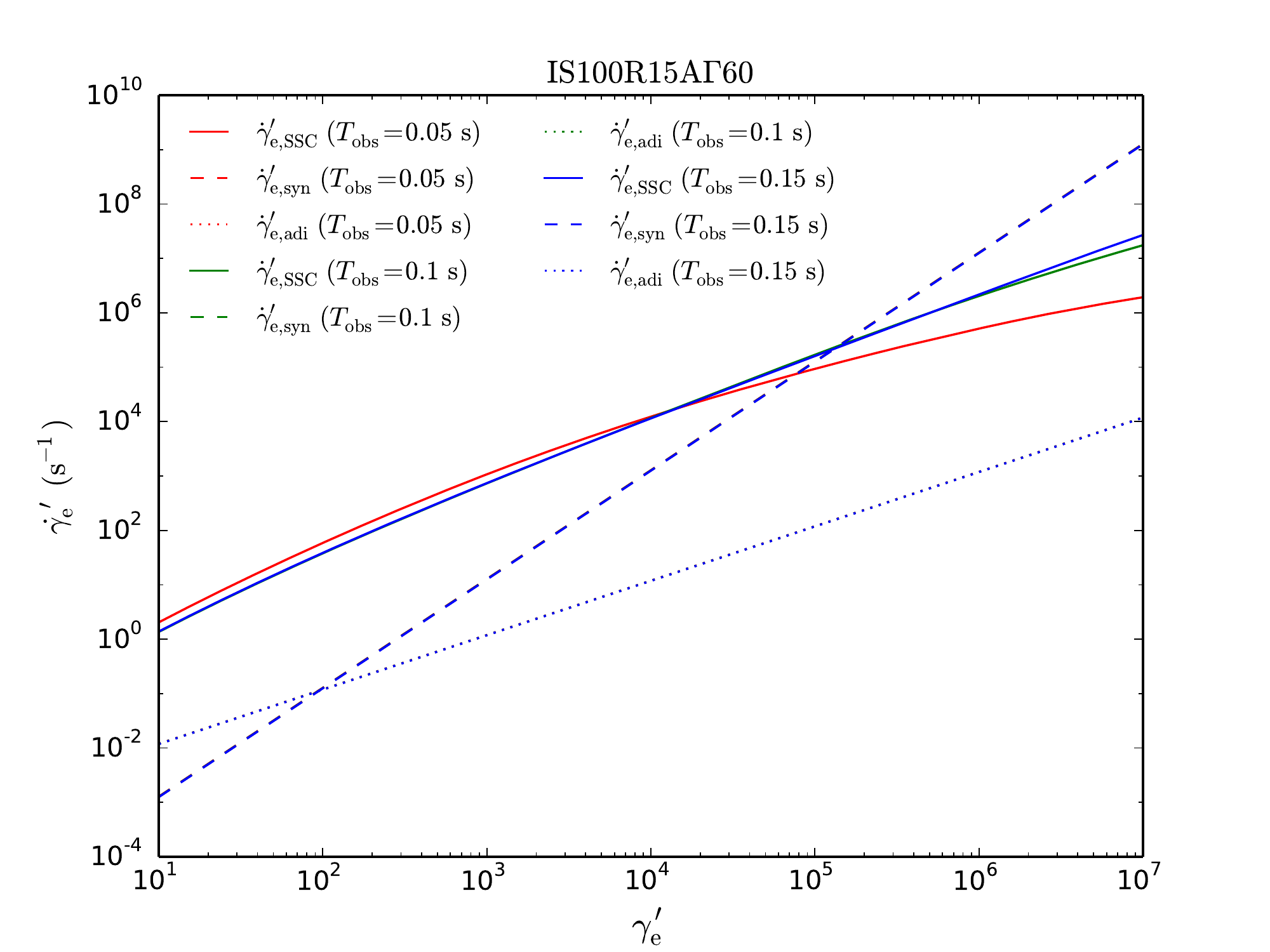}}
    \subfloat{\includegraphics[width=0.25\paperwidth]{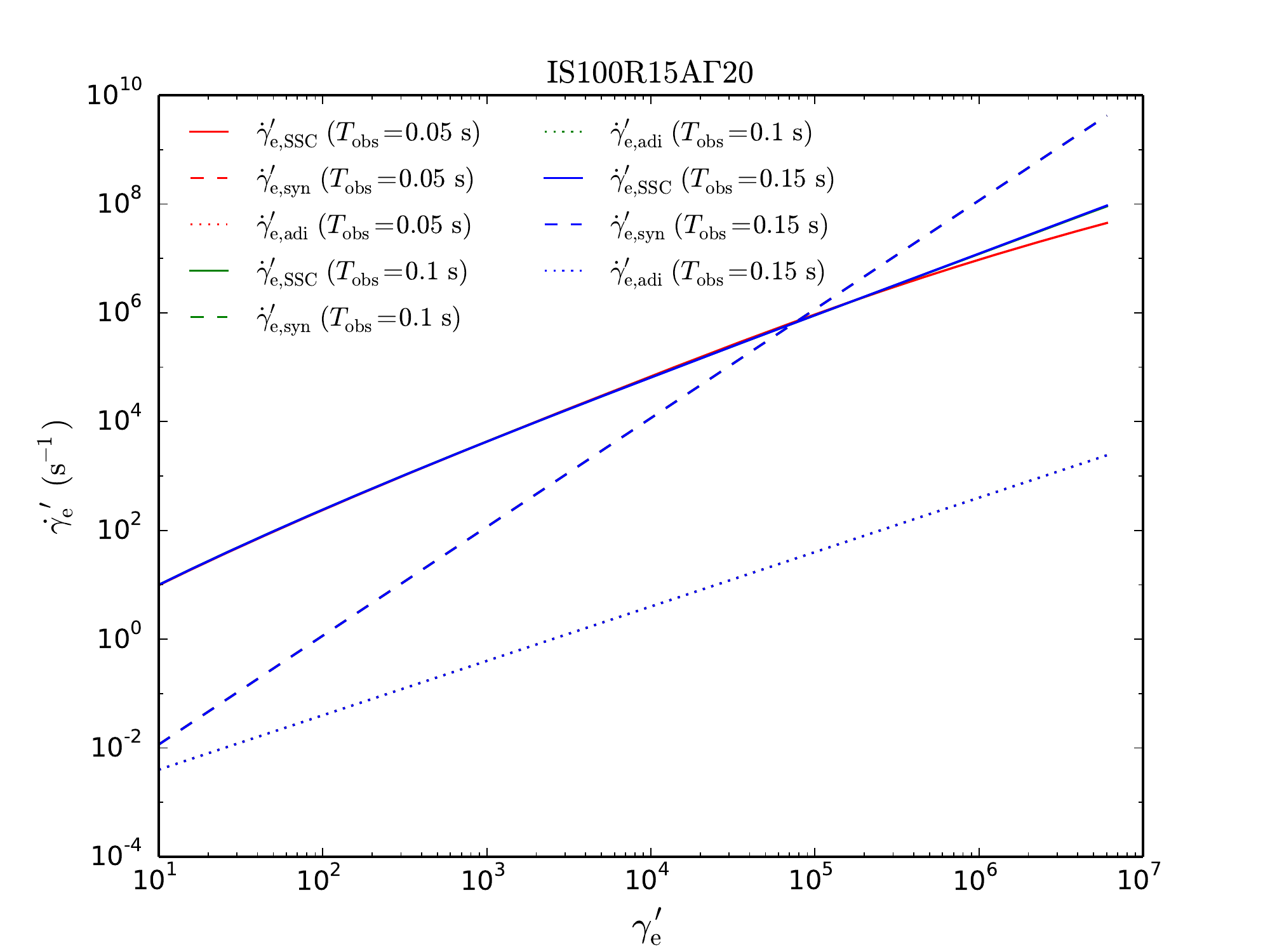}} \\
    \subfloat{\includegraphics[width=0.25\paperwidth]{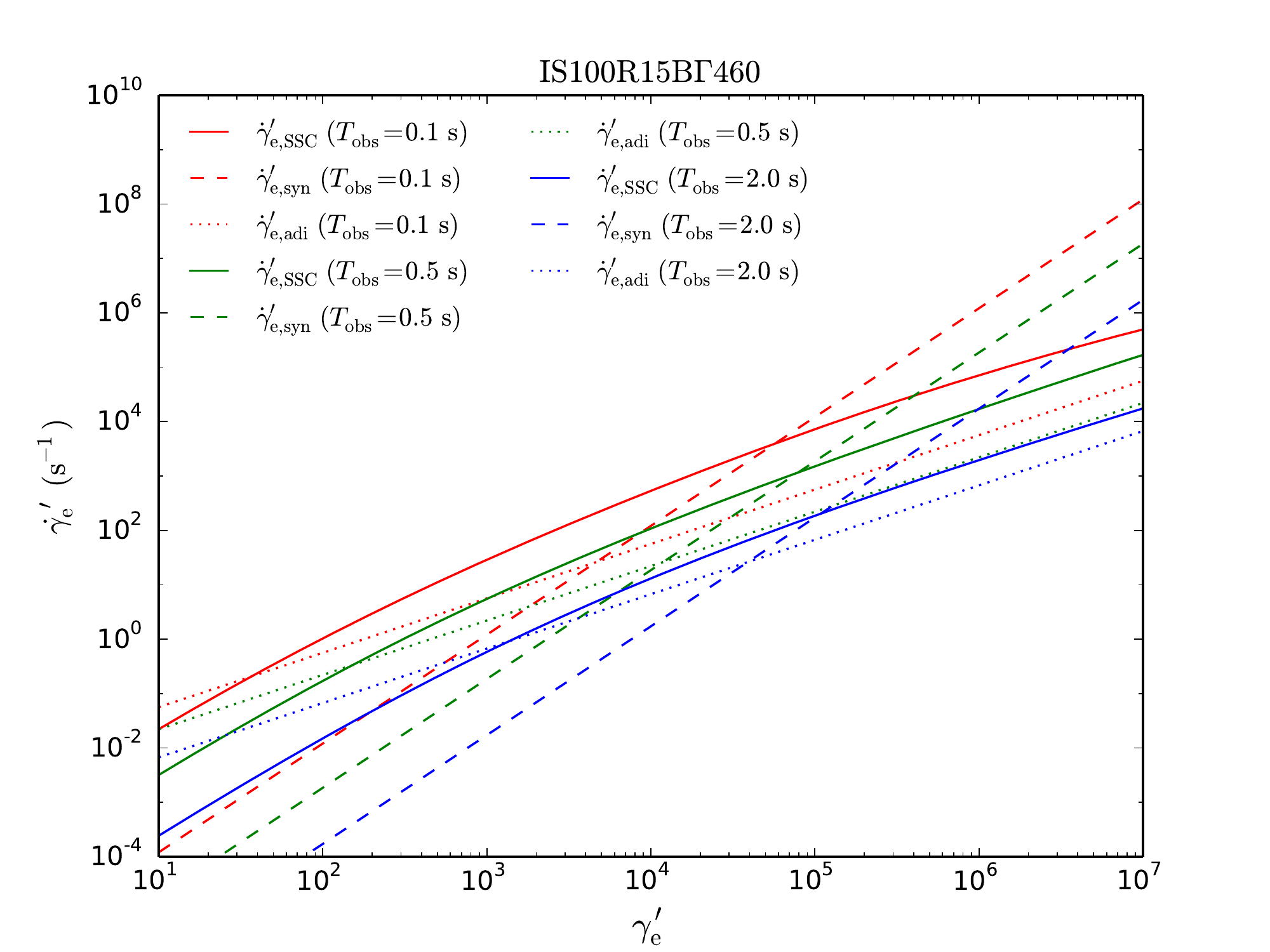}}
    \subfloat{\includegraphics[width=0.25\paperwidth]{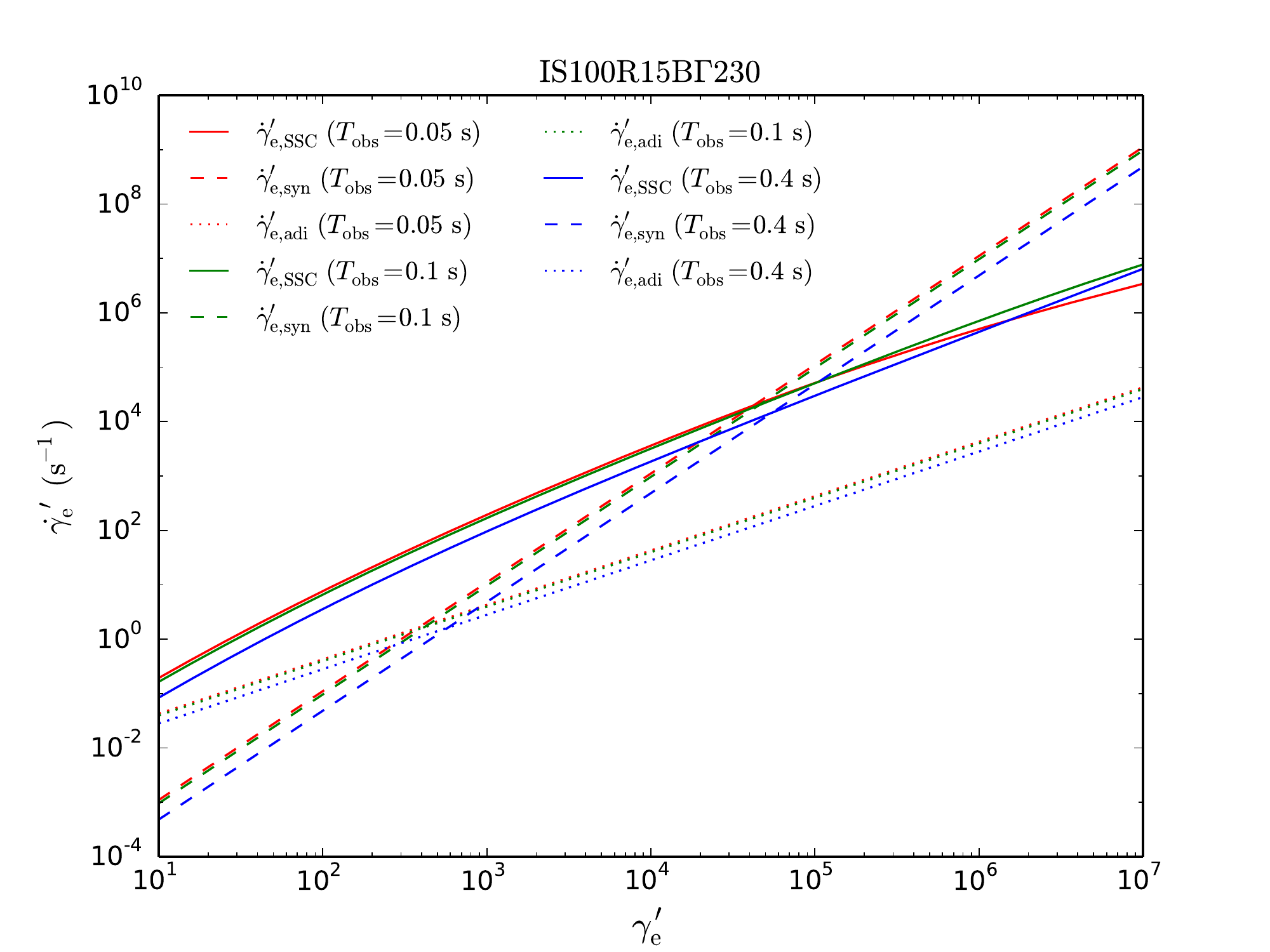}}
    \subfloat{\includegraphics[width=0.25\paperwidth]{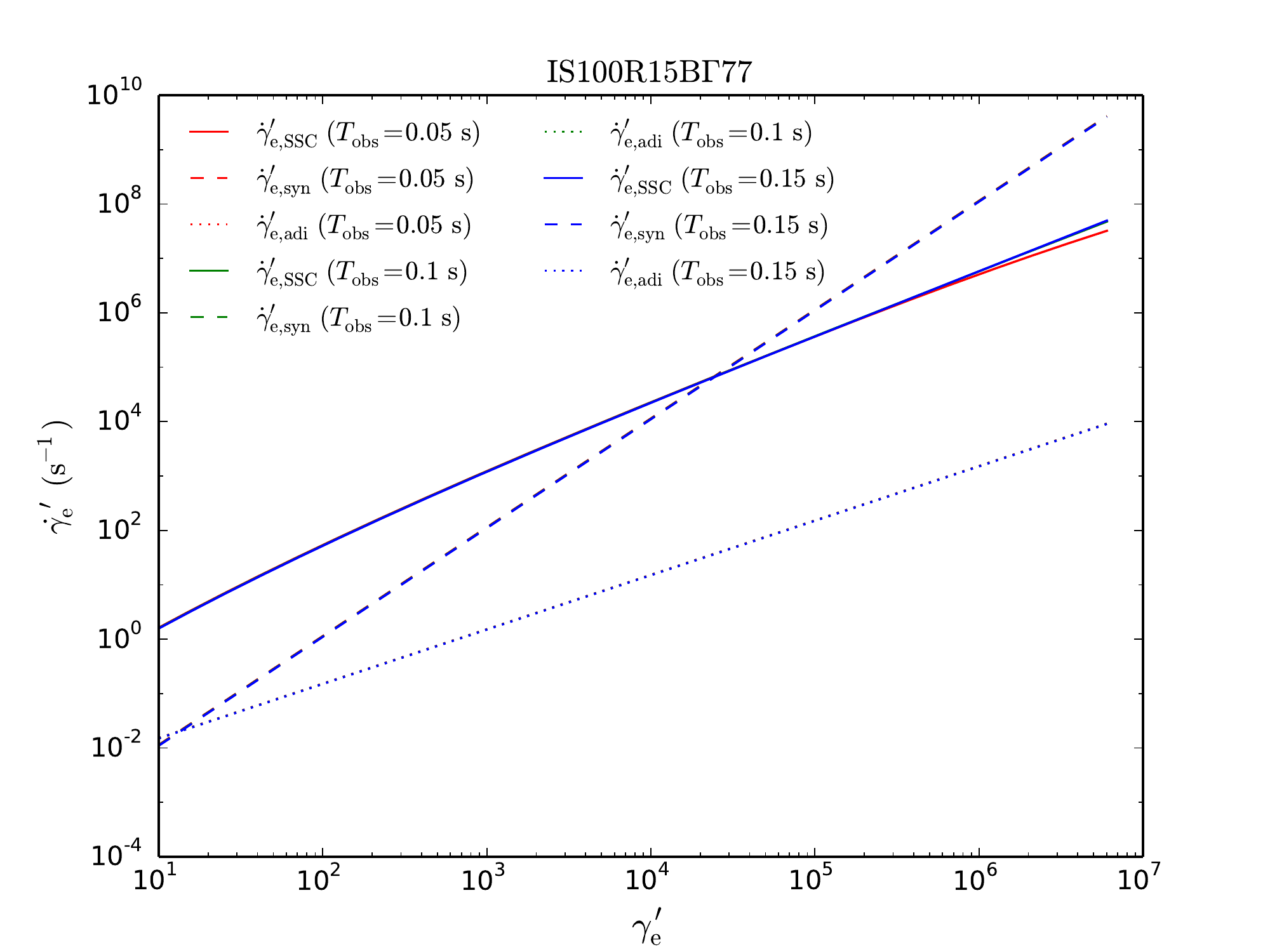}} \\
    \subfloat{\includegraphics[width=0.25\paperwidth]{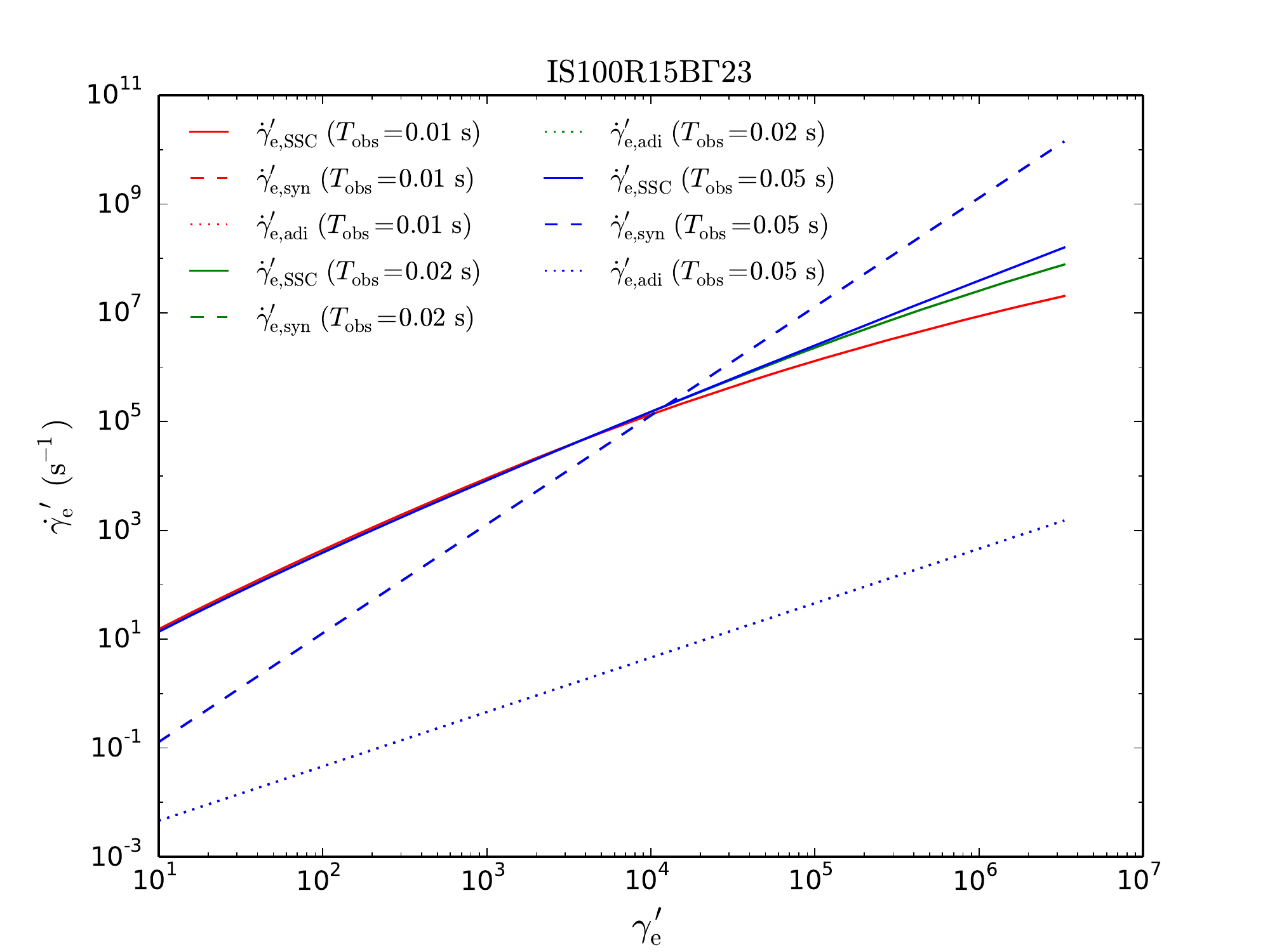}}
    \subfloat{\includegraphics[width=0.25\paperwidth]{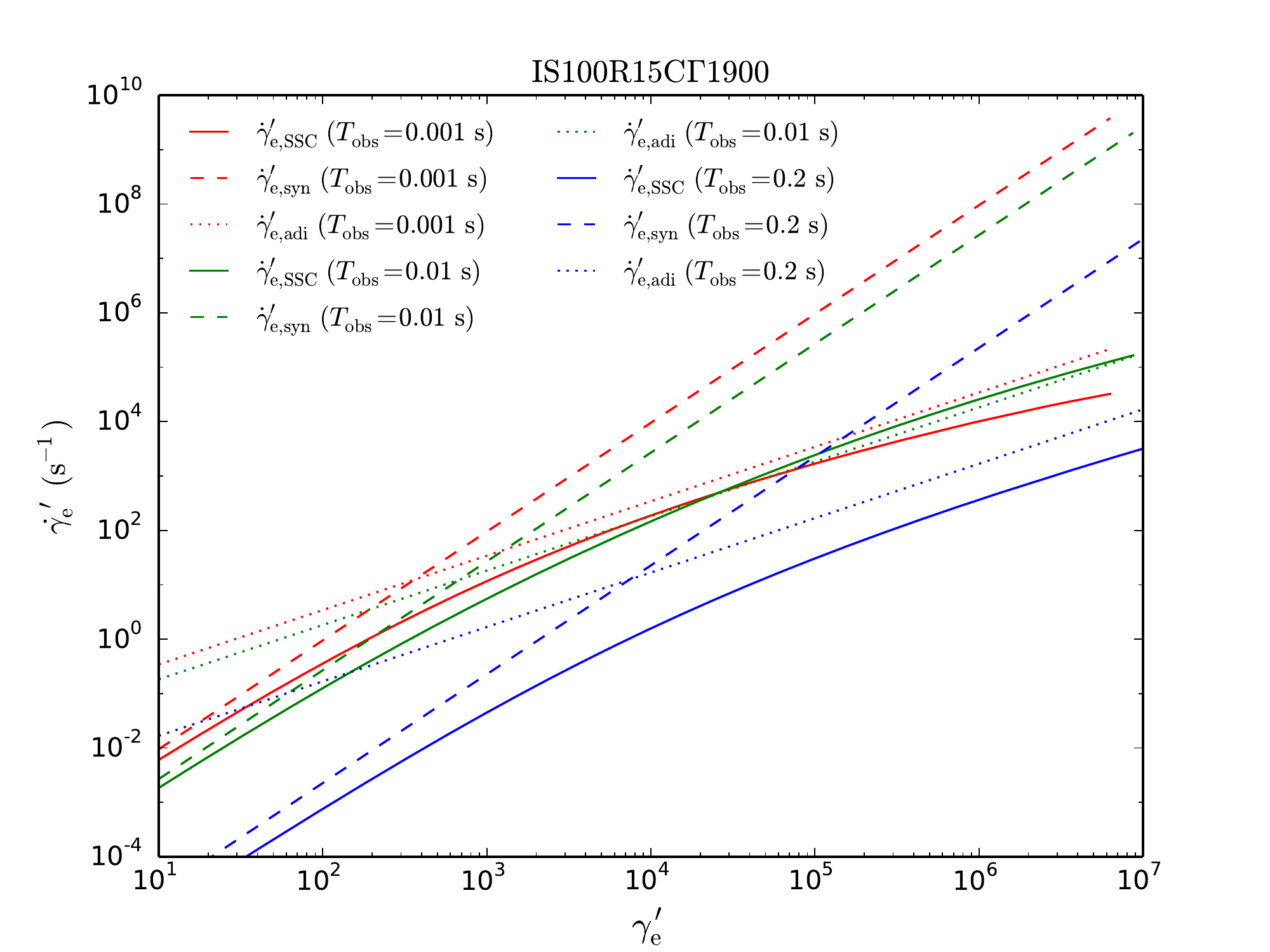}}
    \subfloat{\includegraphics[width=0.25\paperwidth]{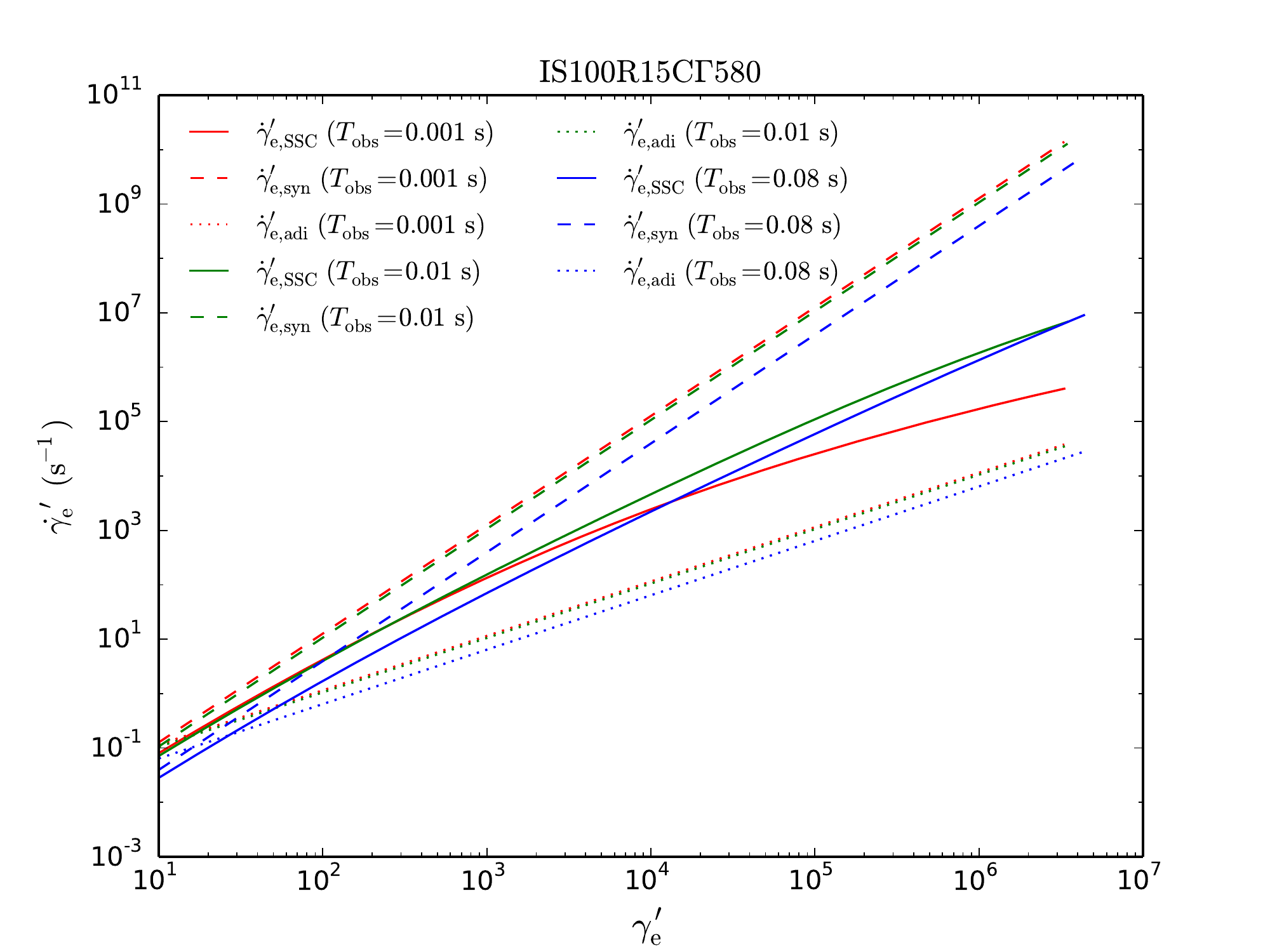}}  \\
    \subfloat{\includegraphics[width=0.25\paperwidth]{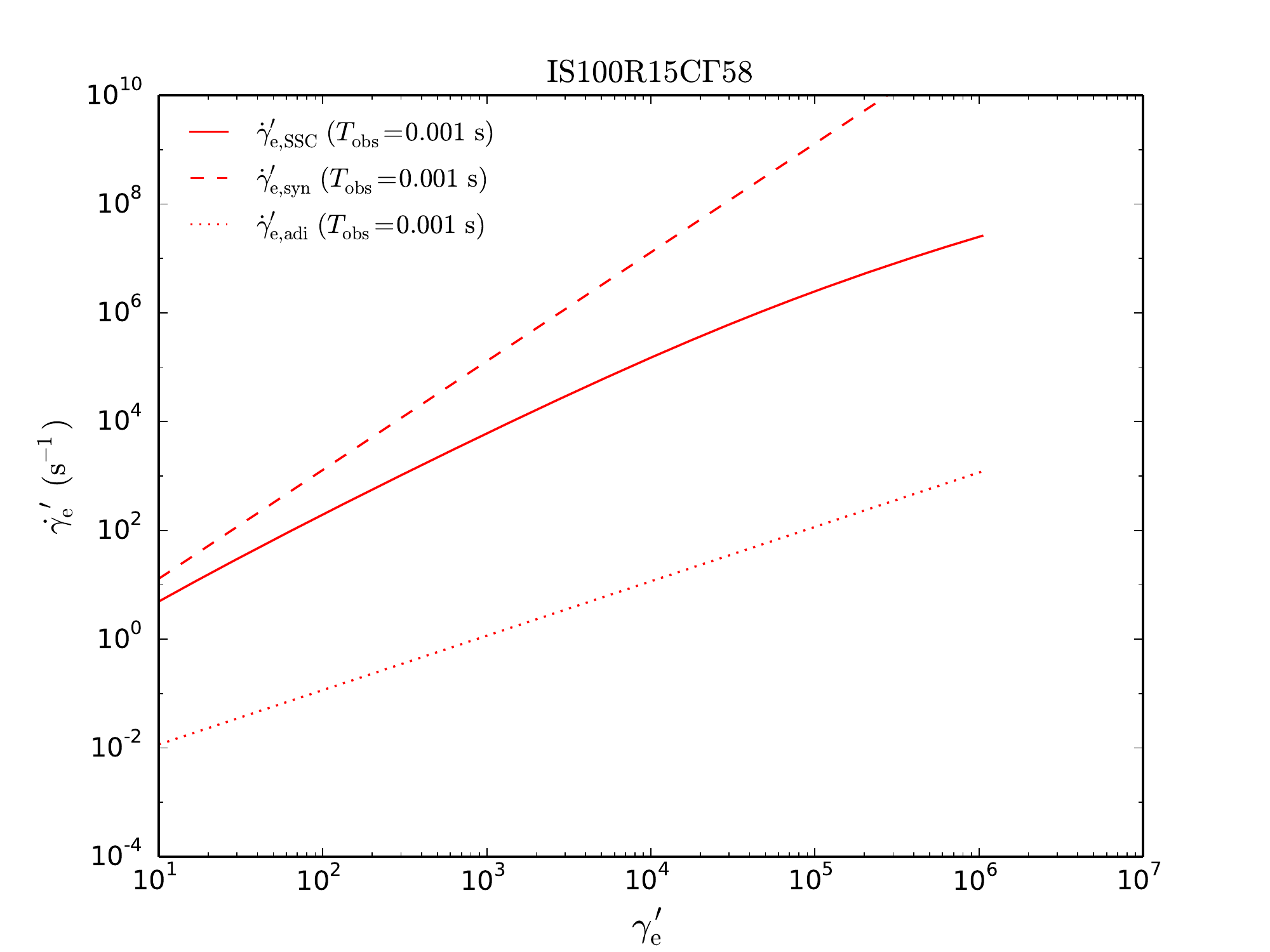}}
    \subfloat{\includegraphics[width=0.25\paperwidth]{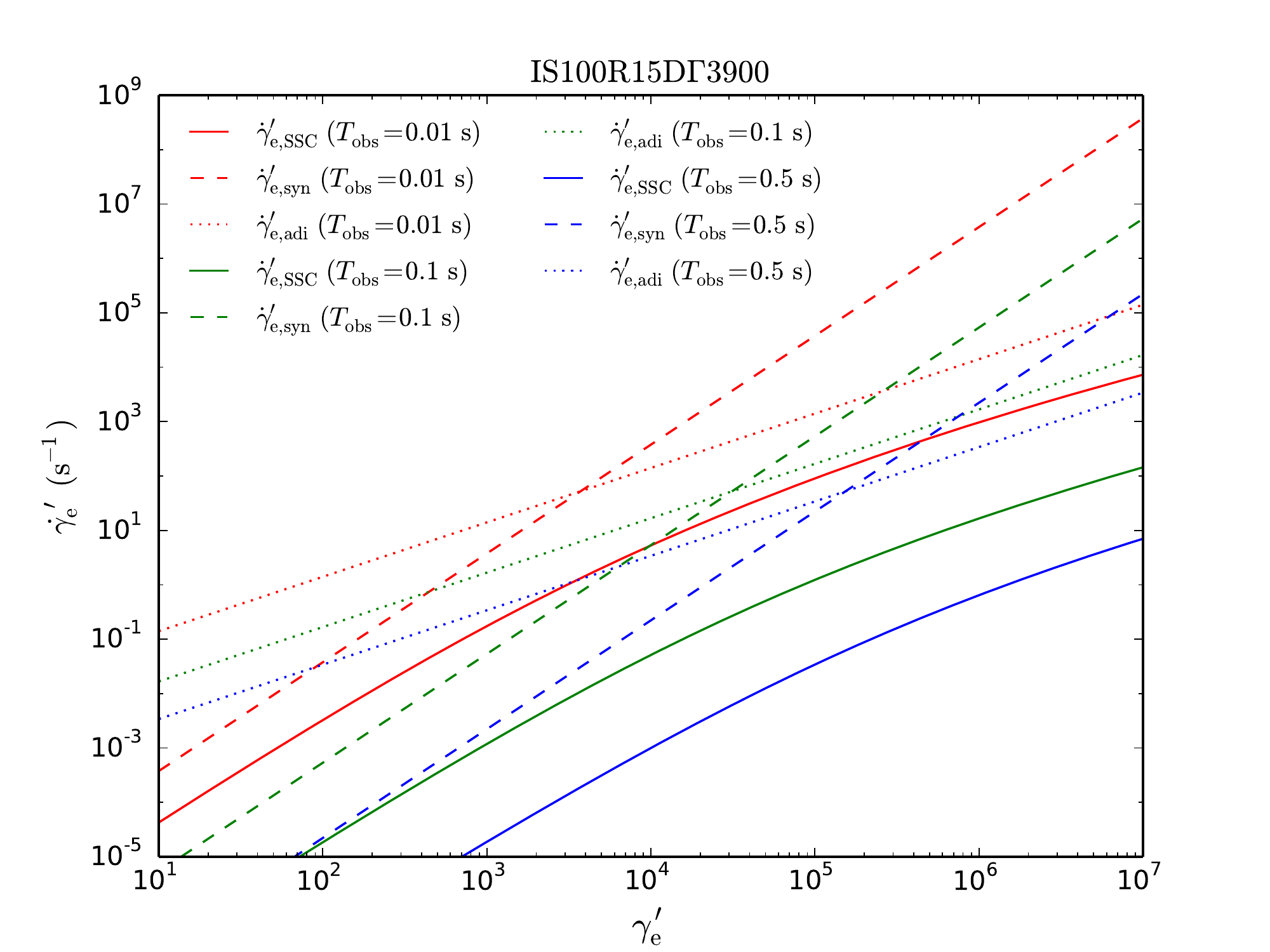}}
    \subfloat{\includegraphics[width=0.25\paperwidth]{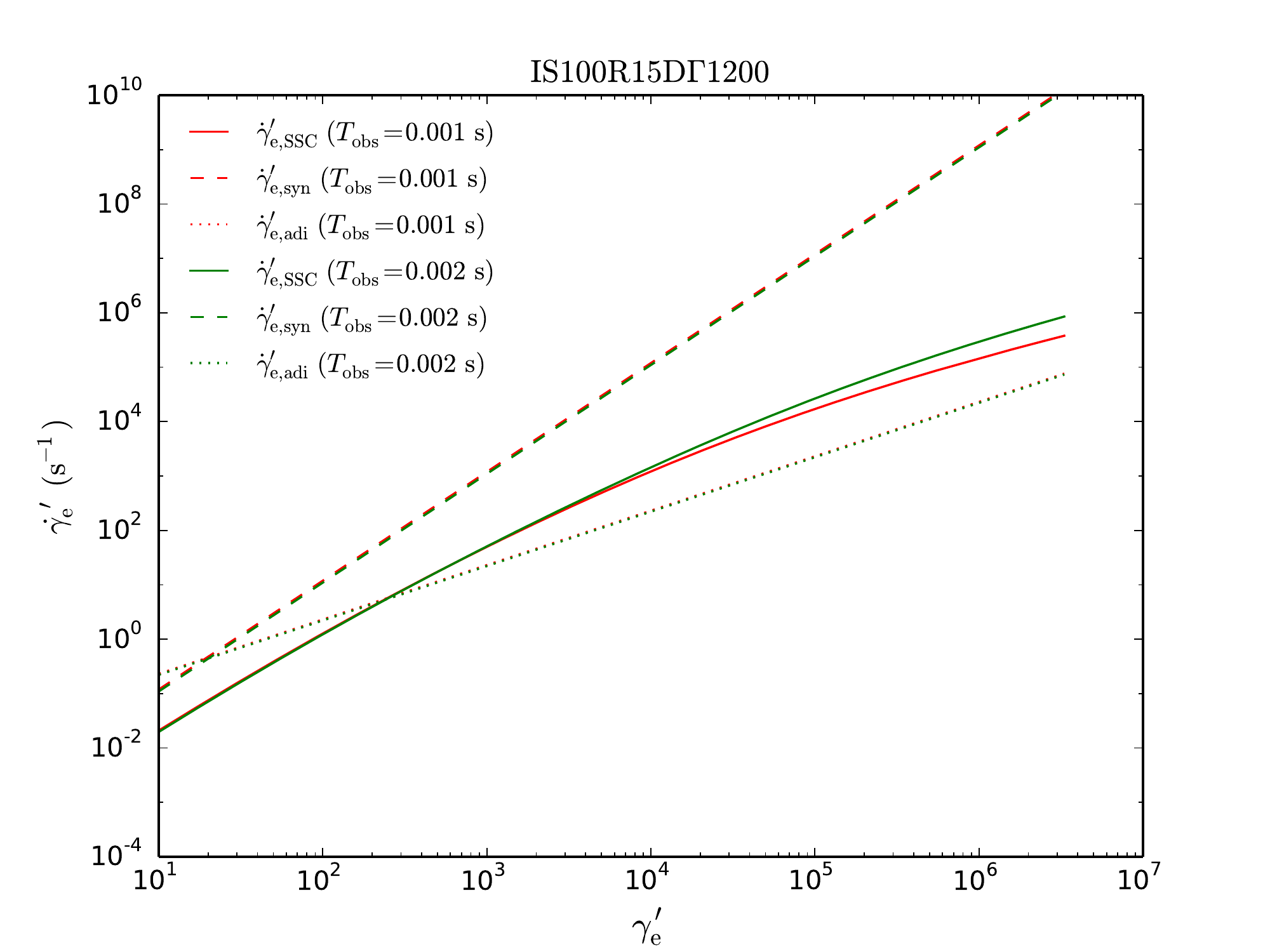}} \\
    \subfloat{\includegraphics[width=0.25\paperwidth]{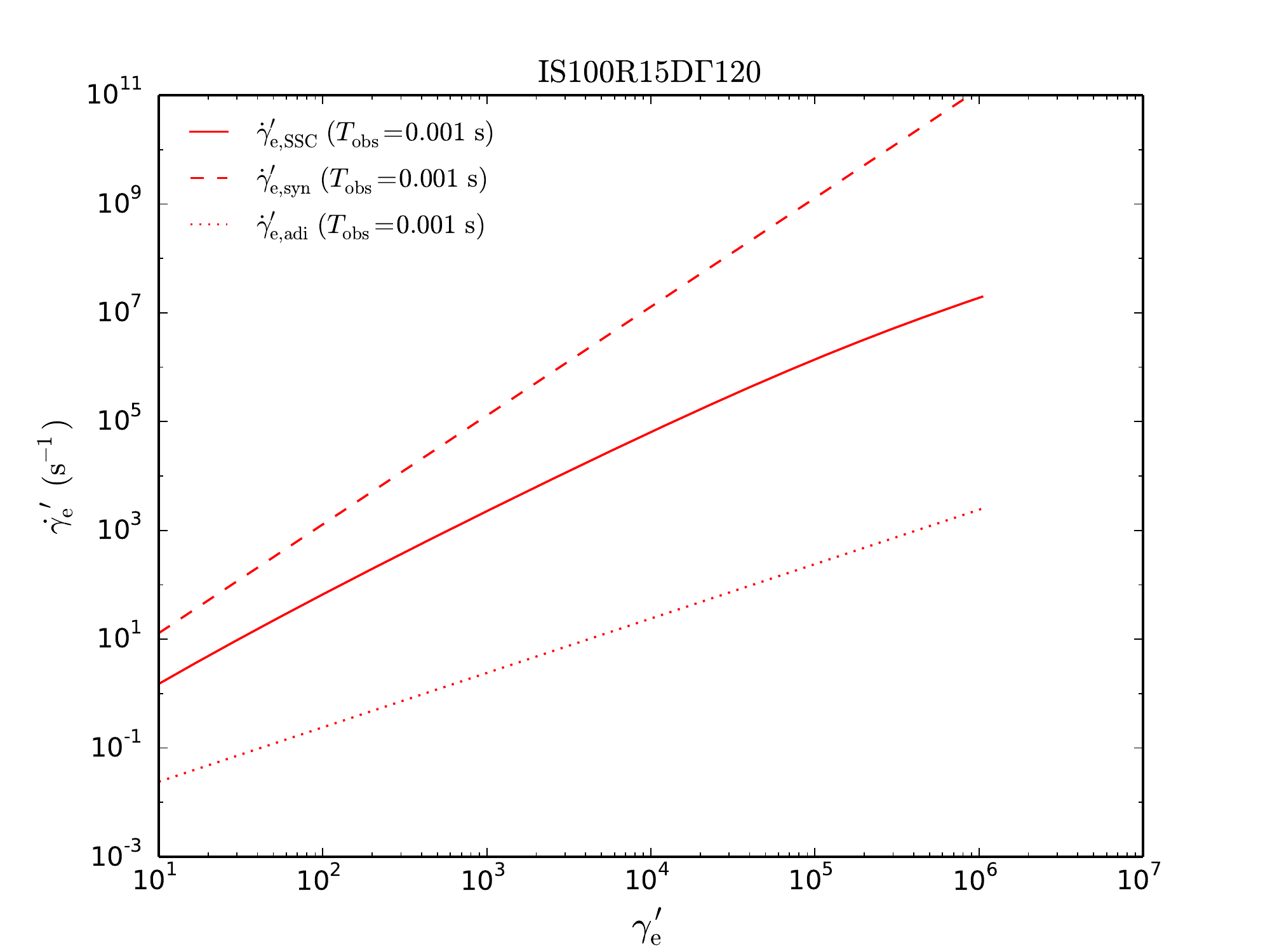}}
    \caption{The co-moving cooling rates of different cooling mechanisms for the electrons with the energy distribution
    presented in Figure \ref{fig:MB-electron}.\label{fig:MB-rate}}
\end{adjustwidth}
\end{figure}

\clearpage

\begin{figure}
\begin{adjustwidth}{-2cm}{-2cm}
\centering
    \subfloat{\includegraphics[width=0.3\paperwidth]{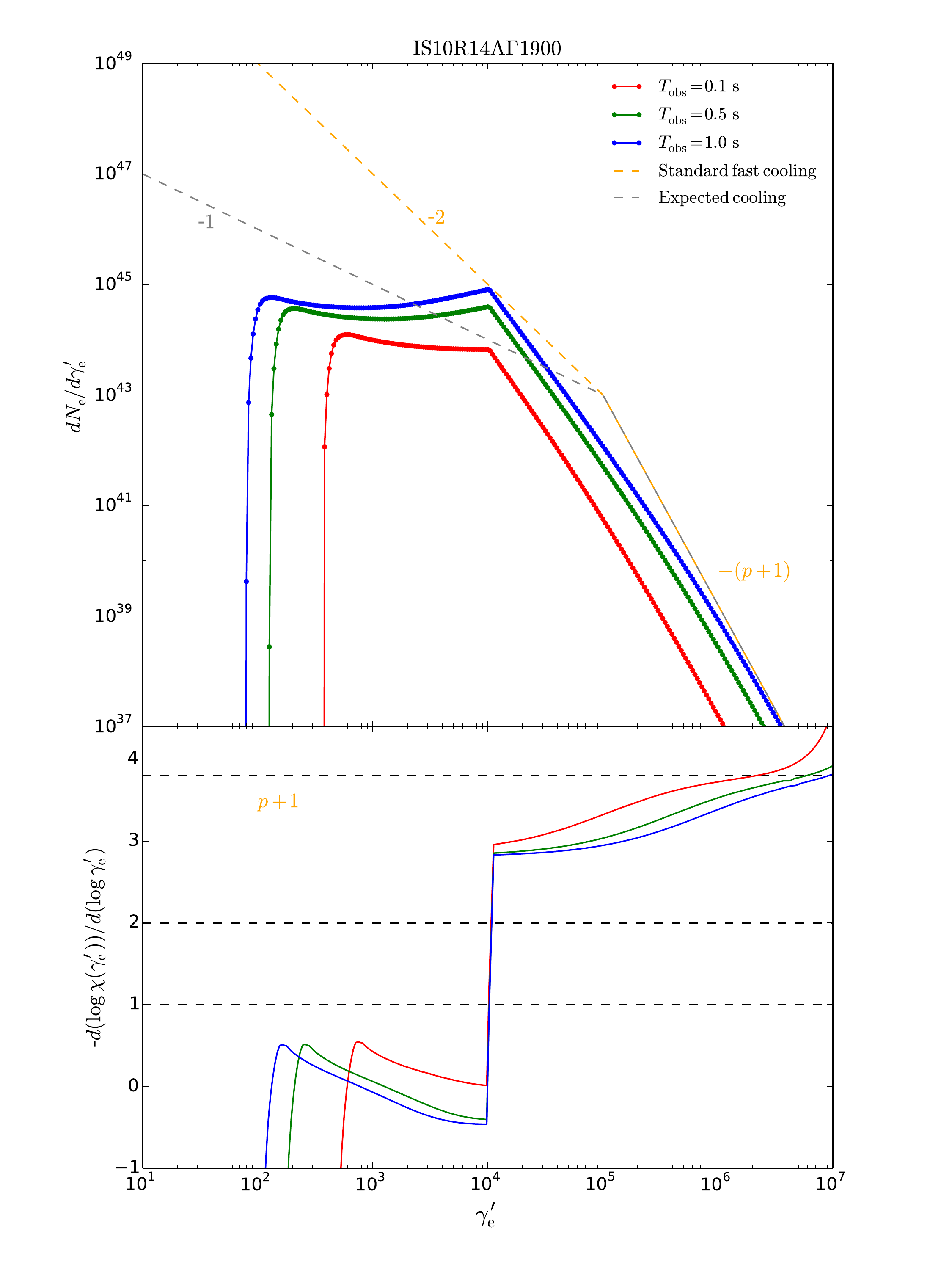}}
    \subfloat{\includegraphics[width=0.3\paperwidth]{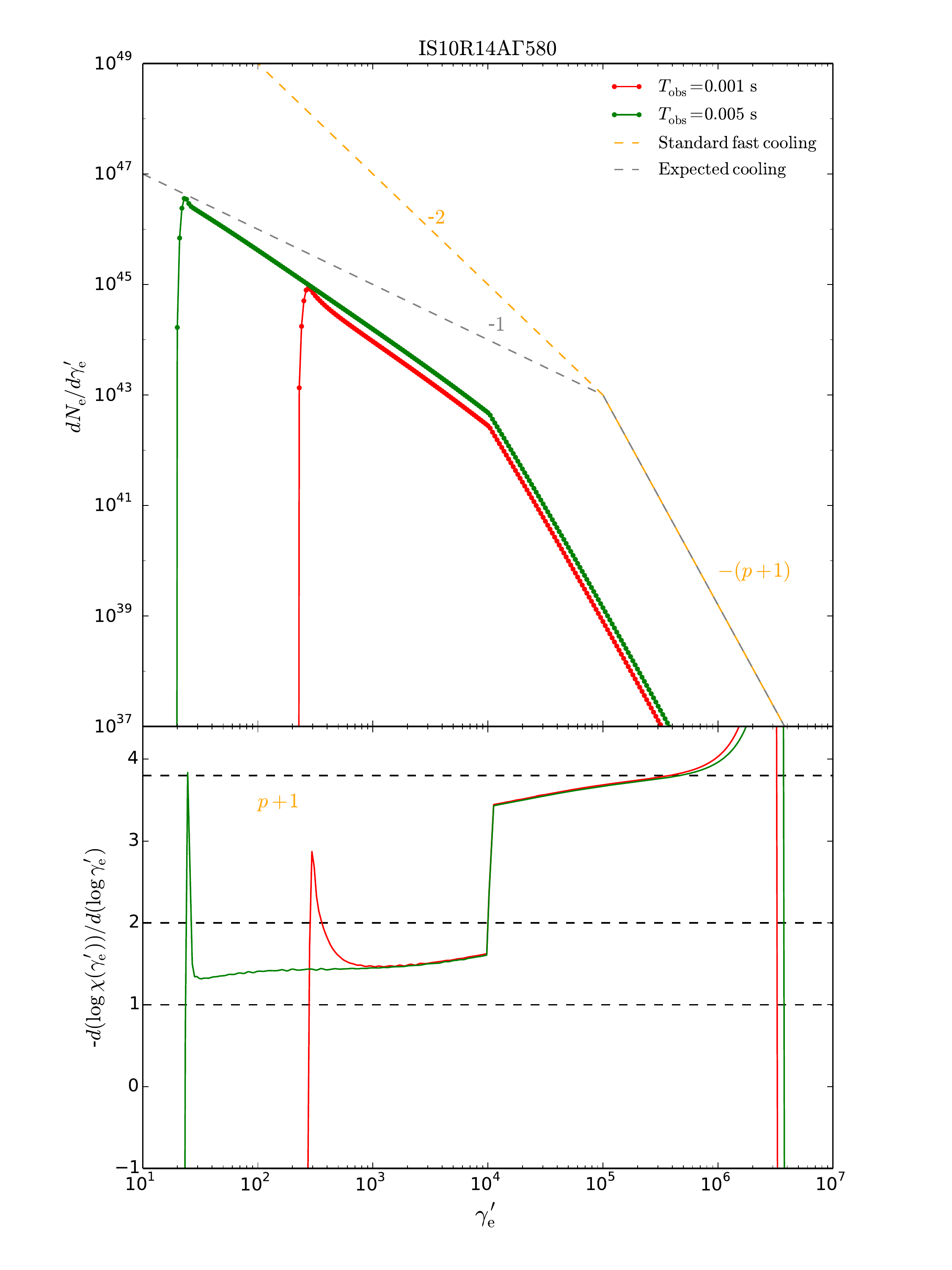}}
    \subfloat{\includegraphics[width=0.3\paperwidth]{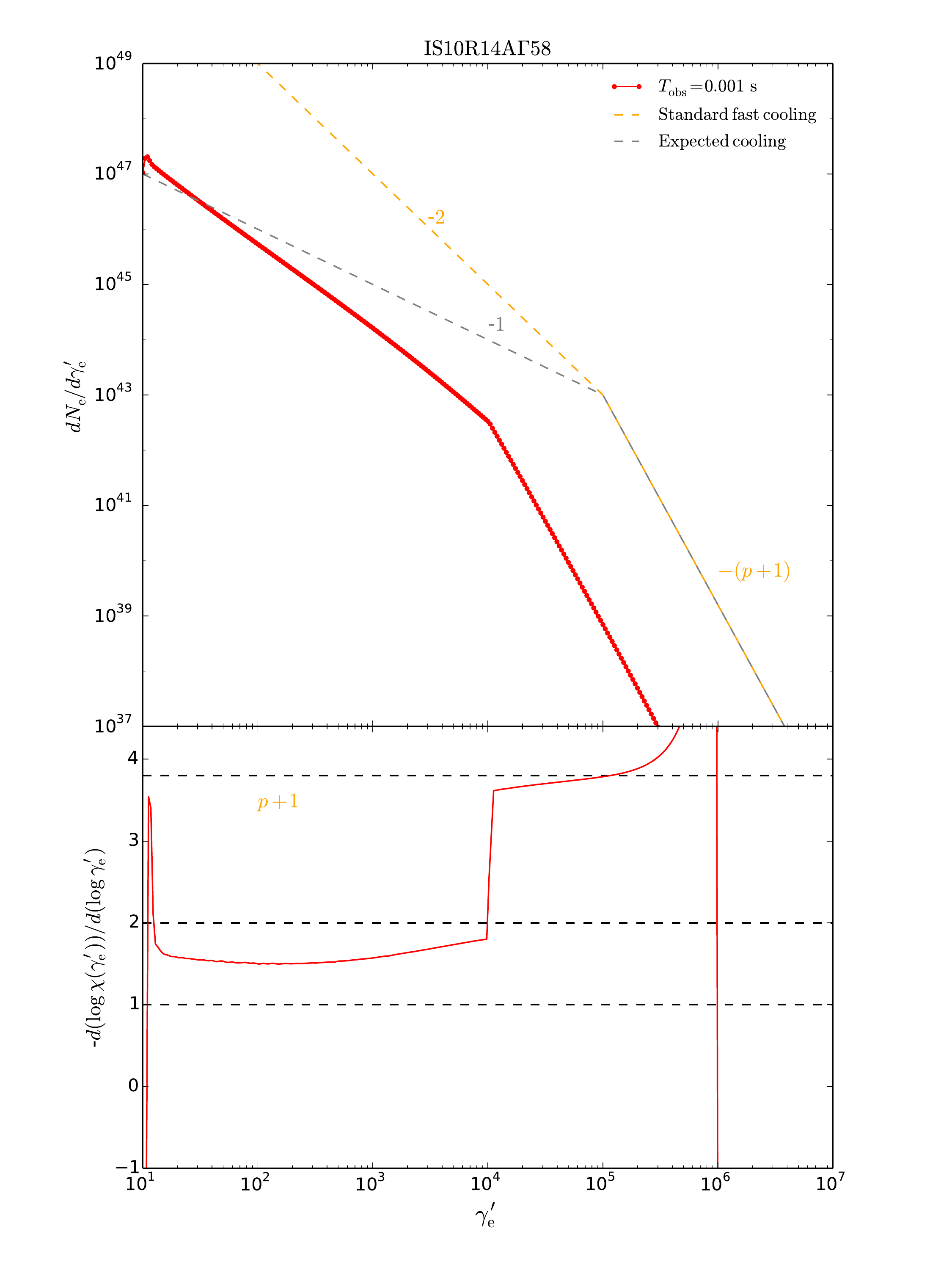}} \\
    \subfloat{\includegraphics[width=0.3\paperwidth]{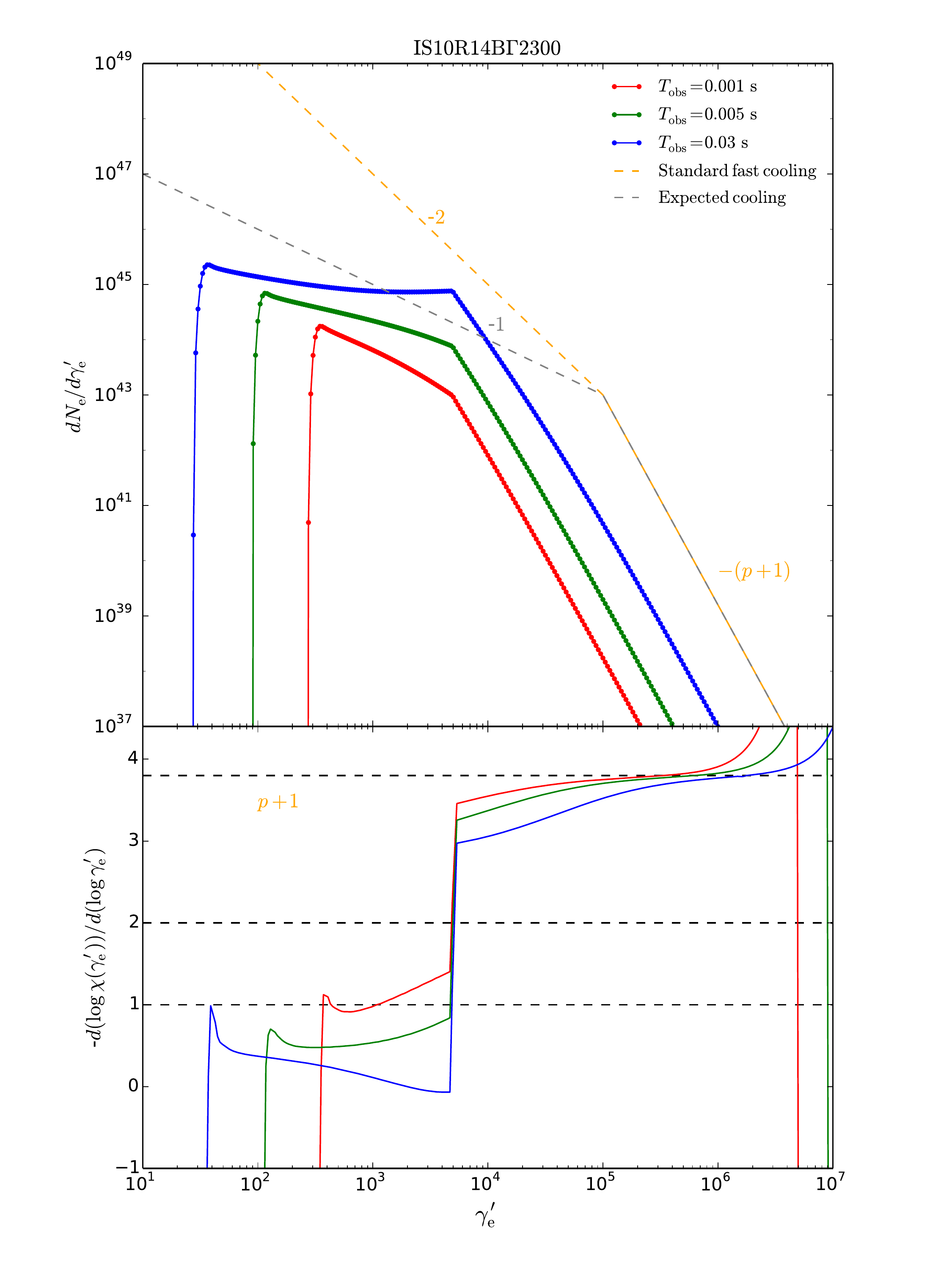}}
    \subfloat{\includegraphics[width=0.3\paperwidth]{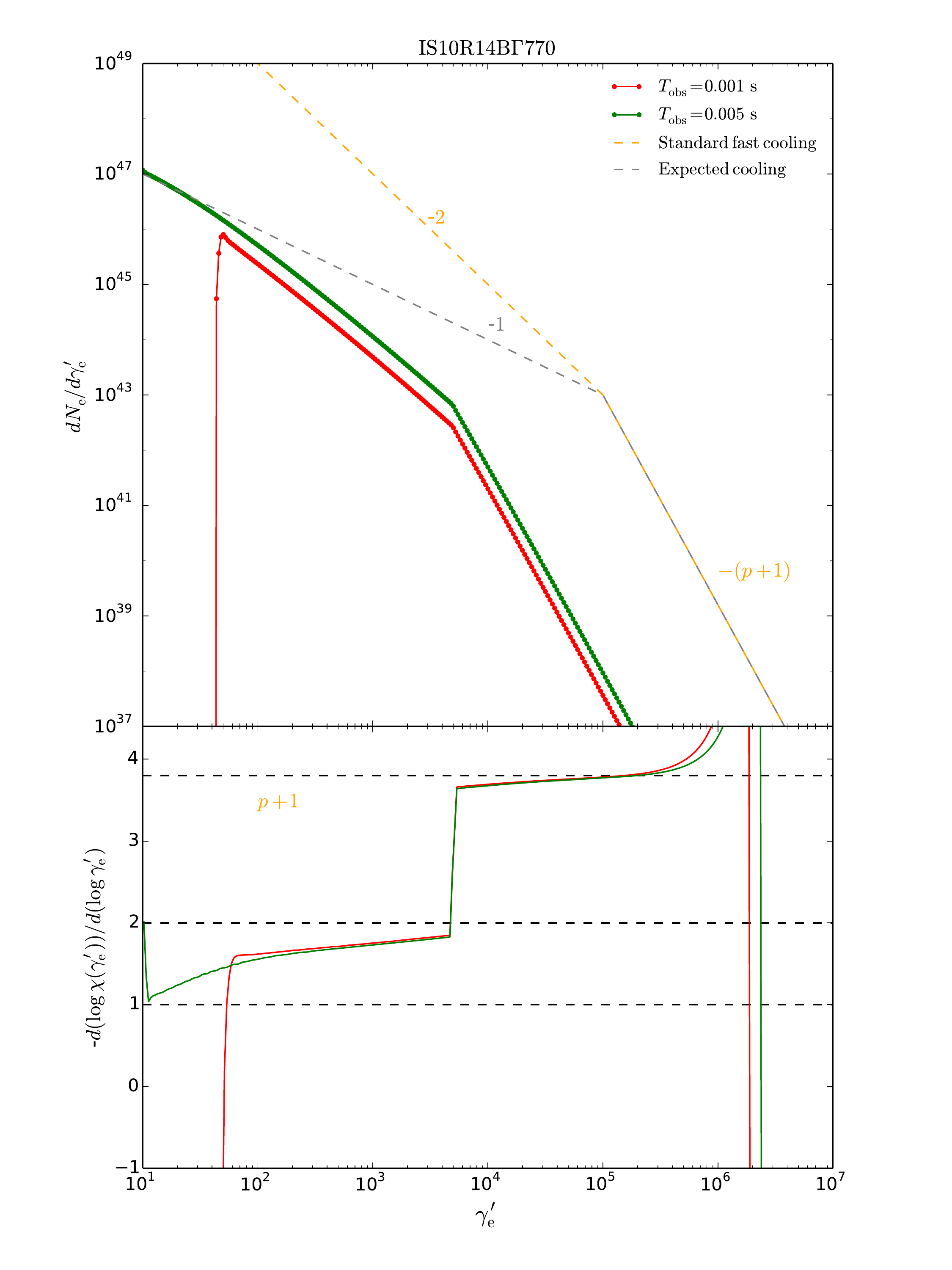}}
    \subfloat{\includegraphics[width=0.3\paperwidth]{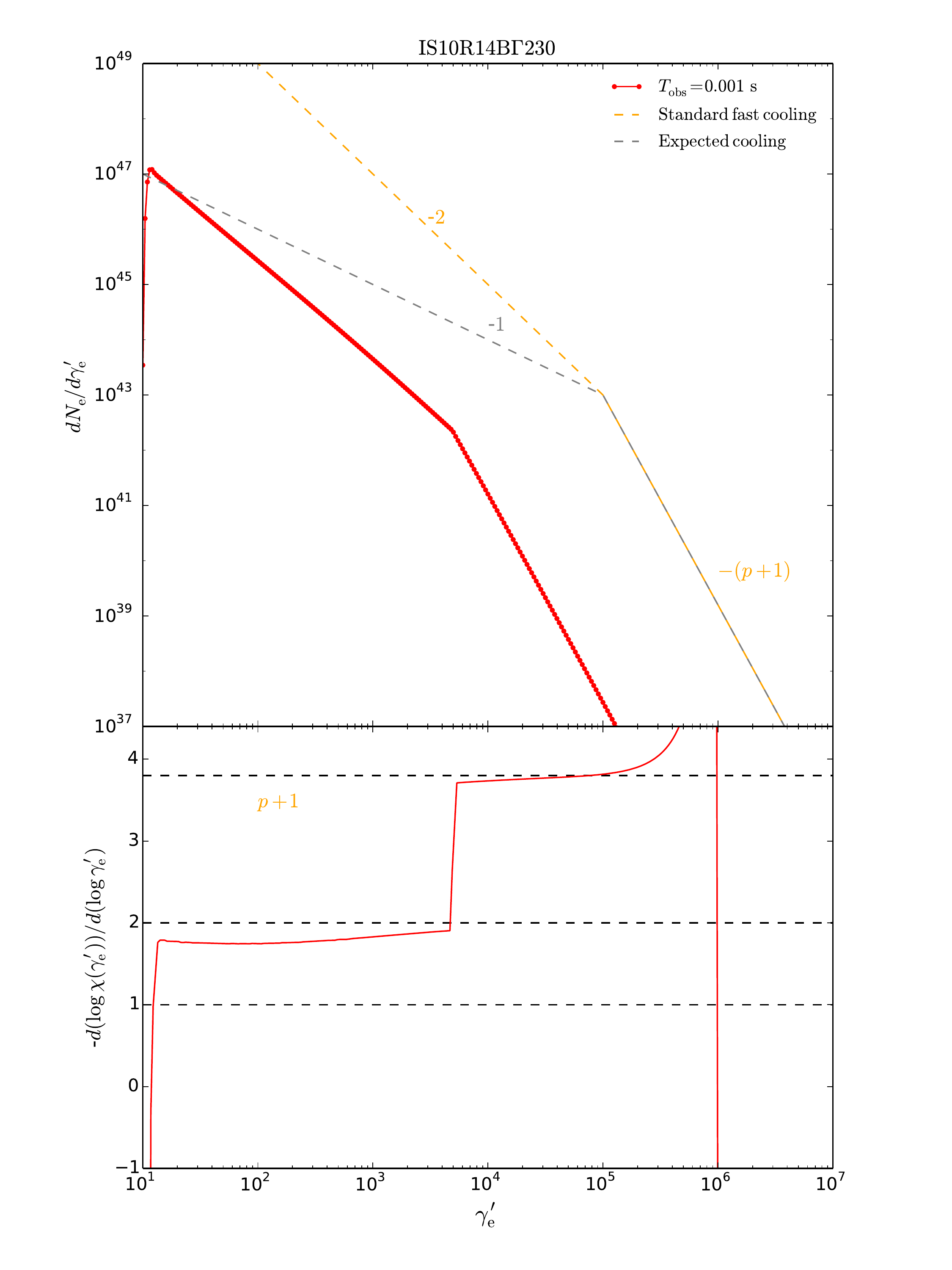}}
    \caption{The evolution of the electron energy spectrum for the six cases in Group IS10R14 (see Table \ref{TABLE:C}).\label{fig:MC-electron}}
\end{adjustwidth}
\end{figure}

\clearpage

\begin{figure}
\begin{adjustwidth}{-2cm}{-2cm}
\centering
    \subfloat{\includegraphics[width=0.3\paperwidth]{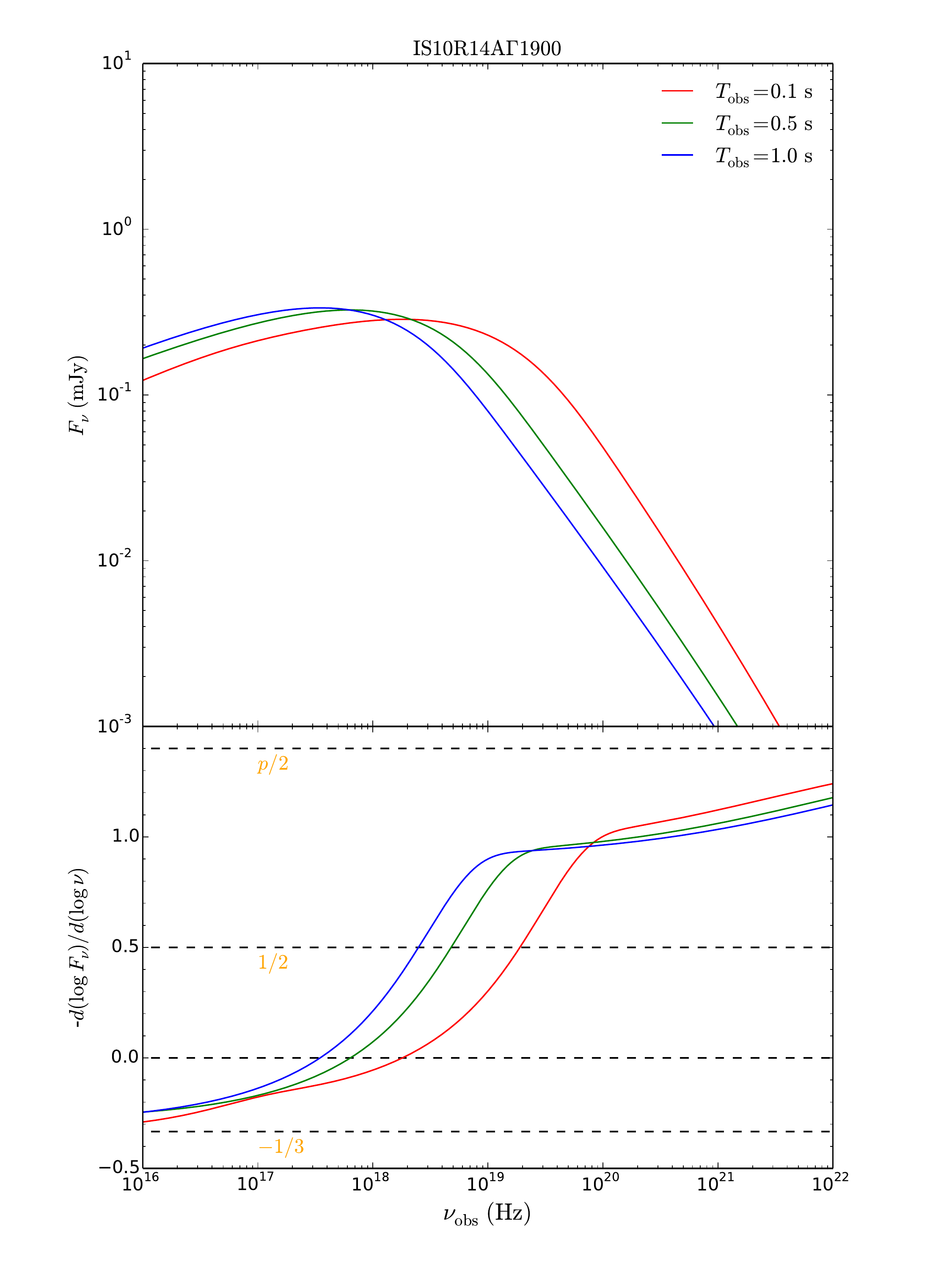}}
    \subfloat{\includegraphics[width=0.3\paperwidth]{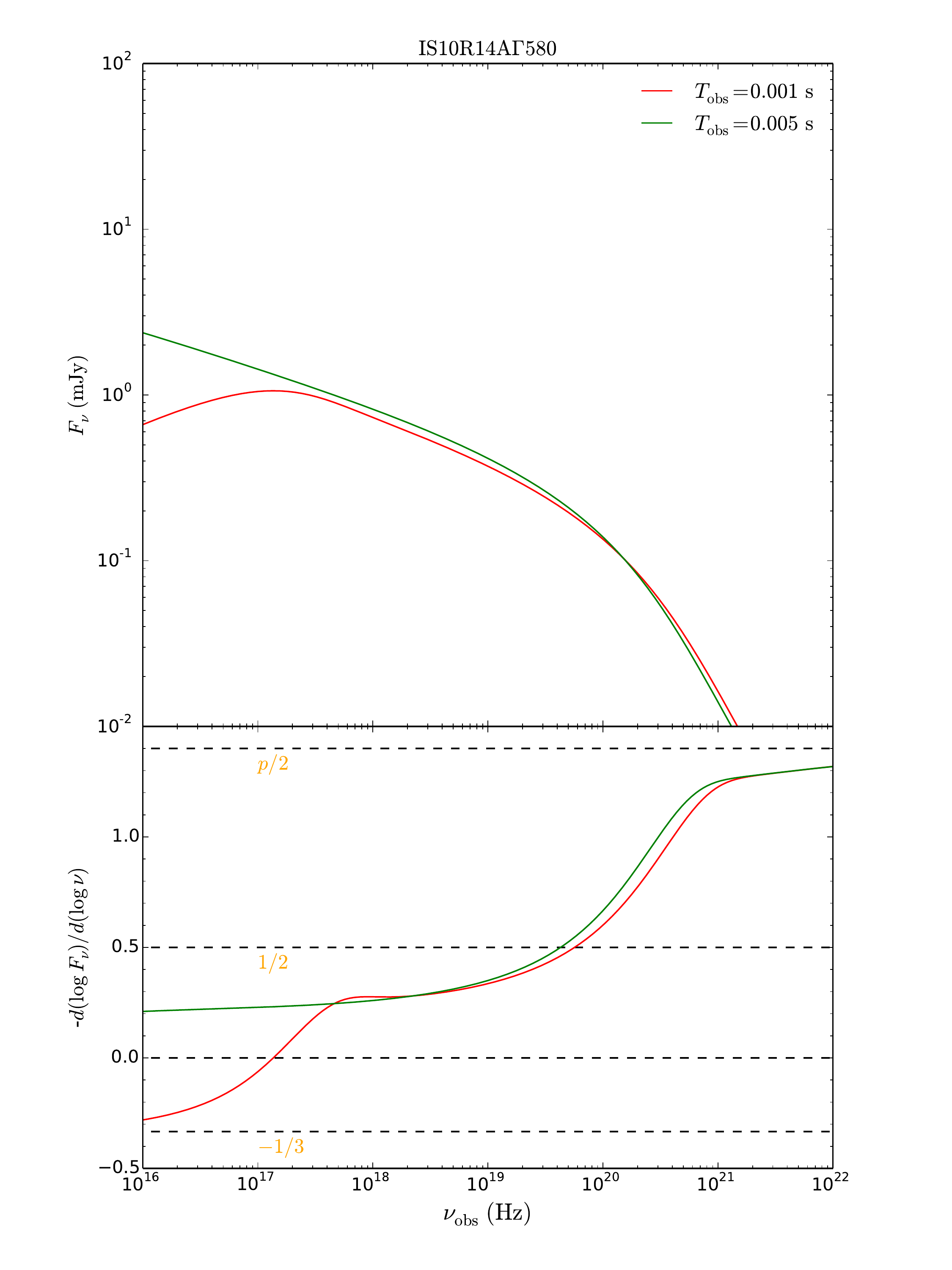}}
    \subfloat{\includegraphics[width=0.3\paperwidth]{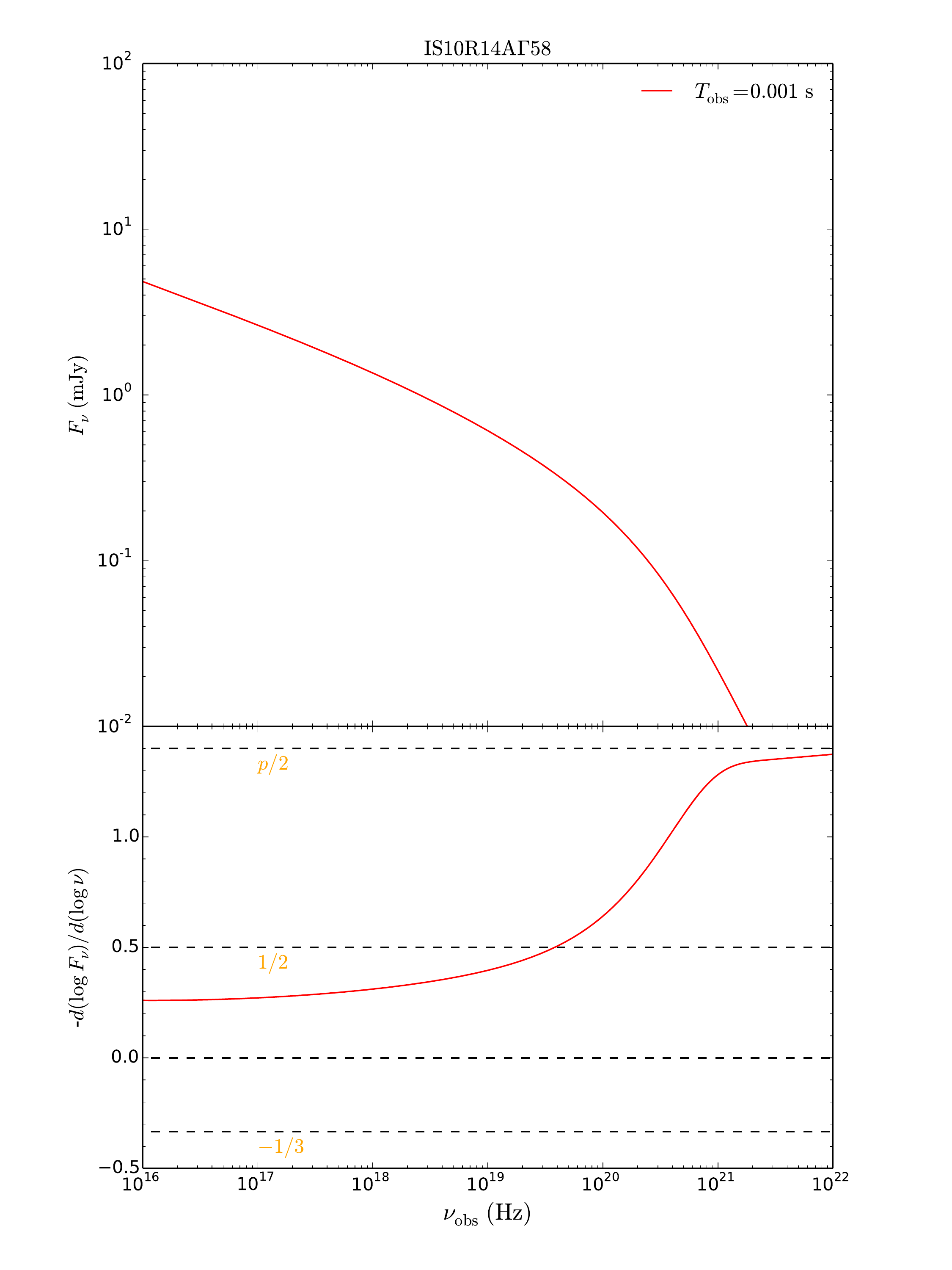}} \\
    \subfloat{\includegraphics[width=0.3\paperwidth]{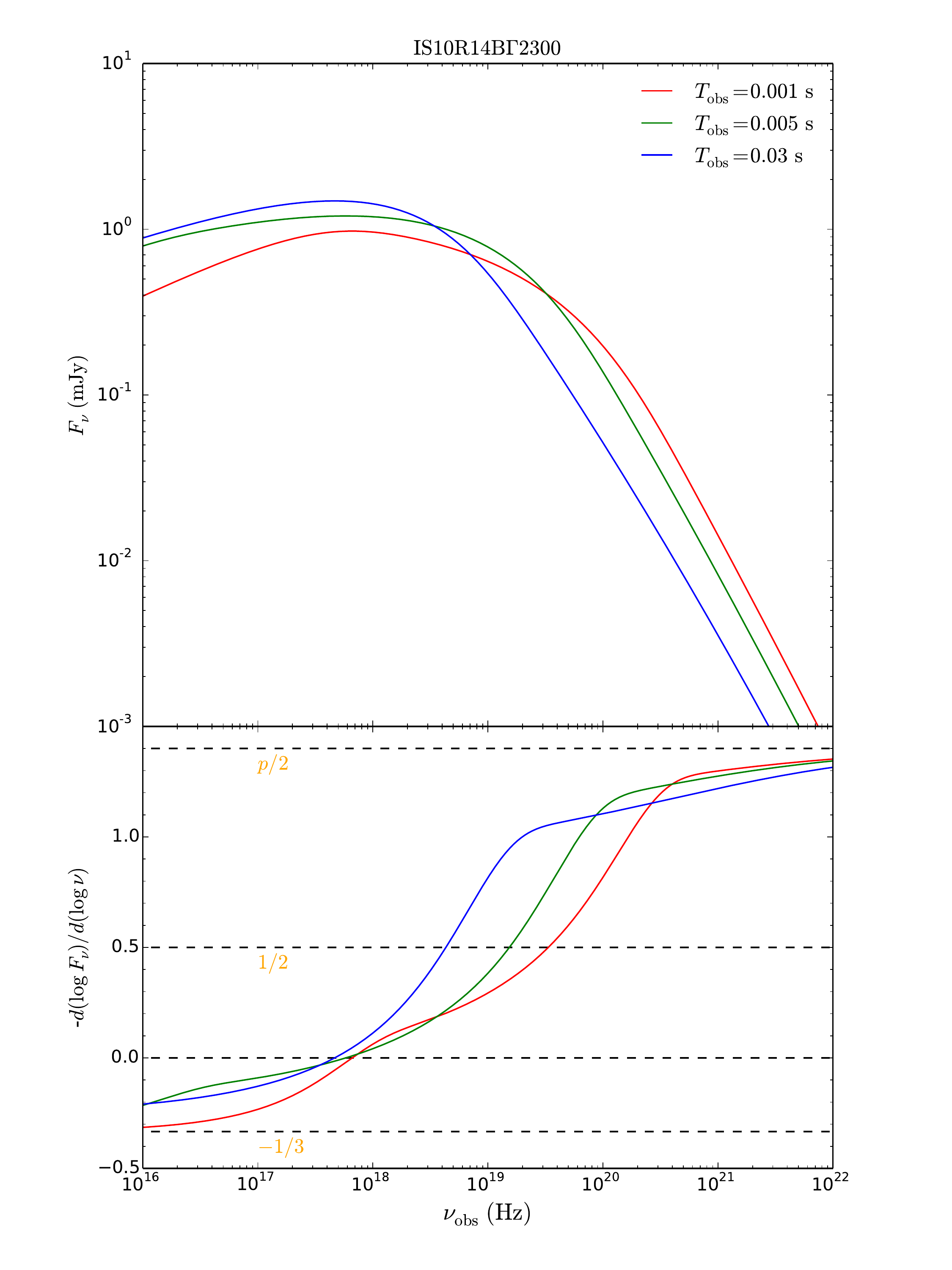}}
    \subfloat{\includegraphics[width=0.3\paperwidth]{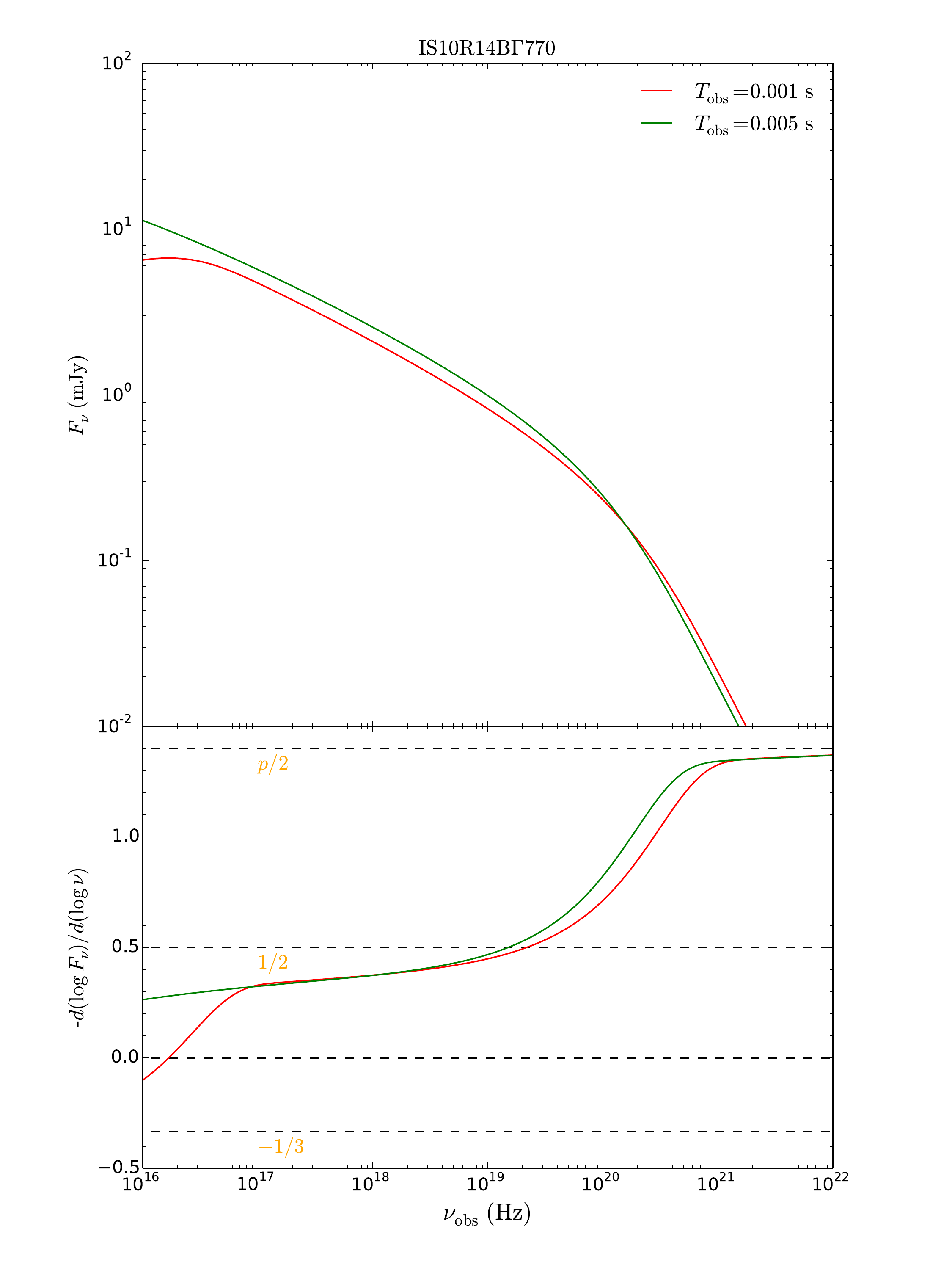}}
    \subfloat{\includegraphics[width=0.3\paperwidth]{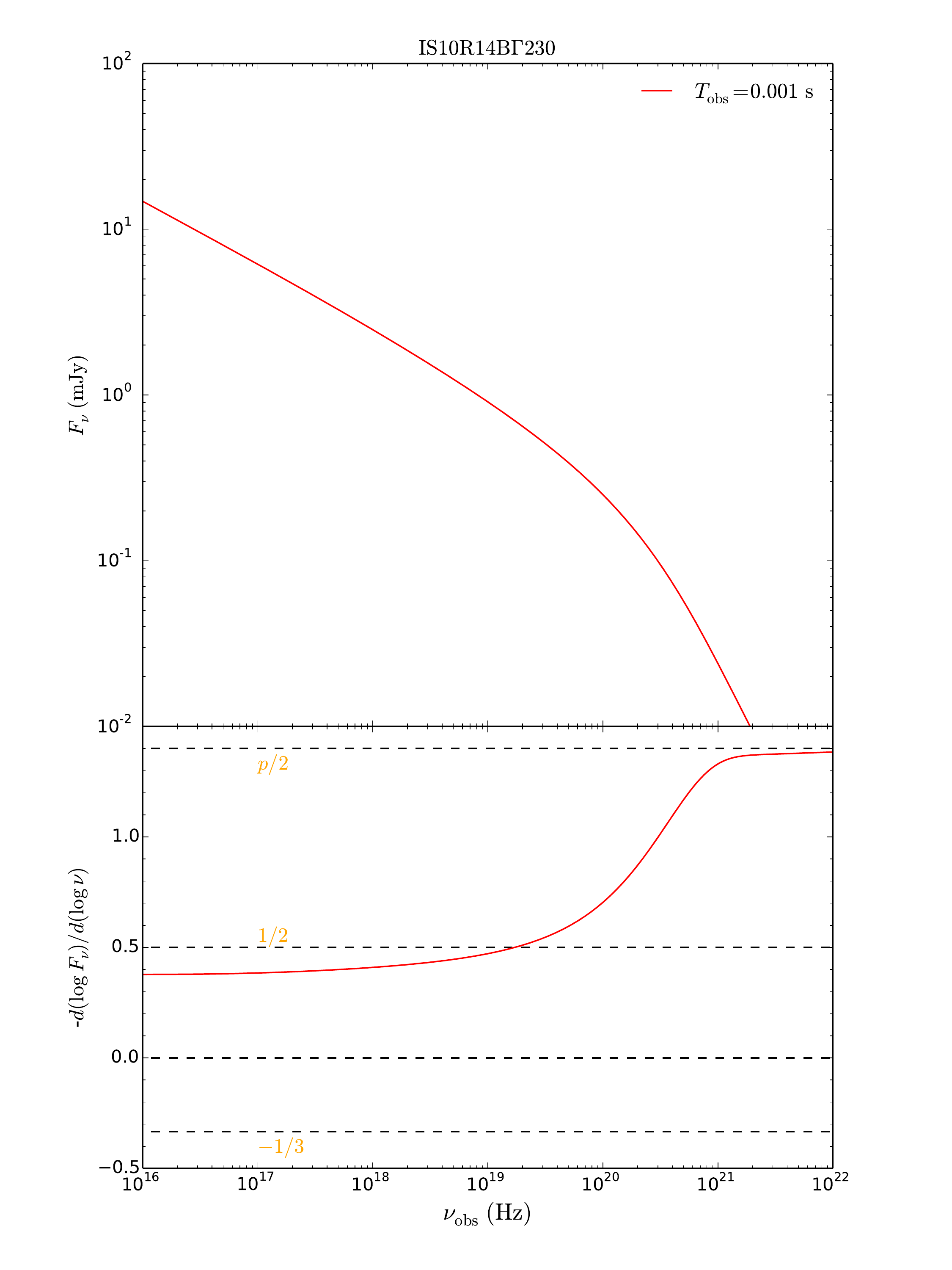}}
    \caption{The corresponding synchrotron flux-density spectra $F_{\nu}$ from the electrons with the energy distribution presented in Figure
    \ref{fig:MC-electron}.\label{fig:MC-spectra}}
\end{adjustwidth}
\end{figure}

\clearpage

\begin{figure}
\begin{adjustwidth}{-2cm}{-2cm}
\centering
    \subfloat{\includegraphics[width=0.3\paperwidth]{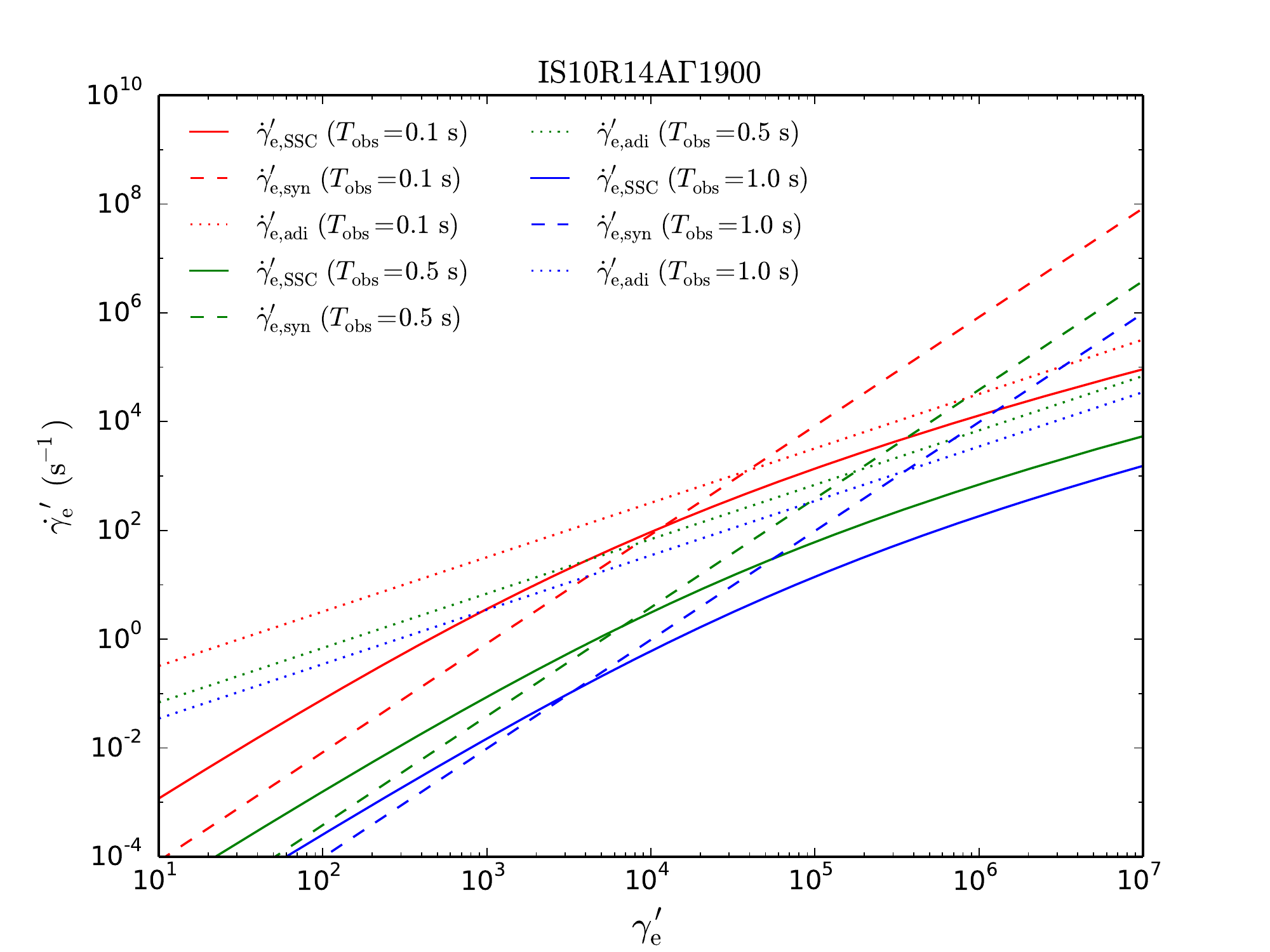}}
    \subfloat{\includegraphics[width=0.3\paperwidth]{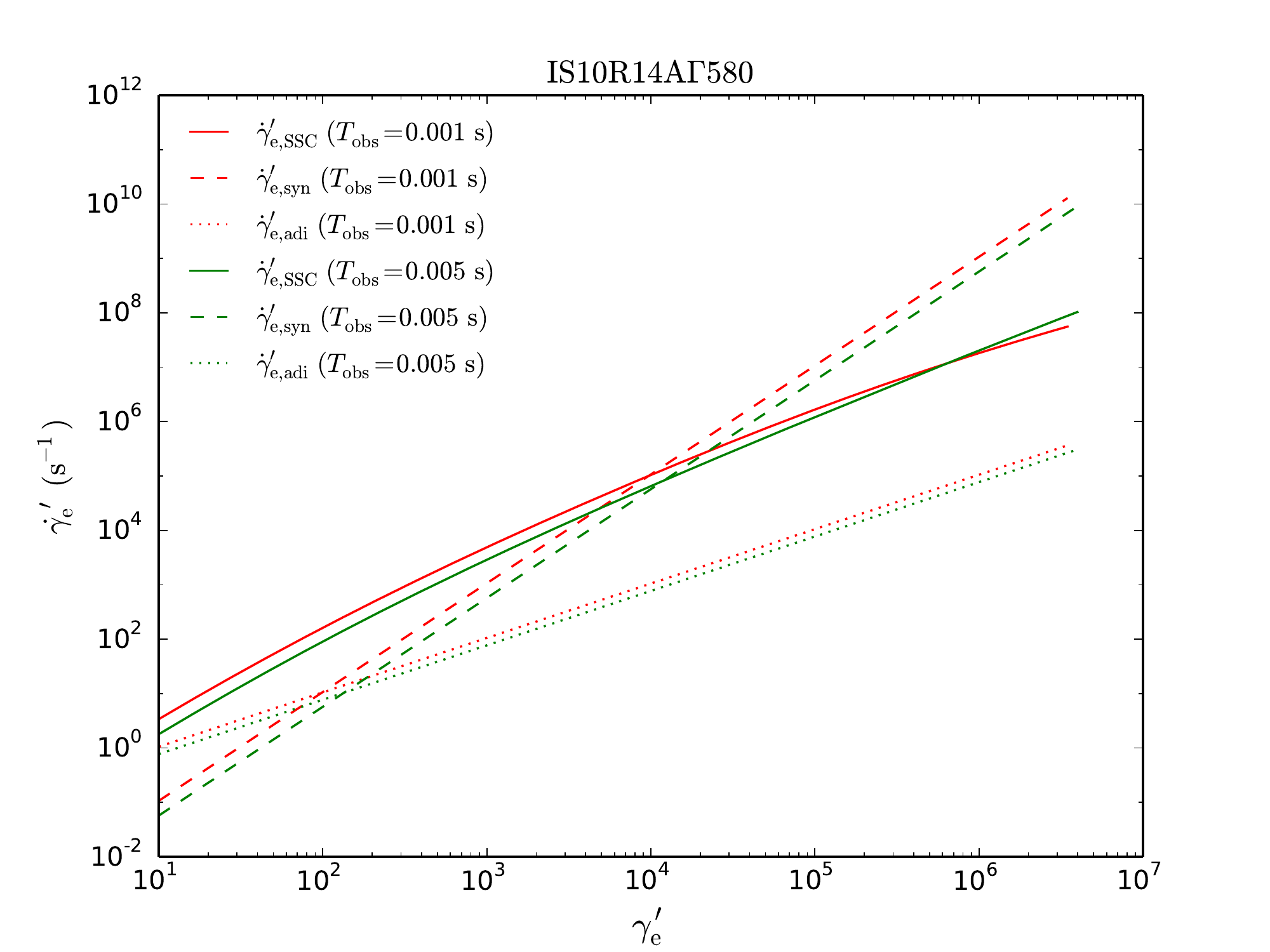}}
    \subfloat{\includegraphics[width=0.3\paperwidth]{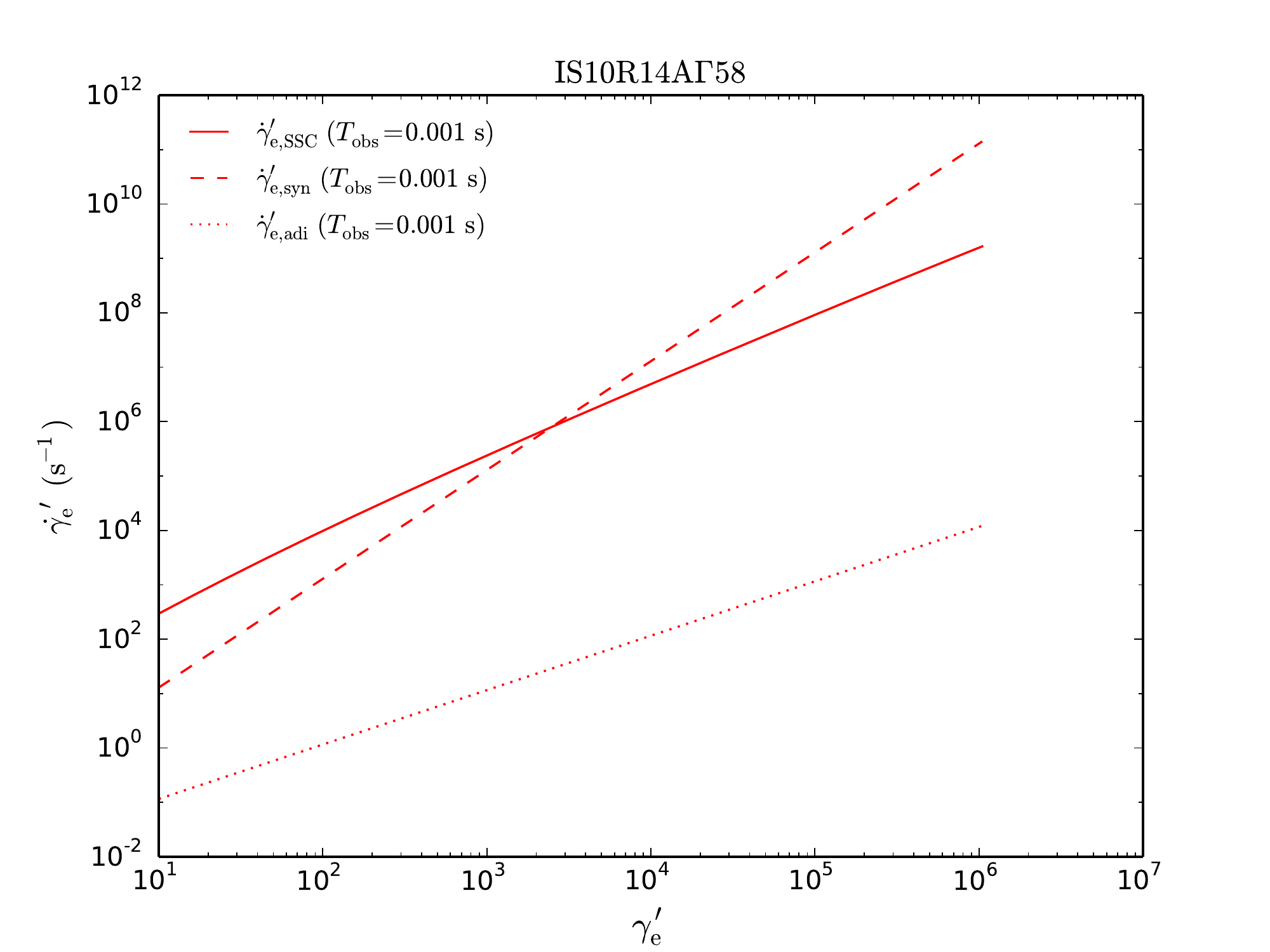}} \\
    \subfloat{\includegraphics[width=0.3\paperwidth]{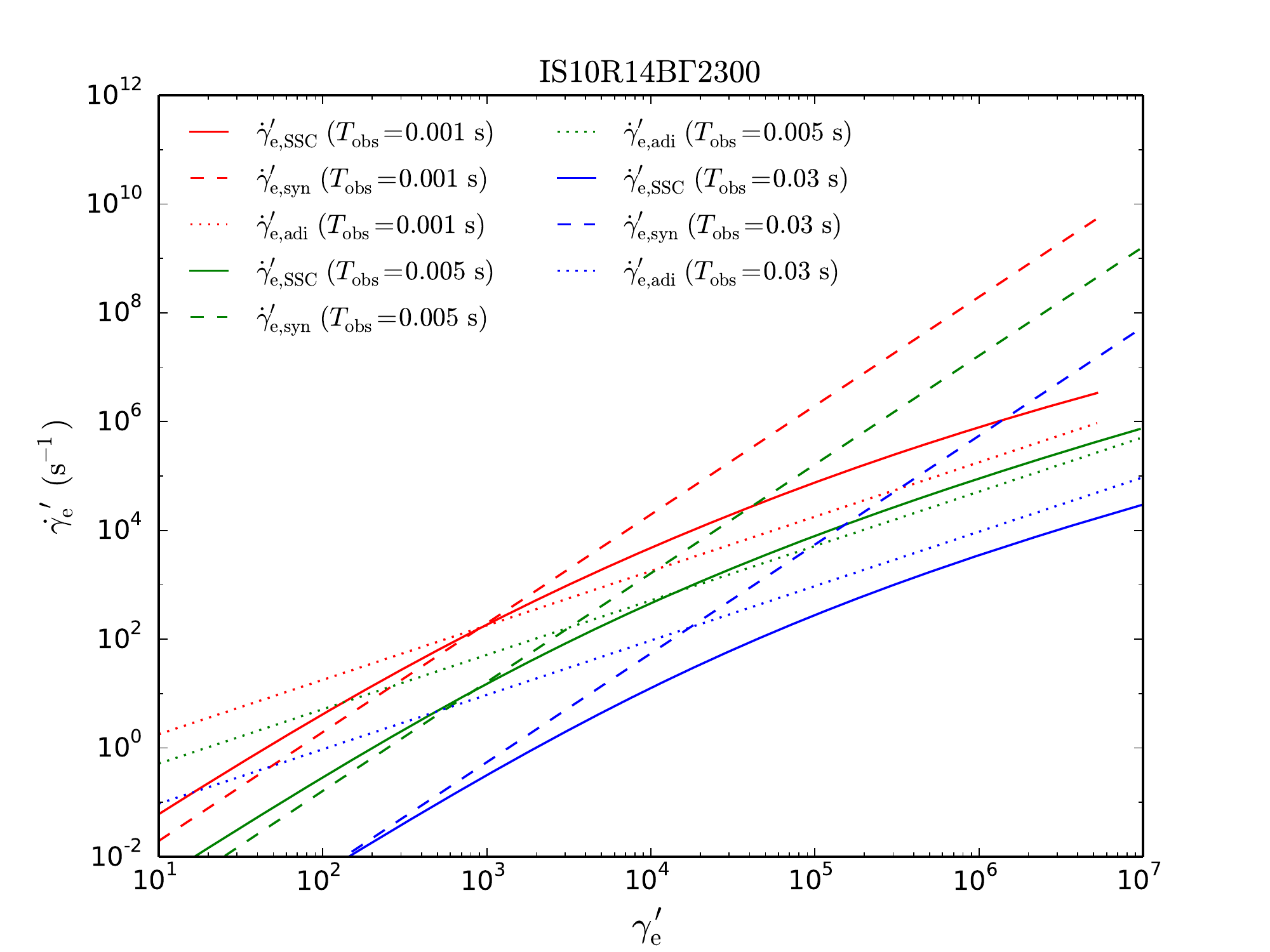}}
    \subfloat{\includegraphics[width=0.3\paperwidth]{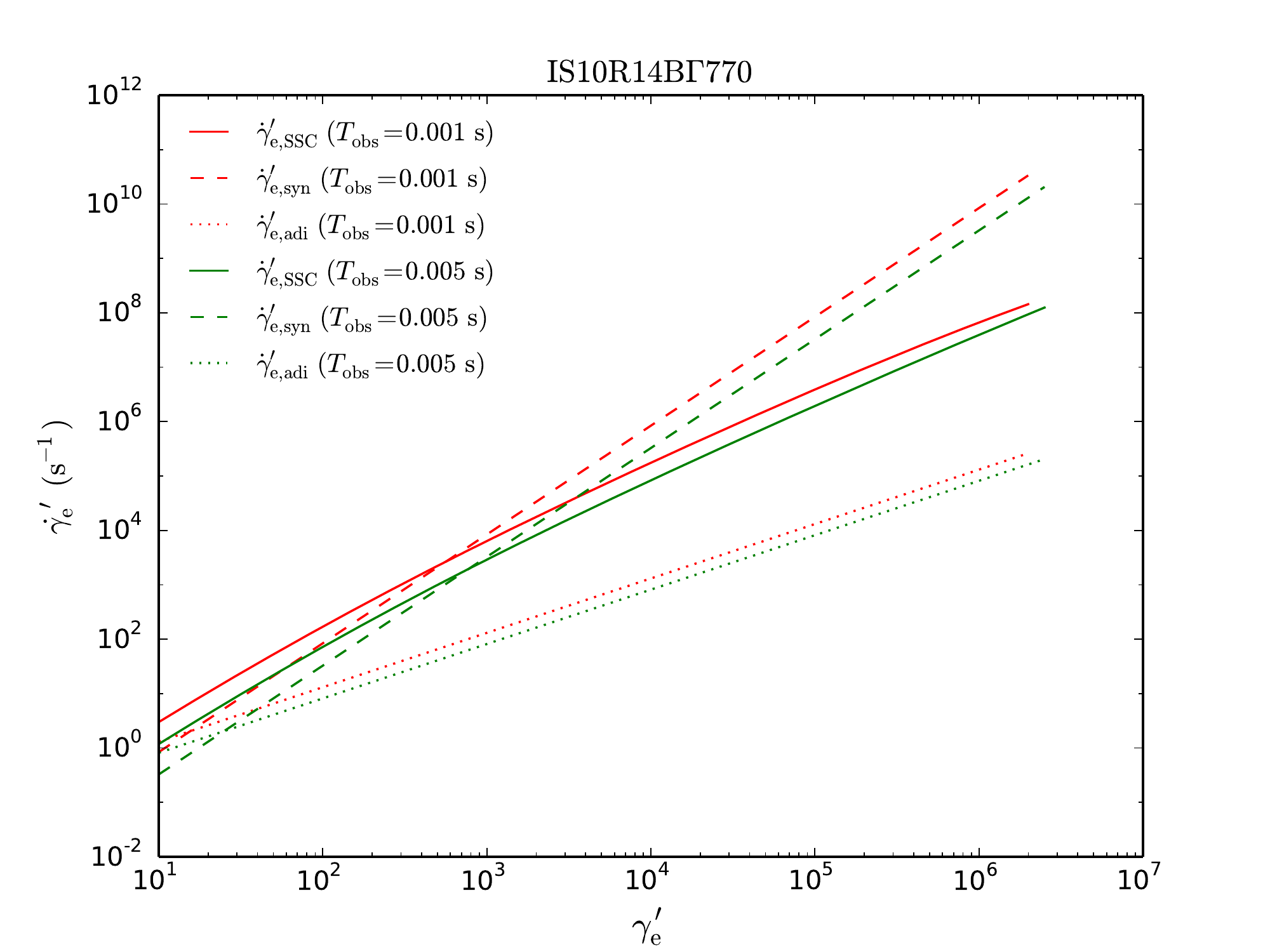}}
    \subfloat{\includegraphics[width=0.3\paperwidth]{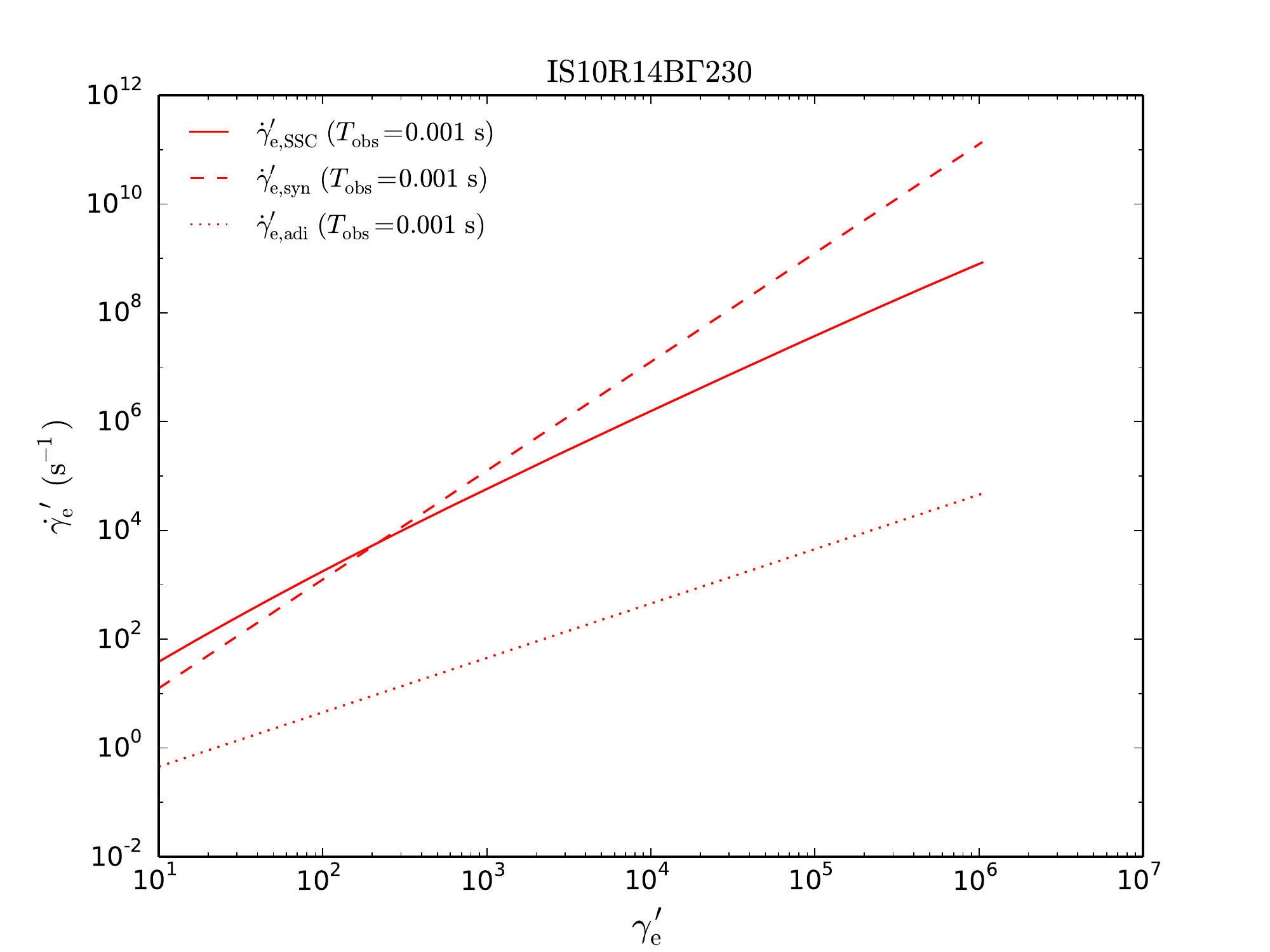}}
    \caption{The co-moving cooling rates of different cooling mechanisms for the electrons with the energy distribution
    presented in Figure \ref{fig:MC-electron}.\label{fig:MC-rate}}
\end{adjustwidth}
\end{figure}

\clearpage

\begin{figure}
\begin{adjustwidth}{-2cm}{-2cm}
\centering
    \subfloat{\includegraphics[width=0.25\paperwidth]{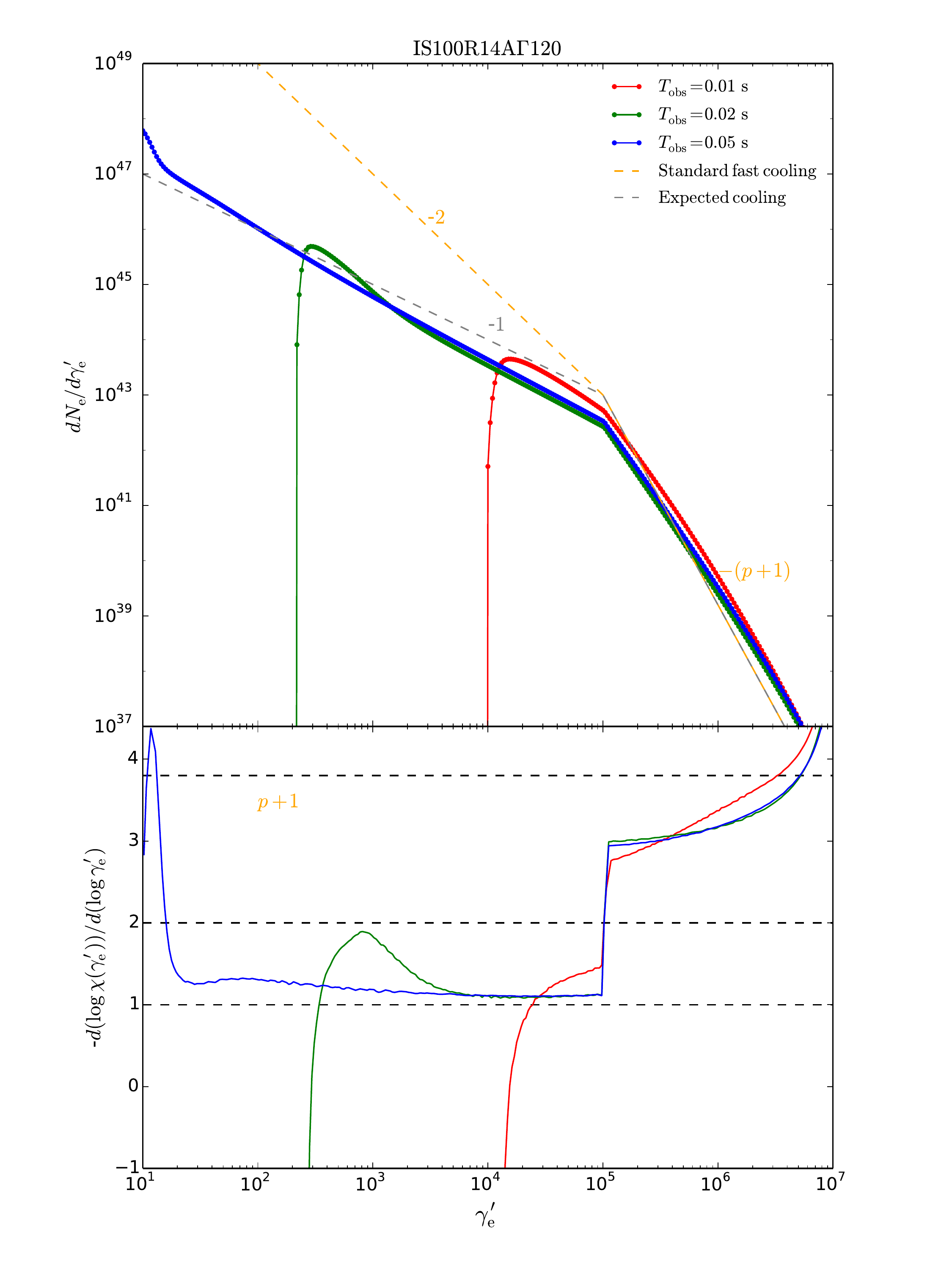}}
    \subfloat{\includegraphics[width=0.25\paperwidth]{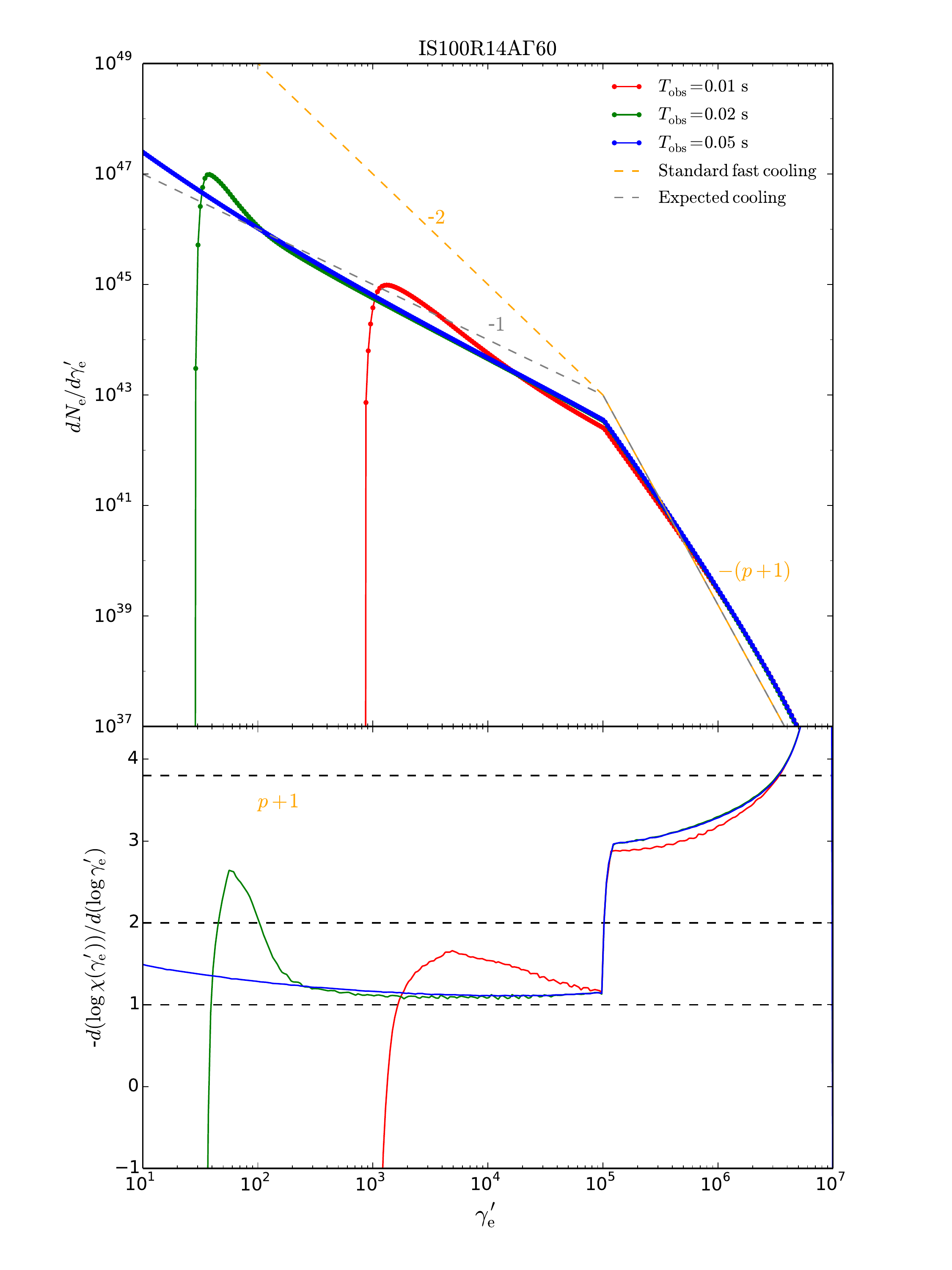}}
    \subfloat{\includegraphics[width=0.25\paperwidth]{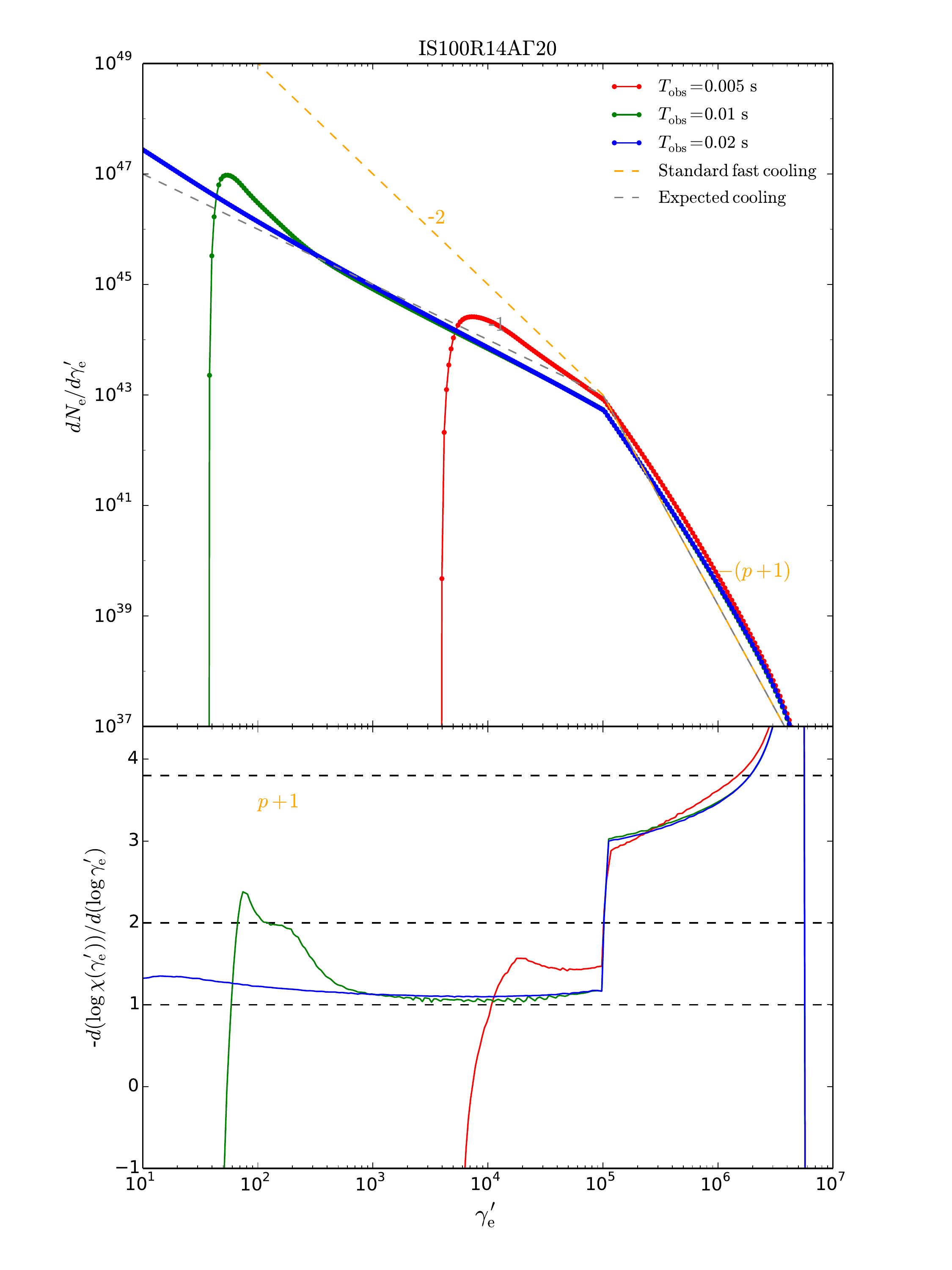}} \\
    \subfloat{\includegraphics[width=0.25\paperwidth]{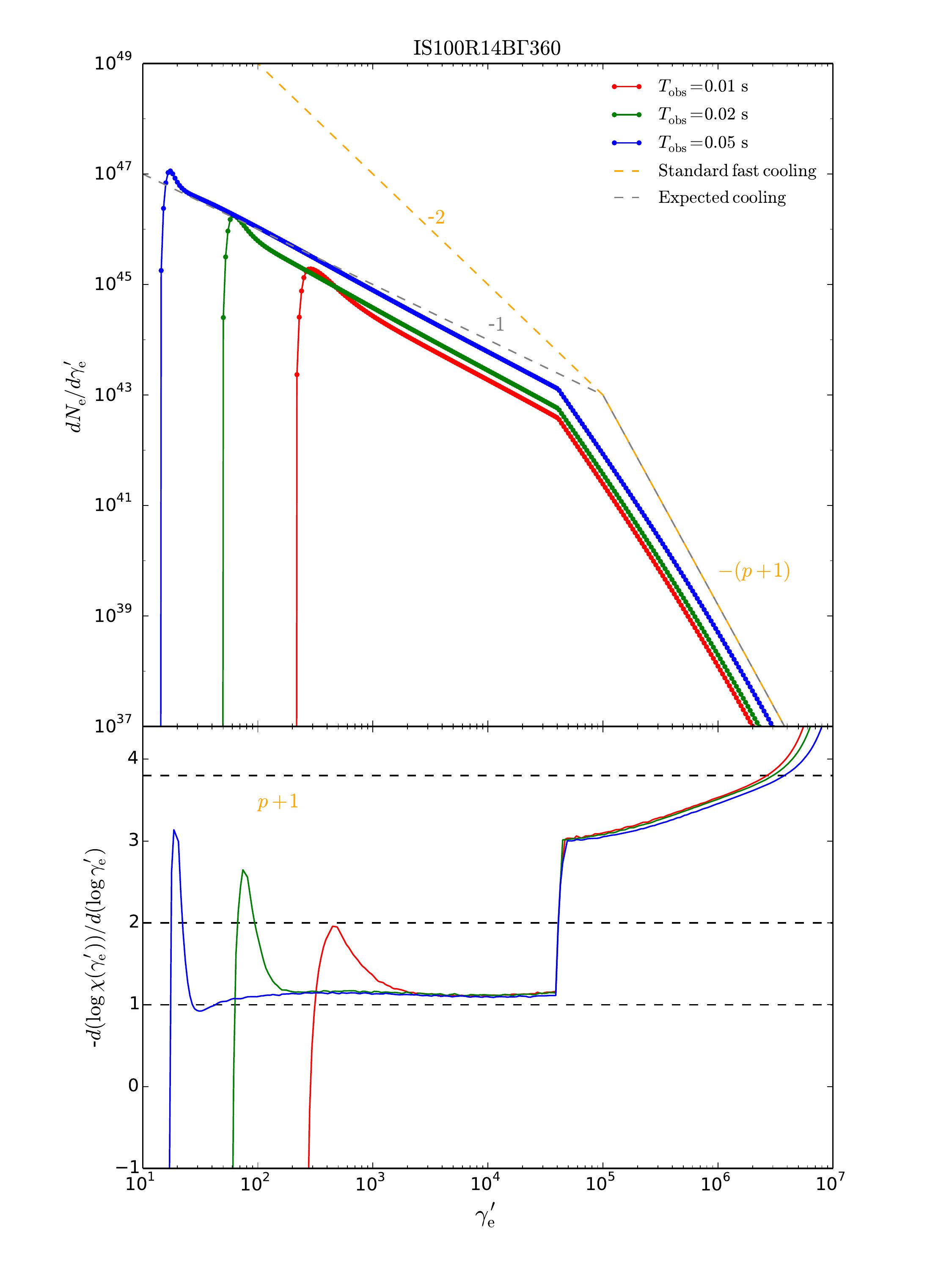}}
    \subfloat{\includegraphics[width=0.25\paperwidth]{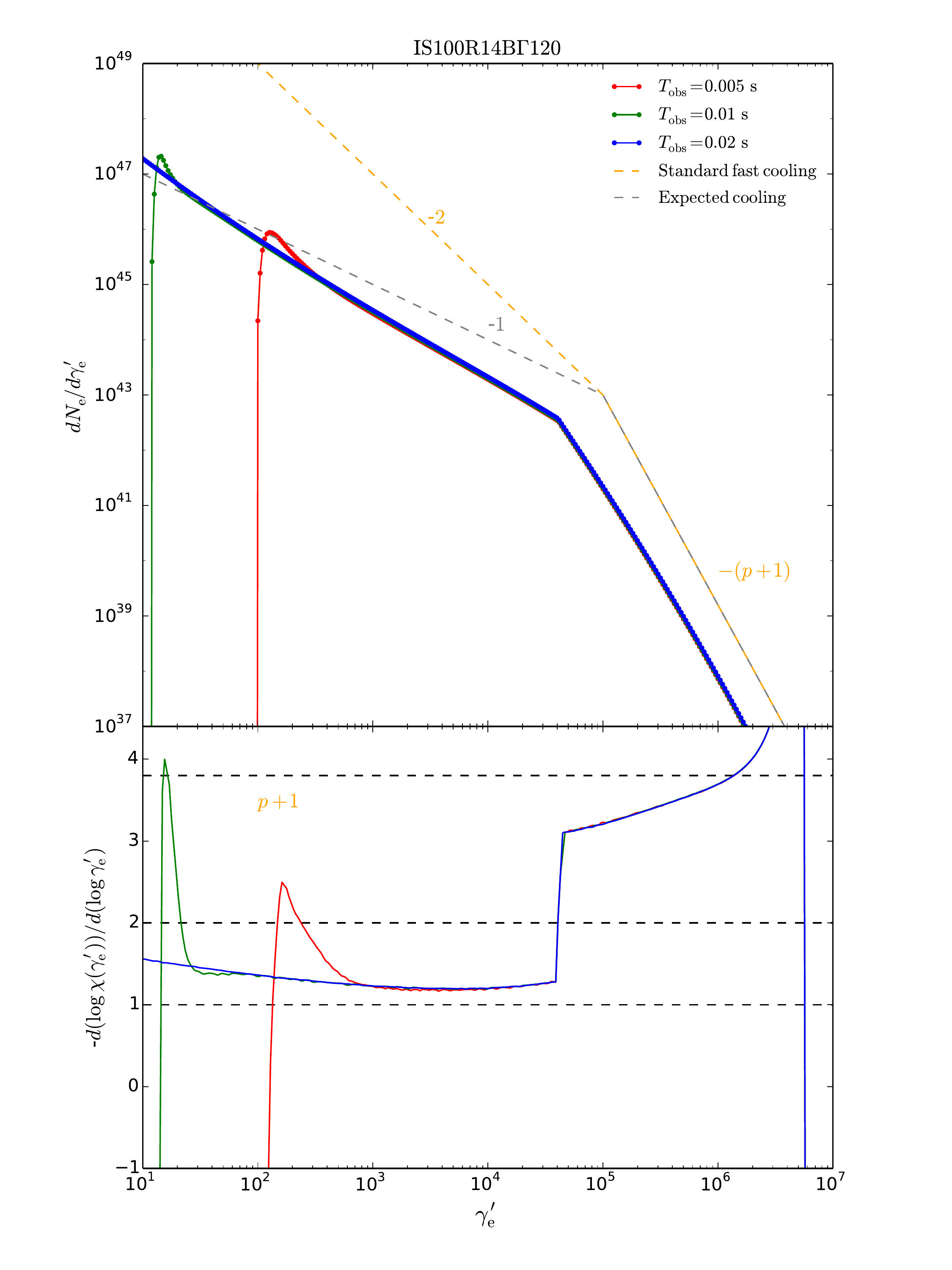}}
    \subfloat{\includegraphics[width=0.25\paperwidth]{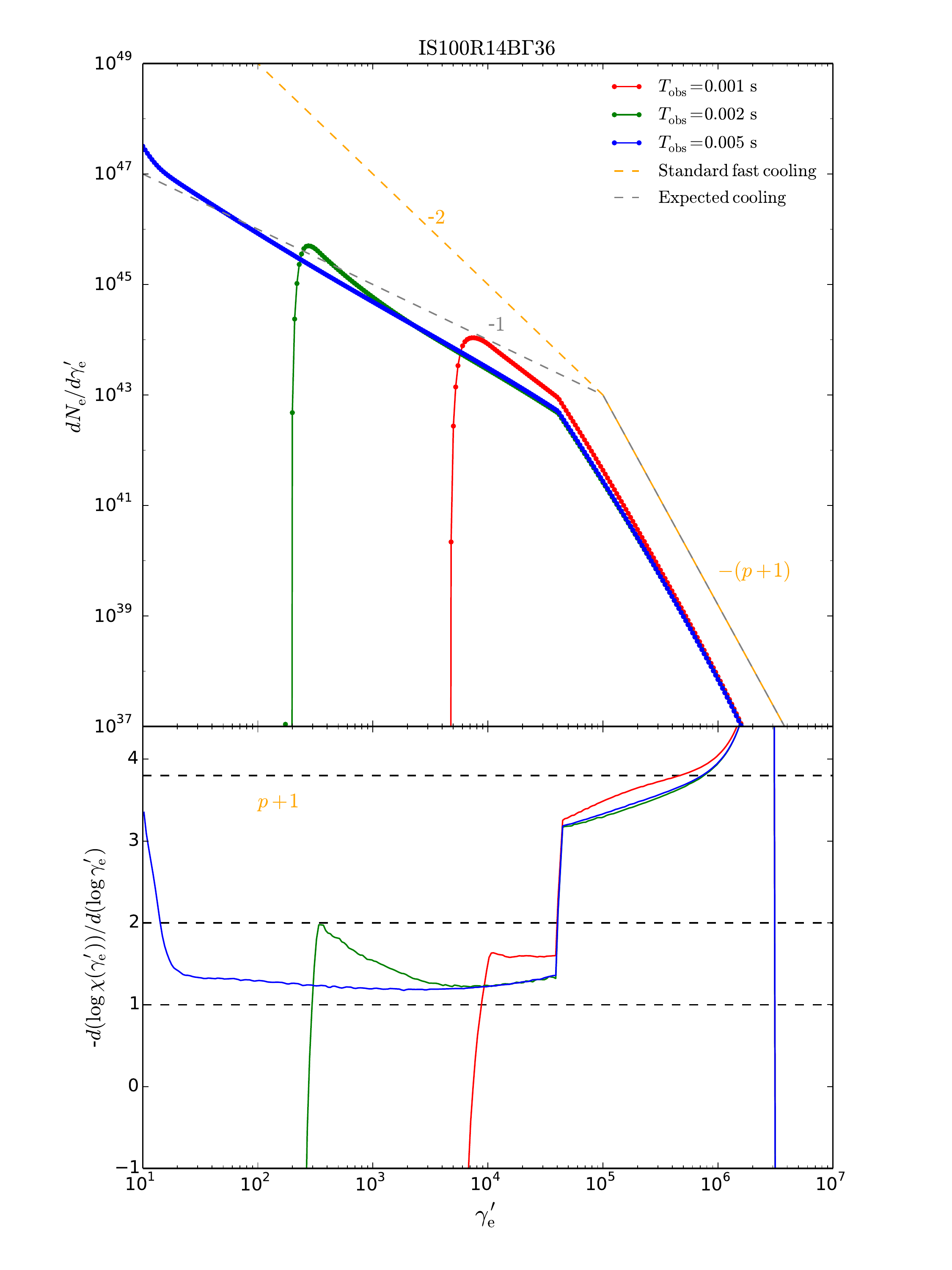}} \\
    \subfloat{\includegraphics[width=0.25\paperwidth]{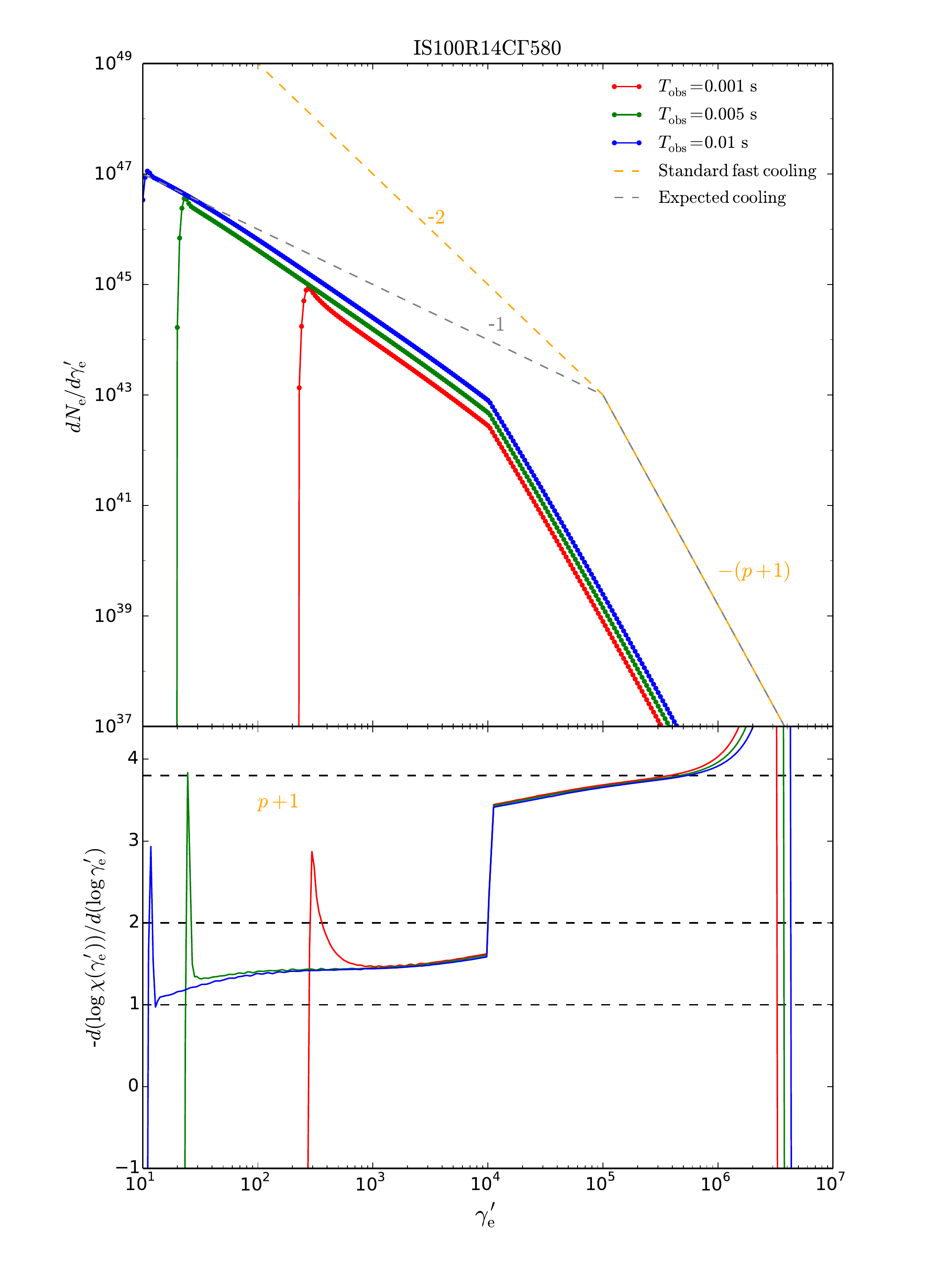}}
    \subfloat{\includegraphics[width=0.25\paperwidth]{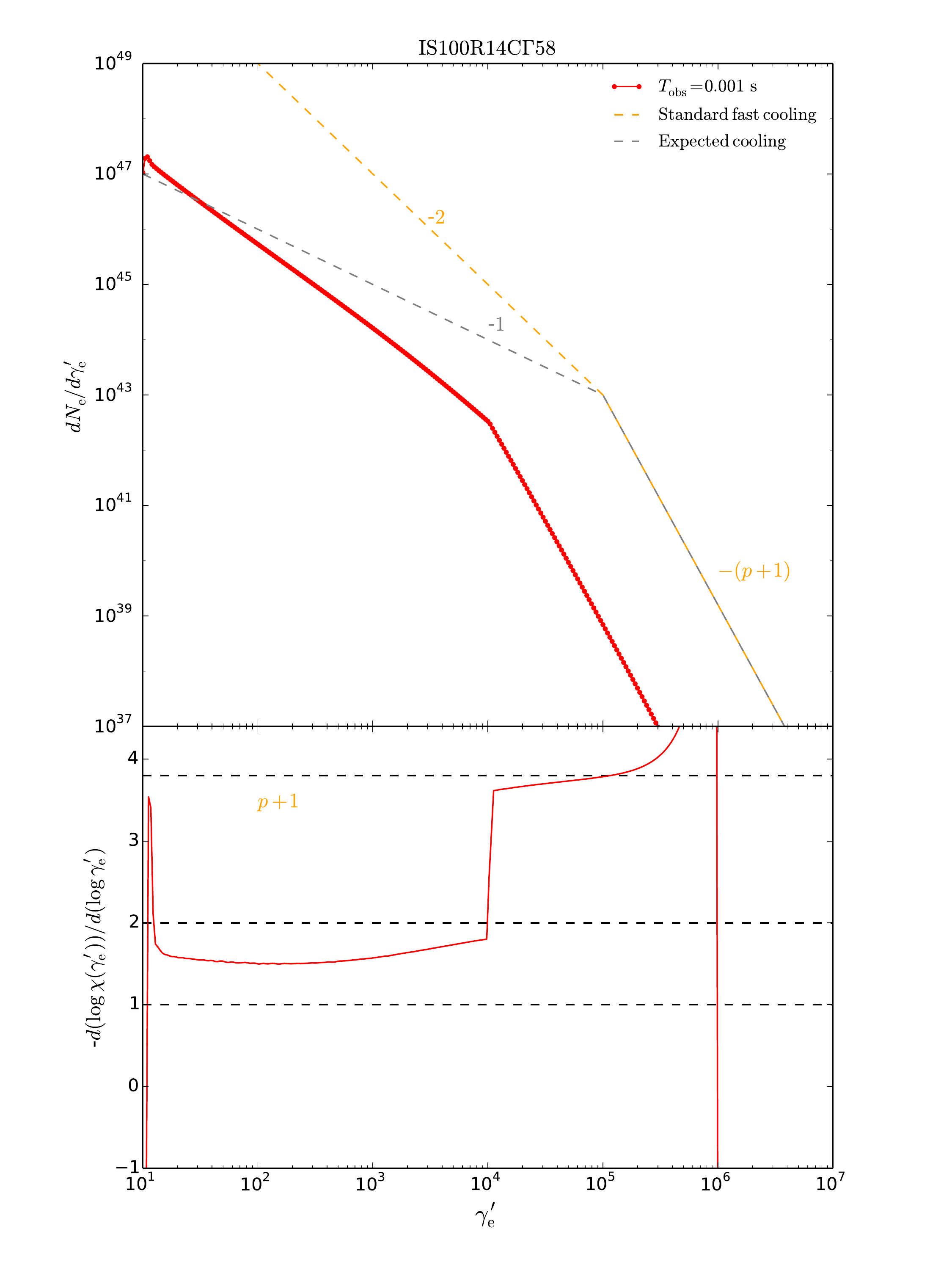}}
    \subfloat{\includegraphics[width=0.25\paperwidth]{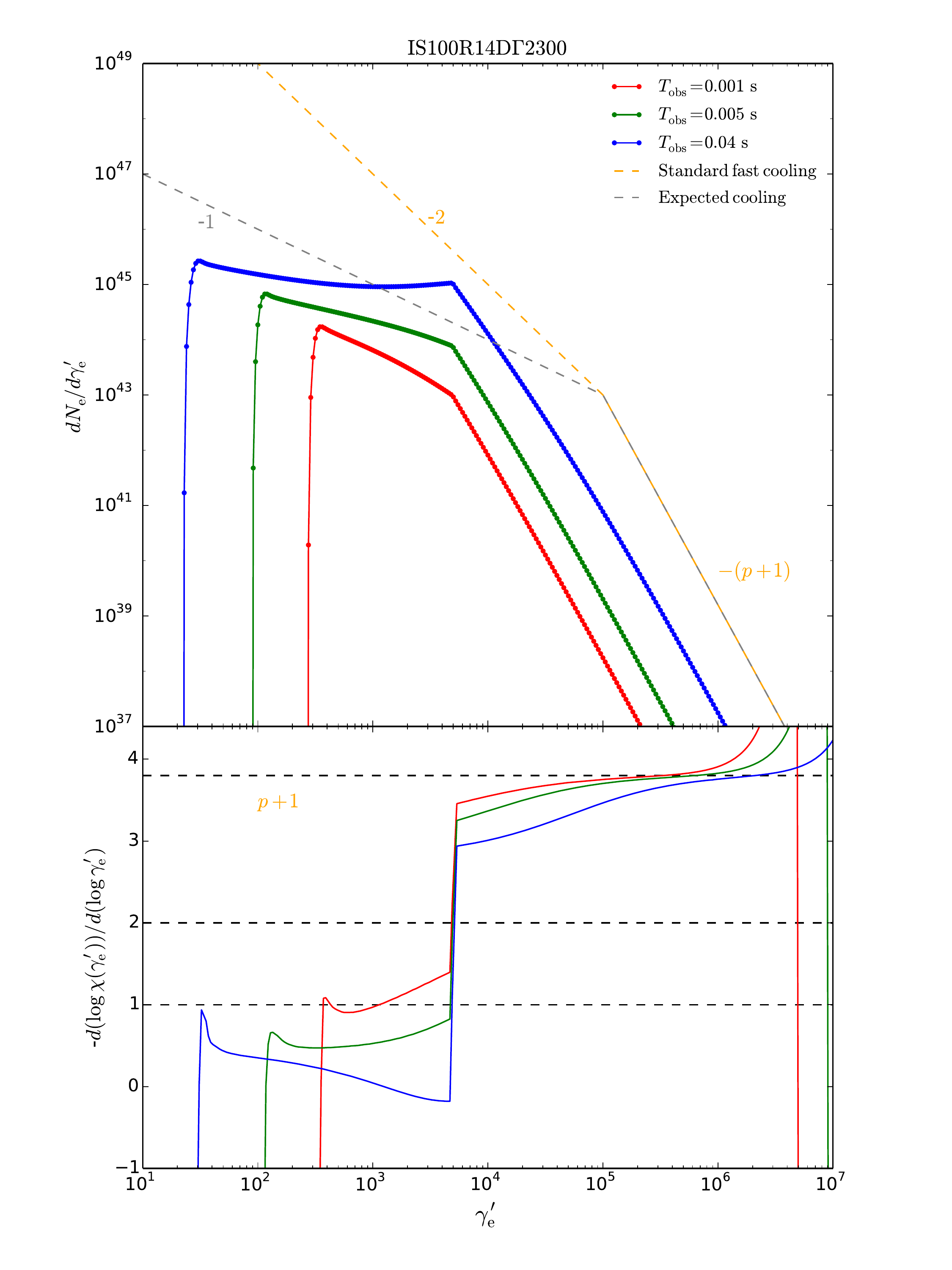}}
    \caption{The evolution of the electron energy spectrum for the ten cases in Group IS100R14 (see Table \ref{TABLE:D}).\label{fig:MD-electron}}
\end{adjustwidth}
\end{figure}

\clearpage

\begin{figure}
\begin{adjustwidth}{-2cm}{-2cm}
\centering
    \subfloat{\includegraphics[width=0.25\paperwidth]{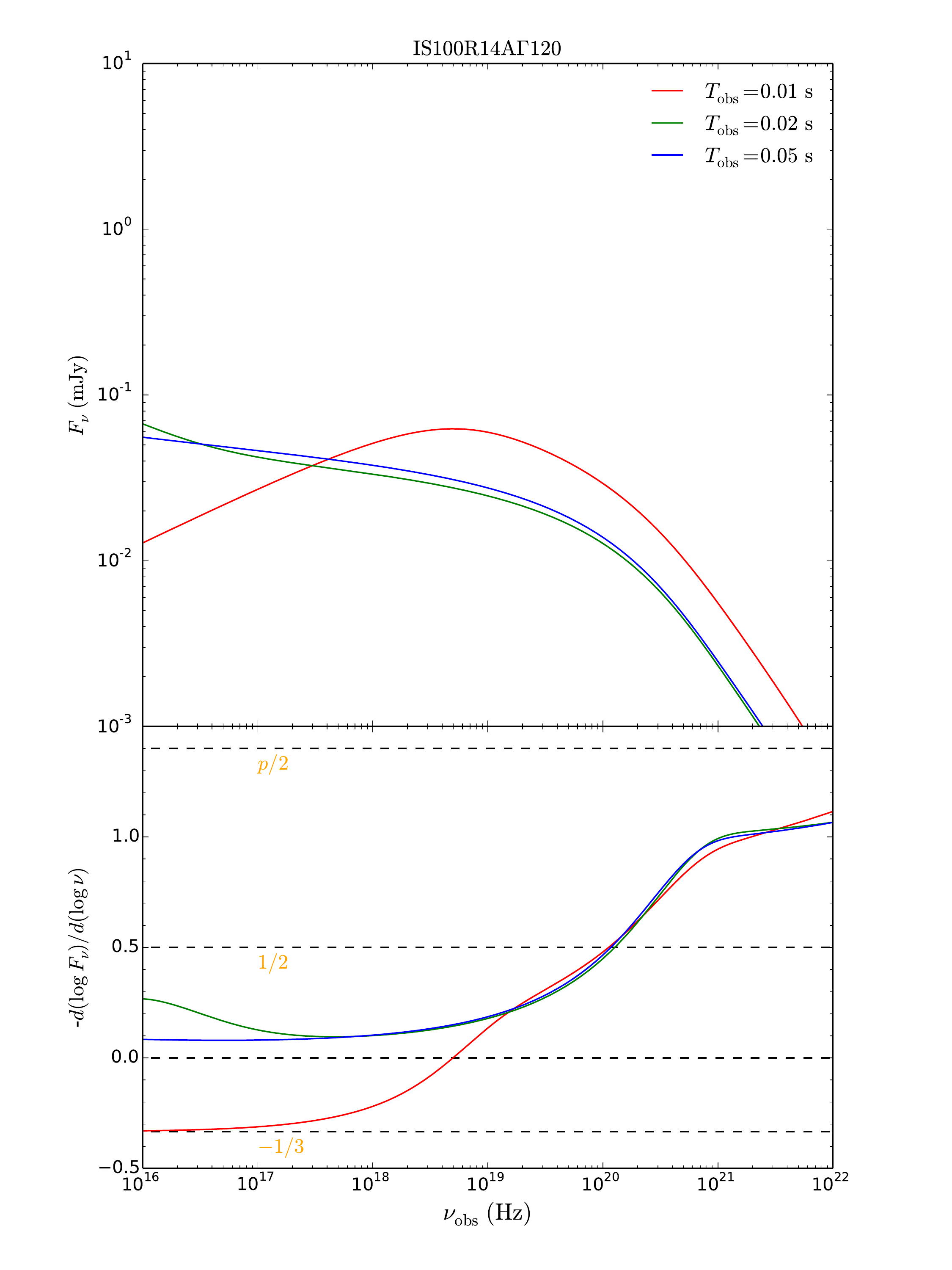}}
    \subfloat{\includegraphics[width=0.25\paperwidth]{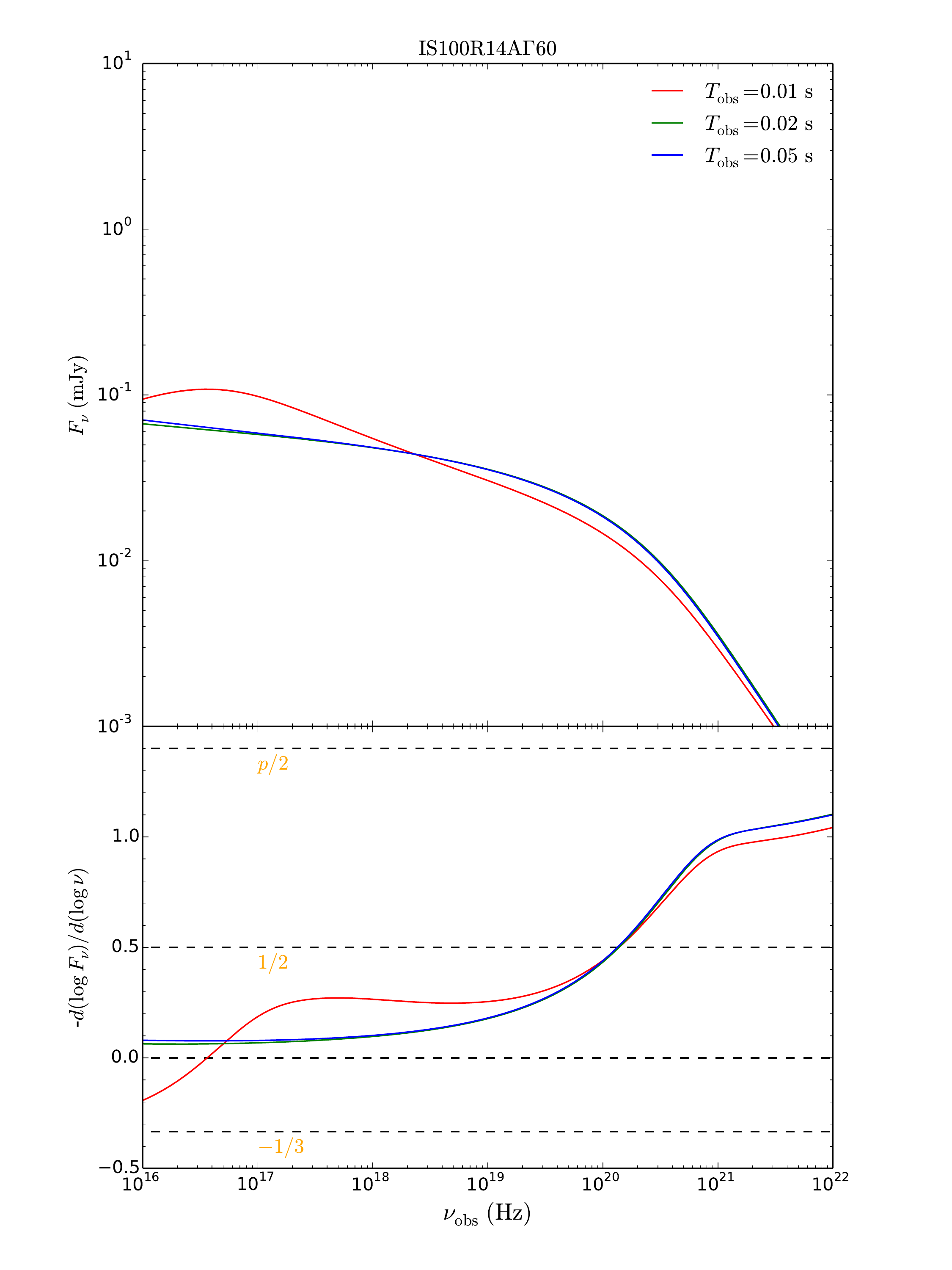}}
    \subfloat{\includegraphics[width=0.25\paperwidth]{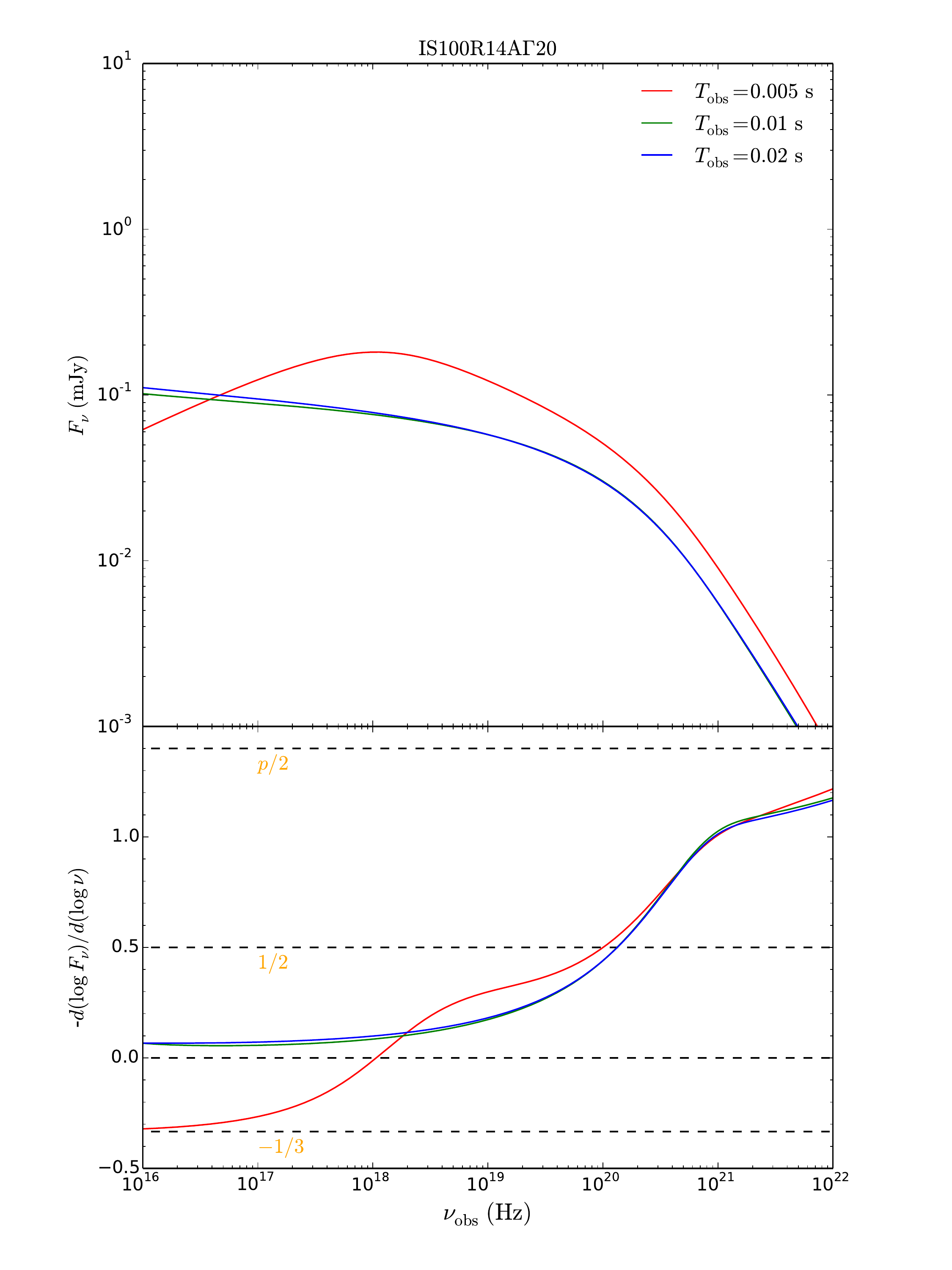}} \\
    \subfloat{\includegraphics[width=0.25\paperwidth]{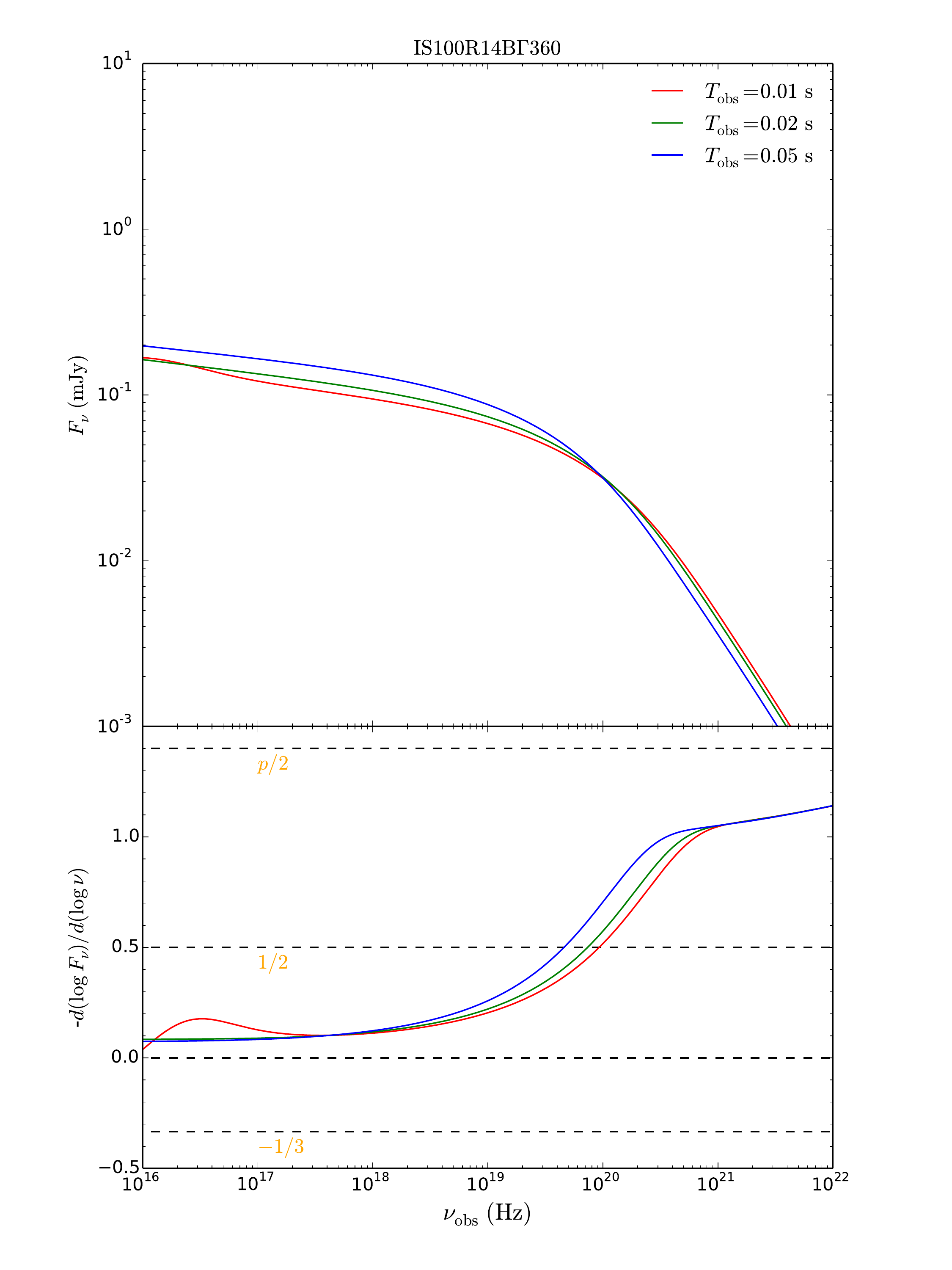}}
    \subfloat{\includegraphics[width=0.25\paperwidth]{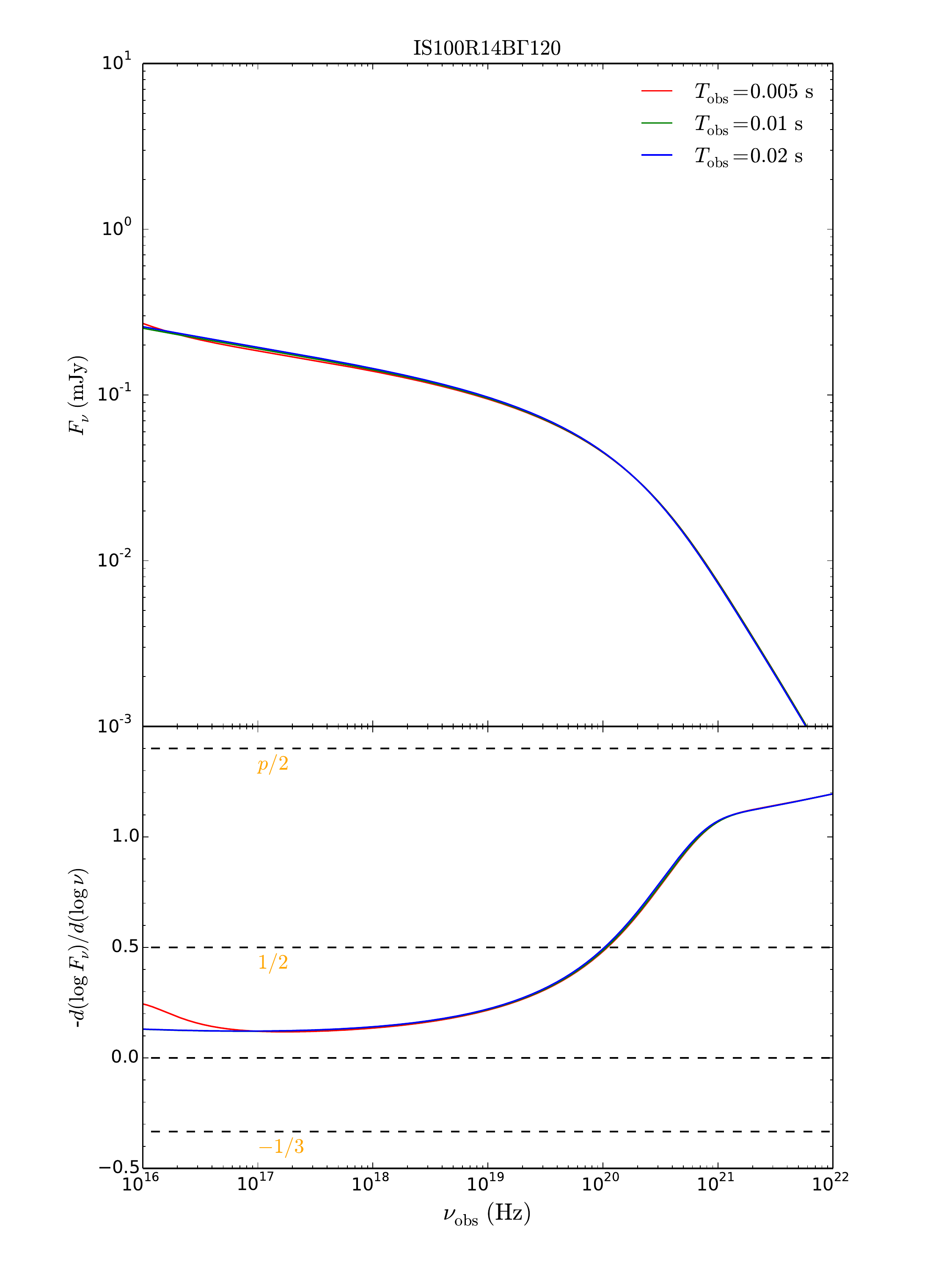}}
    \subfloat{\includegraphics[width=0.25\paperwidth]{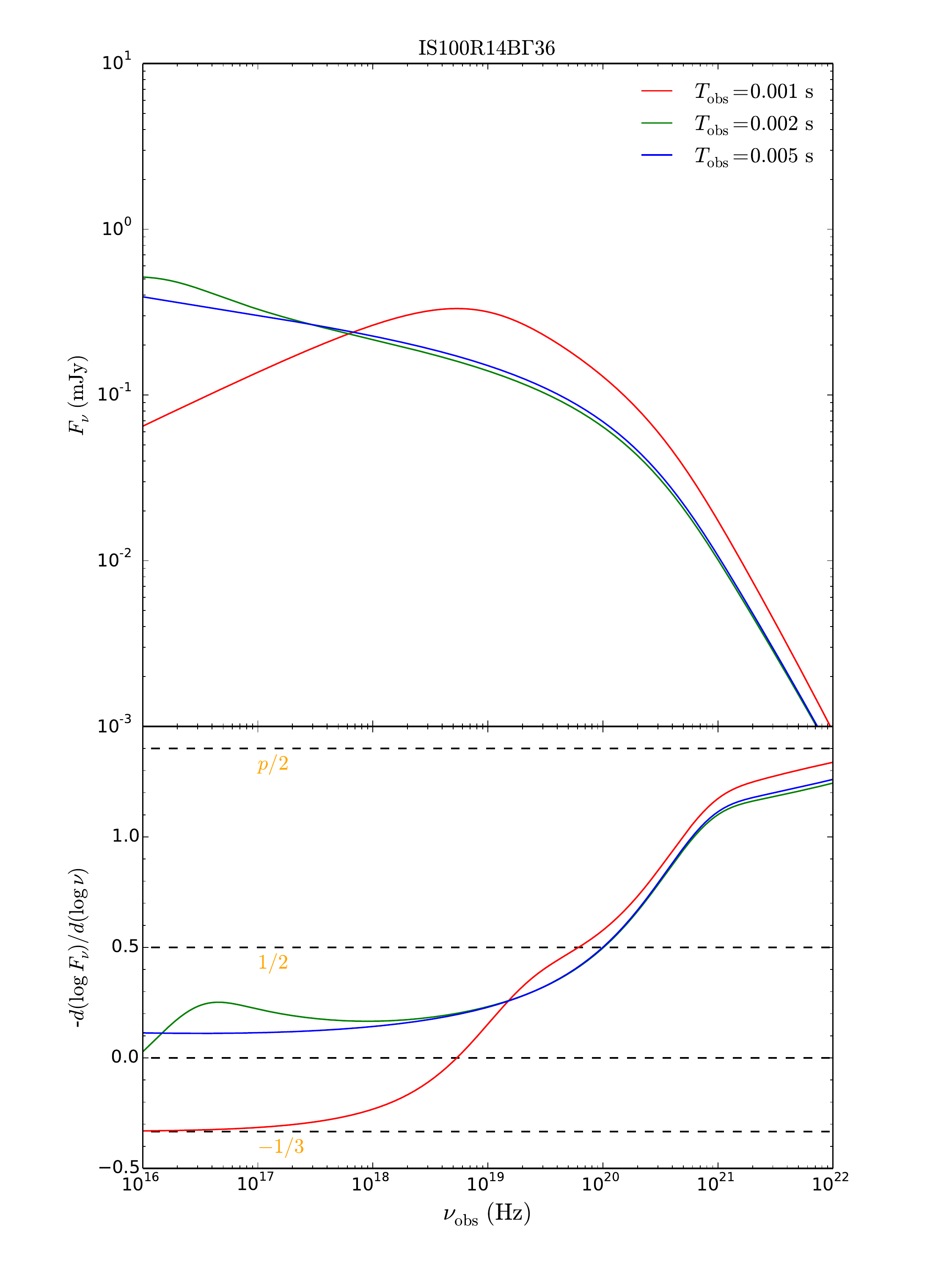}} \\
    \subfloat{\includegraphics[width=0.25\paperwidth]{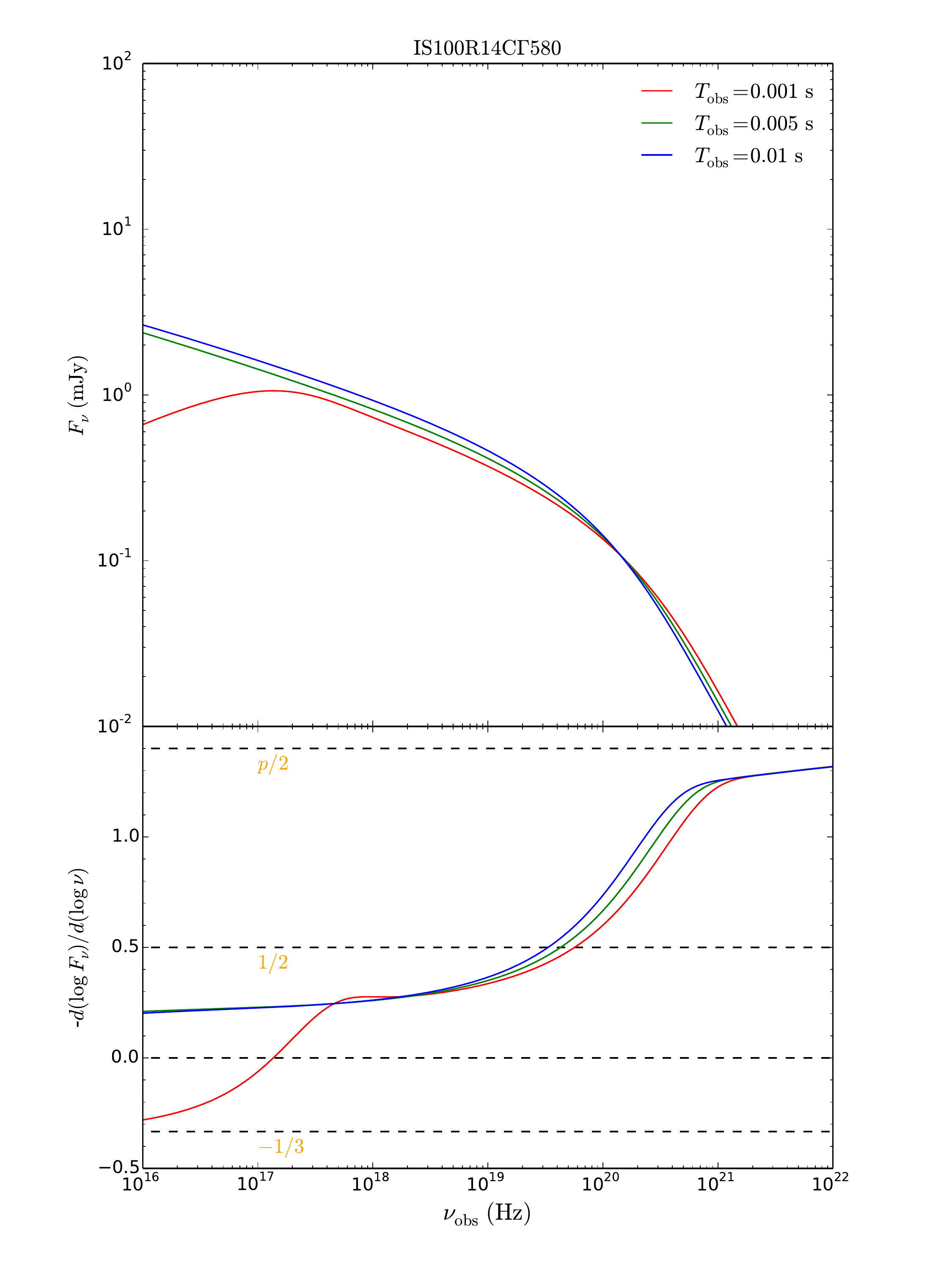}}
    \subfloat{\includegraphics[width=0.25\paperwidth]{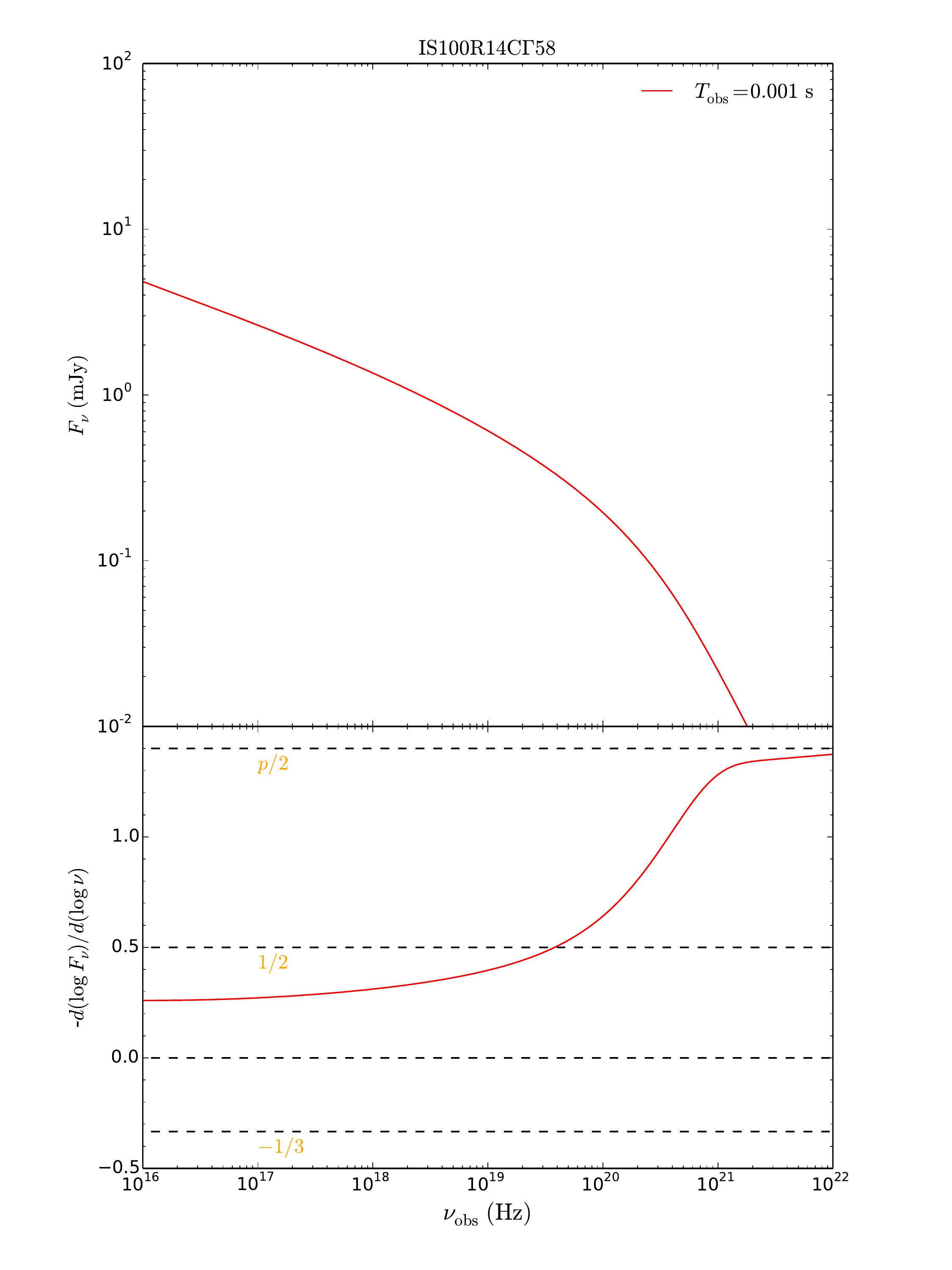}}
    \subfloat{\includegraphics[width=0.25\paperwidth]{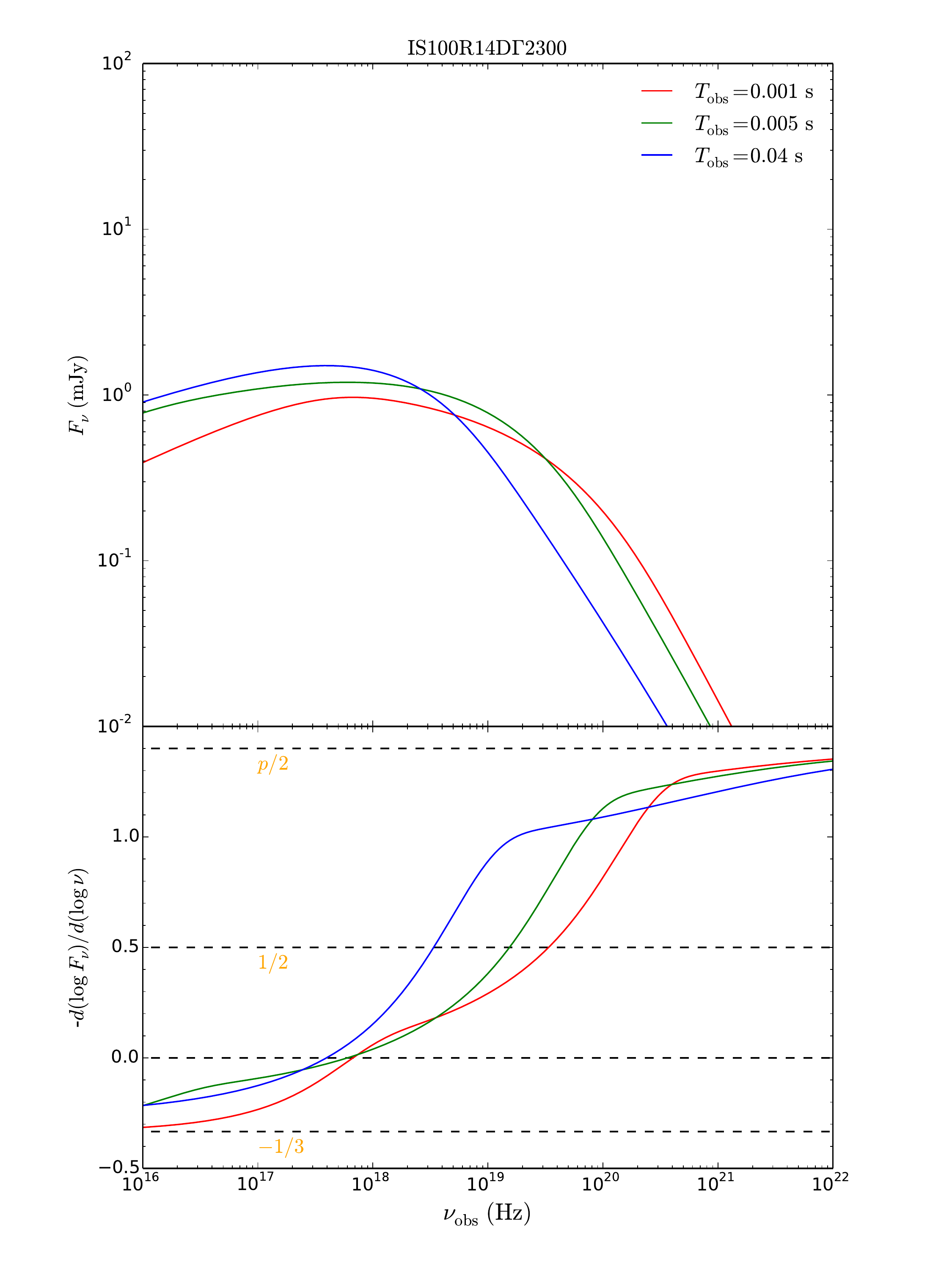}}
    \caption{The corresponding synchrotron flux-density spectra $F_{\nu}$ from the electrons with the energy distribution presented in Figure
    \ref{fig:MD-electron}.\label{fig:MD-spectra}}
\end{adjustwidth}
\end{figure}

\clearpage

\begin{figure}
\begin{adjustwidth}{-2cm}{-2cm}
\centering
    \subfloat{\includegraphics[width=0.25\paperwidth]{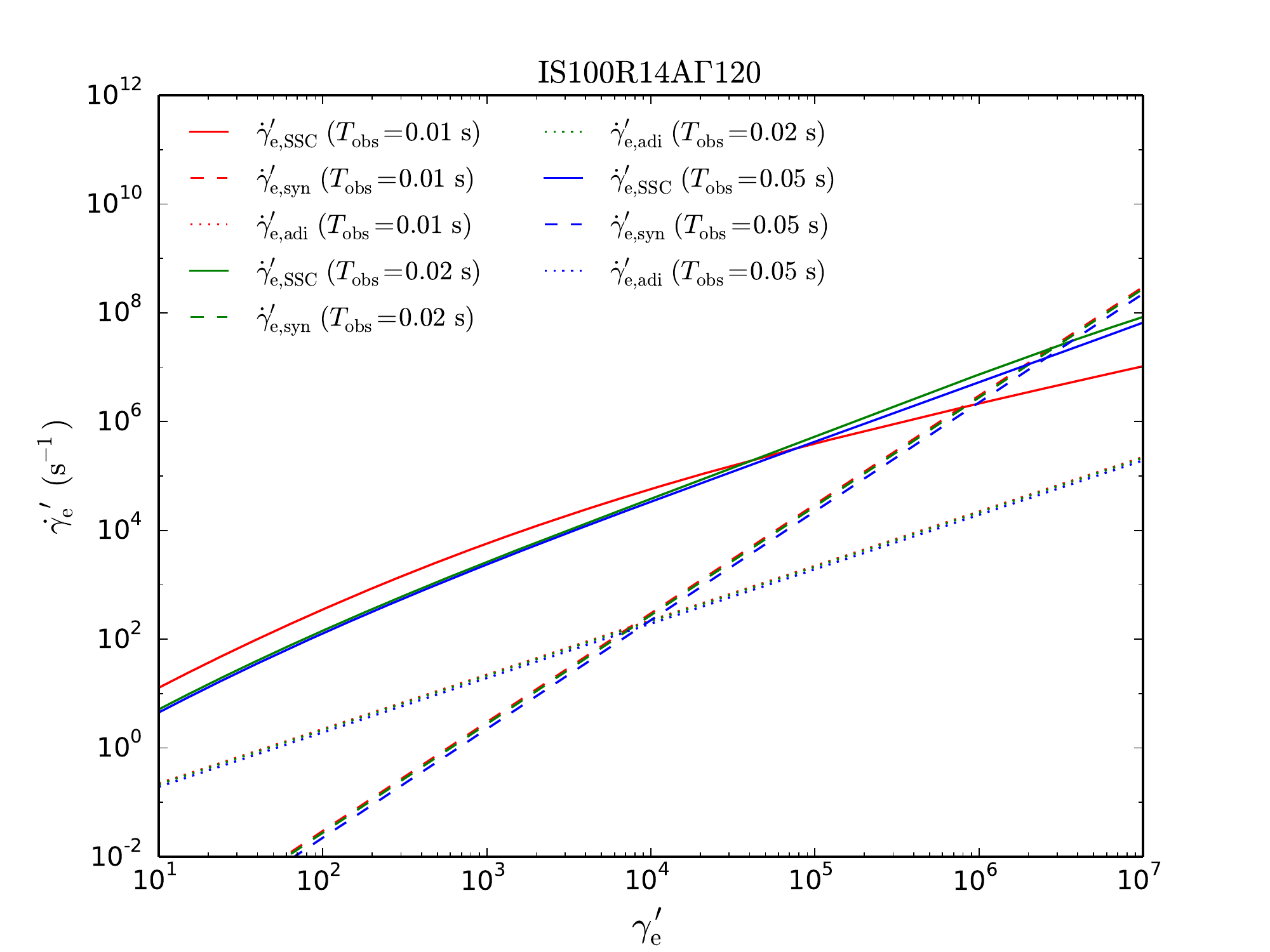}}
    \subfloat{\includegraphics[width=0.25\paperwidth]{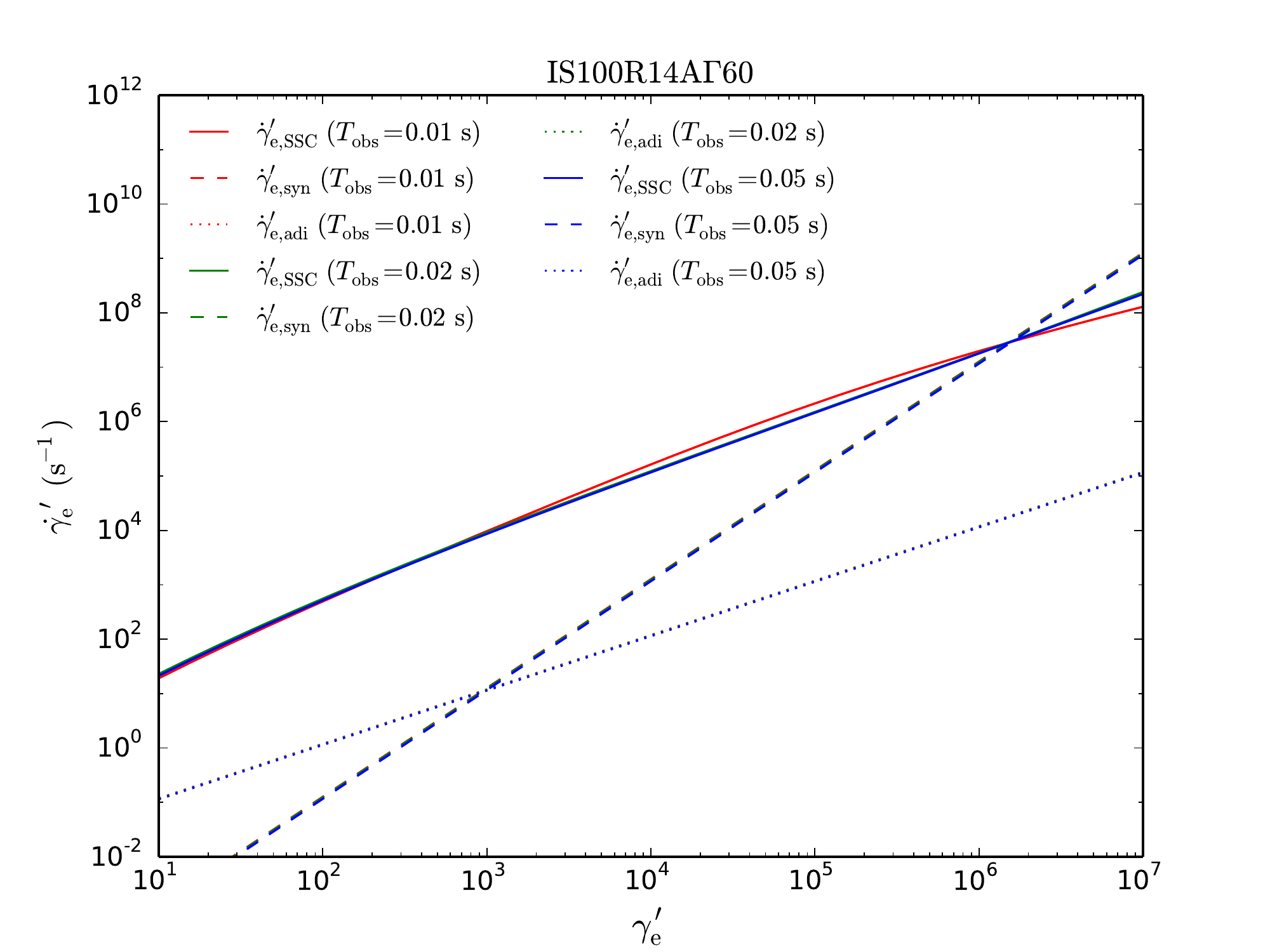}}
    \subfloat{\includegraphics[width=0.25\paperwidth]{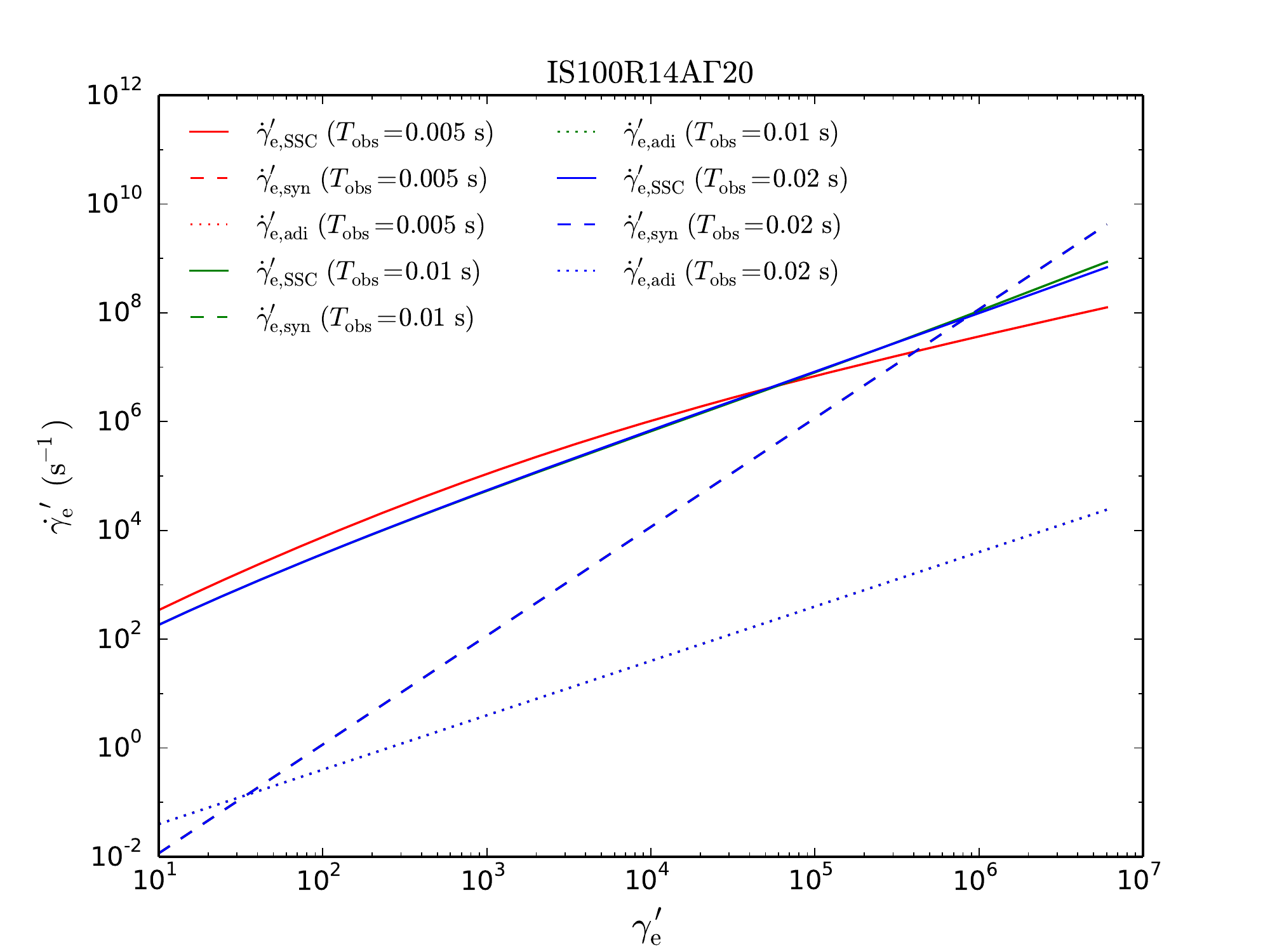}} \\
    \subfloat{\includegraphics[width=0.25\paperwidth]{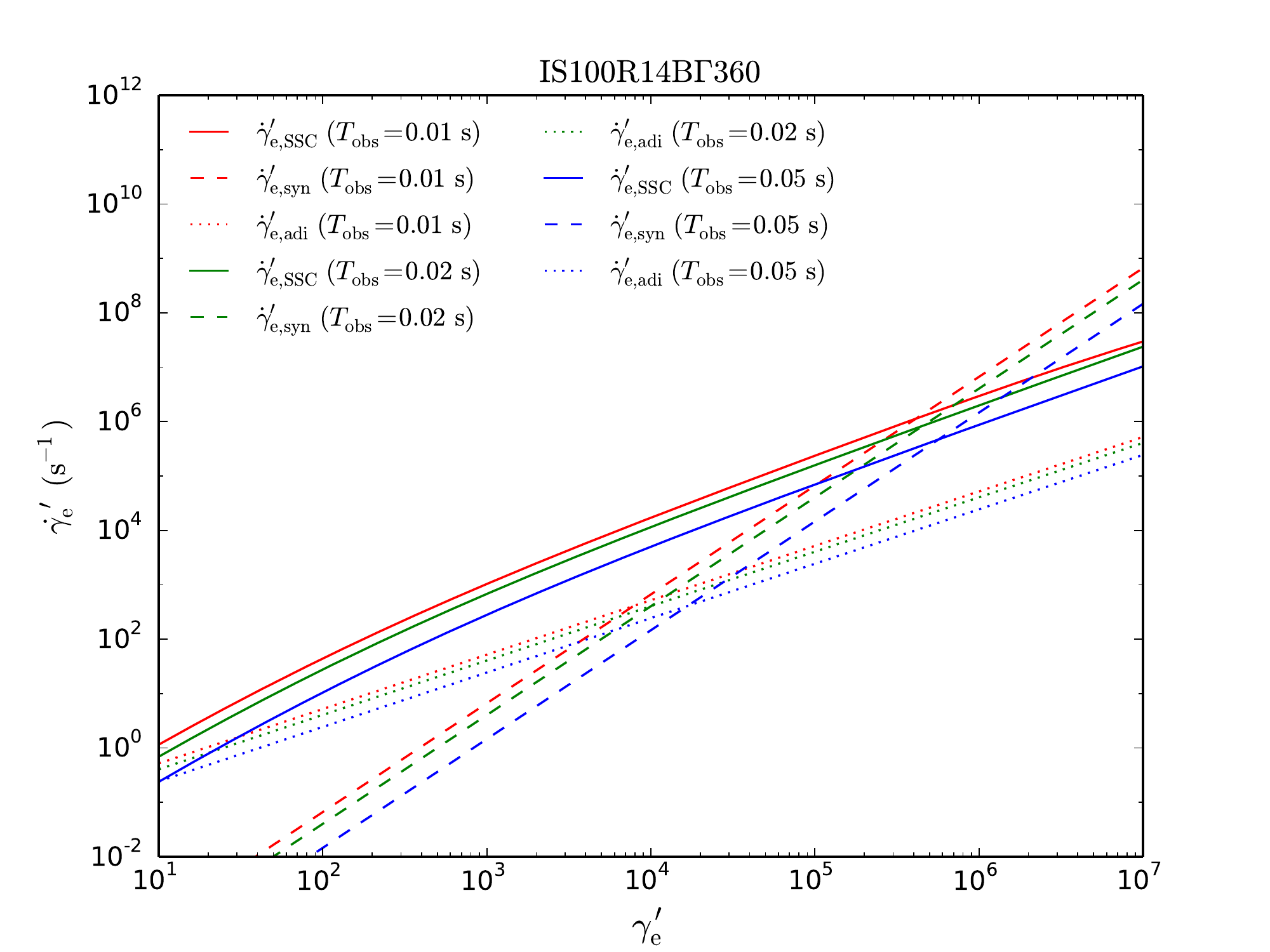}}
    \subfloat{\includegraphics[width=0.25\paperwidth]{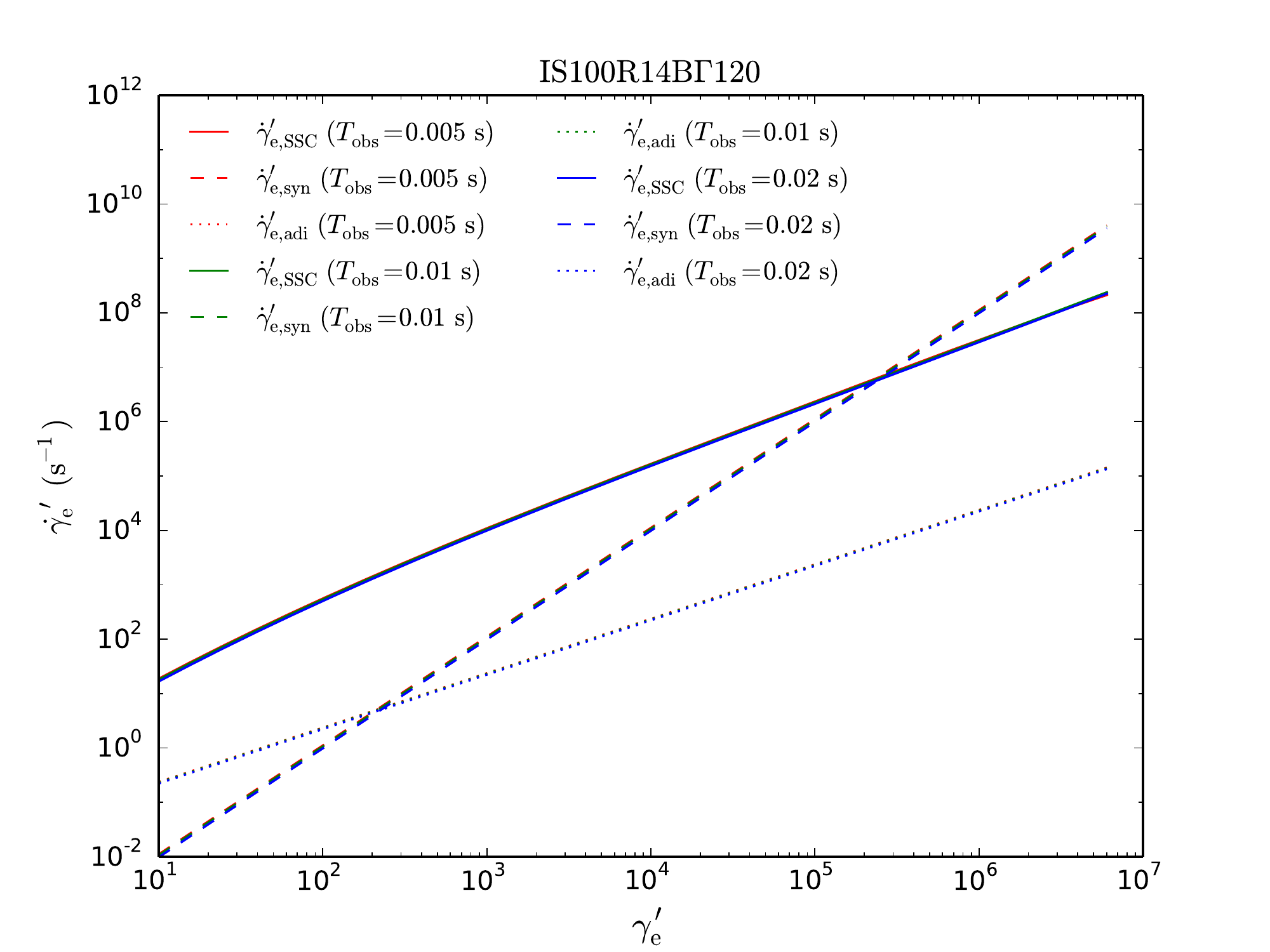}}
    \subfloat{\includegraphics[width=0.25\paperwidth]{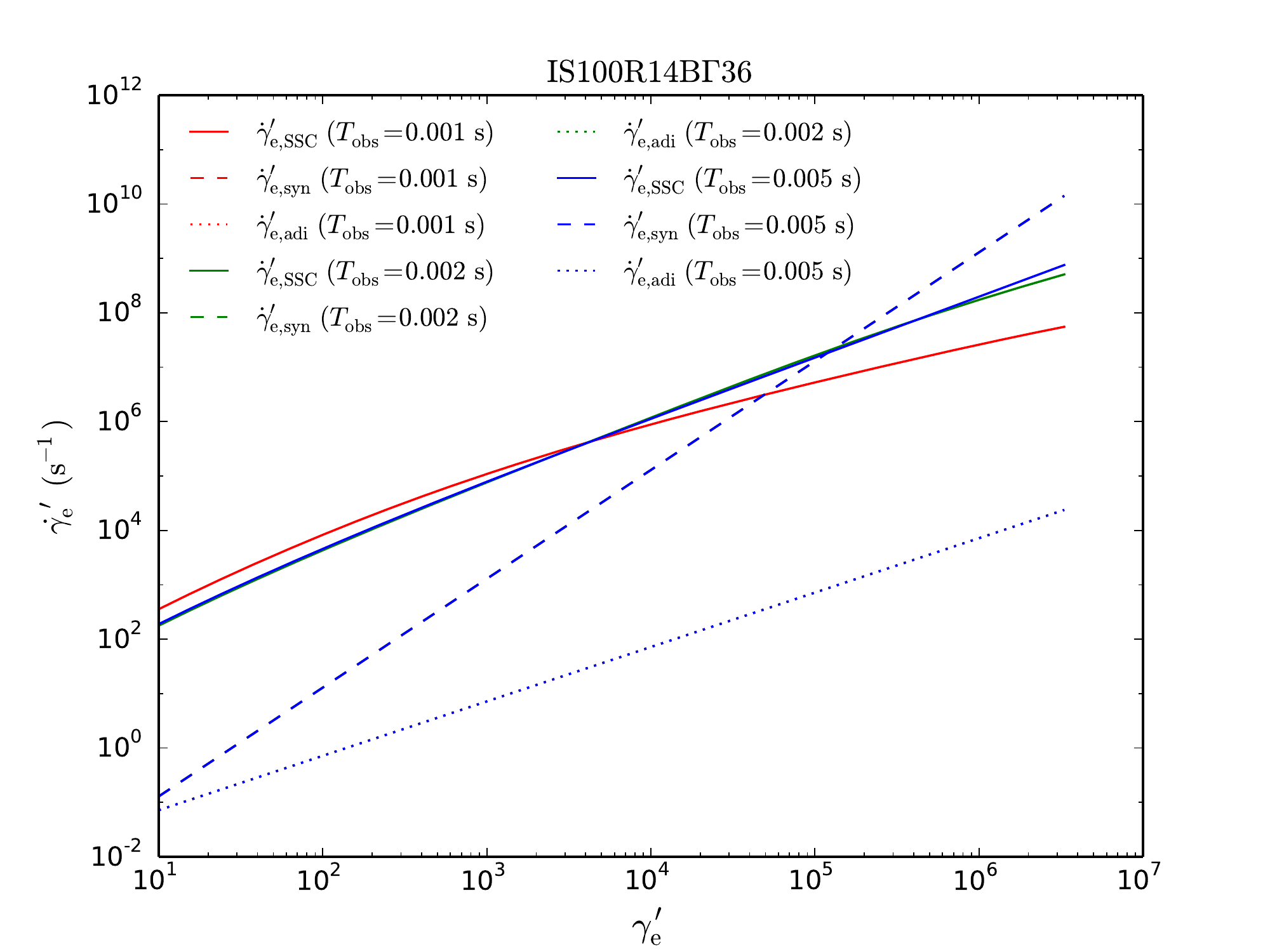}} \\
    \subfloat{\includegraphics[width=0.25\paperwidth]{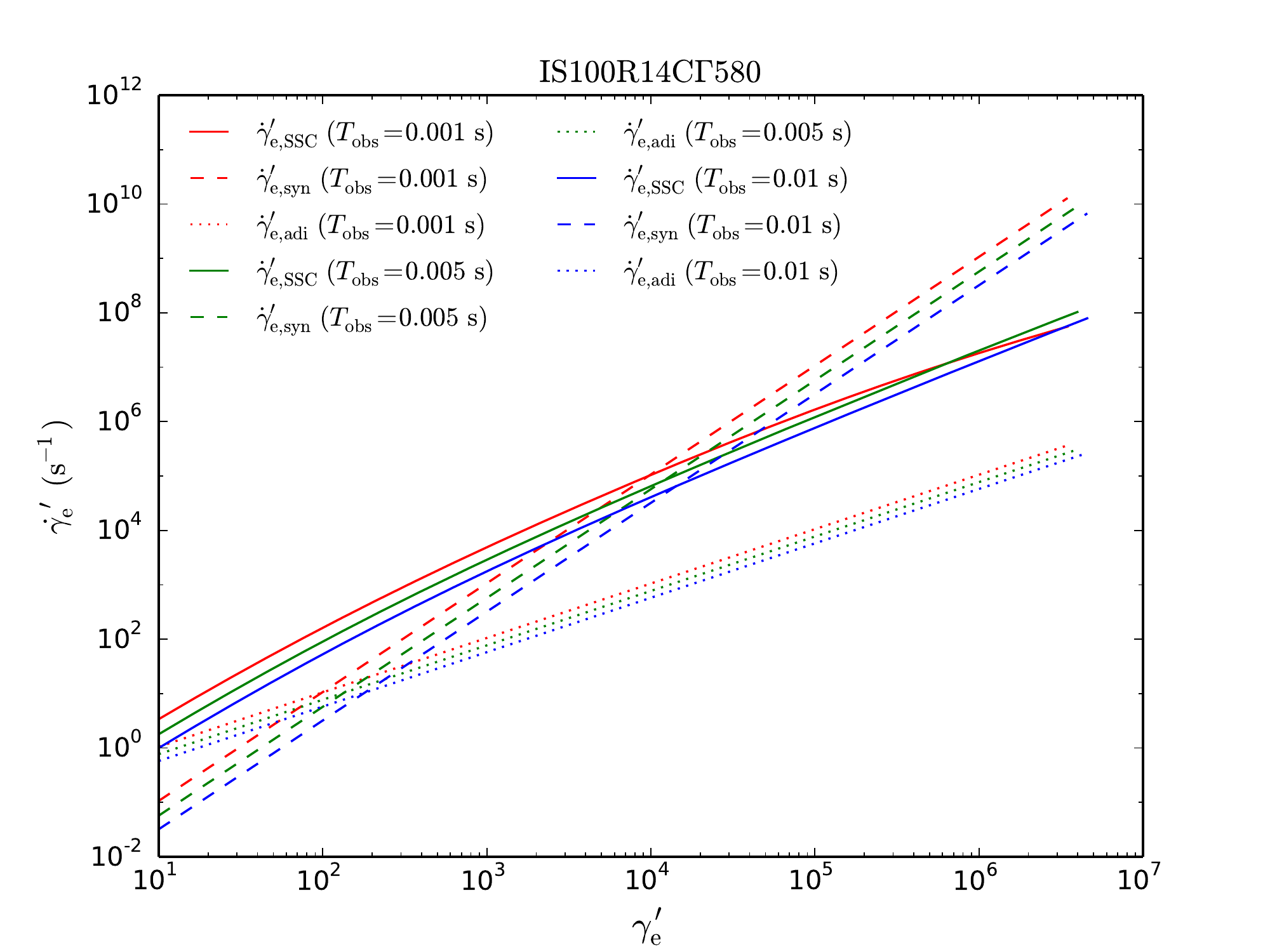}}
    \subfloat{\includegraphics[width=0.25\paperwidth]{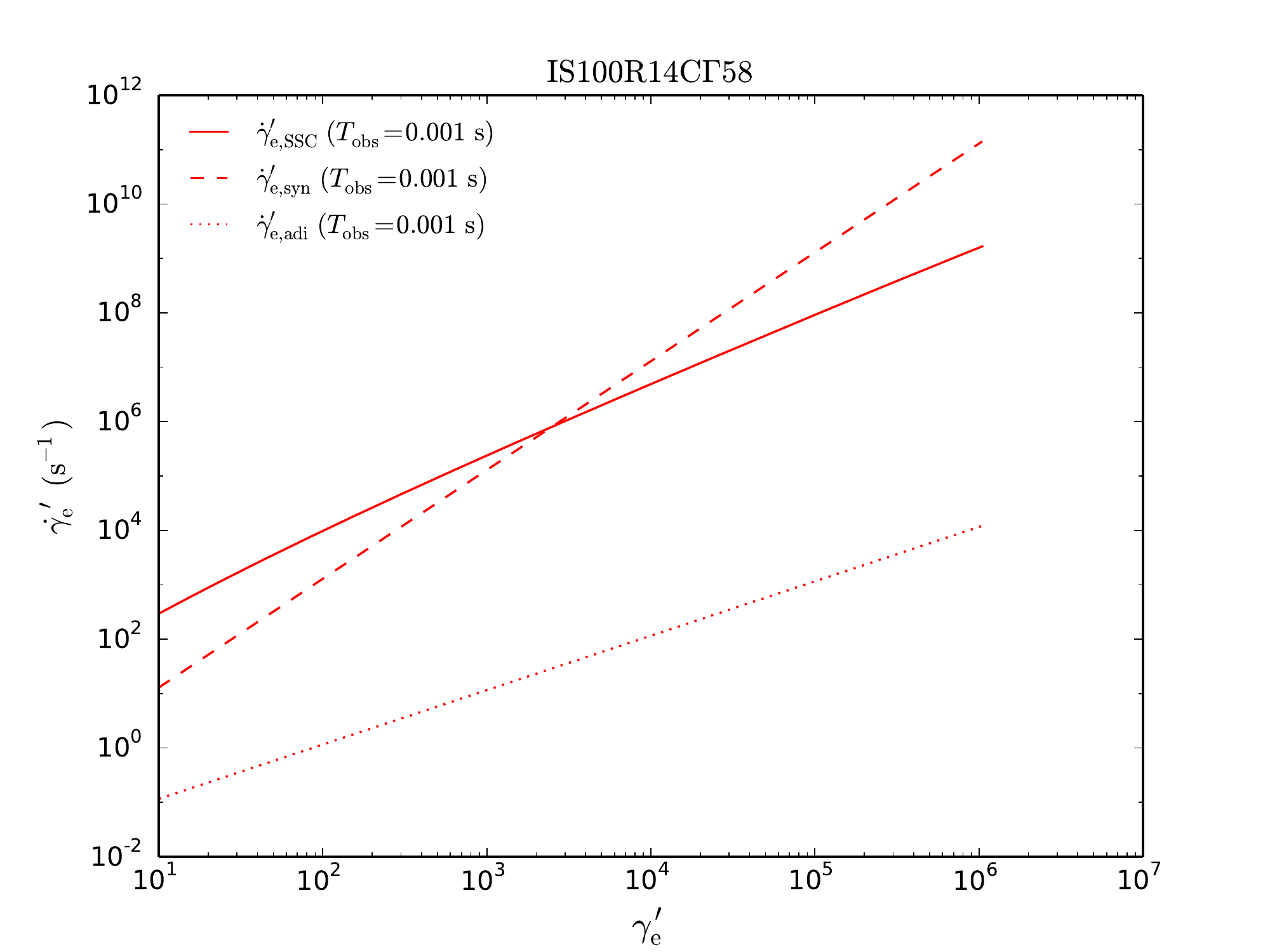}}
    \subfloat{\includegraphics[width=0.25\paperwidth]{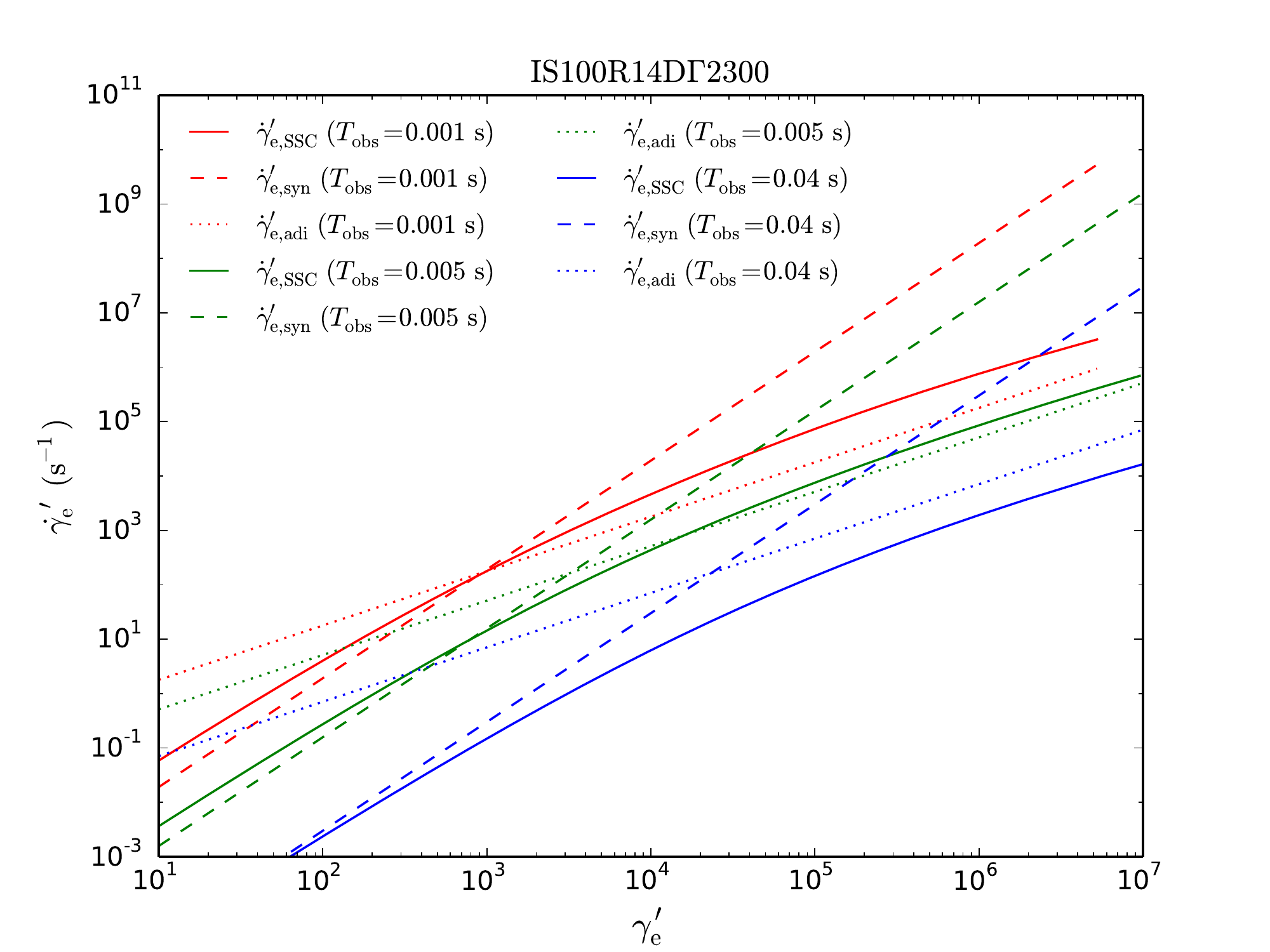}}
    \caption{The co-moving cooling rates of different cooling mechanisms for the electrons with the energy distribution
    presented in Figure \ref{fig:MD-electron}.\label{fig:MD-rate}}
\end{adjustwidth}
\end{figure}

\end{document}